\documentclass[usenatbib]{mn2e}
\usepackage{apjfonts,amsfonts,amsmath,amssymb,bm,ctable,verbatim}

\usepackage[figuresleft]{rotfloat}
\makeatletter
\let\@makecaption=\SFB@makefigurecaption
\makeatother
\usepackage{pdflscape}

\newcommand{\gizmourl}{\href{http://www.tapir.caltech.edu/~phopkins/Site/GIZMO.html}{\url{http://www.tapir.caltech.edu/~phopkins/Site/GIZMO.html}}}

\newcommand{\etal}{et al.}

\newcommand{\acknowledgments}[1]{\begin{small}\section*{Acknowledgments}\end{small}{\noindent #1}\vspace{5pt}}
\newcommand{\datastatement}[1]{\begin{small}\section*{Data Availability Statement}\end{small}{\noindent #1}\vspace{5pt}}

\newcommand{\coloryes}[1]{\textcolor{blue}{#1}}
\newcommand{\colormid}[1]{\textcolor{black}{#1}}
\newcommand{\colorno}[1]{\textcolor{red}{#1}}
\newcommand{\colormtoohigh}[1]{\textcolor{magenta}{#1}}

\makeatletter
\newcommand*{\@rowstyle}{}
\newcommand*{\rowstyle}[1]{\gdef\@rowstyle{#1}\@rowstyle\ignorespaces}% sets the style of the next row
\newcolumntype{=}{>{\gdef\@rowstyle{}}}% resets the row style
\newcolumntype{+}{>{\@rowstyle}}% adds the current row style to the next column
\makeatother
\newcommand{\toohigh}{}

\title[Disk-Burst Tests]{What Causes The Formation of Disks and End of Bursty Star Formation?}

\author[Hopkins \etal]{
\parbox[t]{\textwidth}{
Philip F.~Hopkins$^1$, 
Alexander B.\ Gurvich$^2$, 
Xuejian Shen$^1$, 
Zachary Hafen$^3$, 
Michael Y.\ Grudi{\'c}$^4$, 
Shalini Kurinchi-Vendhan$^1$, 
Christopher C. Hayward$^5$, 
Fangzhou Jiang$^3$, 
Matthew E.\ Orr$^{5,6}$,
Andrew Wetzel$^7$, 
Du\v{s}an Kere\v{s}$^8$, 
Jonathan Stern$^9$, 
Claude-Andr{\'e} Faucher-Gigu{\`e}re$^2$, 
James Bullock$^3$, 
Coral Wheeler$^{10}$, 
Kareem El-Badry$^{11}$, 
Sarah R. Loebman$^{12}$,
Jorge Moreno$^{13}$, 
Michael Boylan-Kolchin$^{14}$, 
Eliot Quataert$^{15}$
}\vspace*{4pt} \\
$^1$ TAPIR, Mailcode 350-17, California Institute of Technology, Pasadena, CA 91125, USA. E-mail:phopkins@caltech.edu \\
$^2$ Department of Physics \& Astronomy and CIERA, Northwestern University, 1800 Sherman Ave, Evanston, IL 60201, USA \\
$^3$ Department of Physics and Astronomy, University of California Irvine, CA 92697, USA \\
$^4$ Carnegie Observatories, 813 Santa Barbara St, Pasadena, CA 91101, USA \\
$^5$ Center for Computational Astrophysics, Flatiron Institute, 162 5th Ave., New York, NY 10010 USA \\ 
$^6$ Department of Physics and Astronomy, Rutgers University, 136 Frelinghuysen Road, Piscataway, NJ 08854, USA \\
$^7$ Department of Physics \& Astronomy, University of California, Davis, CA, USA 95616 \\
$^8$ Department of Physics, Center for Astrophysics and Space Sciences,University of California San Diego, 9500 Gilman Drive, La Jolla, CA 92093, USA \\ 
$^9$ School of Physics \& Astronomy, Tel Aviv University, Tel Aviv 69978, Israel \\ 
$^{10}$ {Department of Physics and Astronomy, California State Polytechnic University, Pomona, Pomona, CA 91768, USA} \\
$^{11}$ Center for Astrophysics, Harvard \&\ Smithsonian, 60 Garden Street, Cambridge, MA 02138, USA \\
$^{12}$ Department of Physics, University of California, Merced, 5200 N. Lake Road, Merced, CA 95343, USA \\
$^{13}$ Department of Physics and Astronomy, Pomona College, Claremont, CA 91711, USA \\ 
$^{14}$ {Department of Astronomy, The University of Texas at Austin, 2515 Speedway, Stop C1400, Austin, Texas 78712-1205, USA} \\
$^{15}$ Department of Astrophysical Sciences, Princeton University, Princeton, NJ 08544, USA 
}

\date{}
\begin{document}
\maketitle

\begin{abstract}
As they grow, galaxies can transition from irregular/spheroidal with ``bursty'' star formation histories (SFHs), to disky with smooth SFHs. But even in simulations, the direct physical cause of such transitions remains unclear. We therefore explore this in a large suite of numerical experiments re-running portions of cosmological simulations with widely varied physics, further validated with existing FIRE simulations. We show that gas supply, cooling/thermodynamics, star formation model, Toomre scale, galaxy dynamical times, and feedback properties do {\em not} have a direct causal effect on these transitions. Rather, both the formation of disks and cessation of bursty star formation are driven by the gravitational potential, but in different ways. Disk formation is promoted when the mass profile becomes sufficiently centrally-concentrated in shape (relative to circularization radii): we show that this provides a well-defined dynamical center, ceases to support the global ``breathing modes'' which can persist indefinitely in less-concentrated profiles and efficiently destroy disks, promotes orbit mixing to form a coherent angular momentum, and stabilizes the disk. Smooth SF is promoted by the potential or escape velocity $V_{\rm esc}$ (not circular velocity $V_{\rm c}$) becoming sufficiently large at the radii of star formation that cool, mass-loaded (momentum-conserving) outflows are trapped/confined near the galaxy, as opposed to escaping after bursts. We discuss the detailed physics, how these conditions arise in cosmological contexts, their relation to other correlated phenomena (e.g. inner halo virialization, vertical disk ``settling''), and observations.
\end{abstract}

\begin{keywords}
galaxies: general --- galaxies: evolution --- galaxies: formation ---  Galaxy: structure --- galaxies: star formation --- ISM: structure
\end{keywords}

\section{Introduction}
\label{sec:intro}

The origin of galaxy morphology and kinematic structure, and its relation to their star formation histories, remain crucial open questions in extragalactic astronomy \citep[for recent reviews, see][]{2015ARA&A..53...51S,dale:sf.feedback.review,2017ARA&A..55...59N,bullock:2017.small.scale.LCDM.challenges,vogelsberger:2020.review.galaxy.form.sims}. Small dwarf galaxies and high-redshift galaxies tend to exhibit irregular, ``clumpy'' morphologies with correspondingly disordered kinematics, and ``bursty'' star formation with large temporal fluctuations about some mean star formation rate \citep{kriek:z2.hubble.sequence,daddi:highz.gal.high.fgas,ceverino:2010.clump.disks.cosmosims,forster-schreiber:2011.hiz.gal.morph,newman:z2.clump.winds,oklopcic:clumpy.highz.gals.fire.case.study.clumps.not.long.lived,sparre.2015:bursty.star.formation.main.sequence.fire,2016ApJ...833...37G,2016ApJ...833..254S,emami:2019.bursty.dynamics.similar.to.fire.in.obs.but.fire.too.rapid.in.intermediate,2019ApJ...884..133F,2021MNRAS.501.4812F}. This is quite different from massive, low-redshift star-forming galaxies which often exhibit morphologically thin and dynamically cold disks with smooth (weakly-fluctuating, i.e.\ non-bursty) galaxy-integrated star formation rates. While numerical simulations now reproduce both regimes qualitatively \citep[see references above and][]{2015MNRAS.454.1886S,crain:eagle.sims,garrisonkimmel:fire.morphologies.vs.dm}, and indeed anticipated the observed high-redshift disordered/bursty phase \citep{fukuda:nuclear.ring,immeli:fragmentation.vs.clump.properties,bournaud:disk.clumps.to.bulge,elmegreen:2009.clump.properties.hudf}, most still struggle to match {\em in detail} the precise morphological mix of galaxies as a function of mass and redshift \citep{2016MNRAS.459..467O,garrisonkimmel:fire.morphologies.vs.dm,peebles:2020.thin.disks.challenging,haslbauer:2022.thin.disks.vs.cdm,2022arXiv220214038C}. More fundamentally, it still remains unclear (even in the simulations) how, when, where, and why the onset of disk formation really occurs.

What does appear clear is that at some point in their history, often between halo masses $M_{\rm halo}\sim 10^{11}-10^{12}\,M_{\odot}$, galaxies in cosmological simulations often undergo a rapid kinematic and morphological transformation. The galaxy mass grows; the potential well deepens; escape velocity, density, and characteristic acceleration scales increase, while the dynamical time decreases; feedback becomes less efficient and more dominated by ``venting''; gas fractions deplete and the Toomre/clump mass scale decreases; star formation becomes less ``bursty'' and more smooth; inflows allow a coherent central mass to form; the inner circumgalactic medium (CGM) halo becomes quasi-virialized, with cooling times longer than dynamical times; and the galaxy forms an increasingly stable, gradually-settling and growing disk. While the details of models and feedback treatments vary (and this introduces changes in e.g.\ exactly where and when such transitions occur), this has now been seen in a wide variety of different simulations and it appears increasingly robust that such a transition does, in fact, occur \citep[references above and][]{2016MNRAS.462.2418S,hafen:2022.criterion.for.thin.disk.related.to.halo.ang.mom,gurvich:2022.disk.settling.fire,2022MNRAS.511.3895F,2022MNRAS.512.3806V}. Moreover growing observational evidence following high-redshift progenitors of Milky-Way mass galaxies or tracing back Galactic history via galactic archeology also appears to support this scenario \citep{hung:fire.sf.gal.kinematic.evol.gas.dispersion,2021MNRAS.508.4484J,2022arXiv220315822Z,2022arXiv220402989C,2022ApJ...928...58L,2022arXiv220304980B}.

From a theoretical point of view, however, it is extremely difficult to disentangle the {\em causal} role of all of these different processes, because they tend to occur at nearly the same cosmic time for a given galaxy, and because many of them are causally linked in multiple different directions (i.e.\ they can ``trigger'' one another). The ``chicken or egg'' question of which came first and {\em why} disks begin to form at a given time (and therefore the most important physics involved) remains elusive. This is crucial for understanding galaxies in their own right, but also for constraining the underlying models: if observations favor disk formation or settling at earlier times than a simulation predicts, then understanding what triggers this process is fundamental to inform any exploration of which physics might be missing from said simulations.

In this paper, we therefore present a series of controlled studies designed -- to the extent possible -- to isolate and explore the different mechanisms above and their effects on whether or not disks can form. These allow us to separate all of the physics above, and to identify the key causal chain of events. We describe our methods in \S~\ref{sec:methods}, discuss the key results in \S~\ref{sec:results}, describe how the studied variations which are important lead to the formation of disks (\S~\ref{sec:diskform}) and/or smooth star formation (\S~\ref{sec:bursty}) and/or thin disk heights (\S~\ref{sec:settling}), briefly consider how this relates to observations of galaxy structure and which physical processes could non-linearly produce the key conditions in \S~\ref{sec:obs}, and summarize our conclusions in \S~\ref{sec:conclusions}.

\begin{footnotesize}
\ctable[caption={{\normalsize Some Simulation Parameters Explored for {\bf m11a} in this paper. Columns show: {\bf (1)} Simulation name. {\bf (2)} Stellar mass $M_{\ast}$ at $z=0$ (in units of $10^{7}\,M_{\odot}$; \colormtoohigh{magenta} denotes experiments producing galaxies far more massive than observed). {\bf (3)} Gas mass $M_{\rm gas}$ (within $<20\,$kpc) at $z=0$ (in $10^{7}\,M_{\odot}$). {\bf (4)} Does the galaxy have a clear gas disk at $z=0$? (\coloryes{Blue}/\colormid{black}/\colorno{red} for yes/ambiguous/no) {\bf (5)} Is the SFR near $z\sim0$ ``smooth''? {\bf (6)} Additional notes.}
\label{table:sims}},center,star
]{=l +l +l +l +l +l}{}{
\hline\hline
Name & $M_{\ast}$ & $M_{\rm gas}$ & {\bf Disk}? & {\bf Smooth SF}? & Other Notes \\
\hline
% m11a_m2e3_fire2_Apr262022_restartdisktest_default
Default & 5.6 & 21 & \colorno{No} & \colorno{No} & Our ``default'' or ``baseline'' simulation.\\ 
\hline\hline
\multicolumn{6}{l}{Modified {\bf Potential} (Point Mass): $a_{\ast} = G\,M_{0}/(r^{2}+\epsilon_{\ast}^{2})$ with $M_{0}$ in $10^{9}\,M_{\odot}$, $\epsilon_{\ast}$ in kpc. (Added to potential per Eqs.~\ref{eqn:a.ext.full}-\ref{eqn:a.ext})}\\
\hline
% m11a_m2e3_fire2_Apr262022_restartdisktest_ptmass_3a
\toohigh $M_{0}=5$ & \colormtoohigh{60} & 14 & \coloryes{\bf Yes} & \colormid{Some (Late)} & $\epsilon_{\ast}=0.2$. Compact ($<0.5$\,kpc) thin gas+stellar disk. SFR dips to zero between bursts.\\
% m11a_m2e3_fire2_Apr262022_restartdisktest_ptmass_7b
$M_{0}=2$ & 14 & 40 & \coloryes{\bf Yes} & \colorno{No} & $\epsilon_{\ast}=0.2$. Compact ($<2\,$kpc) disk, but spiral structure to $10\,$kpc. Bursty outflows.\\
% m11a_m2e3_fire2_Apr262022_restartdisktest_ptmass_1
$M_{0}=1$ & 4.9 & 78 & \coloryes{\bf Yes} & \colorno{No} & $\epsilon_{\ast}=0.2$. Extended, stable, well-ordered disk.\\
% m11a_m2e3_fire2_Apr262022_restartdisktest_ptmass_4
$M_{0}=1$\,S & 6.5 & 76 & \coloryes{\bf Yes} & \colorno{No} & $\epsilon_{\ast}=0.1$. Very ordered early disk which gets thicker later but also builds out to $\sim10\,$kpc.\\
% m11a_m2e3_fire2_Apr262022_restartdisktest_ptmass_5
$M_{0}=1$\,L   & 6.0 & 30 & \coloryes{\bf Yes (some)} & \colorno{No} & $\epsilon_{\ast}=1$.  Intermittent disk with significant angular momentum (AM).\\
% m11a_m2e3_fire2_Apr262022_restartdisktest_ptmass_6
$M_{0}=0.5$ & 6.3 & 42 & \colormid{Yes (weak)} & \colorno{No} & $\epsilon_{\ast}=0.2$. Gas-rich, not well-ordered but large arc with significant AM.\\
% m11a_m2e3_fire2_Apr262022_restartdisktest_ptmass_2
$M_{0}=0.2$ & 5.2 & 43 & \colorno{No} & \colorno{No} & $\epsilon_{\ast}=0.2$. Gas-rich but no real disk.\\
\hline\hline
\multicolumn{6}{l}{Modified {\bf Potential} (Constant $V_{\rm c}$): $a_{\ast} = V_{0}^{2}/(r^{2}+\epsilon_{\ast}^{2})^{1/2}$ with $V_{0}$ in ${\rm km\,s^{-1}}$, $\epsilon_{\ast}=0.2$ in kpc.} \\
\hline
% m11a_m2e3_fire2_Apr262022_restartdisktest_VcConst_7
\toohigh $V_{0}=300$ & \colormtoohigh{120} & 1.0 & \coloryes{\bf Yes} & \coloryes{\bf Yes} & Smooth SFR (30x enhanced). Compact $\sim 1\,$kpc very thin gas+stellar disk.\\
% m11a_m2e3_fire2_Apr262022_restartdisktest_VcConst_1
$V_{0}=200$ & 4.1 & 1.5 & \coloryes{\bf Yes} & \colormid{Some} & Extremely compact ($<0.5\,$kpc), very thin disk. SF smooth after initial burst, bursty later.\\
% m11a_m2e3_fire2_Apr262022_restartdisktest_VcConst_2
$V_{0}=100$ & 12 & 5.1 & \coloryes{\bf Yes} & \coloryes{\bf Yes} & Compact ($<3\,$kpc), thin disk. SF smooth after initial large burst.\\
% m11a_m2e3_fire2_Apr262022_restartdisktest_VcConst_6
$V_{0}=72$ & 4.6 & 42 & \coloryes{\bf Yes} & \colormid{Some (Late)} & Clear disk, large warp (from former polar disk). SF bursty to $z=0.17$ then smooth later.\\ 
% m11a_m2e3_fire2_Apr262022_restartdisktest_VcConst_5
$V_{0}=60$ & 4.9 & 35 & \coloryes{\bf Yes} & \colorno{No} & Medium (3-5 kpc) disk growing, pretty disordered but clear AM.\\
% m11a_m2e3_fire2_Apr262022_restartdisktest_VcConst_3
$V_{0}=50$ & 5.1 & 19 & \colorno{No} & \colorno{No} & Gas-rich, no significant difference in AM versus default conditions.\\
\hline\hline
\multicolumn{6}{l}{Modified {\bf Potential} (Constant $|{\bf a}|$): $a_{\ast}=a_{0}$ with $a_{0}$ in units of $G\,M_{\odot}/{\rm pc^{2}} = 1.4\times10^{-11}\,{\rm cm\,s^{-2}}.$} \\
\hline
% m11a_m2e3_fire2_Apr262022_restartdisktest_AccConst_4
\toohigh $a_{0}=10000$ & \colormtoohigh{29} & 14 & \coloryes{\bf Yes} & \coloryes{\bf Yes} & Compact $<2$\,kpc very thin disk. Initial burst, then smooth (very high) SFR.\\
% m11a_m2e3_fire2_Apr262022_restartdisktest_AccConst_1
$a_{0}=3000$ & 6.3 & 1.4 & \colorno{No} & \coloryes{\bf Mostly} & Gas-Poor. Large initial burst, then smooth SF with some drops. No disk, just one rotating clump.\\
% m11a_m2e3_fire2_Apr262022_restartdisktest_AccConst_2
$a_{0}=1000$ & 7.7 & 4.9 & \colorno{No} & \colormid{Some (Late)} & Gas-Poor. Some initial burst, then smooth SF.\\
% m11a_m2e3_fire2_Apr262022_restartdisktest_AccConst_3
$a_{0}=300$ & 6.2 & 8.6 & \colorno{No} & \colorno{No} & Somewhat less gas-poor. Some initial burst, but not as large as $a_{0}=1000$.\\
% m11a_m2e3_fire2_Apr262022_restartdisktest_AccConst_5
$a_{0}=100$ & 6.5 & 30 & \colorno{No} & \colorno{No} & No morphological disk, but some shell/streamer with some significant AM at $2\,$kpc.\\
\hline\hline
\multicolumn{6}{l}{Modified {\bf Potential} (Constant $\rho$): $a_{\ast}=(4\pi/3)\,G\,\rho_{0}\,r$ with $\rho_{0}$ in units of $m_{p}\,{\rm cm^{-3}} = 1.7\times10^{-24}\,{\rm g\,cm^{-3}}.$} \\
\hline
% m11a_m2e3_fire2_Apr262022_restartdisktest_RhoConst_3 (r=reff)
\toohigh $\rho_{0}=10$ & \colormtoohigh{21} & 6.5 & \colorno{No} & \coloryes{\bf Yes} & Substantial AM in gas halo, but no disk. SF follows burst-then-smooth-decay pattern.\\
% m11a_m2e3_fire2_Apr262022_restartdisktest_RhoConst_2 (r=reff)
$\rho_{0}=5$ & 12 & 21 & \colorno{No} & \coloryes{\bf Yes (Mid)} & No morphological disk (just clump at $2\,$kpc \& significant halo AM). Burst-decay then rising SF.\\
% m11a_m2e3_fire2_Apr262022_restartdisktest_RhoConst_2a (r=rp)
$\rho_{0}=5$a & 11 & 17 & \colorno{No} & \coloryes{\bf Yes (Mid)} & No visual disk. Halo gas has some AM, but cool is clumpy+spherical. Smooth SF after early burst.\\
% m11a_m2e3_fire2_Apr262022_restartdisktest_RhoConst_1 (r=reff)
$\rho_{0}=2$ & 9.6 & 25 & \coloryes{\bf Yes} & \colormid{Some (Late)} & $\epsilon_{\ast}=0.2$. Polar ($\sim 5\,$kpc) disk. Burst-smooth decay SF (with occasional drops).\\
% m11a_m2e3_fire2_Apr262022_restartdisktest_RhoConst_1a (r=rp)
$\rho_{0}=2$a & 6.8 & 12 & \colorno{No} & \colormid{Slight} & Clumpy/disordered gas (some AM). Burst-decay SF but still mostly bursty.\\
% m11a_m2e3_fire2_Apr262022_restartdisktest_RhoConst_4 (fixed r=rp)
$\rho_{0}=1$ & 5.0 & 32 & \colormid{Yes (weak)} & \colormid{Slight (Late)} & Inner disk with polar outer disk.  Weak burst-decay SF with bursts slightly weaker later.\\
% m11a_m2e3_fire2_Apr262022_restartdisktest_RhoConst_5 (fixed r=rp)
$\rho_{0}=0.3$ & 4.9 & 8.7 & \colorno{No} & \colorno{No} & Couple gas clumps in center but no disk. Weaker burst-then-decay SF retaining fluctuations.\\
\hline\hline
\multicolumn{6}{l}{Modified {\bf Potential} (Increasing $\rho\propto r$): $a_{\ast}=q_{0}\,r^{2}$ with $q_{0}$ in units of $10^{6}\,G\,M_{\odot}/{\rm kpc}^{4}$.} \\
\hline
% m11a_m2e3_fire2_Apr262022_restartdisktest_X0Const_1
\toohigh $q_{0}=120$ & \colormtoohigh{37} & 5.0 & \colorno{No} & \coloryes{\bf Yes} & Significant gas in center but not disky. Burst-smooth decay SF.\\
% m11a_m2e3_fire2_Apr262022_restartdisktest_X0Const_2
\toohigh $q_{0}=30$ & \colormtoohigh{17} & 18 & \colorno{No} & \colormid{Slight (Late)} & Weaker burst-slow decay, burstiness calms in decay. Some AM and gas in center but no disk.\\
% m11a_m2e3_fire2_Apr262022_restartdisktest_X0Const_3
$q_{0}=7.5$ & 9.6 & 30 & \colormid{Somewhat} & \coloryes{\bf Yes (Late)} & Gas center clumpy but slight disky structure. Burst-slow decay SF, burstiness calms later.\\
% m11a_m2e3_fire2_Apr262022_restartdisktest_X0Const_4
$q_{0}=2$ & 7.8 & 10 & \colorno{No} & \colorno{No} & Some gas in center but little AM, no disk. Still bursty.\\
% m11a_m2e3_fire2_Apr262022_restartdisktest_X0Const_5
$q_{0}=0.5$ & 6.0 & 40 & \colorno{No} & \colorno{No} & Some gas AM, gas-rich but clumpy/irregular morphology, no disk. Bursty outflows.\\
\hline\hline
\multicolumn{6}{l}{Modified {\bf Radiative Cooling/Heating}: $\Lambda_{\rm rad} \rightarrow \eta_{\rm cool}\,\Lambda_{\rm rad}$. (Further variations or more extreme, $\eta_{\rm cool} \rightarrow \infty$ cases, in \S~\ref{sec:bursty.ruled.out} \&\ Appendix~\ref{sec:additional.params})} \\
\hline
% m11a_m2e3_fire2_Apr262022_restartdisktest_modTCool_4
$\eta_{\rm cool}=0.001$ & 1.9 & 26 & \colorno{No} & \colorno{Quenched} & One last burst from gas already in disk then nothing. Little halo gas AM.\\
% m11a_m2e3_fire2_Apr262022_restartdisktest_modTCool_1
$\eta_{\rm cool}=0.01$ & 1.9 & 49 & \colorno{No} & \colorno{Quenched} & Last burst then nothing. Small halo gas AM, but nothing at all disky. \\
% m11a_m2e3_fire2_Apr262022_restartdisktest_modTCool_2
$\eta_{\rm cool}=0.1$ & 2.2 & 22 & \colorno{No} & \colorno{No} & Few big bursts then eventually cuts off at $z=0.3$. Little halo gas AM.\\
% m11a_m2e3_fire2_Apr262022_restartdisktest_modTCool_3
$\eta_{\rm cool}=0.3$ & 2.8 & 24 & \colorno{No} & \colorno{Burstier} & Notably more bursty than default. Few bursts then SF drops to nil, then repeats.\\
% m11a_m2e3_fire2_Apr262022_restartdisktest_modTCool_5
\toohigh $\eta_{\rm cool}=10$ & \colormtoohigh{36} & 150 & \colorno{No} & \colorno{No} & Bursty, SFR elevated 10x. Lots of gas in center but clumpy/irregular. \\
\hline\hline
\multicolumn{6}{l}{Modified {\bf Stellar Feedback}: Multiply feedback strengths/rates, $(\dot{P}_{\rm fb},\,\dot{E}_{\rm fb}) \rightarrow \eta_{\rm fb}\,(\dot{P}_{\rm fb},\,\dot{E}_{\rm fb})$. (Further variations noted in \S~\ref{sec:bursty.ruled.out} \&\ Appendix~\ref{sec:additional.params})} \\
\hline
% m11a_m2e3_fire2_Apr262022_restartdisktest_FBx0pt1 (SNe_Energy_Renormalization=StellarMassLoss_Energy_Renormalization=RP_Local_Momentum_Renormalization=HIIRegion_fLum_Coupled=PhotonMomentum_Coupled_Fraction=0.1)
\toohigh $\eta_{\rm fb}=0.1$ & \colormtoohigh{370} & 55 & \coloryes{\bf Yes (Late)} & \colormid{Mixed (Late)} & SFR rises 100x. Bursty with no disk until $V_{\rm c} \gtrsim 125\,{\rm km\,s^{-1}}$ in center ($z\sim0.3$) then disky \&\ smooth.\\
% m11a_m2e3_fire2_Apr262022_restartdisktest_FBx0pt1a (mod Ndot via calculate_relative_light_to_mass_ratio_from_imf)
\toohigh $\eta_{\rm fb}=0.1$a & \colormtoohigh{210} & 72 & \coloryes{\bf Yes (Late)} & \colormid{Mixed (Late)} & Disk forms after centrally-peaked $V_{\rm c}$ with $M_{\ast} \gtrsim 40$ forms, then SFR smooths after $M_{\ast} \gtrsim 120$.\\
% m11a_m2e3_fire2_Apr262022_restartdisktest_FBx0pt3  (SNe_Energy_Renormalization=StellarMassLoss_Energy_Renormalization=RP_Local_Momentum_Renormalization=HIIRegion_fLum_Coupled=PhotonMomentum_Coupled_Fraction=0.3)
\toohigh $\eta_{\rm fb}=0.3$ & \colormtoohigh{110} & 100 & \colorno{No} & \colorno{No} & Some AM but no disk, irregular gas clumps. Elevated SFR but equally bursty.\\
% m11a_m2e3_fire2_Apr262022_restartdisktest_FBx0pt3a (mdot Ndot via calculate_relative_light_to_mass_ratio_from_imf)
\toohigh $\eta_{\rm fb}=0.3$a & \colormtoohigh{23} & 37 & \colorno{No} & \colorno{No} & Irregular/clumpy gas, no disk. SFR elevated, with bursty SF and outflows.\\
% m11a_m2e3_fire2_Apr262022_restartdisktest_FBx3
$\eta_{\rm fb}=3$ & 2.2 & 17 & \colorno{No} & \colorno{No} & SFR drops then mostly self-quenches after a couple feedback bursts.\\
% m11a_m2e3_fire2_Apr262022_restartdisktest_FBx10
$\eta_{\rm fb}=10$ & 1.9 & 6.6 & \colorno{No} & \colorno{No} & SFR drops rapidly, bursty at very low level. Gas poor, no disk.\\
% m11a_m2e3_fire2_Apr262022_restartdisktest_FBRadx30SNex0pt1 (rad fb/luminosites x30, SNe rate x0.1)
\toohigh $\eta_{\rm fb}=0.1$R & \colormtoohigh{39} & 100 & \colorno{No} & \colorno{No} & Radiative FB boosted 30x, SNe lowered 0.1x. Gas clumpy, no disk. SF+outflows bursty.\\
% m11a_m2e3_fire2_Apr262022_restartdisktest_FBRadx30_sameRSNe (rad fb/luminosites x30)
$\eta_{\rm fb}=30$R & 13 & 84 & \colorno{No} & \colorno{Slightly} & Radiative FB boosted 30x. Gas-rich (minimal halo gas AM), but irregular. Slightly less bursty.\\
% m11a_m2e3_fire2_Apr262022_restartdisktest_FBSNePtermx10 (terminal momentum x10 [all else equal])
$\eta_{\rm fb}=10$P & 3.3 & 47 & \colorno{No} & \colorno{Burstier} & Terminal momentum $p_{\rm term}$ x10. Irregular/spheroidal (no disk). Very bursty SF.\\
% m11a_m2e3_fire2_Apr262022_restartdisktest_FBSNePtermx0pt1 (terminal momentum x0.1 [all else equal])
\toohigh $\eta_{\rm fb}=0.1$P & \colormtoohigh{16} & 40 & \colorno{No (weak)} & \colorno{No} & $p_{\rm term}$ x0.1. Significant gas AM but no morphological disk. SF elevated but remains bursty.\\
\hline\hline
\multicolumn{6}{l}{Modified {\bf Star Formation}: Threshold SF Density goes from $1000\,{\rm cm^{-3}} \rightarrow n_{\rm crit}\,{\rm cm^{-3}}$; SFR per $t_{\rm ff}$ in self-gravitating gas from unity to $\eta_{\rm sf}.$} \\
\hline
% m11a_m2e3_fire2_Apr262022_restartdisktest_SFE
$\eta_{\rm sf}=0.01$ & 3.9 & 20 & \colorno{No} & \colorno{No} & $n_{\rm crit}=1$; Somewhat more bursty (vs.\ default). Gas in center very clumpy, no disk.\\
% m11a_m2e3_fire2_Apr262022_restartdisktest_SFRho
$n_{\rm crit}=1$ & 4.0 & 16 & \colorno{No} & \colorno{Burstier} & $\eta_{\rm sf}=1$; Significantly more bursty. Gas in center but no AM and just clumps.\\
\hline\hline
\multicolumn{6}{l}{Modified {\bf Gas Fractions/Mass}: Gas Mass Inside ISM \&\ CGM multiplied by $\eta_{\rm gas}$.} \\
\hline
% m11a_m2e3_fire2_Apr262022_restartdisktest_fgas0pt1
$\eta_{\rm gas}=0.1$ & 6.7 & 52 & \colorno{No} & \colorno{No} & SFR initially dips, then large inflow with more gas re-excites. No disk. Still bursty.\\
% m11a_m2e3_fire2_Apr262022_restartdisktest_fgas0pt1a (out to 200 kpc)
$\eta_{\rm gas}=0.1$a & 2.1 & 10 & \colorno{No} & \colorno{No (weak)} & SFR very bursty, then brief smooth phase then bursty again. Bit of gas in center, not at all disky.\\
% m11a_m2e3_fire2_Apr262022_restartdisktest_fgas0pt3
$\eta_{\rm gas}=0.3$ & 4.9 & 25 & \colorno{No} & \colorno{No} & Gas diffuse and irregular, no coherent disk. SF similar to default just lower mean.\\
% m11a_m2e3_fire2_Apr262022_restartdisktest_fgas0pt3a (out to 200 kpc)
$\eta_{\rm gas}=0.3$a & 2.8 & 13 & \colorno{No} & \colorno{No} & SFR gentle declines by amount expected. Gas in center but diffuse/irregular.\\
% m11a_m2e3_fire2_Apr262022_restartdisktest_fgas3
$\eta_{\rm gas}=3$ & 13 & 34 & \colorno{No} & \colorno{No} & SFR elevated as expected. Gas in center just incoherent clumps, no disk.\\
% m11a_m2e3_fire2_Apr262022_restartdisktest_fgas10
\toohigh $\eta_{\rm gas}=10$ & \colormtoohigh{43} & 140 & \colorno{No (weak)} & \colorno{No} & Gas center clumpy/irregular, no disk (AM dominated by one clump). SFR elevated but bursty.\\ 
\hline
}
\end{footnotesize}

\section{Methods}
\label{sec:methods}

\subsection{Simulation Overview}

The ``default'' simulations and numerical methods here have been extensively utilized in previous work, so we only briefly summarize them here. The simulations were run with {\small GIZMO}\footnote{A public version of {\small GIZMO} is available at \gizmourl} \citep{hopkins:gizmo}, in its meshless finite-mass MFM mode (a mesh-free finite-volume Lagrangian Godunov method). Gravity is solved with fully-adaptive Lagrangian force softening. All include the physics of cooling, star formation, and stellar feedback from the FIRE-2 version of the Feedback in Realistic Environments (FIRE) project \citep{hopkins:2013.fire}, described in detail in \citet{hopkins:fire2.methods}. Gas cooling is followed from $T=10-10^{10}\,$K (including e.g.\ metal-line, molecular, fine-structure, dust, photo-electric, photo-ionization cooling/heating, and accounting for self-shielding and both local radiation sources and the meta-galactic background). We follow 11 distinct abundances accounting for turbulent metal diffusion as in \citet{colbrook:passive.scalar.scalings,escala:turbulent.metal.diffusion.fire}. Gas is converted to stars using a sink-particle prescription if and only if it is locally self-gravitating at the resolution scale \citep{hopkins:virial.sf}, self-shielded/molecular \citep{krumholz:2011.molecular.prescription}, Jeans-unstable, and denser than $n_{\rm crit,\,sf}>1000\,{\rm cm^{-3}}$. Each star particle is then evolved as a single stellar population with IMF-averaged feedback properties calculated following \citet{starburst99} for a \citet{kroupa:2001.imf.var} IMF and its age and abundances. We explicitly treat mechanical feedback from SNe (Ia \&\ II) and stellar mass loss (from O/B and AGB stars) as discussed in \citet{hopkins:sne.methods}, and radiative feedback including photo-electric and photo-ionization heating and UV/optical/IR radiation pressure with a five-band radiation scheme as discussed in \citet{hopkins:radiation.methods}. The simulations are cosmological ``zoom-in'' runs with a high-resolution region of size $\sim$\,Mpc surrounding the primary halo of interest \citep{onorbe:2013.zoom.methods}. Details of all of these numerical methods are in \citet{hopkins:fire2.methods}.

\subsection{Initial Conditions}

In order to have a controlled experiment, we consider in this paper a series of ``controlled restarts,'' as in \citet{orr:ks.law,su:discrete.imf.fx.fire}. Specifically, we re-start a fully cosmological simulation from some chosen time (for our default case, redshift $z_{0}=1$, or scale-factor $a_{z}\equiv1/(1+z) = a_{z,0}=1/2$), which we initially ran to that redshift with the ``base'' physics described above, and re-run it to $z=0$ ($a_{z}=1$) with each physics variation.

For this case study, we will initially focus on one particular initial condition/high-resolution volume, with the primary (most massive) galaxy in the volume denoted ``{\bf m11a}.'' This galaxy is a dwarf galaxy extensively studied in previous FIRE papers \citep[see e.g.,][and references below]{hopkins:fire2.methods,hopkins:cr.mhd.fire2,el.badry:jeans.modeling.dwarf.coherent.oscillations.biases.mass,elbadry:fire.morph.momentum,elbadry:HI.obs.gal.kinematics}. At $z=0$ in our default FIRE-2 simulations, it has a virial mass of $M_{\rm vir} \approx 4\times10^{10}\,M_{\odot}$, virial radius $R_{\rm vir} \approx 90\,$kpc, and galaxy stellar mass of $\approx 6 \times10^{7}\,M_{\odot}$. In many previous studies, we have shown this is a ``typical'' dwarf at this mass scale in many respects including its position on the size-mass, stellar-halo mass, baryonic Tully-Fisher, and mass-metallicity relations \citep{hopkins:fire2.methods,chan:fire.udgs,elbadry:fire.morph.momentum,elbadry:HI.obs.gal.kinematics,emami:2021.fire.testing.bursty.sf.size.fluct.corr}, as well as more detailed internal properties such as its morphology and kinematics \citep{elbadry:fire.morph.momentum,elbadry:HI.obs.gal.kinematics}, dark matter density profile shape \citep{chan:fire.udgs}, merger history \citep{fitts:mergers.in.dwarf.form}, CGM statistics \citep{hopkins:2020.cr.outflows.to.mpc.scales}, outflow properties \citep{pandya:2021.loading.factors.of.fire}, abundance ratios \citep{gandhi:2022.metallicity.dependent.Ia.rates.statistics.fire}, and more. 

Importantly, in all these studies, we robustly find that {\bf m11a} does not form a gaseous or stellar disk at any cosmological time (it is a star-forming dIrr-type galaxy), and exhibits ``bursty'' star formation at all times (see \S~\ref{sec:results} below). Moreover, we find this is robust across dozens of FIRE-2 studies where we have varied many different physics and numerics including gravitational force softening and simulation mass resolution \citep{hopkins:fire2.methods}; the strength and detailed prescriptions for stellar mechanical feedback \citep{hopkins:sne.methods}; methods of treating cooling and radiation \citep{hopkins:radiation.methods};  inclusion of magnetic fields, anisotropic conduction and viscosity, and/or cosmic rays \citep{hopkins:cr.mhd.fire2}; the version of FIRE (and corresponding stellar evolution tracks) between FIRE-2 and FIRE-3 \citep{hopkins:fire3.methods}; detailed star formation prescriptions \citep{orr:ks.law}; and inclusion of AGN feedback \citep{wellons:2022.smbh.growth}.

However, part of our motivation for this study is the recognition that this particular galaxy {\em can}, under the right conditions, form a disk. Specifically, in experiments presented in \citet{shen:2021.dissipative.dm.dwarfs.fire} where we replaced our default collisionless cold dark matter assumption with a dissipative dark matter model, we saw that in models which produced steep central ``cusps'' in the dark matter (very different from the flat-central-density cores that robustly form in CDM), {\bf m11a} (and some other FIRE galaxies which otherwise do not form disks) did in fact robustly form a well-ordered disk. This therefore allows us to ask the question, under what conditions could this galaxy form a disk? 
 
We chose the time of restart ($z=1$) because the stellar mass is $1.9\times10^{7}\,M_{\odot}$, so the system will still form most ($\gtrsim 60\%$) of its final mass at later times, and there is an order-unity redshift change and sufficient time for new equilibria to be reached. But this is sufficiently late in the galaxy history that there is already a well-defined stellar ``galaxy'' with $\sim1-2$\,kpc extent (important for the ``center'' of the potential we impose to be defined at all), and the epoch of initial collapse/rapid assembly has concluded so almost all growth occurs in-situ past this point, making it ideal for this kind of controlled restart. But we have checked that restarting at almost any times between $z\sim 0.2-3$ give very similar qualitative conclusions (just with more/less time for the galaxy to ``adapt'' to the modified conditions). 

As noted below, for a more limited set of comparisons we will consider additional galaxies, for example {\bf m10q} and {\bf m12i} at order-of-magnitude smaller and larger mass scales, respectively, restarted at different times, as well as other galaxies with the same mass scale as {\bf m11a} (e.g.\ {\bf m11b}). But our procedure is identical in each case.

\begin{figure*}
	\includegraphics[width=0.45\columnwidth]{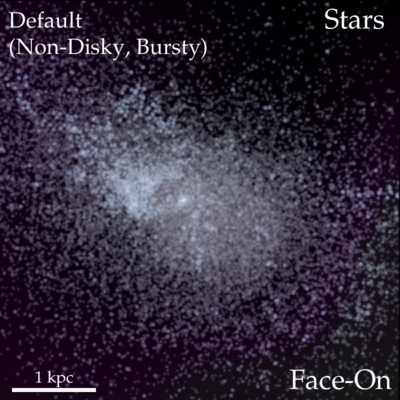}
	\includegraphics[width=0.45\columnwidth]{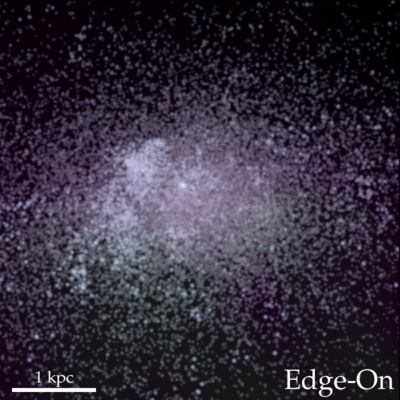}
	\includegraphics[width=0.45\columnwidth]{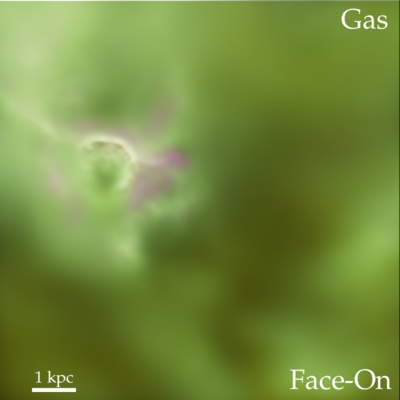} 
	\includegraphics[width=0.45\columnwidth]{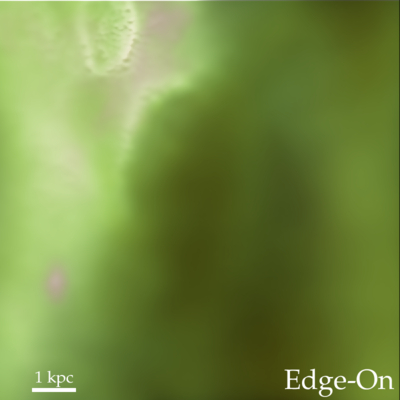} 
	\\
	\includegraphics[width=0.45\columnwidth]{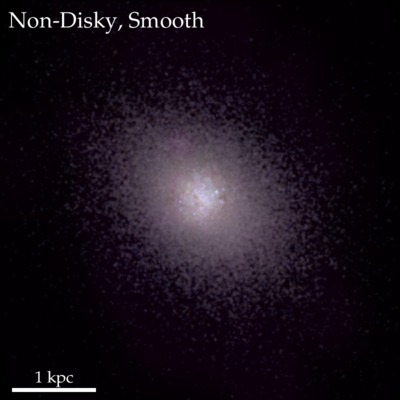}
	\includegraphics[width=0.45\columnwidth]{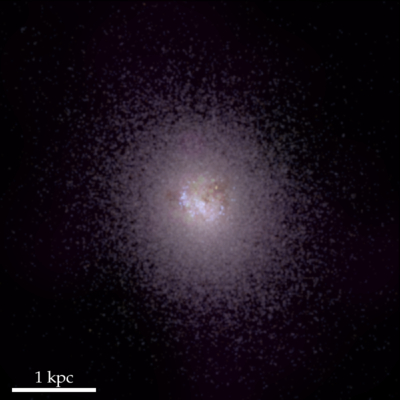}
	\includegraphics[width=0.45\columnwidth]{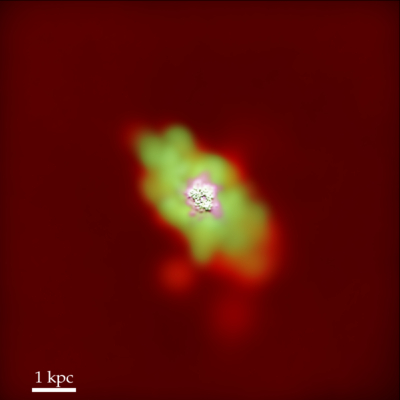} 
	\includegraphics[width=0.45\columnwidth]{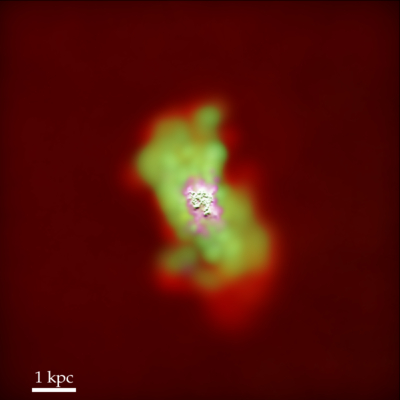} 
	\\
	\includegraphics[width=0.45\columnwidth]{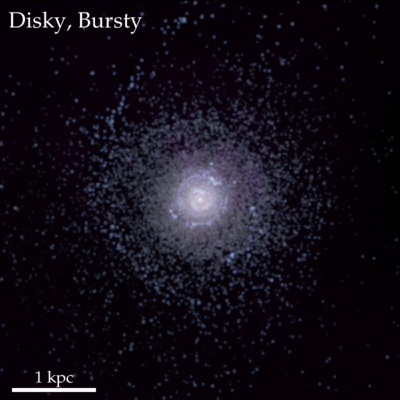}
	\includegraphics[width=0.45\columnwidth]{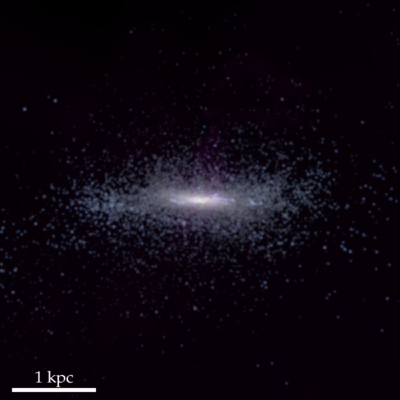}
	\includegraphics[width=0.45\columnwidth]{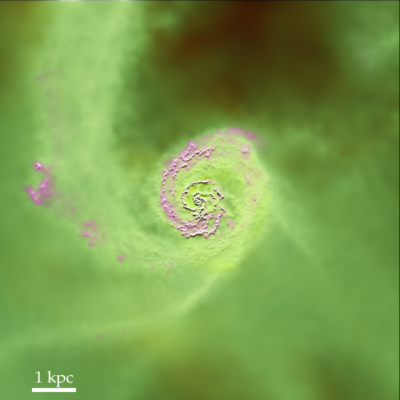} 
	\includegraphics[width=0.45\columnwidth]{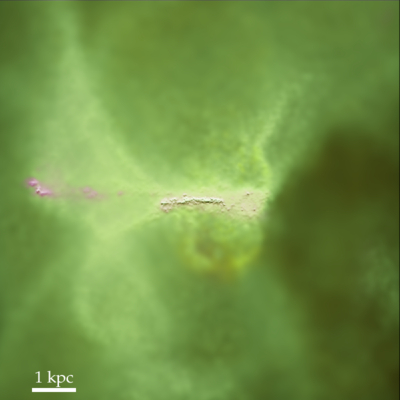} 
	\\
	\includegraphics[width=0.45\columnwidth]{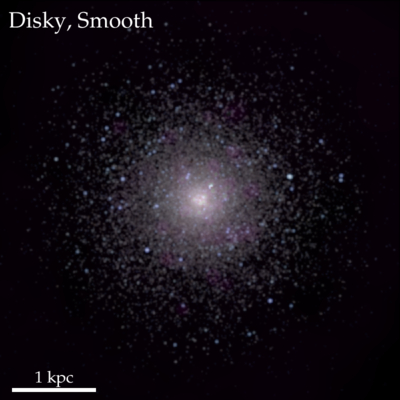}
	\includegraphics[width=0.45\columnwidth]{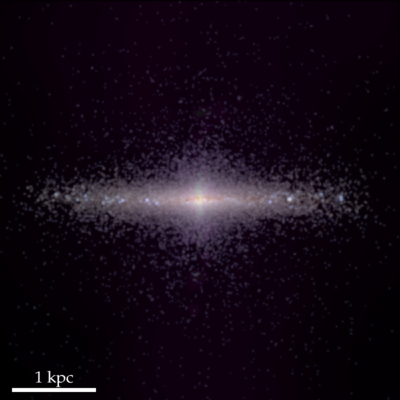}
	\includegraphics[width=0.45\columnwidth]{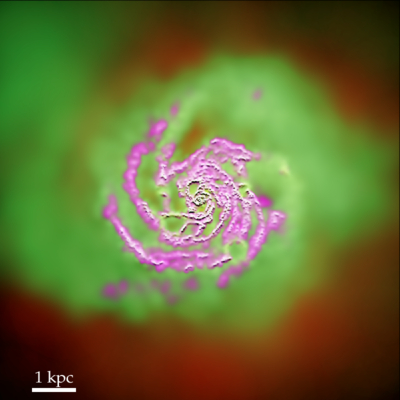} 
	\includegraphics[width=0.45\columnwidth]{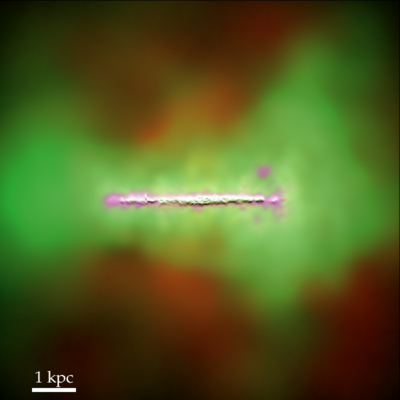} 
	\\
	\vspace{-0.1cm}
	\caption{Images of representative test dwarf galaxy simulations from Table~\ref{table:sims} at $z=0$. 
	We show mock $ugr$ HST composites with the pink alpha layer showing H$\alpha$ emission, projected face-on ({\em first from left}) and edge-on ({\em second}) to the total angular momentum vector, and three-band gas images showing different phases (cold gas at $\ll 10^{4}\,$K, {\em magenta}; warm at $\sim10^{4}$\,K, {\em green}; hot at $\gtrsim 10^{5}\,$K, {\em red}; typical densities for the circum-galactic warm and hot phases are $\sim 10^{-2}\,{\rm cm^{-3}}$ and $\sim {\rm a\ few}\, 10^{-4}\,{\rm cm^{-3}}$, respectively) face-on ({\em third}) and edge-on ({\em fourth}). 
	Rows show different simulations.
	{\em Top:} Default: This produces a system which is spheroidal (not disky) and has bursty SF.
	{\em Second from Top:} $\rho_{0}=10$: Adding a more extended deep potential produces results which are spheroidal but with a smooth SFH.
	{\em Third:} $M_{0}=2$: Adding a concentrated central potential/mass profile produces a disky system with bursty SF.
	{\em Bottom:} $V_{0}=100$: This features a deep potential and centrally-concentrated mass profile which is disky with smooth SF.
	\label{fig:images}\vspace{-0.3cm}}
\end{figure*}

\subsection{Core Physics Variations Considered}

The list of varied idealized test simulations which we primarily study is given in Table~\ref{table:sims}, together with many of their properties. We describe the variations here. Throughout the text, we also briefly discuss additional tests which combine variations below or consider slightly more complicated special cases.

\subsubsection{Modified Potential/Acceleration/Velocity/Dynamical Times}

Following \citet{garrisonkimmel:fire.subhalo.destruction}, when we restart, we insert a special ``source'' collisionless particle at the center of the galaxy of interest. Specifically, we insert a particle with $100\times$ the normal single star-particle mass at the potential minimum of the galaxy at $z=z_{0}$. The mass is chosen so that it is massive enough to avoid artificially $N$-body scattering out of the galaxy center, but low enough mass to introduce negligible perturbations to the galaxy mass profile. In the restarts we can then optionally add a position-dependent term to the gravitational force, which we define as:
\begin{align}
\label{eqn:a.ext.full} {\bf a}_{\rm grav,ext}({\bf x}) &\equiv -a_{\ast}(|{\bf x}-{\bf x}_{0}|)\,\frac{{\bf x}-{\bf x}_{0}}{|{\bf x}-{\bf x}_{0}|}\,e^{-\frac{|{\bf x}-{\bf x}_{0}|}{r_{\rm 0}}}\,\left[1 - e^{-\frac{a_{z}-a_{z,0}}{\Delta a_{z}}} \right]
\end{align}
where ${\bf x}_{0}$ is the position of the central particle defined above, $r_{0} = 6\,$kpc is chosen to roughly correspond to the outer radius of the galaxy (and serves to ensure the added term acts only in the galactic region and remains finite/bounded), $a_{\rm z}$ is the scale factor at $z$, and the $\Delta a_{z} \approx 0.05$ term avoids pathological behaviors by allowing this additional term to ``smoothly'' ramp on. Note we have varied $r_{0}$ from $5-20\,$kpc, and $\Delta a_{z}$ from $0.05-0.2$ in a subset of the simulation tests and find these do not change any of our significant conclusions. The physics of interest for each test is contained in the function $a_{\ast}$, for which we consider:
\begin{align}
\label{eqn:a.ext} 
a_{\ast}(r) &= 
\begin{cases}
0 \hfill ({\rm Default}) \\ 
\frac{G\,M_{0}}{r^{2}+\epsilon_{\ast}^{2}}  \hfill ({\rm Point\ Mass}) \\ 
\frac{V_{0}^{2}}{(r^{2}+\epsilon_{\ast}^{2})^{1/2}}\ \ \ \ \   \hfill ({\rm Constant\ V_{\rm c}}) \\ 
a_{0} \hfill ({\rm Constant\ |{\bf a}|}) \\ 
\frac{4\pi\,G\,\rho_{0}\,r}{3} \hfill ({\rm Constant\ \rho}) \\ 
q_{0}\,r^{2} \hfill ({\rm Increasing\ \rho \propto r}) 
\end{cases}
\end{align}
i.e.\ $a_{\ast}(r) \propto r^{\alpha}$ with $\alpha=(-2,\,-1,\,0,\,1,\,2)$, corresponding to spherically-symmetric mass distributions with $\rho(r) \propto r^{\beta}$ with $\beta = (< -3,\,-2, -1, 0,\, 1)$. 

\subsubsection{Modified Cooling Physics}

We also consider cases where we arbitrarily multiply the radiative cooling+heating rates by a universal constant, i.e.\ $\Lambda_{\rm rad} \rightarrow \eta_{\rm cool}\,\Lambda_{\rm rad}$ with $\eta_{\rm cool} = (0.001,\,0.01,\,0.1,\,0.3,\,1,\,3,\,10)$, which, for fixed initial conditions, directly translates to multiplying the ratio of cooling to dynamical time by $\eta_{\rm cool}^{-1}$. 

As described below we have also considered a set of variants where we replace $\Lambda_{\rm rad}$ entirely with a piecewise constant function (with a single value $\Lambda_{0}$ above/below $10^{4}$\,K); we find this gives identical conclusions to our $\eta_{\rm cool}$ experiments provided the ``typical'' cooling rate is similar ($\Lambda_{0} \sim \langle \eta_{\rm cool}\,\Lambda_{\rm rad}(T,\,...) \rangle$).

We note that more physically-motivated changes to the cooling functions, involving changes to tabulated cooling rates, ionization chemistry, the meta-galactic UV background, treatments of dust and molecular hydrogen, etc., have been explored in \citet{wheeler:ultra.highres.dwarfs,hopkins:fire3.methods}. In those studies, these variations had little effect on the burstiness or diskiness of this system. But the effect of the $\eta_{\rm cool}$ values we choose here is (intentionally) much larger than those physically-interesting variations.

\subsubsection{Modified Star Formation \&\ Stellar Feedback}

We follow \citet{hopkins:fire2.methods} and consider variations where (after restart) we uniformly multiply all feedback ``rates'' (e.g.\ SNe and stellar mass-loss and radiative heating, etc.) rates by a factor $\eta_{\rm fb} = (0.1,\,0.3,\,1,\,3,\,10)$. We also consider models where we modify the energy injection (for mechanical feedback and radiative heating) or momentum flux (for radiation pressure) by a factor $\eta^{\prime}_{\rm fb} = (0.01,\,0.1,\,1,\,30)$, keeping the rates fixed (so feedback is just as ``frequent'' but with less or more energy ``per event''). 

We have also considered experiments where we separately modify the radiative, SNe feedback, and/or stellar mass-loss (O/B and AGB) feedback terms (as opposed to multiplying all of them together). And we considered models where we modify the ``terminal momentum'' $p_{\rm term}$, which determines the ratio of thermal versus kinetic energy coupled from SNe (and the momentum coupled when the SNe cooling radii are unresolved; see \citealt{hopkins:sne.methods}). None of these change our conclusions from our $\eta_{\rm fb}$ runs.

For star formation, we consider variations in the density threshold for star formation, $n_{\rm crit,sf} = (1,\,1000)\,{\rm cm^{-3}}$ (i.e.\ changing it from our default value of $1000\,{\rm cm^{-3}}$, which selects dense gas, to a much lower value applicable to almost all the cool ISM). We also consider variations to the star formation efficiency per free-fall time: a gas cell that meets all of the necessary criteria for star formation (described above) forms stars at a rate $\dot{\rho}_{\ast} = \eta_{\rm sf}\,\rho_{\rm molecular} / t_{\rm freefall}(\rho)$, where we consider $\eta_{\rm sf} = (0.01,\,1)$ (i.e.\ changing it from our default unity value to artificially much ``slower'' star formation). 

Again, more physically-motivated changes to the above, exploring different numerical schemes for coupling SNe/mechanical feedback to the gas, or coupling and propagating radiation in the simulations, as well as adding/removing different feedback terms in more detail (SNe Types Ia and/or II, O/B or AGB mass-loss, radiation in different individual bands, radiative heating, photoionization, radiation pressure, etc.) individually, or adopting different stellar evolution tracks and therefore input rates/spectra for the different feedback channels, are extensively explored in \citet{hopkins:fire2.methods,hopkins:sne.methods,hopkins:radiation.methods,hopkins:fire3.methods}. Similarly, more physical variations to the star formation criteria (modifying or adding/removing a virial/self-gravity, Jeans, molecular, inflow/outflow, temperature, density criteria), or star formation efficiency (considering different observationally-motivated values and/or dynamic models where this is variable) are explored in \citet{hopkins:fire2.methods,orr:ks.law,chan:fire.udgs,elbadry:fire.morph.momentum}. Again, none of these (modulo removing feedback mechanisms entirely, akin to our $\eta_{\rm fb} \ll 1$ models) qualitatively produce differences in diskiness or burstiness so we again are choosing much larger systematic variations to survey than the physically-motivated changes explored in these studies.

\subsubsection{Gas Supply/Fractions}

We vary the gas supply or gas fraction within and around the galaxy by simply multiplying all gas masses in the initial conditions by a factor 
$F_{\rm gas} = 1 - (1-\eta_{\rm gas})\,\exp{(-|{\bf x}-{\bf x}_{0}|^{2}/2\,r_{\rm F})}$, where we consider $\eta_{\rm gas}=(0.1,\,0.3,\,1,\,3,\,10)$, and $r_{\rm F}=40$\,kpc or $r_{\rm F}=200$\,kpc (runs denoted ``$\eta_{\rm gas}=0.1$a'' or ``$0.3$a,'' etc.). We find it makes no difference to our conclusions if we do this in the restarted initial conditions or have the mass change ``ramp'' up.

\subsection{Analysis}

In our analysis below, some quantities are relatively easy to define: we measure the stellar mass, gas mass, and star formation rates within a sphere of $20\,$kpc. For this stellar mass and SFR, this is larger than the main galaxy and the contribution of satellites is negligible, so the choice of sphere size has little effect. For the gas, the values do depend on radius and this choice is somewhat arbitrary, but our qualitative conclusions are independent of where we define this measurement (it is simply a convenient reference point). When we show circular velocity curves we simply define:
\begin{align}
\label{eqn:Vc} V_{\rm c}^{2}(r) \equiv a_{\rm spherical}\,r = \frac{G\,M_{\rm enc}(<r)}{r} + a_{\rm grav,\,ext}(r)|\,r\ , 
\end{align}
(where $a_{\rm grav,\,ext}$ is the added term in Eq.~\ref{eqn:a.ext}), 
i.e.\ we neglect the (very small, for our purposes) corrections to $V_{\rm c}$ from non-spherical terms in the potential. Images are mock HST-like $ugr$ composites or multi-temperature composites computed by post-processing the simulations self-consistently with the stellar population models and gas/dust extinction models in-code, following \citet{hopkins:lifetimes.letter,hopkins:2013.fire}. Star formation histories are archeological (SFH of stars in the galaxy at $z=0$), although because the stars are primarily formed in-situ, this is very similar to a SFH following the ``main progenitor.'' 

``Diskiness'' and ``Burstiness'' in Table~\ref{table:sims} are qualitative terms, so need some clarification. When we refer to ``diskiness'' we generally are referring to the visual morphology of the gas (and stellar, if we explicitly refer to a stellar disk), though we will show cases with formally-measured $V_{\rm rot}/\sigma$, $H/R$, and $j/j_{\rm c}$, to support this. Likewise, we rely on visual classification of the SFHs to label simulations as ``bursty'' or not in Table~\ref{table:sims}. But in the large majority of cases, the classification is un-ambiguous, and we specifically identify cases where it is ambiguous (either because it appears ``in between'' in some sense, or because different indicators disagree). More extensive exploration of different labeling schemes and justification for our simple classification is given in  \citet{sparre.2015:bursty.star.formation.main.sequence.fire,yu:2021.fire.bursty.sf.thick.disk.origin,gurvich:2022.disk.settling.fire,2021MNRAS.501.4812F}.

\section{Results \&\ Discussion}
\label{sec:results}

Table~\ref{table:sims} summarizes many of the key properties of the simulations at $z=0$. We see, as discussed above, that while our ``default'' simulation and many variants produce no disk and remain ``bursty,'' we do simulate a number of variants which produce either disks and/or smooth SFHs (see lines with ``yes'' label in blue).

Fig.~\ref{fig:images} shows the visual morphologies at $z=0$ of several of these galaxies, with and without disks. We can clearly see the disks in the cases labeled as ``disky,'' and clearly see that those labeled not disky have a typical dIrr-like morphology (often with some gas, but it is in a clumpy/bursty/shell/fragment configuration, with no preferred geometry).

\begin{figure}
	\includegraphics[width=0.95\columnwidth]{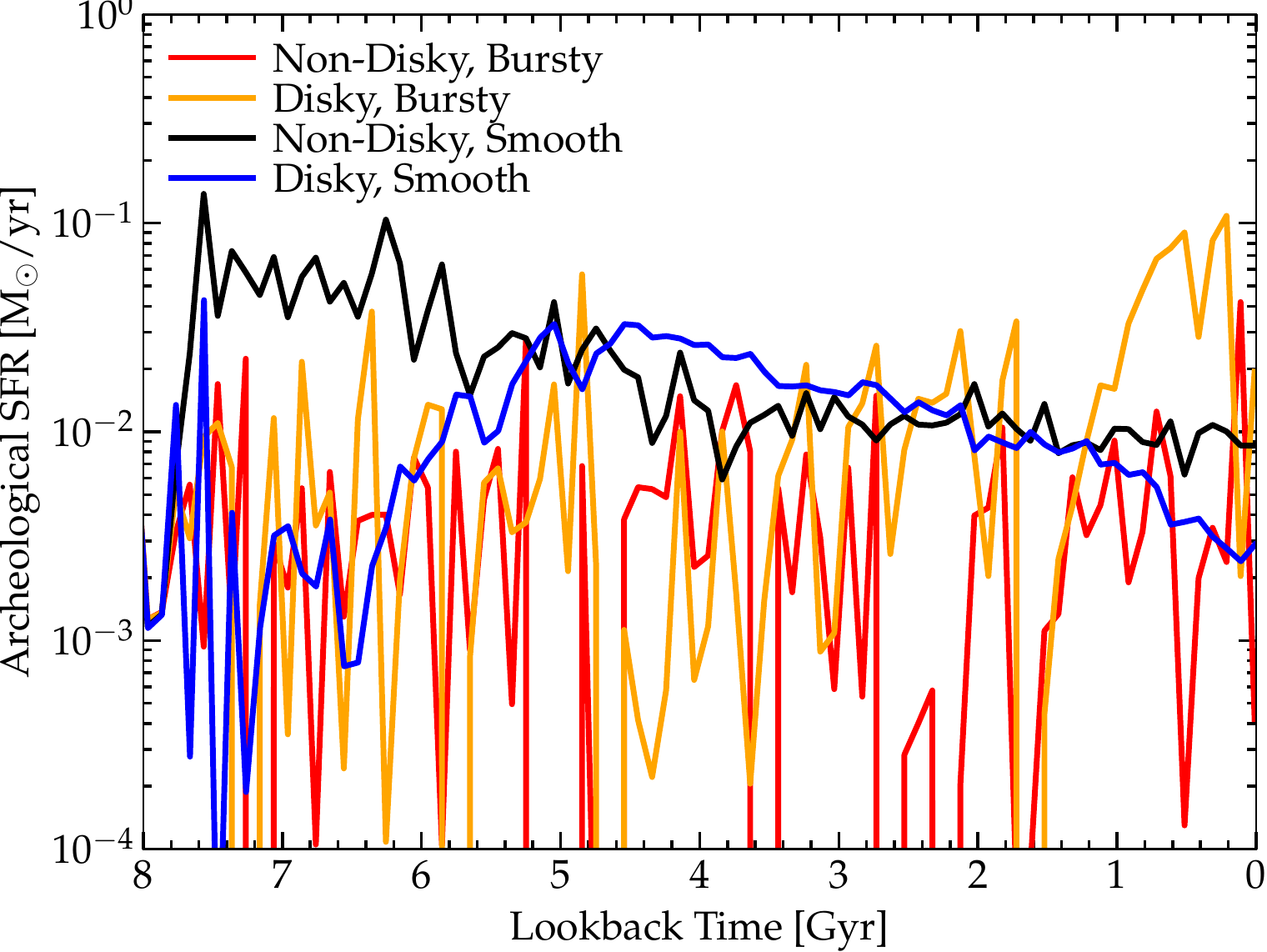}
	\vspace{-0.1cm}
	\caption{SFRs versus lookback time of the galaxies from Fig.~\ref{fig:images}. We restrict to the times after the experiments were started (since they are identical by definition before). We see a clear separation in the bursty versus smooth SF histories.
	\label{fig:sfr}}
\end{figure}

Fig.~\ref{fig:sfr} shows the SFHs for the galaxies in Fig.~\ref{fig:images}, labeling those which we call ``smooth''. While different physics changes produce systematically higher or lower SFRs (reflected in the integrated stellar masses formed given in Table~\ref{table:sims}), the difference between bursty and smooth star formation at any normalization is visually obvious.

\begin{figure}
	\includegraphics[width=0.95\columnwidth]{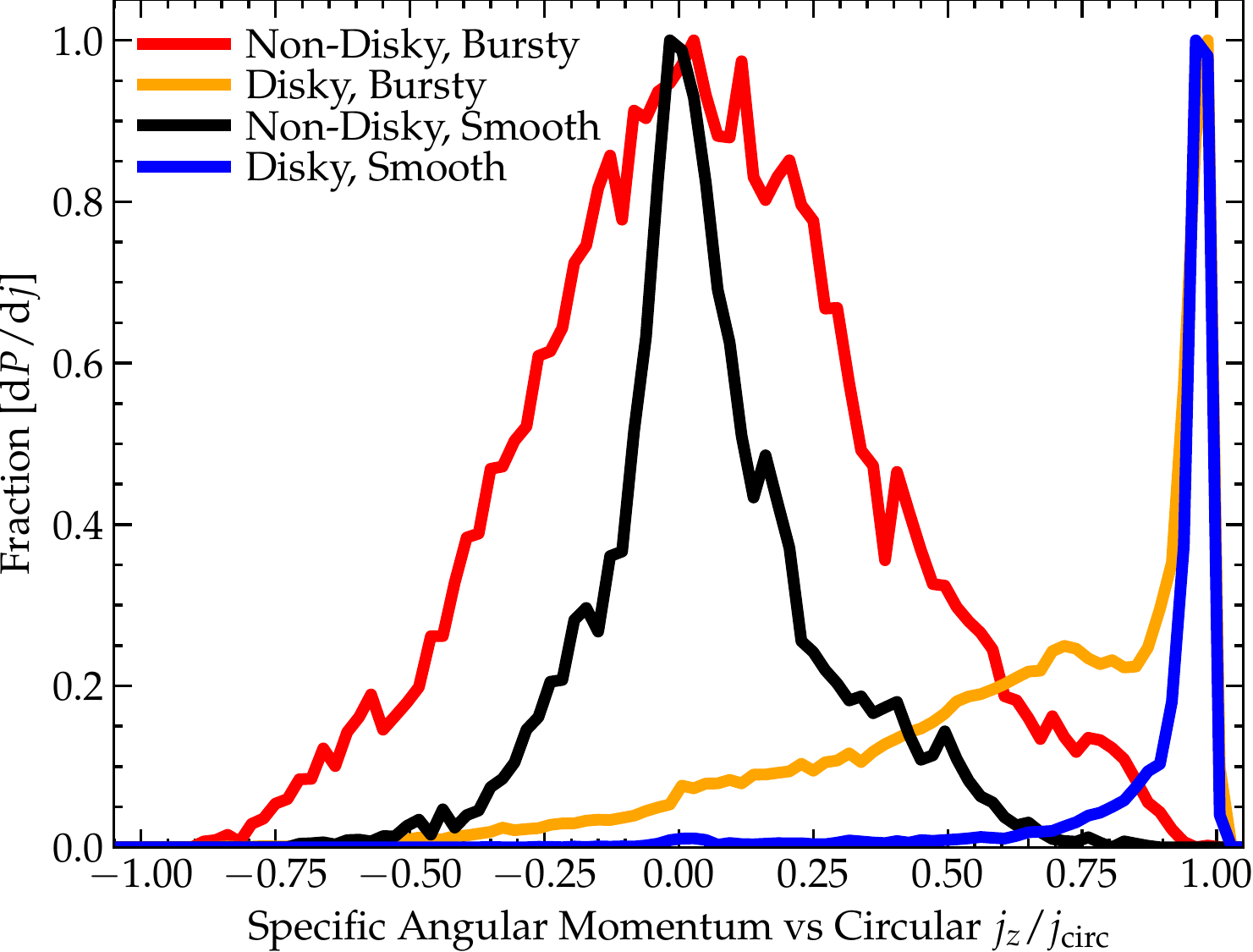}
	\vspace{-0.1cm}
	\caption{Distribution of specific angular momentum of all gas within $<10\,$kpc of the galaxy center, relative to that of a circular orbit with the same energy, in the runs from Fig~\ref{fig:images}. The disk separation is quantitatively clear (with the ``disky, smooth'' example being a ``very thin'' disk).
	\label{fig:jc}}
\end{figure}

\begin{figure}
	\includegraphics[width=0.95\columnwidth]{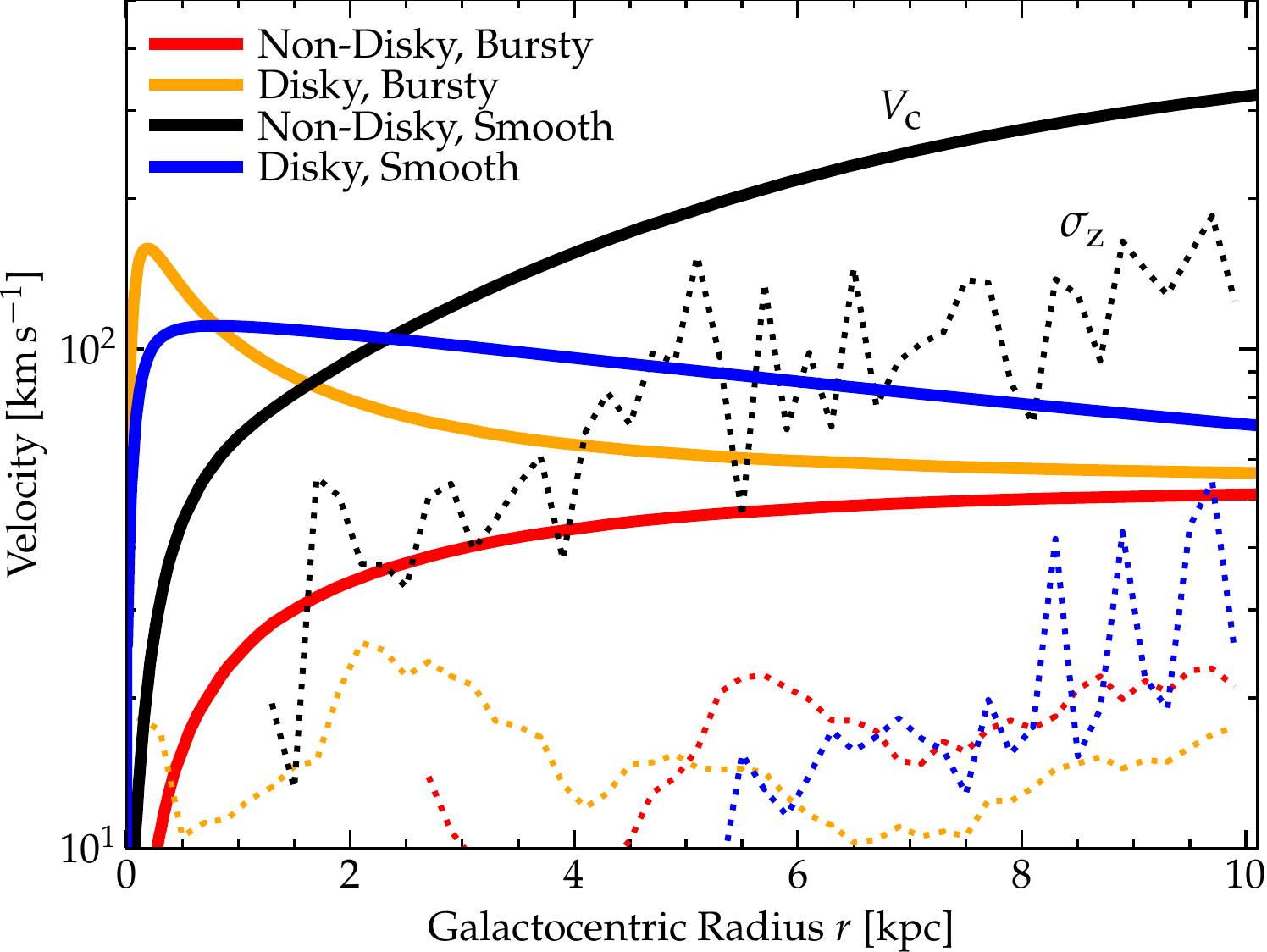}
	\vspace{-0.1cm}
	\caption{Circular velocity $V_{\rm c}$ (Eq.~\ref{eqn:Vc}; {\em solid}) and vertical gas velocity dispersion $\sigma_{z}$ ({\em dotted}) measured in radial annuli, in the runs from Fig~\ref{fig:images}. The disky cases exhibit centrally-concentrated $V_{\rm c}$ profiles.
	\label{fig:Vrot}}
\end{figure}

Figs.~\ref{fig:jc} \&\ \ref{fig:Vrot} quantify the angular momentum content and the ratio of rotation velocity to velocity dispersion in more detail, for the same simulations. Note that the results shown are robust to the precise time chosen to view the simulation: time-averaged versions or examples at any redshift between $z=0 - 0.5$ are similar. These are chosen to illustrate the fact that the classification assigned to the galaxies is qualitatively robust to this labeling scheme.\footnote{As can be inferred from direct comparison between Fig.~\ref{fig:images} and Fig.~\ref{fig:jc}, it also does not significantly alter our conclusions regarding the ``diskiness'' and angular momentum distribution if we select all gas or just HI or star-forming gas (e.g.\ exclude hot phases and/or outflows), or vary the radius interior to which Fig.~\ref{fig:jc} is measured.}

\subsection{The Structure of The Gravitational Potential Is Key}

From a cursory examination of Table~\ref{table:sims} and Figs.~\ref{fig:images}-\ref{fig:Vrot}, it is already clear (as we will show in detail below) that the most important physics controlling both diskiness and burstiness is the form of the gravitational potential/acceleration. However, it is striking that different changes in the potential manifest differently for disk formation and the cessation of bursty star formation. In brief, while the most dramatic modifications to the potential can produce both diskiness and smooth SFHs, it is clear that in general, the more centrally-concentrated potential/mass-profile modifications (the point-mass-like or constant-$V_{\rm c}$ models) are much more easily able to promote disk formation (but often with bursty SFHs), while the more extended potential terms (the constant-density or rising-density/$q_{0}$ models) are more effective at promoting smooth SF. We will discuss both of these in greater detail below in \S~\ref{sec:diskform}-\ref{sec:bursty}, but first wish to discuss other variations which do not so directly influence diskiness or burstiness.

\subsection{Disk Formation and Reduced ``Burstiness'' Are Not Necessarily Coupled}
\label{sec:decoupling}

Before discussing specific physics variations, one thing which is immediately clear, as noted above, that the formation of a disk is {\em not} necessarily tied to the transition of star formation into the smooth (i.e. ``not bursty'') mode, and vice versa. While our default simulation and many variants are both (a) not disky and (b) bursty in SF, and we have a couple of variants which are both (a) disky and (b) smooth in SF, we actually see that {\em most} of the tests which produce clear well-ordered gaseous disks still feature bursty star formation, and conversely {\em most} of the tests which feature relatively smooth SF (notably reduced burstiness) do not produce gas disks! 

In typical cosmological simulations, the emergence of disks and transition to smooth SF are tightly coupled in time \citep{ma:radial.gradients,ma:2016.disk.structure,angles.alcazar:particle.tracking.fire.baryon.cycle.intergalactic.transfer,orr:stacked.vs.bursty.sf.fire,elbadry:fire.morph.momentum,elbadry:HI.obs.gal.kinematics,garrisonkimmel:fire.morphologies.vs.dm,hung:fire.sf.gal.kinematic.evol.gas.dispersion,yu:2021.fire.bursty.sf.thick.disk.origin,stern21_ICV,trapp:2021.radial.transport.fire.sims,hafen:2022.criterion.for.thin.disk.related.to.halo.ang.mom}. This has led, naturally, to speculation that there is a direct causal link between the two: perhaps SF is smooth if and only if it occurs in a well-ordered disk, perhaps because stellar feedback could ``vent'' in the polar direction without disrupting the whole disk and causing a cessation of SF, or perhaps because of differences in the associated Toomre mass \citep{cafg:bursty.sf.toymodel}. 

However, we see no such one-to-one relation. It is clear, that the {\em local} causal agent of smooth SF and disk formation can be different. Moreover, a disky configuration is {\em neither necessary nor sufficient} for smooth star formation, and smooth star formation is {\em neither necessary nor sufficient} to promote disk formation.

\subsubsection{Possible Exception: A Very Thin Disk May Often Imply Sufficient Conditions For Smooth Star Formation}

That said, one caveat is that all of our {\em very thin} disks in the {\bf m11a} experiments above (e.g.\ the models with $V_{0} \gtrsim 100$, or $a_{0} \gtrsim 10^{4}$, or $M_{0} \gtrsim 5$), produce at least some period of smooth star formation later in their history (after much of the initial gas mass has been exhausted by star formation and well after the disk is established), provided we restrict to the star formation occurring within the disk.\footnote{The model with $V_{0}=200$ has a period of smooth star formation followed by some bursts at very late times, but it has nearly gas-exhausted at this time and so some of this very late-time burstiness is likely due to the very low gas content; see \citealt{orr:ks.law} for discussion and other examples in this limit.} Thus it may be the case that whatever conditions are necessary for a very thin disk to form are also sufficient for the star formation within that disk to be ``smooth.'' We discuss this further below when we discuss the broader criteria for smooth SF, but note that there is at least one exception we study ({\bf m11b}) below, which forms a very thin disk with bursty SF.

\begin{figure}
	\includegraphics[width=0.95\columnwidth]{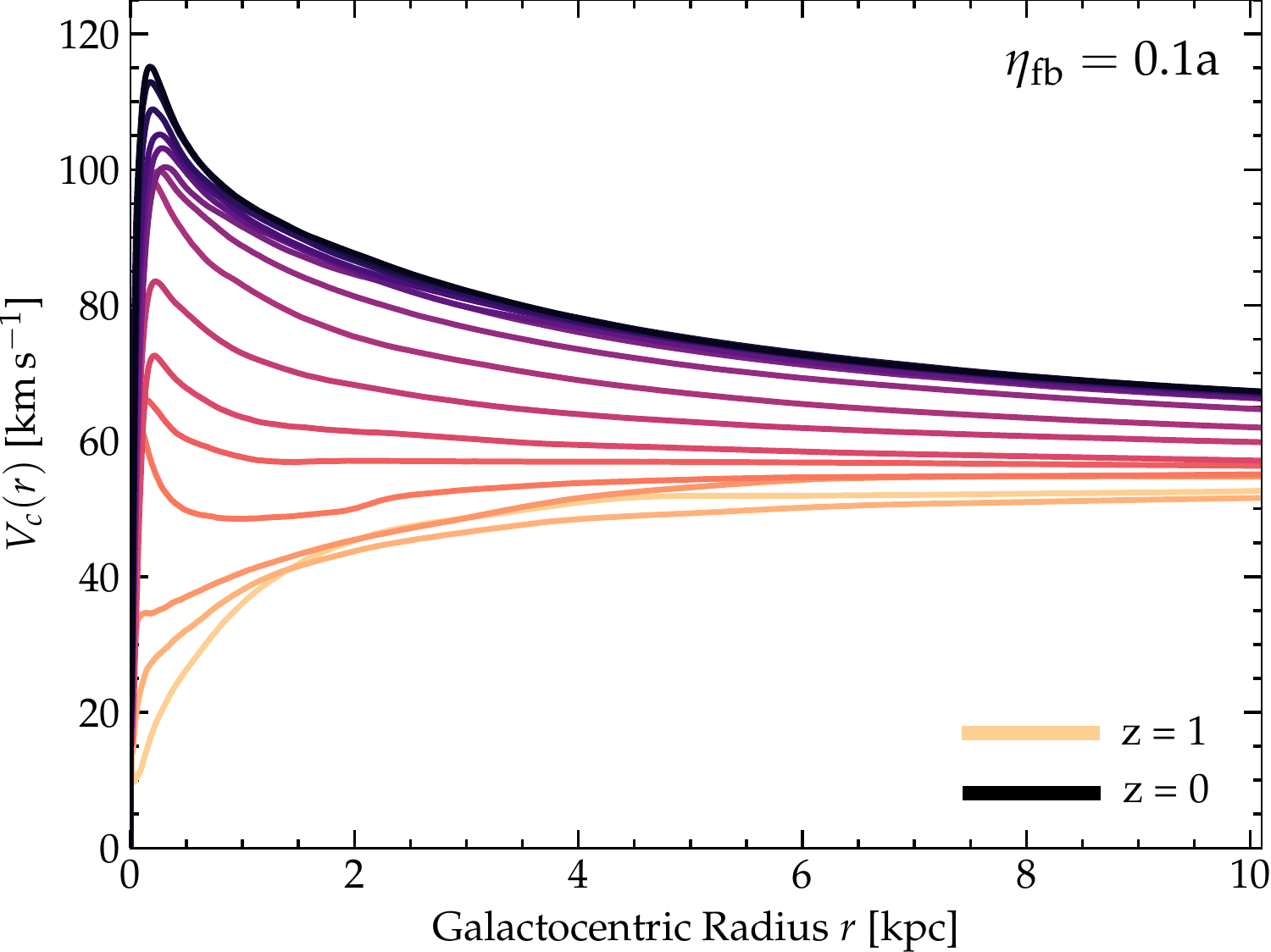}
	\includegraphics[width=0.95\columnwidth]{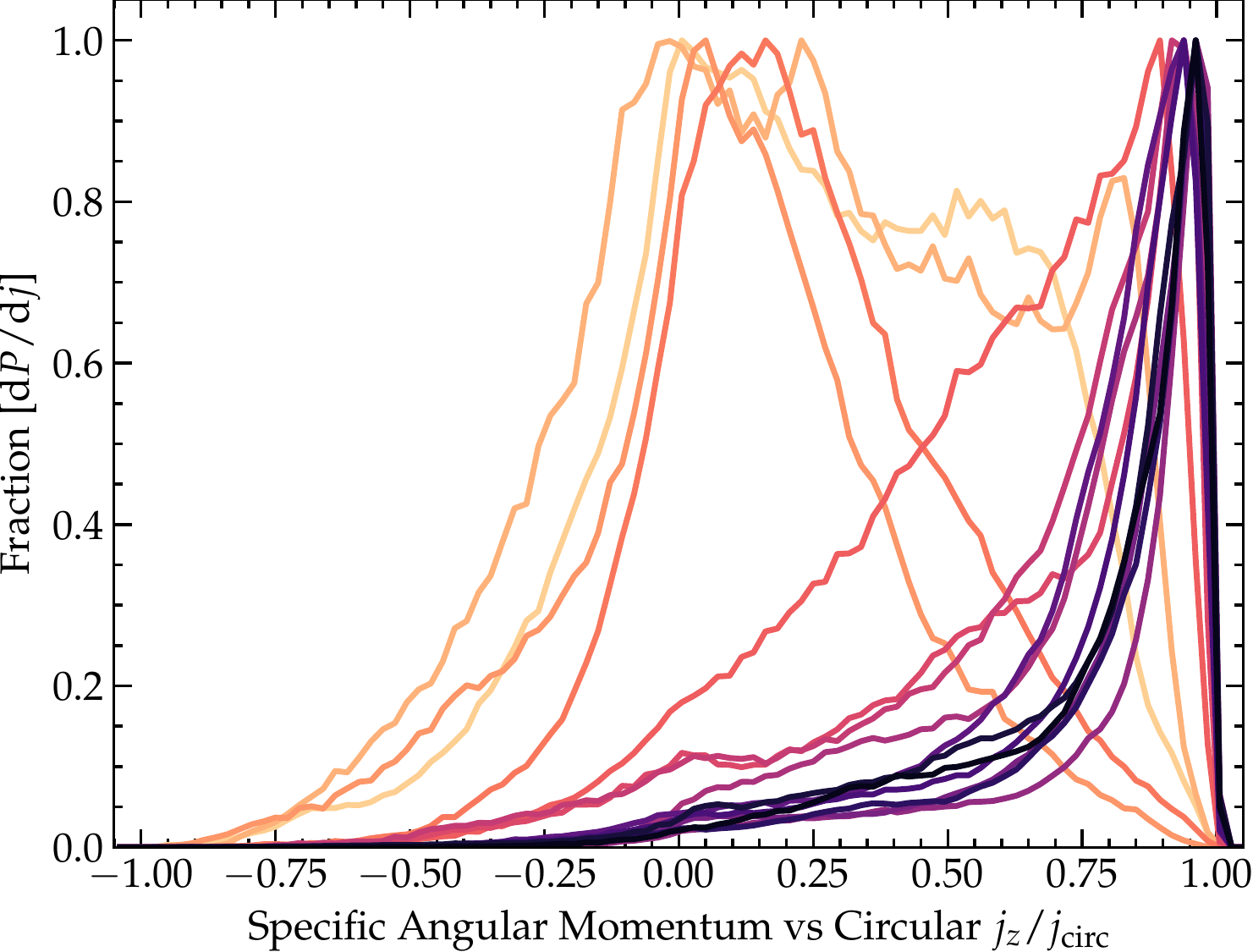}
	\vspace{-0.1cm}
	\caption{Time evolution of the rotation curve (as Fig.~\ref{fig:Vrot}) and angular momentum (as Fig.~\ref{fig:jc}) of the $\eta_{\rm fb}=0.1$a run with very weak stellar feedback: different line colors correspond to snapshots roughly uniformly spaced in proper time between $z=1$ (lighter) when we begin the experiment and $z=0$ (darker). The mass profile becomes more and more centrally-concentrated owing to efficient star formation (given the very weak feedback) in the center; coherent disky structure only builds up after the potential is sufficiently centrally-concentrated.
	\label{fig:weak.fb}}
\end{figure}

\subsection{Stellar Feedback and Star Formation Criteria Do Not Directly Cause Diskiness or Burstiness}
\label{sec:sffb.fx.weak}

It is clear that making stellar feedback stronger or weaker (either modifying the {\em rates} of events, e.g.\ number of SNe, or {\em energy} of events, e.g.\ explosion energy/momentum per SNe), and/or changing the star formation model (making star formation less/more efficient per freefall time, and/or changing the densities at which gas is allowed to turn into stars) has very weak effects on both the ``burstiness'' of star formation and the ``diskiness'' of the resulting galaxies. None of our simulations with varied feedback or star formation physics directly produces any significant enhancement in ``diskiness'' of gas or stars (neither visually nor measured in any way we can quantify it), and none produces significantly smoother star formation -- in fact, one variant (lowering the critical SF density to $n_{\rm crit}=1\,{\rm cm^{-3}}$) produces somewhat more bursty behavior. We stress that this has already been shown repeatedly in previous studies (see discussion above), which systematically varied feedback and star formation models and assumptions to a much greater extent than we consider in this study here. We simply include it here for the sake of completeness, and because it illustrates some important and potentially non-intuitive physics.

It might seem at first glance as if our extremely-weak feedback runs ($\eta_{\rm fb} \lesssim 0.1$ models in Table~\ref{table:sims}) are an exception to this, as they form disks and have smooth star formation at late cosmic times. However, examination of their time-history and rotation curves in Fig.~\ref{fig:weak.fb} immediately shows that this is a non-linear effect akin to our ``concentrated mass profile'' models. After restart, these simulations {\em do not} exhibit disks nor smooth SF. However given the weak feedback, they form a huge number of stars in each ``burst'' before self-regulating. This builds up an extremely large, compact central stellar mass -- already in severe violation of any reasonable observational constraints from the stellar mass function or satellite abundances \citep{behroozi:2012.abundance.matching.sfhs,moster:2013.abundance.matching.sfhs,brook:2014.mgal.mhalo.local.group,torrey:abundance.matching.evolution,sgk:2016.mgal.mhalo.lowmass.scatter,read:2017.shallow.mgal.mhalo,nadler:2020.abundance.matching.w.lmc.fx}. This leads to a $V_{\rm c}(r)$ curve which becomes centrally-peaked. Only when the mass profile is both (a) sufficiently centrally concentrated, and (b) exceeds $V_{\rm c} \gg 60\,{\rm km\,s^{-1}}$ does any disky structure start to form, and only after this peak reaches $\gtrsim 100\,{\rm km\,s^{-1}}$ (with the escape velocity from small radii exceeding $\sim 200\,{\rm km\,s^{-1}}$) does the SFR become smooth. These are consistent with the values where we see the same behaviors in our idealized experiments (which do not necessarily produce such enormous SFRs and stellar masses in violation of the observations). So these simulations actually support the hypothesis that it is the potential, {\em not} feedback directly, that influences the diskiness and burstiness.

\subsubsection{Connection to Previous Work: Self-Regulation Is Critical}

From some earlier work, one might imagine that ``burstiness'' and ``lack of a gas disk'' are, in some sense, related to either (a) feedback being ``too strong'' (blowing apart the galaxy), or (b) star formation being ``too efficient'' or ``too strongly clustered'' (in space/time) such that single clumps of stars (as compared to a ``steady drip'' of star formation) dominate the dynamics. And indeed this has been suggested explicitly in some studies \citep{roskar:2014.stellar.rad.fx.approx.model,kim:disk.self.reg}. And indeed this could be true at some level if ``all else were equal'' -- if e.g. one took a perfectly fixed gas initial condition in a fixed external potential and then instantaneously subjected it to a large feedback force, it would be unbound, while if it was subjected to weak feedback, it would collapse. The latter is more or less the experiment in \citealt{2015MNRAS.454..238W,kim.ostriker:sne.momentum.injection.sims}, and we consider directly analogous cases in the text below.

However, the problem is that ``all else is {\em not} equal'' in any dynamical system. Many studies, including those cited above and \citet{wise:2008.first.star.fb,schaye:2010.cosmo.sfh.sims,ostriker:2010.molecular.reg.sf,hopkins:rad.pressure.sf.fb,hopkins:2013.fire,krumholz:2011.rhd.starcluster.sim,trujillo-gomez:2013.rad.fb.dwarfs,federrath:2014.low.sfe,grudic:sfe.cluster.form.surface.density,agertz:sf.feedback.multiple.mechanisms,kimm.cen:escape.fraction,rosdahl:2015.galaxies.shine.rad.hydro}, have shown that star formation is {\em self-regulating}. If one allows the gas to actually evolve naturally in a multi-phase, self-gravitating system, then in a self-gravitating gas cloud or disk, if there is ``less feedback per star formed,'' the system will continue to collapse under self-gravity and form more and more stars until enough stars form to provide sufficient feedback to disrupt the structure. Thus, {\em feedback self-regulates to the same total strength}. What changes is merely the star formation efficiency. We see this directly in our tests: the mass of stars formed after the ``restart'' is inversely proportional to the feedback strength $\eta_{\rm fb}$. Similarly, if we change the density threshold for star formation it does nothing to alter the global balance: forcing a low star formation rate in dense gas merely means more gas will ``pile up'' at high densities before feedback can self-regulate it, and gas at those high densities will get denser and denser before being exhausted or disrupted, so the details of the phase distribution of gas (e.g.\ amount of gas at the highest densities relative to that at intermediate densities) will change, but to leading order the global feedback properties are unchanged (again, this is shown in much more detail in previous studies, see \citealt{hopkins:rad.pressure.sf.fb,hopkins:dense.gas.tracers,hu:2022.public.gizmo.multiphase.ism.thermochemistry.sims}). Changing the ``form'' of feedback, e.g.\ more events with less energy per event, or changing the feedback ``budget'' to reduce energy in SNe and place more into radiative or cosmic ray feedback, also has remarkably little effect on these conclusions, as shown explicitly in \citet{hopkins:cr.mhd.fire2}. Basically, on galactic scales for all but the smallest ultra-faint dwarf galaxies (for those, see discussion in \citealt{su:discrete.imf.fx.fire}), feedback is already in the ``continuum limit'' and regulated by global energy/momentum balance \citep{pandya:2021.loading.factors.of.fire}. This also means more ``indirect'' variations, like changing the IMF, supernova mass distribution or energy distribution versus mass, etc, all have negligible effects here \citep[see][]{hopkins:fire3.methods}.

\subsubsection{Feedback And Other Physics Can Be Important Indirectly}

Of course, this is not to say that feedback cannot have a quantitative effect on diskiness or burstiness, especially {\em indirectly}. For example, many studies have shown that ignoring radiative feedback from stars leads to quantitatively enhanced ``burstiness'' because the weaker ``early'' feedback disrupting nascent GMCs allows them to collapse further, producing more-strongly-clustered star formation and therefore SNe \citep{kannan:photoion.feedback.sims,rosdahl:2015.galaxies.shine.rad.hydro,kimm:lyman.alpha.rad.pressure,emerick:rad.fb.important.stromgren.ok,hu:2017.rad.fb.model.photoelectric,emerick:rad.fb.important.stromgren.ok,hopkins:radiation.methods}. However, in all these studies, this does not fundamentally change whether the galaxy is in the ``bursty'' or ``smooth'' regime as we define it here (just the magnitude of that burstiness). 

Another obvious non-linear way that feedback can indirectly influence burstiness or diskiness is through processes like those described above for our weak-feedback simulations: feedback can and does, over cosmic times, play a key role in determining what the actual shape of the potential is. So of course, feedback is very important in this sense. {\em Our point is simply that it is not the proximate causal agent of burstiness or diskiness.} There are also many processes which can disrupt or destroy disks (e.g.\ galaxy mergers) -- and it has been known for decades that models without stellar feedback run from early cosmic times ($z\gtrsim 100$) will turn all the baryons into stars rapidly at high redshifts, so galaxies assemble via collisionless mergers that tend to destroy stellar disks (even if those disks did initially form; see e.g.\ \citealt{white:1991.galform,barneshernquist92,katz:treesph,somerville99:sam,governato04:resolution.fx,maller:sph.merger.rates,mashchenko:2008.dwarf.sne.fb.cusps,naab:dry.mergers,naab:etg.formation,keres:fb.constraints.from.cosmo.sims}). We will discuss some of these subsequent processes below but stress that here we focus exclusively on what causes the initial formation of disks.

%\clearpage

\subsection{Gas Supply, Toomre Masses, and Thermal/Cooling Physics Do Not Control Diskiness or Burstiness}
\label{sec:thermo}

We find that arbitrarily varying the gas supply/fraction ($\eta_{\rm gas}$) and/or gas cooling rates ($\eta_{\rm cool}$) has no qualitative effect on the formation of a disk or transition to smooth SF in the simulated galaxies. 

Note that since most of our simulations varying $\eta_{\rm cool}$ were run with otherwise ``default'' physics, so remain bursty and do not form disks, we also wished to re-test this with (a) simulations that clearly form a disk, and (b) simulations which exhibit smooth SF, to test the converse (whether changing the cooling rate could prevent this from occurring). We therefore re-ran variants of our $M_{0}=2$ and $\rho_{0}=10$ models, replacing our usual complicated cooling function with a simple toy model $\Lambda\rightarrow \Lambda_{\rm toy}$ where $\Lambda_{\rm toy} = 10^{-25}\,{\rm erg\,cm^{3}\,s^{-1}}$ for $T<10^{4}$\,K and $\Lambda_{\rm toy} = (10^{-23},\,10^{-22},\,10^{-21})\,{\rm erg\,cm^{3}\,s^{-1}}$ for $T\ge 10^{4}\,$K. The variants of $\rho_{0}=10$ all show ``smooth'' SF after the restart (despite the changes in $\Lambda$), and the variants of $M_{0}$ all show disky structure (though the disk mass varies substantially). The gas supply changes, as expected: the more efficient-cooling models produce significantly larger stellar/baryonic galaxy masses and non-linearly more concentrated/deeper potentials by $z=0$.  

As discussed below, we have also repeated many of these experiments in Milky Way-mass galaxies, with the same conclusions.

\subsubsection{Disk Properties May be Influenced, But Not Whether It Can Initially Form}

The fact that gas supply and/or cooling has little effect is less surprising for the initial disk formation question. While some models have argued that whether the halo is virialized or accretion proceeds in the rapid ``cold mode'' ($t_{\rm cool} \lesssim t_{\rm dyn}$) or slow ``hot mode'' ($t_{\rm cool} \gtrsim t_{\rm dyn}$) can be important for how thin the galactic disk is and how much angular momentum is present \citep{keres:hot.halos,keres:cooling.revised,keres:cooling.clumps.from.broken.filaments,keres:fb.constraints.from.cosmo.sims,dekelbirnboim:mquench,ceverino:2010.clump.disks.cosmosims,bett:2011.halo.spin.flips,torrey:2011.arepo.disks,2017ApJ...835..289S,stewart:scylla.halo.accretion.vs.codes,stern21_ICV,hafen:2022.criterion.for.thin.disk.related.to.halo.ang.mom}, it is important to note that these models and simulations were making a point about the thickness, mass, or amount of angular momentum (or resulting size, given said angular momentum) of disks {\em assuming such disks can form at all}, not about the binary question of whether or not a disk can form {\em in the first place} \citep[see e.g.][]{brook:2010.low.ang.mom.outflows}. 

One might naively assume that ``more gas = more disk'' (i.e.\ more rapid cooling or more gas supply would provide more disk), but again this assumes said gas can form a disk at all. Similarly, many studies have noted that gas disks can rapidly ``re-form'' after destructive events such as galaxy mergers provided there is sufficient gas supply \citep{hopkins:disk.heating,hopkins:disk.survival.cosmo,hopkins:disk.survival,hopkins:m31.disk,hopkins:inflow.analytics,hammer:hubble.sequence.vs.mergers,koda:disk.survival.prescriptions,hoffman:mgr.orbit.structure.vs.fgas,stewart:disk.survival.vs.mergerrates,puech:2012.disk.survival}. But again, by definition these studies began from pre-existing disks so the necessary conditions for disk formation were implicitly already met, so this is a quantitative question of how much mass could be supplied for said disk. So of course, provided a disk can form, more gas and more rapid cooling may enhance its mass, but this does not answer how it first forms. 

Moreover, as we discuss at length above (\S~\ref{sec:sffb.fx.weak}), feedback is self-regulating: we see that if we increase the gas supply, more stars form but more gas is also blown out/ejected; conversely, if we decrease the gas supply only within some finite radius, then feedback is weaker so more gas flows in from large radii until a similar balance is achieved. And of course, with no gas supply or cooling (e.g.\ our lowest $\eta_{\rm cool}$ runs, which entirely quench star formation after one final burst of the pre-existing dense gas already in the disk at the initial conditions), no gas disk can form, but this is trivial.

\subsubsection{Changing the Cooling Rate and/or Gas Supply Cannot Alone ``Break Out'' of Burst-Quench Cycles}

It is perhaps slightly more surprising that varying the cooling rate and/or gas supply does not seem capable of pushing the galaxies into the smooth mode of SF. It has been argued that the halo moving into the ``hot'' mode, or the inner halo virializing and moving from a gas cooling time $t_{\rm cool} \lesssim t_{\rm freefall}$ (with typical gas temperatures $T \ll T_{\rm vir}$) to a quasi-cooling flow solution with $t_{\rm cool} \gtrsim t_{\rm freefall}$ (and $T \sim T_{\rm vir}$), is very well-correlated in time with the transition from ``bursty'' to ``disky'' and could represent the causal agent driving this transition \citep{yu:2021.fire.bursty.sf.thick.disk.origin,stern21_ICV,hafen:2022.criterion.for.thin.disk.related.to.halo.ang.mom,gurvich:2022.disk.settling.fire}. It has also been argued that the transition could represent the characteristic most unstable or ``Toomre'' mass for a disk with $Q\sim 1$, which scales as $M_{\rm char}[r]/M_{\rm gas}[<r] \sim (M_{\rm gas}[<r]/M_{\rm enc}[<r])^{2}$ \citep[for a more rigorous discussion for supersonically turbulent systems demonstrating this still holds even for $Q\ne 1$, see][]{hopkins:excursion.ism,hopkins:frag.theory}, becoming small compared to the galaxy gas mass so that the star formation is dominated by many independent small clouds which can ``average out'' and fail to expel the entire ISM \citep{cacciato:2011.analytic.disk.instab.cosmo.evol,huertas.company:2016.surface.density.vs.quenching.morph,cafg:bursty.sf.toymodel}. But when we artificially force any ratio we desire of $t_{\rm cool}/t_{\rm freefall}$ by varying $\eta_{\rm cool}$,\footnote{Using the criterion defined specifically in \citet{stern21_ICV} for $t_{\rm cool}^{(s),\,0.1\,R_{\rm vir}}/t_{\rm ff}$ at $0.1\,R_{\rm vir}$, our $\eta_{\rm cool}$ variations span $0.01 \lesssim t_{\rm cool}^{(s),\,0.1\,R_{\rm vir}}/t_{\rm ff} \lesssim 100$.}
or any gas fraction (hence Toomre/characteristic fragment mass) varying $\eta_{\rm gas}$, we do not see any values which generate ``smooth'' star formation without changing the potential.

First, regarding the cooling efficiency $\eta_{\rm cool}$ and ``hot'' versus ``cold'' mode halos, there are several reasons this does not necessarily produce smooth SF in the experiments here (all discussed in more detail in \S~\ref{sec:bursty} below). 
(1) Even if inflow were perfectly ``smooth'' in time, this would only produce smooth SFRs if there were no processes {\em internal to the galaxy} which could produce SFR fluctuations. 
(2) We show below in our more detailed experiments that suppressing bursty SF is closely related to suppressing ``overshoot'' in local fluctuations of the ISM SFR, and confining the ensuing outflows so they remain trapped in the disk. For a fixed gas supply, the effects of a ``hot halo'' might be expected to be relatively modest: even assuming a spherical, virialized gas halo the theoretical maximum enhancement in the energy required to push most of the ISM+inner CGM gas (the mass-loading required to create a significant ``burst-quench'' cycle) out to some radius (or equivalently, the minimum effective Bernoulli parameter) is just a factor $\sim 2$ (relative to a vacuum), much smaller than other variations we consider. 
(3) Even in spherical symmetry with a pressurized halo and an initially ``smooth'' inflow rate (and no mechanism to generate fluctuations in the disk for a fixed gas content at some radius $r$), nothing prohibits ``global breathing modes'' where gas shells coherently oscillate in and out of the galaxy, and in fact most analytic models of feedback-induced gas outflow burst-quench cycles causing e.g.\ dark matter cores explicitly invoke such models \citep[see e.g.][]{pontzen:2011.cusp.flattening.by.sne,teyssier:2013.cuspcore.outflow,brook:2015.dm.signals}. In other words, gas could recycle in bursty fashion ``interior to'' the CGM. 
(4) Changing $\eta_{\rm cool}$ in any cooling-flow or precipitation-type model can indeed change the overall mass supply (as we see in the SFRs and stellar masses); but (as argued above) since feedback is self-regulating, this alone has little effect on the dynamics of the gas within the galaxy or its ability to ``overshoot'' in bursts and be ejected in quench cycles. 
(5) If $\eta_{\rm cool}$ is small enough (e.g.\ our $\eta_{\rm cool} \lesssim 0.01$ runs), the entire halo gas supply will have negligible cooling during a Hubble time, so after the galaxy consumes or ejects via feedback the small initial dense gas mass in the disk, star formation is quenched, rather than proceeding steadily.\footnote{In these runs we find an essentially adiabatic gas halo, which does form some coherent halo gas angular momentum as expected from its spin \citep[e.g.][]{momauwhite:disks}, but obviously no disk.}

Additionally, it is important to note that independent of $\eta_{\rm cool}$, the accretion onto the halo {\bf m11a} is always quasi-spherical,\footnote{This does not mean that accretion cannot be ``clumpy,'' let alone that there cannot be large pressure and/or density fluctuations in the halo. That can and does occur even in classical, quasi-spherical ``hot mode'' halos owing to the thermal instability, and of course the pressure fluctuations can be larger if there are trans or super-sonic non-coherent motions in the halo gas. But the accretion will still be statistically quasi-spherical.} as expected (for a more detailed illustration in ``default'' simulations at these masses, see e.g.\ \citealt{hafen:2018.cgm.fire.origins,hafen:2019.fire.cgm.fates}). ``Cold mode'' accretion ($t_{\rm cool} \lesssim t_{\rm ff}$) is sometimes conflated with geometric arguments about filamentary accretion. But the first studies of cold/hot accretion \citep[see e.g.][]{keres:hot.halos,keres:cooling.revised,dekel:cold.streams} clearly showed that the halo accretion geometry is a function primarily of the large-scale structure, such that halos with masses less than the turnover mass in the dark matter halo Schechter function ($M_{\rm halo} \lesssim 10^{13}\,M_{\odot}$, at $z\sim 0$) will accrete quasi-spherically, and only those more massive halos preferentially in ``nodes'' at the cosmic web will accrete in filamentary structures (hence ``cold flows in hot halos'' at low redshifts). Thus we do not anticipate any effect on this structure from $\eta_{\rm cool}$.

So we turn to the second question: why does changing the gas supply (either by directly modifying the gas mass inside the halo, or altering the cooling rate) not produce smooth star formation, as might be expected via the Toomre-mass argument? We see five reasons this argument would not apply to the tests here: (1) This pre-supposes a disk, which does not form in these runs (a spheroid would regulate to quasi-hydrostatic equilibrium, giving a different scaling). (2) It implicitly assumes that an isothermal sound speed $\sim 10\,{\rm km\,s^{-1}}$ (which is a ``floor'' for WNM/WIM gas both in the simulations and observed dwarfs; see \citealt{leroy:2008.sfe.vs.gal.prop}) is insufficient to maintain $Q\sim 1$, so the value of e.g.\ $\sigma/V_{\rm c} \sim H/R$ is determined by super-sonic turbulence self-regulating the vertical disk structure \citep{ostriker.shetty:2011.turb.disk.selfreg.ks,hopkins:fb.ism.prop}. For the galaxy here this requires $\Sigma_{\rm gas} \gg 10\,M_{\odot}\,{\rm pc^{-2}}$ (similar to the standard self-shielding threshold; see \citealt{kmt:molecular.fraction.local.quantities}). Otherwise, $H/R$ is set by this ``floor'' temperature (i.e.\ simple photo-ionization physics) and thus the ratio $M_{\rm char}/M_{\rm gas}$ is independent of the gas supply. (3) Self-regulating star formation (discussed above) negates much of the variation in gas supply. If we modify cooling rates, feedback self-adjusts to maintain a given mass given the potential depth. If we reduce the gas supply in the CGM, out to say $\sim 50\,$kpc, then the immediately lower SFRs (hence weaker feedback) lead to more efficient cooling/inflow of gas from even larger radii, as seen in the surprisingly large gas masses in Table~\ref{table:sims} for our $\eta_{\rm gas}\lesssim 0.3$ runs. We can suppress this by reducing the gas mass all the way out to the turnaround/splashback radius ($\sim 200\,$kpc, our $\eta_{\rm gas}\lesssim 0.3$a runs), which may be responsible for a very transient ``smooth'' SF phase in our most extreme simulation ($\eta_{\rm gas}=0.1$a), but even then there is some negation and a return to bursty SF. (4) The galaxy radius interior to which SF occurs can adjust to maintain $Q\sim 1$, not just the scale-height at a fixed radius -- essentially the Toomre-mass hypothesis we are discussing here assumes gas is ``frozen'' at some galacto-centric radius. (5) This model also ignores the role of {\em global} modes influencing burstiness. If the burstiness is connected to a global ``breathing'' mode with gas flowing in and out coherently, as argued for both analytically \citep{pontzen:2011.cusp.flattening.by.sne} and in simulations \citep{governato:2012.dwarf.form.sims,elbadry.2015:core.transformation.stellar.kinematics.gradients.in.dwarfs,christensen:2016.baryon.cycle}, then the SFR will always rise coherently when the gas mass is large in the center, regardless of the number of ``sub-clumps'' into which it is divided.

\begin{figure}
	\includegraphics[width=0.95\columnwidth]{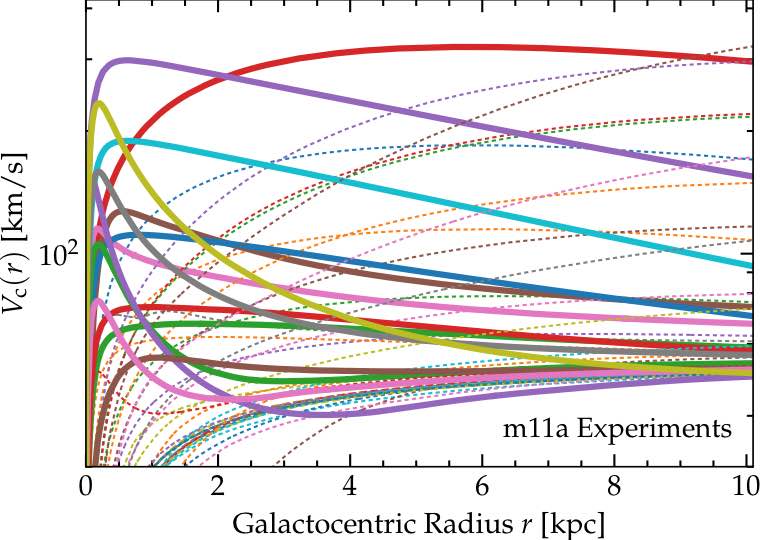}
	\includegraphics[width=0.95\columnwidth]{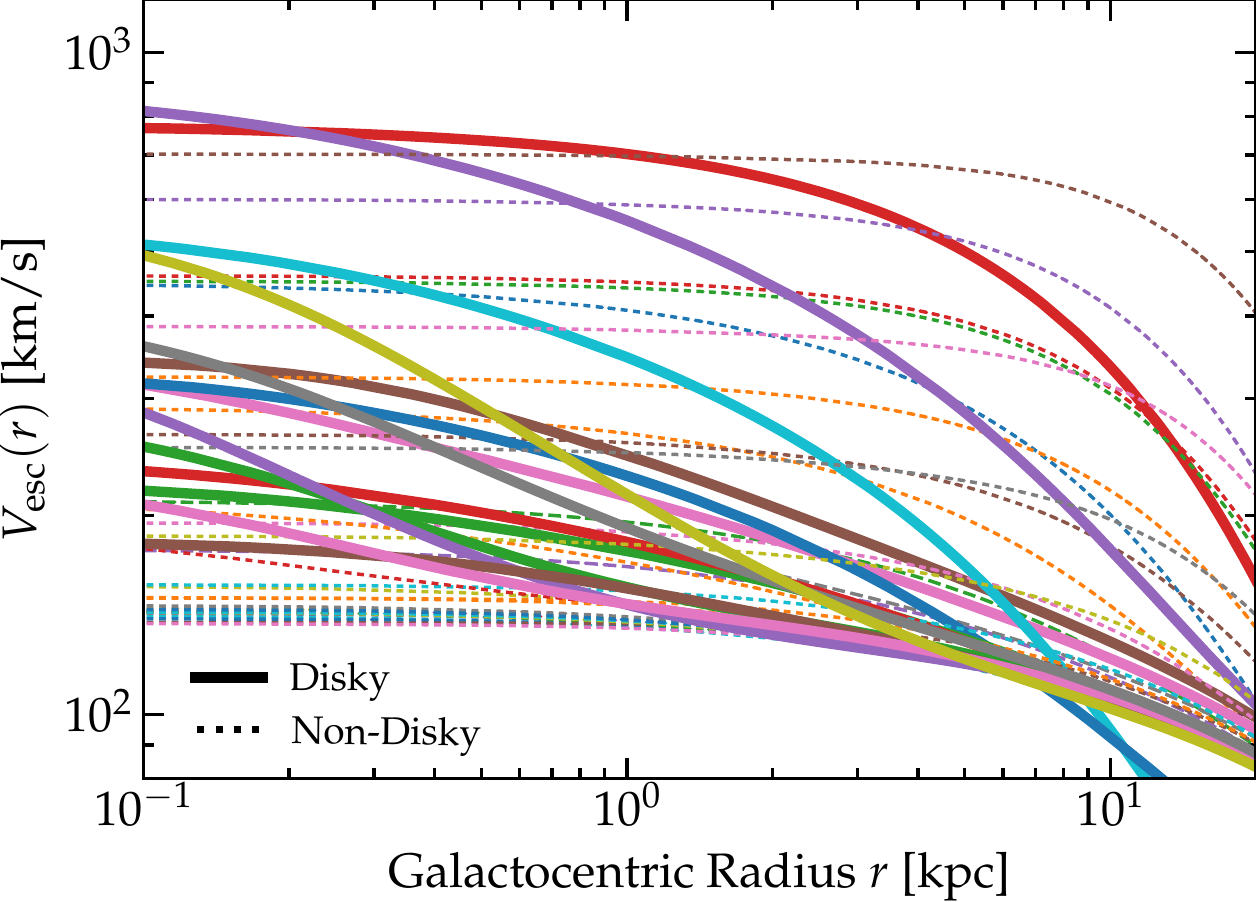}
	\vspace{-0.1cm}
	\caption{{\em Top:} $V_{\rm c}$ profile ({\em top}, as Fig.~\ref{fig:Vrot}) for all simulations in Table~\ref{table:sims}, coded by whether they form a clear disk ({\em thick solid}) or no disk ({\em thin dashed}). There is no obvious trend with $V_{\rm max}$ (the maximum value of $V_{\rm c}$), but there is a clear trend with {\em shape} of the profile, where the disky systems exhibit more centrally-concentrated mass profiles (see \S~\ref{sec:diskform}).
	{\em Bottom:} Escape velocity $V_{\rm esc}$ (Eq.~\ref{eqn:vesc}) profile, in the same style. There is no clear trend of diskiness with $V_{\rm esc}$.
	\label{fig:Vc.vs.disk}}
\end{figure}

\subsection{Other Physics and Numerics Variations}
\label{sec:other.physics}

As noted in \S~\ref{sec:methods}, this galaxy ({\bf m11a}) has been simulated with a wide array of physics and numerics variations in other papers: we briefly review those results here, insofar as they influence whether the galaxy forms a disk and/or develops smooth SF.

Almost all of the previously-studied variations produce no significant change in the ``diskiness'' or ``burstiness'' of the galaxy. This includes: varying the numerical mass resolution by a factor of $\sim 100$; varying the gravitational force softening for collisionless species by factors of $\sim 10$; varying the numerical scheme by which SNe and other mechanical feedback is deposited; varying the strength of SNe feedback; replacing our default FIRE radiation treatment with an explicit M1-radiation hydrodynamics integrator; considering additional bands for radiative feedback (e.g.\ X-ray, He ionizing), or removing radiative feedback entirely; changing the meta-galactic UV background; changing the detailed cooling physics, including/excluding explicit treatment of molecular hydrogen, and changing the stellar evolution tracks, in the FIRE-3 \citep{hopkins:fire3.methods} versus FIRE-2 version of FIRE; varying the sub-grid model for star formation (its efficiency, and the density/virial/molecular/inflow criteria); adding supermassive black holes with various prescriptions for their growth and feedback; including magnetic fields, anisotropic Spitzer-Braginskii conduction and viscosity; explicitly integrating cosmic ray feedback from SNe with various different assumptions for the cosmic ray scattering rates/transport coefficients; adding elastic self-interacting (eSIDM) physics \citep{hopkins:fire2.methods,hopkins:sne.methods,hopkins:radiation.methods,hopkins:cr.mhd.fire2,hopkins:fire3.methods,chan:fire.udgs,el.badry:jeans.modeling.dwarf.coherent.oscillations.biases.mass,elbadry:fire.morph.momentum,elbadry:HI.obs.gal.kinematics,orr:ks.law,gandhi:2022.metallicity.dependent.Ia.rates.statistics.fire,wellons:2022.smbh.growth}.

As discussed below, we have also run a series of more detailed experiments to ask more specific follow-up questions: these include simulations where we vary the cooling function shape, vary the cooling function only in CGM gas (not in the ISM), vary the form of SNe coupling to the surrounding gas, vary which particles can inject SNe in order to set SNe rates ``by-hand,'' vary the terminal momentum of SNe ejecta, vary the rates of SNe and their specific energy, vary the timing of SNe after stars form, vary the star formation criteria more extensively, and {\em simultaneously} vary $\eta_{\rm gas}$ and $\eta_{\rm cool}$ or $\eta_{\rm fb}$ and $\eta_{\rm cool}$. We have also run a limited number of the tests above in re-starts of both lower and higher-mass halos. As we describe below, none of these tests change any of our conclusions.

Given what we have seen in our study thus far, the reason these changes did not qualitatively alter the diskiness or burstiness of this galaxy is fairly obvious. As discussed in each of the papers cited, almost all the variations above produced fairly weak changes to either the star formation history or galaxy potential of {\bf m11a}. Of course, non-linear effects like those discussed in \S~\ref{sec:sffb.fx.weak} can appear -- if we turn off all feedback entirely (akin to our low-$\eta_{\rm fb}$ models), the potentials are radically different, but the models are also ruled out immediately by observations. In practice all the effects above, within the reasonable physically-allowed range, had weak effects.

The only exception was our study in \citet{shen:2021.dissipative.dm.dwarfs.fire} noted in \S~\ref{sec:methods}, with dissipative self-interacting dark matter (dSIDM), which formed disks in some runs. In all of those cases, the dark matter had a sufficiently high cross-section and dissipation rate such that steep central ``cusps'' formed in the dark matter with mass profiles $\rho \propto r^{-1.5}$, significantly steeper even than NFW halos, let alone the ``cored'' $\rho \propto r^{0}$ profiles that tend to form in the default CDM runs. Our experiments in this paper immediately make it clear why those simulations produced disks: they featured a centrally-concentrated mass profile, close to our ``constant-$V_{\rm c}$'' experiments (depending on the specifics of the dark matter model from \citealt{shen:2021.dissipative.dm.dwarfs.fire}). In fact, the disks formed therein are quite similar in extent and morphology to our idealized experiments with $V_{0} \sim 60-70\,{\rm km\,s^{-1}}$, and the galaxies therein also continued to exhibit bursty SF, as our experiments here did as well. Similar hints have been seen in other studies varying different physics with entirely distinct codes and numerical methods \citep{christensen:2014.spiral.morph.in.dwarfs.vs.model.ism.sf}.

\begin{figure}
	\includegraphics[width=0.95\columnwidth]{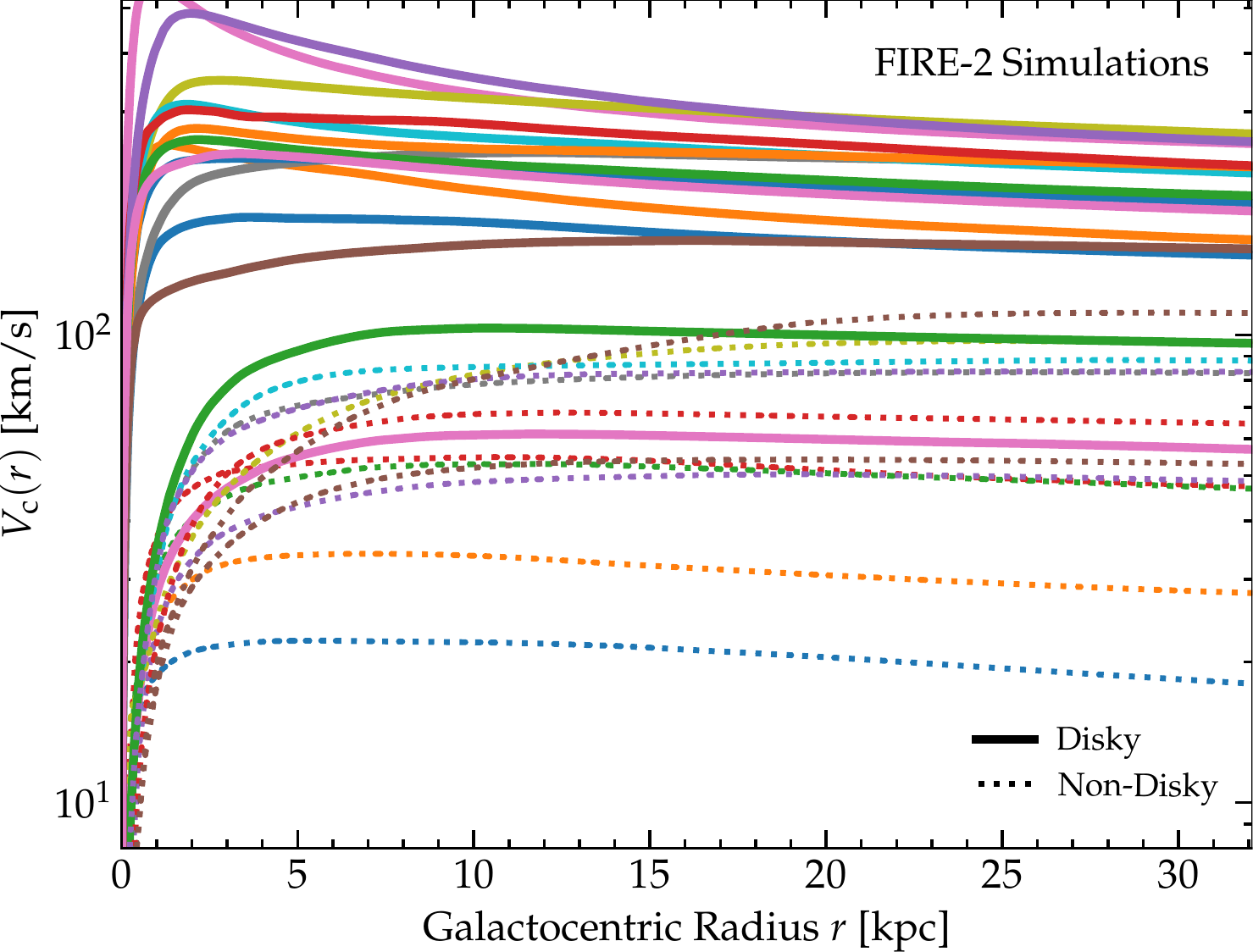}
	\includegraphics[width=0.98\columnwidth]{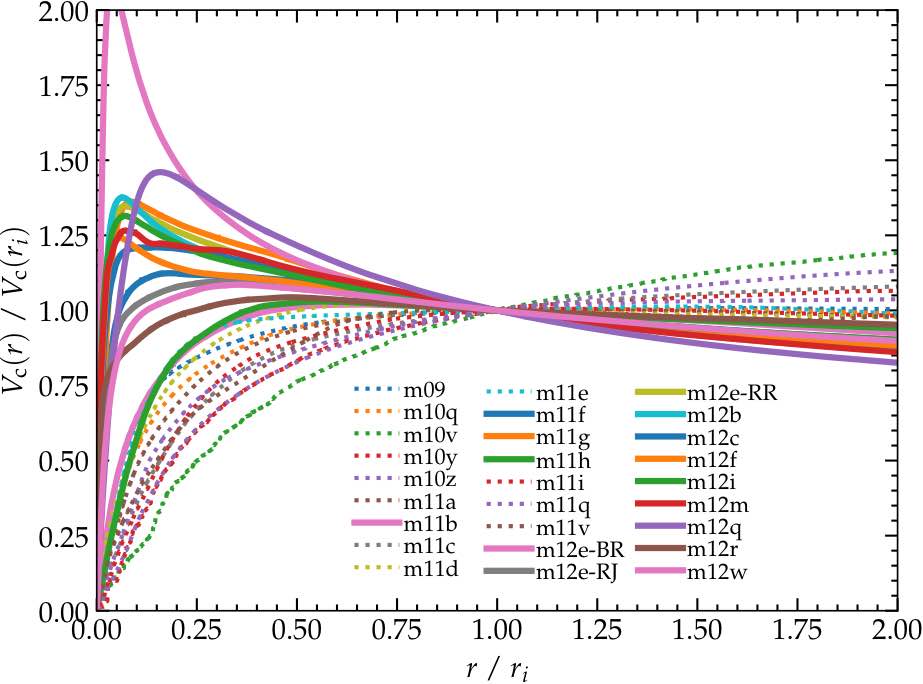}
	\vspace{-0.1cm}
	\caption{Circular velocity curves for all the ``default physics'' FIRE-2 simulations in \citet{hopkins:cr.mhd.fire2} at $z=0$, labeled by whether they form a clear disk as Fig.~\ref{fig:Vc.vs.disk}. These span halo masses $\sim 10^{9}-10^{12.5}\,M_{\odot}$ and a wide range of environments and formation histories. 
	{\em Top:} We show $V_{\rm c}(r)$ in absolute units, where we see that all the very massive galaxies form centrally-concentrated rotation curves (owing to efficient star formation), and disks, but there are is some overlap in disk and non-disk formation at intermediate $V_{\rm c}(r)$, but it is difficult to assess ``how concentrated'' the $V_{\rm c}$ curves are on this scale.
	{\em Bottom:} Same, but normalizing to focus on the $V_{\rm c}$ curve {\em shape} by plotting $V_{\rm c}(r)/V_{\rm c}(r_{i})$ versus $r/r_{i}$, where $r_{i}$ is chosen somewhat arbitrarily to be 3 times the half-cold-neutral HI gas mass radius (using e.g.\ 2-3 times the stellar effective radius or other choices give similar results). Here we see more clearly a division between systems with concentrated mass profiles down to radii at least smaller than this radius (of order the gas circularization radius), which form disks, and systems which are less concentrated and do not (see \S~\ref{sec:diskform}). Note none of these simulations include AGN feedback so none ``quench'' or form ellipticals.
	\label{fig:Vc.fire.disks}}
\end{figure}

\begin{figure*}
	\includegraphics[width=0.48\textwidth]{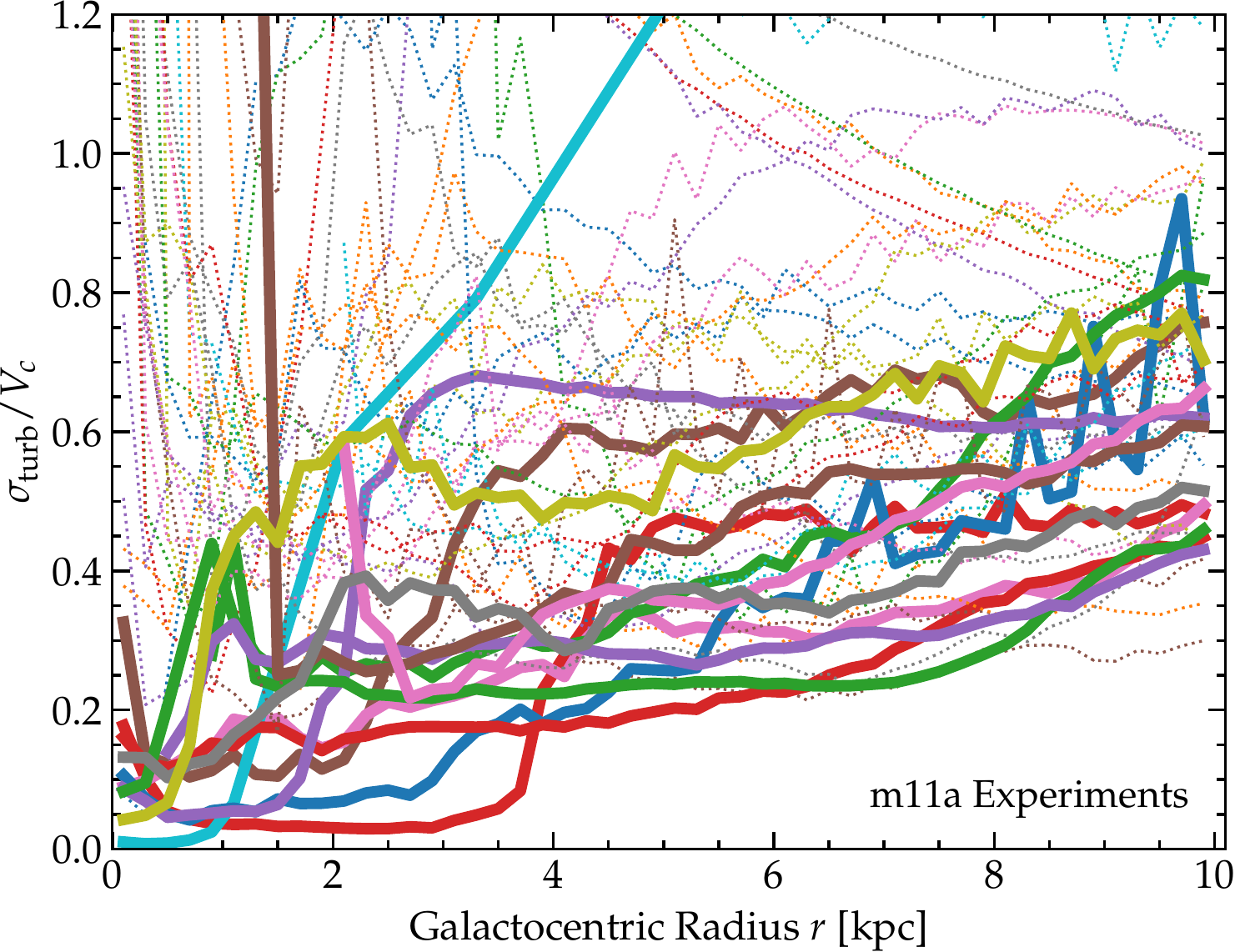} 
	\hspace{0.1cm}
	\includegraphics[width=0.48\textwidth]{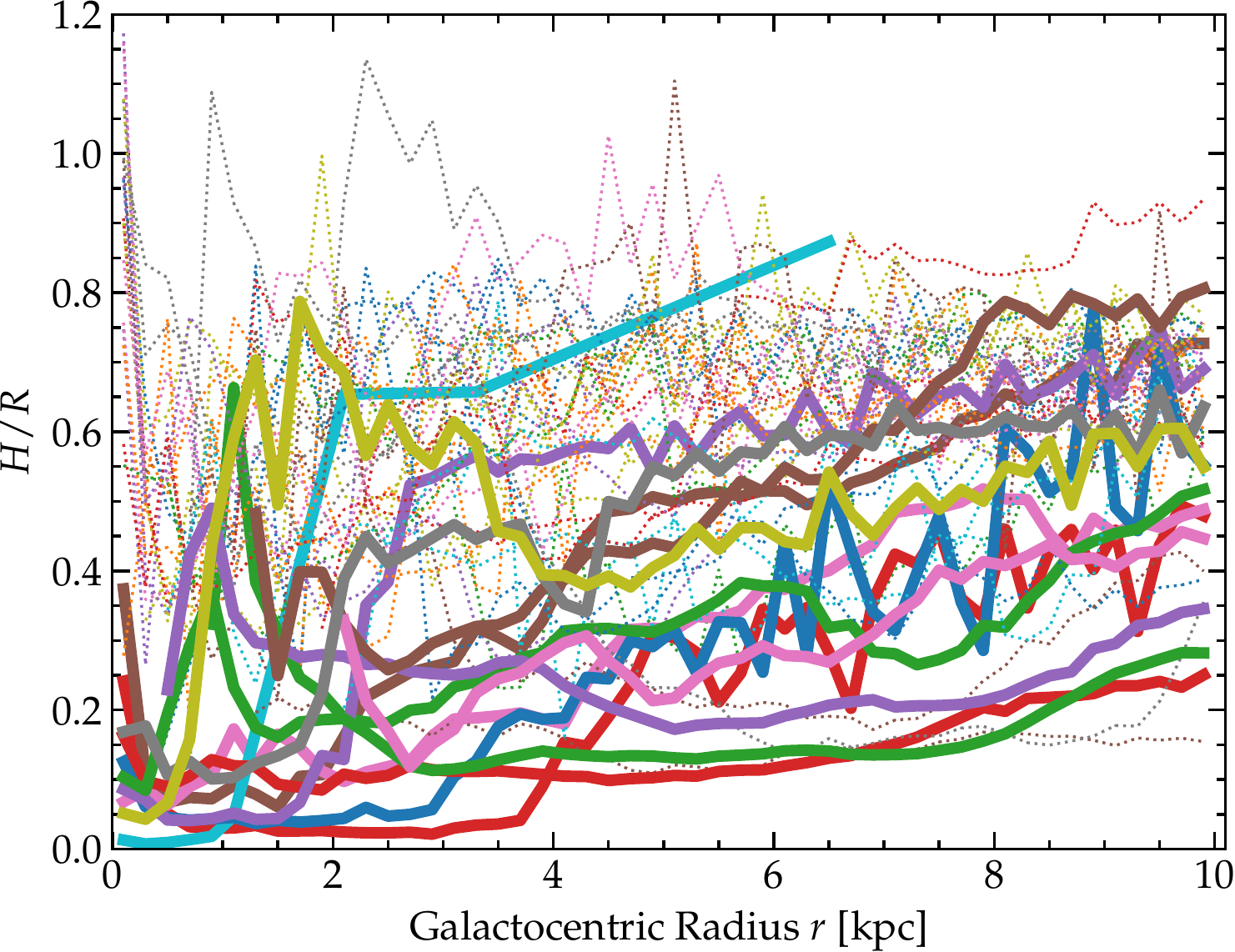} \\
	\includegraphics[width=0.49\textwidth]{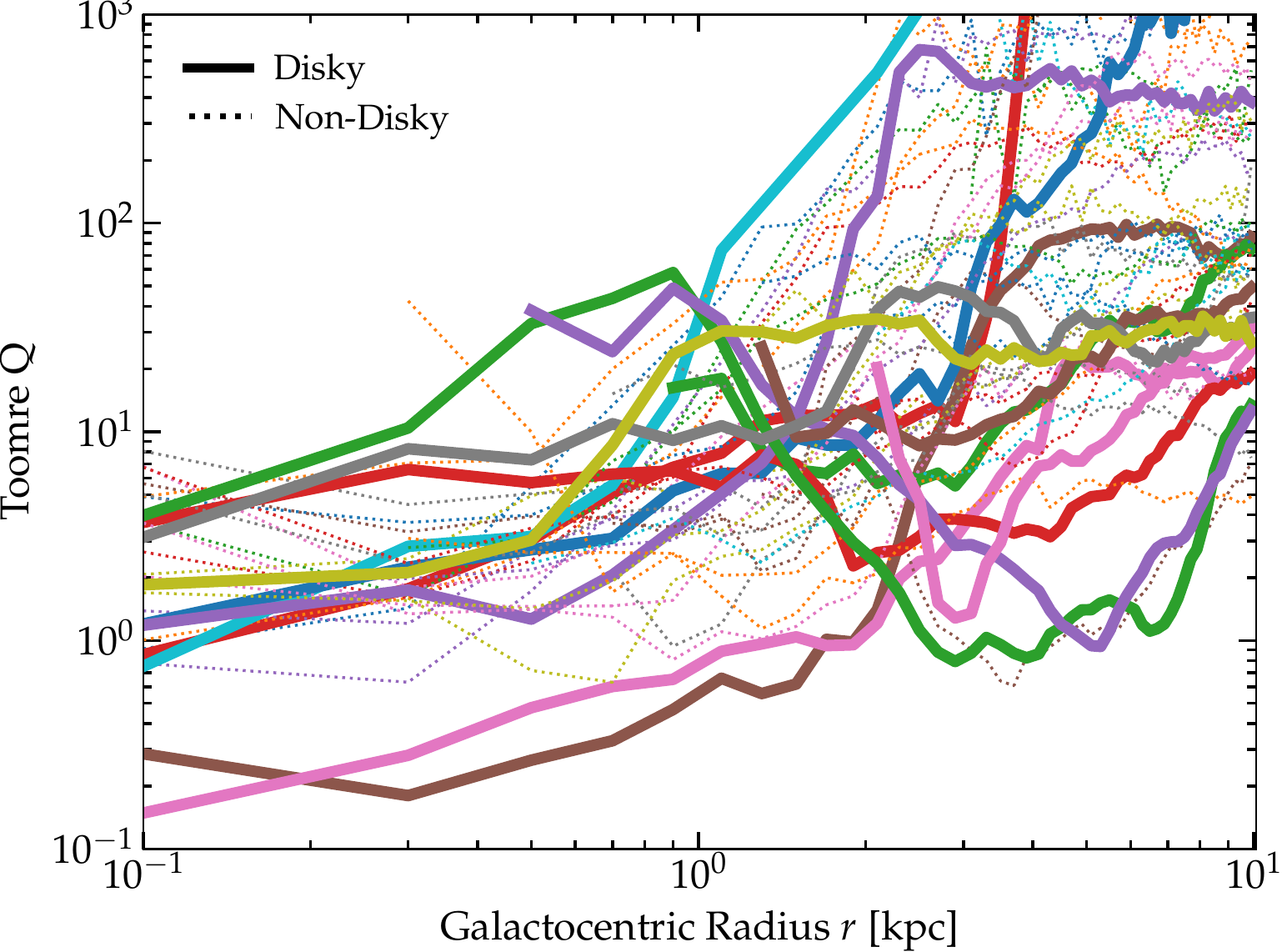}
	\includegraphics[width=0.49\textwidth]{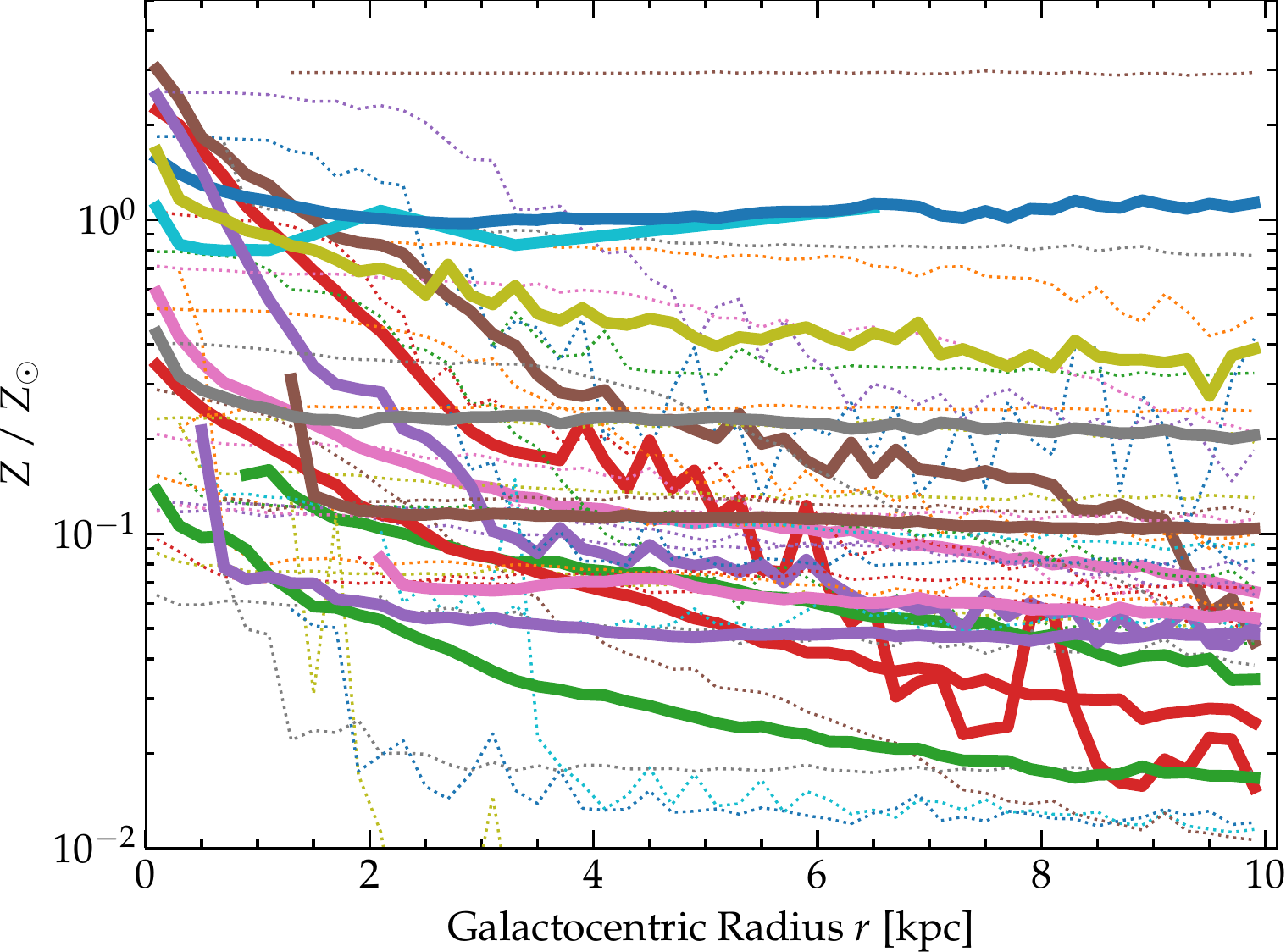}
	\vspace{-0.1cm}
	\caption{Additional profiles of the disky ({\em thick}) versus non-disky ({\em thin}) systems. 
	{\em Top Left:} Vertical turbulent velocity dispersion $\sigma_{z}/V_{\rm c}$ (as Fig.~\ref{fig:Vrot}).
	{\em Top Right:} Scale height $H/R$ in annuli.
	{\em Bottom Left:} Toomre $Q$ parameter (gas-only $Q=\sigma_{\rm gas}\,\kappa/\pi\,G\,\Sigma_{\rm gas}$; accounting for a two-component gas+stellar disk gives similar behavior but somewhat lower $Q$).
	{\em Bottom Right:} Mean gas-phase metallicity.
	The disky systems tend to be kinematically colder and thinner as expected, and have $Q\sim 1$ very broadly around the effective radii of the {\em star-forming} disk ($\sim 1$\,kpc) -- and much larger at larger $r$. There is no trend of diskiness in $Q$ or in metallicity (which, like our $\eta_{\rm cool}$ experiments, influences the cooling rate onto the disks).
	\label{fig:other.disky.profiles}\vspace{-0.3cm}}
\end{figure*}

\section{Centrally-Concentrated Mass Profiles Promote Disk Formation}
\label{sec:diskform}

As noted above, what appears robust in our study is that more centrally-concentrated mass profiles promote disk formation, while more extended and deeper potentials promote smooth SF. Here we further test these hypotheses and discuss the physical mechanisms by which they operate.

\subsection{Validation In Idealized Tests and Additional FIRE Simulations}

First, we will attempt to further validate the hypothesis that a centrally-concentrated mass profiles promotes disk formation.

\subsubsection{Our Default Idealized Tests}

Fig.~\ref{fig:Vc.vs.disk} plots the circular velocity profiles of all our simulations, coded by whether or not they produce a disk. We can immediately see that all the models which are intentionally centrally-concentrated by design (e.g.\ $M_{0} \gtrsim 1$ or $V_{0} \gtrsim 60$) produce disks, but also every other model which produces disks (even if it did not involve direct modifications to the potential or involved one of the less concentrated mass profile models) {\em non-linearly} produces a centrally-concentrated $V_{\rm c}$ profile. Usually, this occurs as described in \S~\ref{sec:sffb.fx.weak} above -- via the deeper potential or weaker feedback leading non-linearly to substantial star formation and adiabatic concentration of the dark matter halo. 

Fig.~\ref{fig:weak.fb} validates this in more detail: as described above, we see explicitly that for the simulations which are not explicitly initialized with a centrally-concentrated mass profile, it is only when the potential becomes centrally-concentrated to the same level as our $V_{0} \gtrsim 60$ simulations that a disk actually begins to form.

\subsubsection{Comparison to the FIRE Simulation Suite (and Other Simulations)}

In Fig.~\ref{fig:Vc.fire.disks} we test this further by considering the entire suite of default FIRE-2 simulations presented in \citet{hopkins:cr.mhd.fire2}. Specifically, these simulations use the ``default'' physics here, but now consider a large ensemble of halos. For each zoom-in volume, for simplicity, we only consider the best-resolved (most massive) galaxy in the high-resolution zoom-in region at $z=0$. We plot the circular velocity curve around each galaxy center as in \citet{hopkins:fire2.methods}, and label each galaxy by whether it does or does not have a gaseous disk. For this labeling we use the specific quantitative criterion defined in \citet{elbadry:fire.morph.momentum}, as this gives a label to all galaxies (including dwarfs) and is specifically targeted at gaseous disks, but as discussed above the classification is fairly insensitive to the quantitative scheme used for labeling. 

We see in Fig.~\ref{fig:Vc.fire.disks} that indeed the {\em shape} of the potential is a key indicator of whether or not the system has a (gaseous) disk. Specifically, the absolute value of the potential or, equivalently, $V_{\rm c}$ at some $r$, is not a particularly good indicator of whether or not a system will form a disk -- something we would not be able to test using our idealized simulations of {\bf m11a} alone. However, every simulation with a disk shows a relatively ``flat'' (shallow-slope) or centrally-rising rotation curve down to radii of order (or smaller than) the circularization radius or effective radius of the gaseous disk. Conversely, the simulations which never form a gaseous disk essentially all feature rotation curves which decline towards small radius compared to the circularization radius. We show a more quantitative example of this below, but simply note for now the clear separation (with no ``overlap'') in the plot here.

This is particularly interesting as it is able to predict outliers among the FIRE simulations. For example, the galaxy {\bf m11b} -- which we study in more detail below -- tends to ubiquitously form a well-ordered gaseous disk (across many FIRE-2 variant simulations), and is one of, if not the single, lowest-mass halo ($\sim 4\times10^{10}\,M_{\odot}$, similar in mass to {\bf m11a}) to do so in the ``default'' setup \citep[see discussion in][]{elbadry:fire.morph.momentum,elbadry:HI.obs.gal.kinematics,chan:fire.udgs,chan:2021.cosmic.ray.vertical.balance,ji:fire.cr.cgm,ji:20.virial.shocks.suppressed.cr.dominated.halos,2021arXiv210905034K,hafen:2022.criterion.for.thin.disk.related.to.halo.ang.mom}. As discussed in those papers, this galaxy does not have a virialized or ``hot'' halo (see also Figs.~5, 9, \&\ 10 in \citealt{hopkins:cr.mhd.fire2}), a high effective surface density or acceleration or escape velocity scale, a low gas fraction, a high halo spin, a particularly ``early'' or ``late'' halo formation history, or other criteria commonly invoked to consider whether a system might form a disk \citep[see][]{garrisonkimmel:fire.morphologies.vs.dm}. On the other hand, galaxy {\bf m11d} does not form a disk in most FIRE-2 realizations, despite having a halo and stellar mass ($M_{\rm halo} \sim 4\times 10^{11}\,M_{\odot}$, $M_{\ast} \sim 4\times10^{9}\,M_{\odot}$), formation time, spin, and other properties very similar to a number of other halos (e.g.\ {\bf m11h}, {\bf m11f}, {\bf m11g}) which all do form disks \citep[see discussion in][]{hafen:2022.criterion.for.thin.disk.related.to.halo.ang.mom}. But a cursory glance at their rotation curves shows that for {\bf m11d}, $V_{\rm c}$ is rising with $r$ out to a radius of $\sim 30\,$kpc (i.e.\ the central potential is very extended), well outside the gas circularization radius, while for {\bf m11b}, the rotation curve is decreasing with $r$ by $r\gtrsim 10$\,kpc which is still well interior to its gaseous disk size/circularization radius. 

Briefly, while a detailed analysis is outside the scope of our study here, it is worth noting that this proxy appears to apply reasonably well to other numerical simulations using different codes, numerical methods, treatments of star formation and the ISM, and feedback, such as those in \citet{christensen:2014.spiral.morph.in.dwarfs.vs.model.ism.sf}. There, it is clear that while the physics varied by the authors (including different cooling and star formation physics in fully cosmological simulations re-run from $z \gtrsim 100$) can non-linearly influence the rotation curve shapes (as expected), all of the runs which form a clear disk morphology have more concentrated $V_{\rm c}$ profiles than those which do not form disks. In fact, the separation between the two groups of rotation curve profiles in that paper, while a smaller sample, appears to fall neatly into the same distinction as shown in Fig.~\ref{fig:Vc.fire.disks}.

\subsubsection{Other Variables Tested}

Fig.~\ref{fig:other.disky.profiles} plots the radial profiles of some other properties of the disky systems, including their turbulent vertical velocity dispersion, scale height, Toomre $Q$ parameter of the gas alone, and gas-phase metallicity. Almost by definition, the disky systems tend to have somewhat smaller $H/R$ and $\sigma_{\rm turb}/V_{\rm c}$, but we clearly see that none of these properties is particularly {\em predictive} of the presence or absence of a disk. We have also compared a number of other properties, including: the orbital frequency $\Omega$, absolute value of $V_{\rm c}$ at different physical radii $r$, escape velocity $V_{\rm esc}$, absolute gas surface density $\Sigma_{\rm gas}$, total acceleration scale $V_{\rm c}^{2}/r$, gas thermal sound speed, gas fraction, the critical dynamical-gas-fraction criterion from \citet{orr:2021.bubble.breakout.model}, the optical depth of the disk ($\propto \Sigma_{\rm gas}\,Z$), and more. None of these variables appears to provide any additional useful discriminating power between ``disky'' and ``non-disky'' simulations.

\subsection{Tests at Different Mass Scales} 
\label{sec:disk.test.vs.mass}

\begin{figure}
	\hspace*{0.11cm}
	\includegraphics[width=0.45\columnwidth]{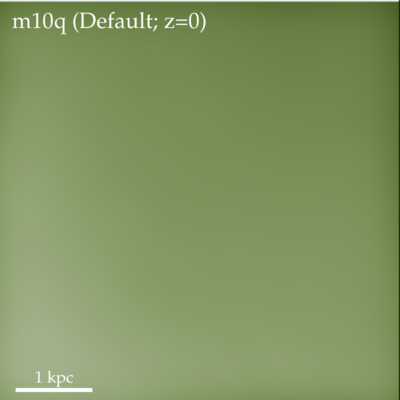}
	\includegraphics[width=0.45\columnwidth]{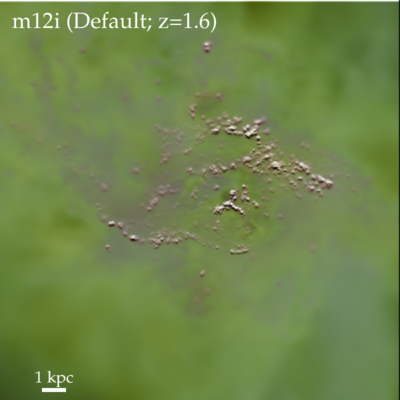}
	\hspace*{0.4cm}
	\includegraphics[width=0.45\columnwidth]{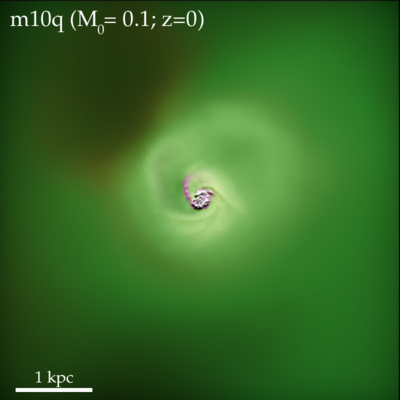}
	%\hspace*{0.14cm}
	\hspace*{0.24cm}
	\vspace{0.1cm}
	\includegraphics[width=0.45\columnwidth]{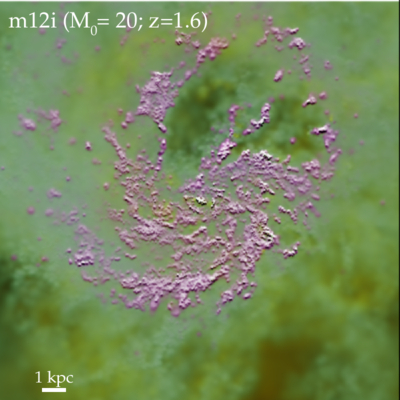}\\
	\includegraphics[width=0.51\columnwidth]{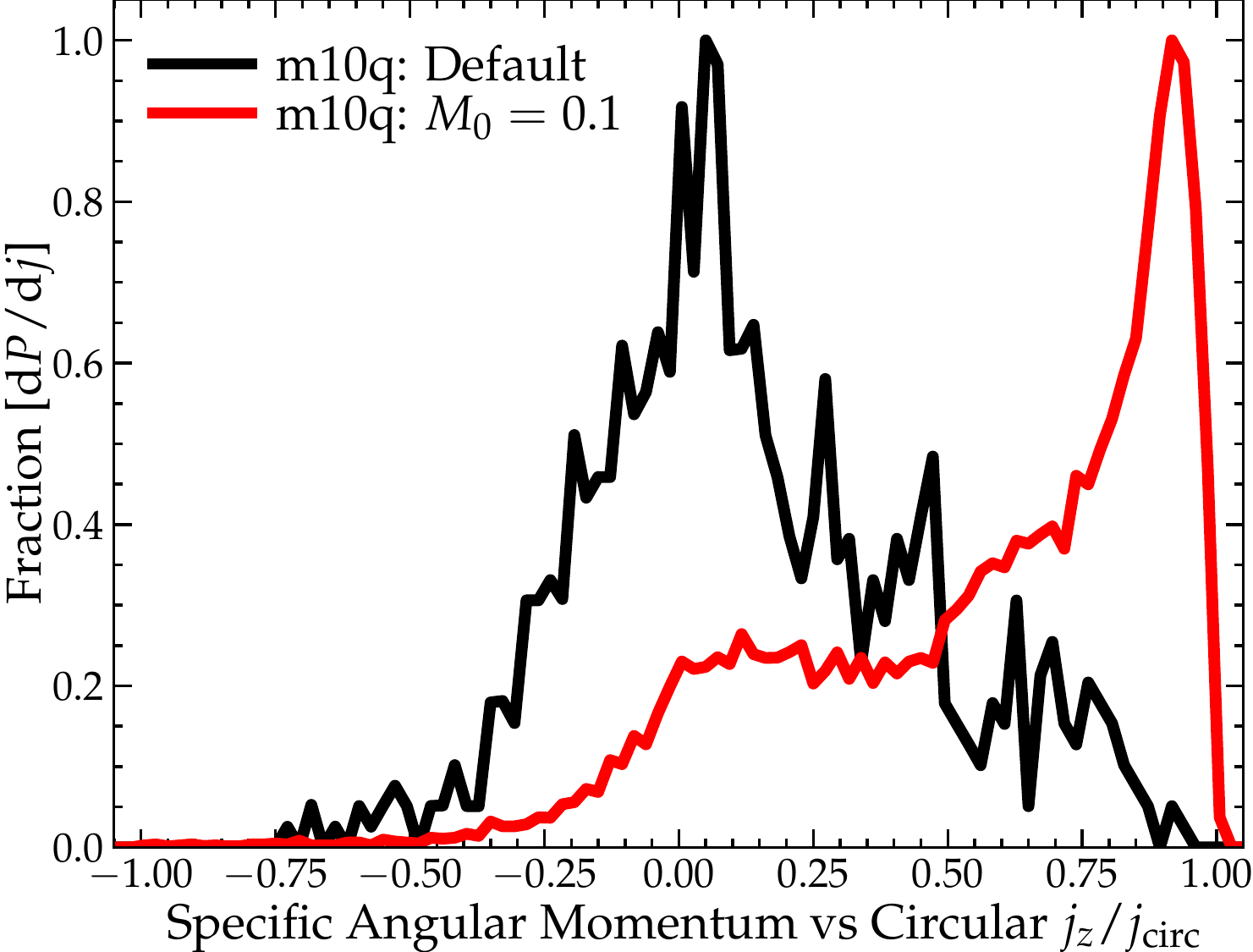}
	\hspace{-0.25cm}
	\includegraphics[width=0.51\columnwidth]{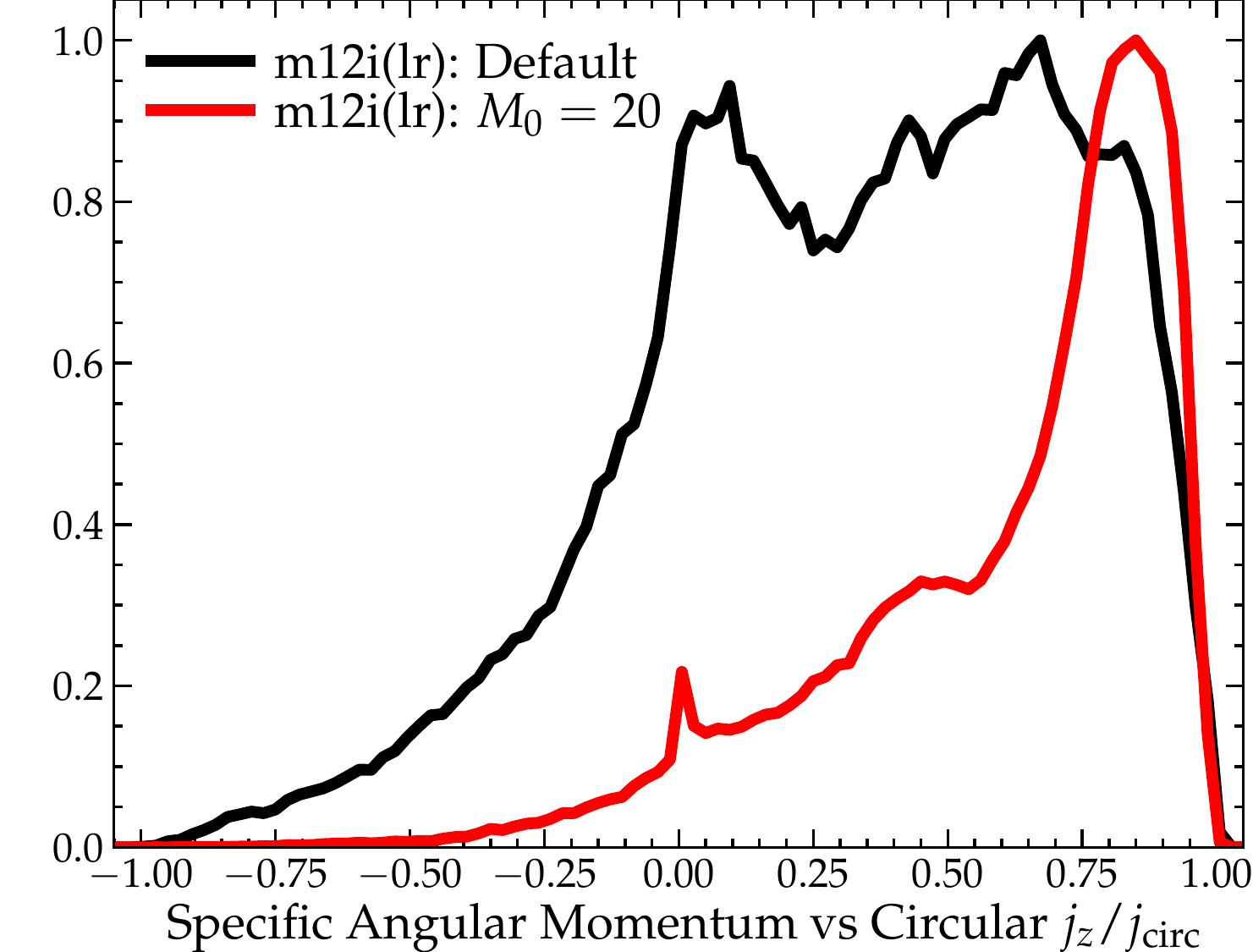}
	\vspace{-0.4cm}
	\caption{Gas images (face-on, as Fig.~\ref{fig:images}) and gas circularity distributions (as Fig.~\ref{fig:jc}) for simulation tests at different mass scales: re-starting {\bf m10q} ({\em left}; $z=0$ halo mass $=8\times10^{9}\,M_{\odot}$, stellar mass $=3\times10^{6}\,M_{\odot}$) at $z=1$ and {\bf m12i} ({\em right}; $z=0$ halo mass $=1\times10^{12}\,M_{\odot}$, stellar mass $=8\times10^{10}\,M_{\odot}$) at $z=2$. We compare the default runs ({\em top} images and black lines) to re-starts akin to our {\bf m11a} experiments, adding a centrally concentrated potential with $M_{0}=0.1$ (for {\bf m10q}) and $M_{0}=20$ (for {\bf m12i}) sufficient to make $V_{\rm c}$ centrally-rising. We show {\bf m10q} at $z=0$; the $M_{0}=0.1$ run forms a disk where no default run does so (there is only diffuse warm halo gas). We show {\bf m12i} at $z=1.6$, since the default run does form a thick disk by later times $z\sim1$ (and some coherent angular momentum is clearly visible already) -- but we clearly see a disk has already formed at this time in the $M_{0}=20$ run.
	\label{fig:disk.altmass.tests}}
\end{figure}

To further test our hypothesis, we repeat our idealized tests which led to disk formation in {\bf m11a} on two very different halos at different mass scales. Specifically, we select halos {\bf m10q} and {\bf m12i} from the FIRE suite \citep{hopkins:2013.fire}. The galaxy {\bf m10q} is an early-forming small classical dwarf with halo mass $<10^{10}\,M_{\odot}$ and stellar mass a couple times $10^{6}\,M_{\odot}$, run with mass resolution $250\,M_{\odot}$, which always forms a dwarf spheroidal without any disky structure or angular momentum whatsoever in ``default'' FIRE runs. We re-start it at $z=1$ as we did for {\bf m11a} adding a point-mass-like potential with $M_{0}=0.1$ (chosen as this is a similar fraction of the total mass inside the galaxy effective radius as our $M_{0}=1$ run for {\bf m11a}). {\bf m12i} is a late-forming Milky Way-mass galaxy at $z=0$, which has been studied extensively in previous FIRE simulation papers \citep{wetzel.2016:latte}. This galaxy eventually does form a disk in the ``default'' run, but this does not occur until relatively late times, $z \lesssim 1$ \citep{ma:2016.disk.structure}. We therefore restart the simulation at $z=2$,\footnote{At $z=2$ the stellar (halo) mass is $\sim 6\times10^{9}\,M_{\odot}$ ($\sim 3\times10^{11}\,M_{\odot}$), the potential is extended with $V_{\rm c}$ rising with $r$ out to $\sim 25\,$kpc.} before a disk forms in the ``default'' run, and run it to $z=1$, here adding a potential term with $M_{0}=20$. 

Fig.~\ref{fig:disk.altmass.tests} shows the visual morphologies of the runs and their angular momentum content: we clearly see that the addition of this centrally-concentrated mass profile promotes disk formation in both cases. For {\bf m10q}, at the time of restart ($z=1$), this has formed a larger fraction of stars than {\bf m11a} (it forms relatively early), and it spends most of the time between $z=0-1$ with a low-density warm gas halo\footnote{In Fig.~\ref{fig:disk.altmass.tests}, the gas densities in the ``inner CGM'' (outside the central $\sim 1\,$kpc but inside a few kpc) in the default {\bf m10q} run are depleted (typical $\sim 3-7 \times 10^{-5}\,{\rm cm^{-3}}$, notably lower than in {\bf m11a} in Fig.~\ref{fig:images}), while for the run with the centrally-concentrated potential they range from $\sim 10^{-4} -10^{-2}$ (with most of the mass in this phase around $\sim 0.3-1\times 10^{-2}\,{\rm cm^{-3}}$, so the pressure is order-of-magnitude similar to the virial pressure). For {\bf m12q}, the densities and pressures are correspondingly larger as expected.} occasionally cooling to form a burst of stars, so the disk is relatively small in size and mass (limited by the gas supply), but obvious visually. The SF remains extremely bursty to $z=0$. For {\bf m12i}, we see that even though the default run is beginning to form some coherent angular momentum by $z=1.6$, the time shown, as the central galaxy is rapidly becoming more massive so the potential is already becoming quite strongly concentrated, this is substantially accelerated by the added potential term.

\subsection{A Case Study of a Bursty-SF Thin Disk with an Extended Core and Rigid-Body Rotation: {\bf m11b}}
\label{sec:m11b}

\begin{figure*}
	\includegraphics[width=0.245\textwidth]{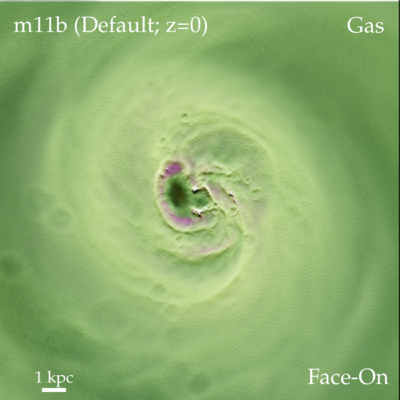}
	\includegraphics[width=0.245\textwidth]{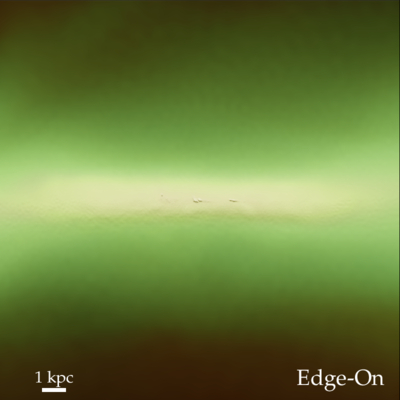}
	\includegraphics[width=0.245\textwidth]{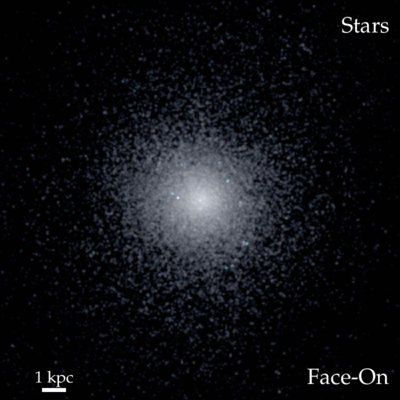}
	\includegraphics[width=0.245\textwidth]{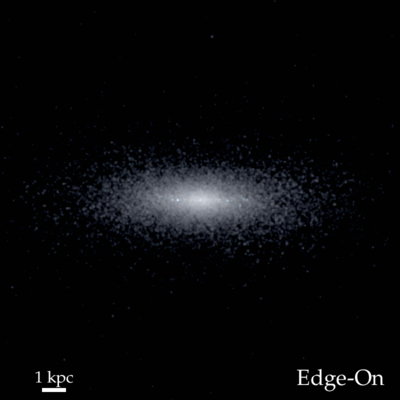} 
	\vspace{0.2cm}\\
	\includegraphics[width=0.33\textwidth]{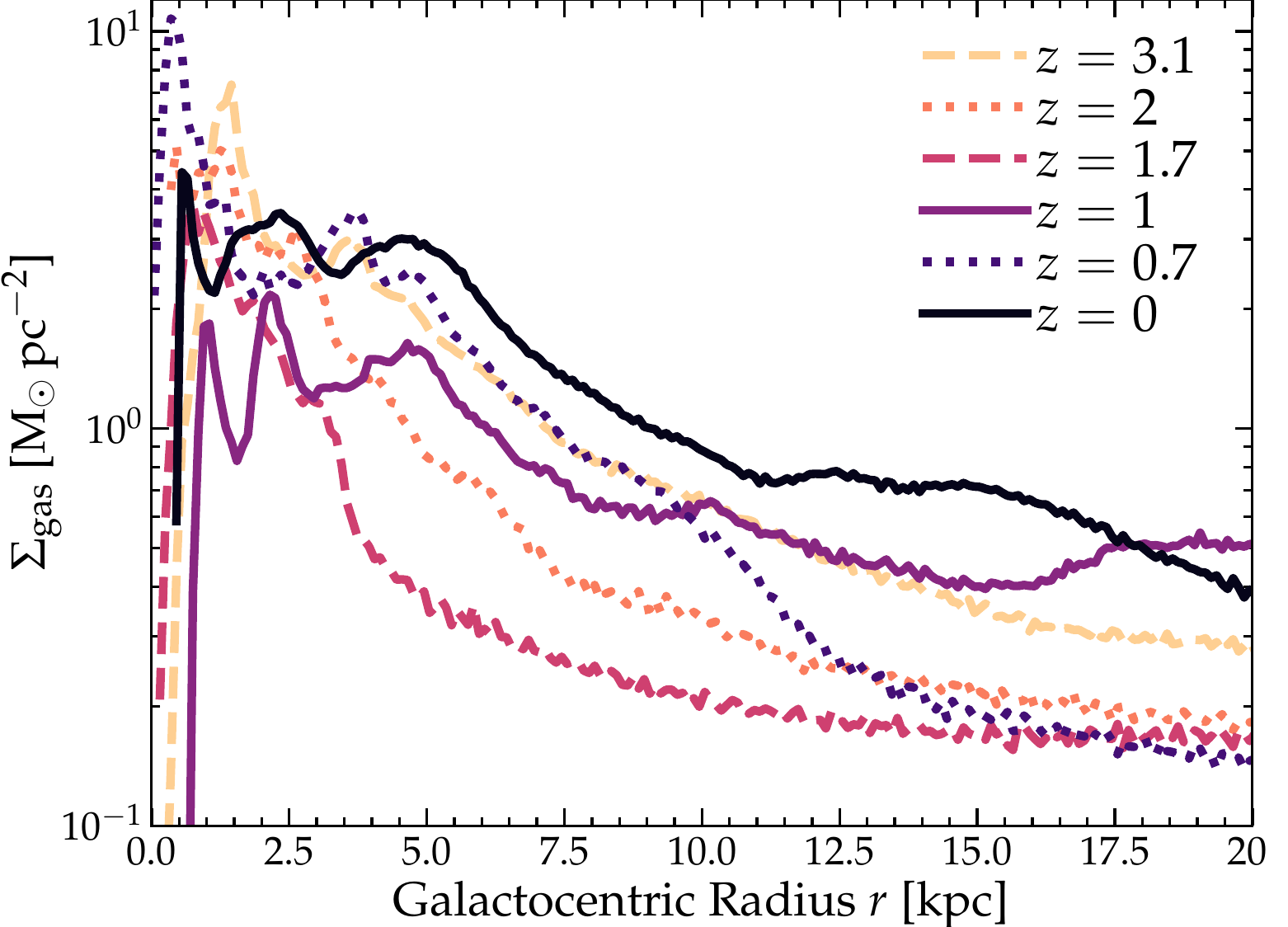} 
	\includegraphics[width=0.33\textwidth]{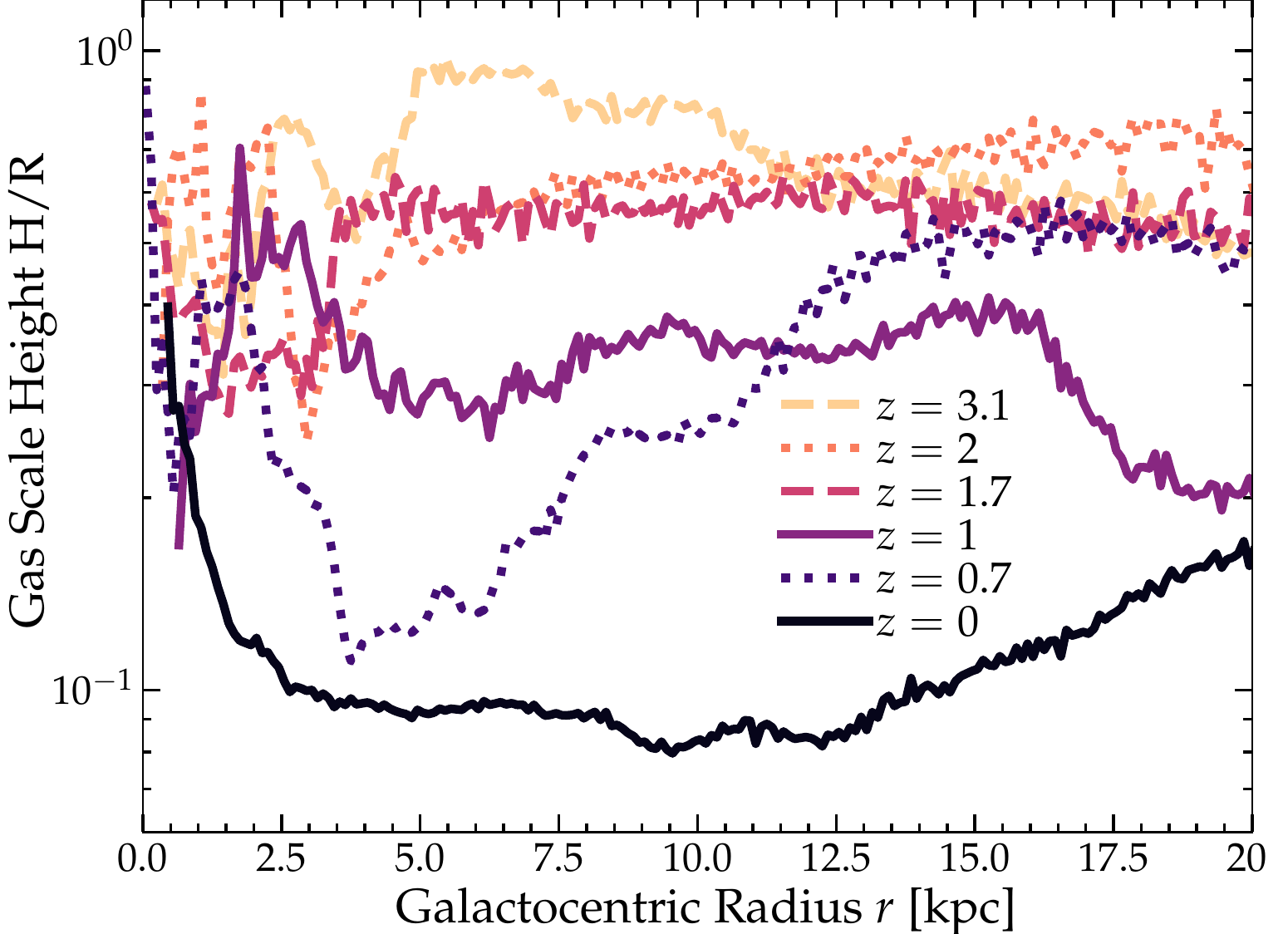}
	\includegraphics[width=0.33\textwidth]{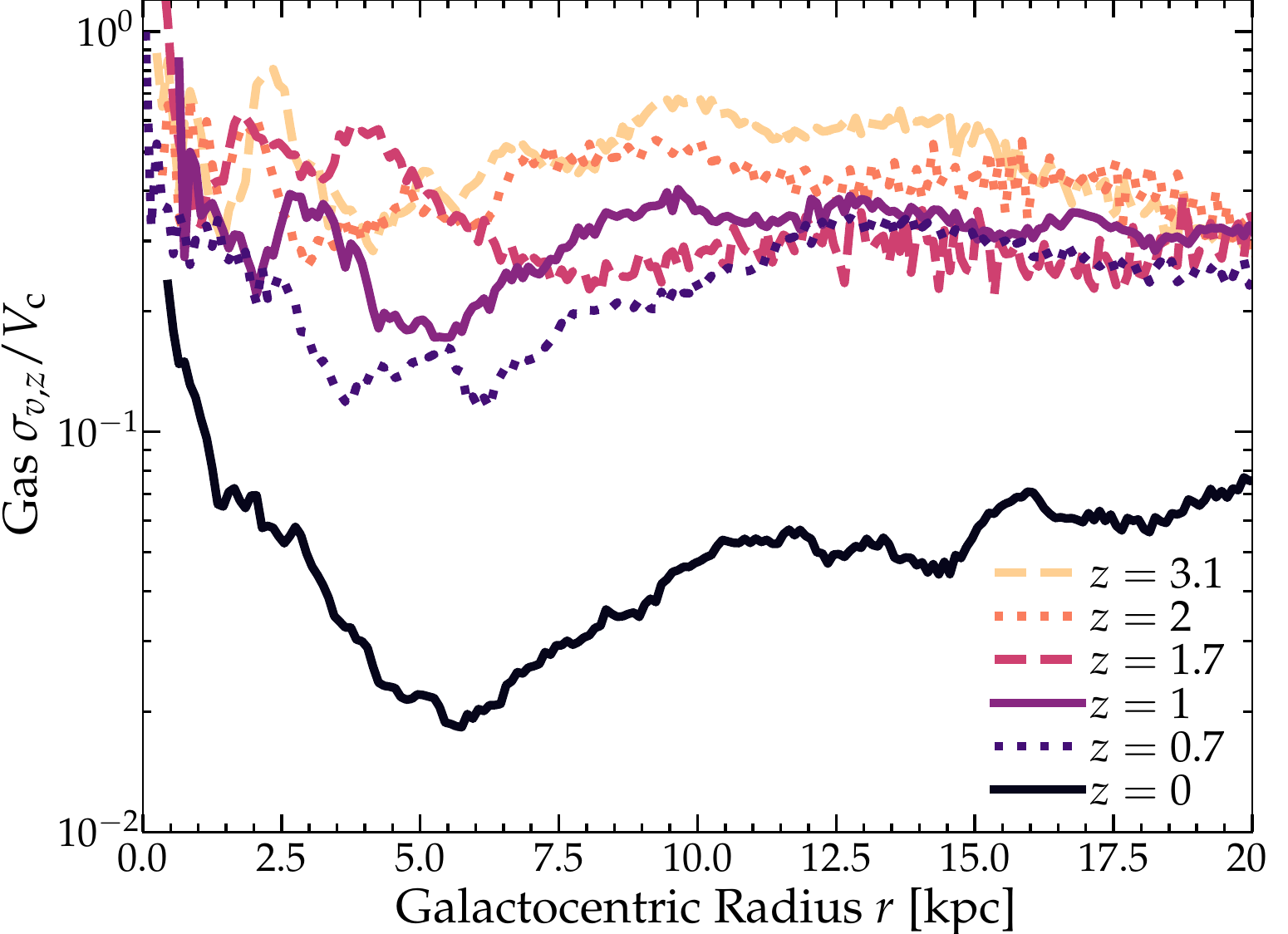} \\
	\includegraphics[width=0.33\textwidth]{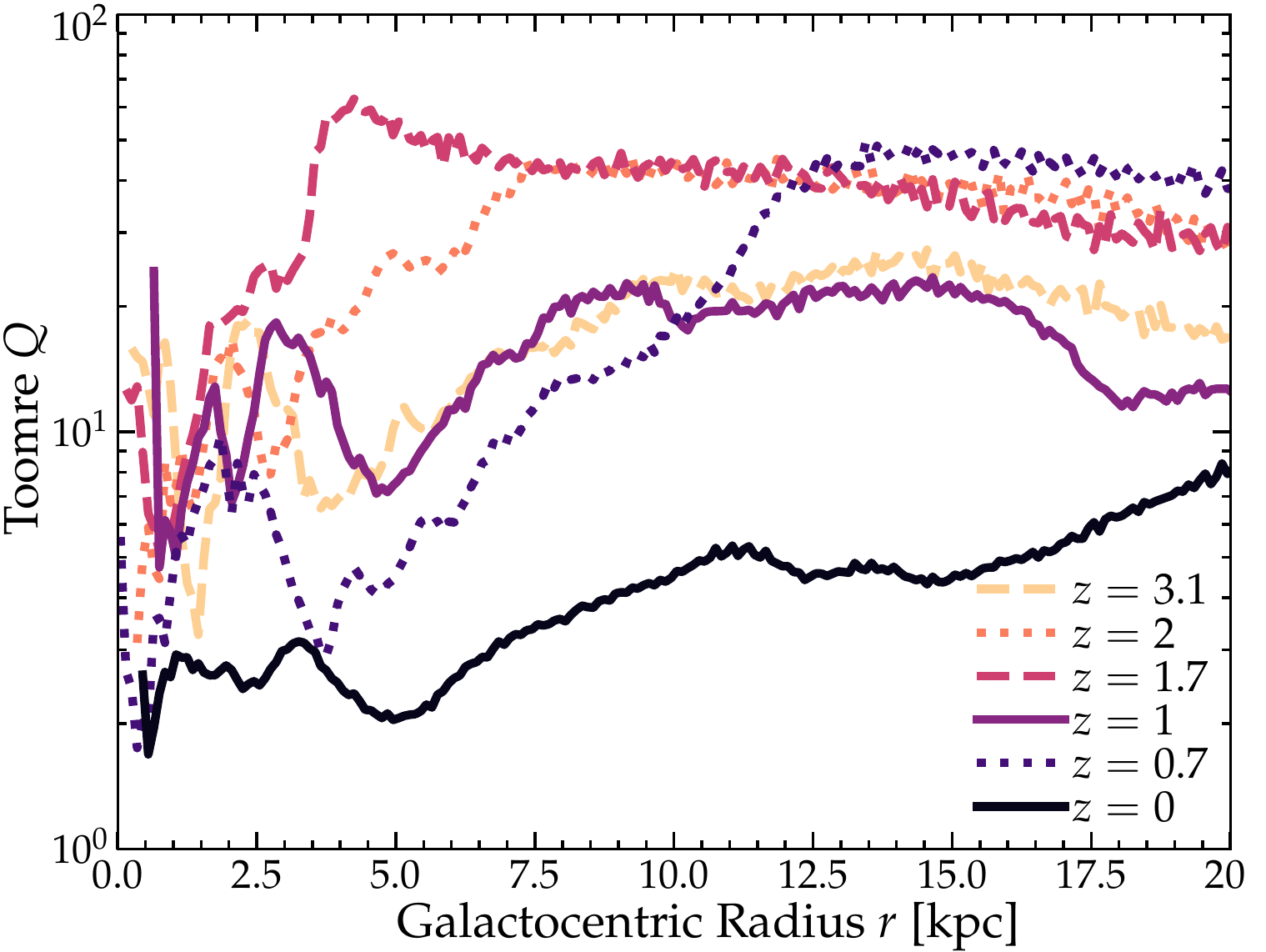} 
	\includegraphics[width=0.33\textwidth]{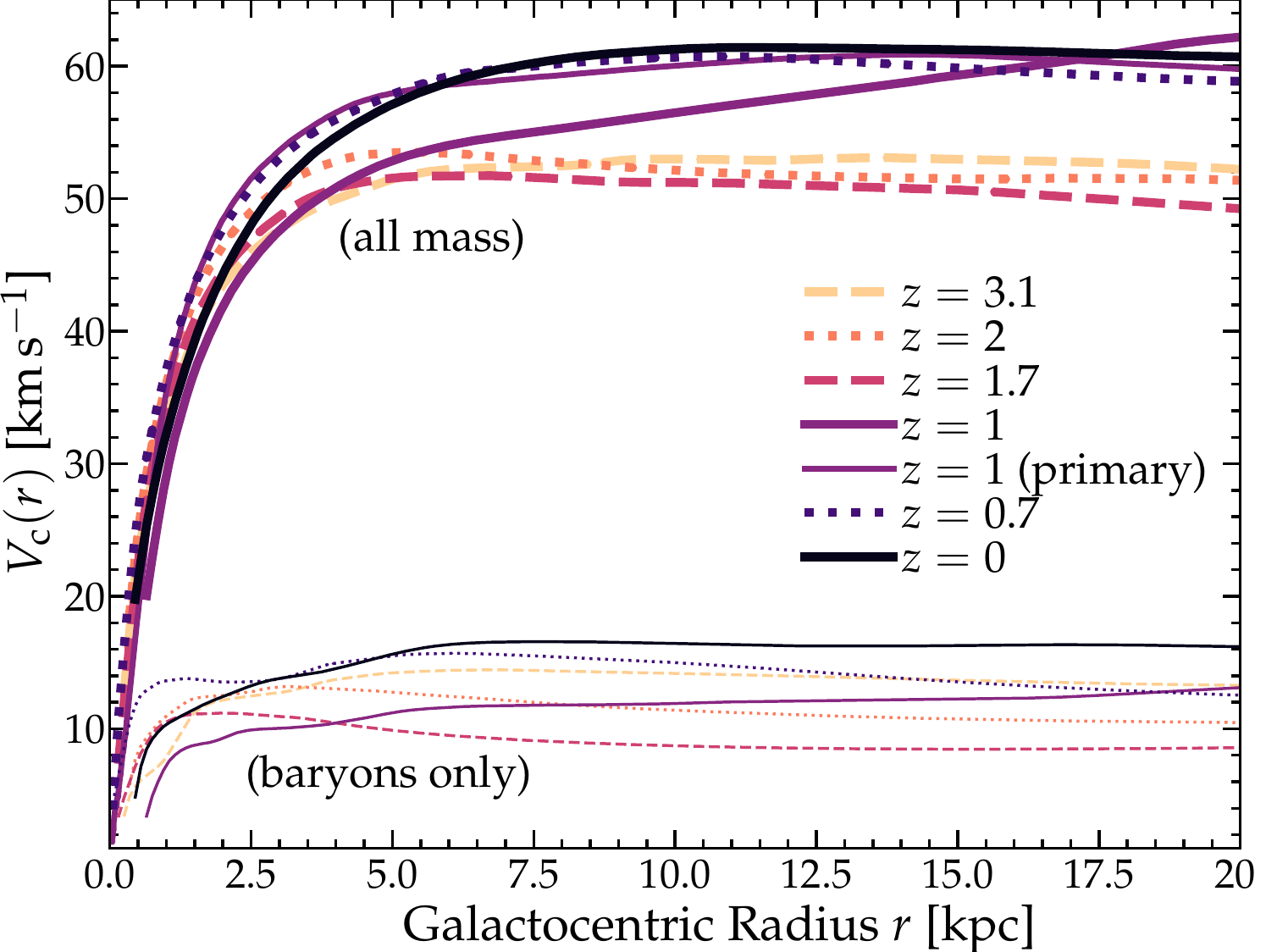} 
	\includegraphics[width=0.33\textwidth]{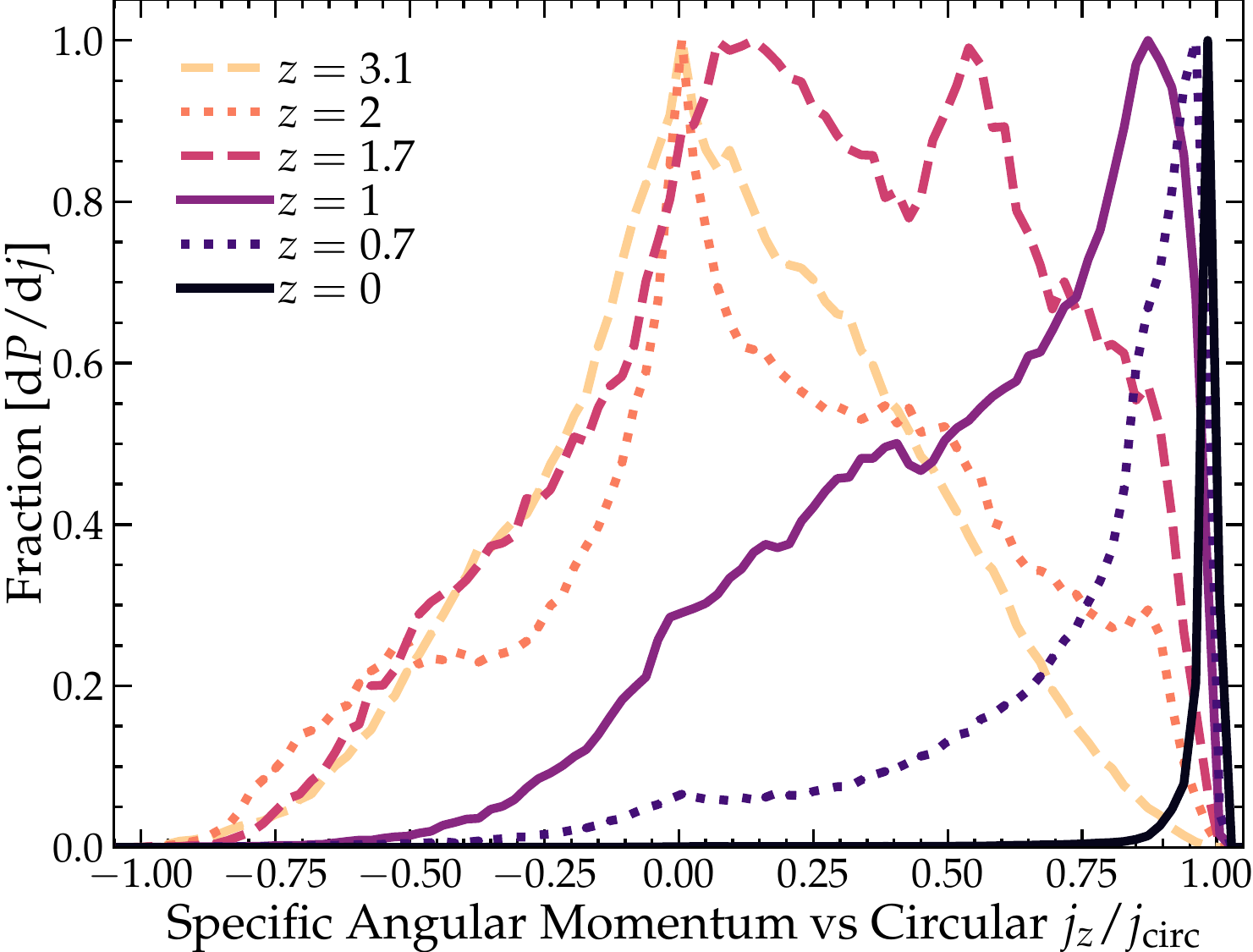} \\
	\includegraphics[width=0.33\textwidth]{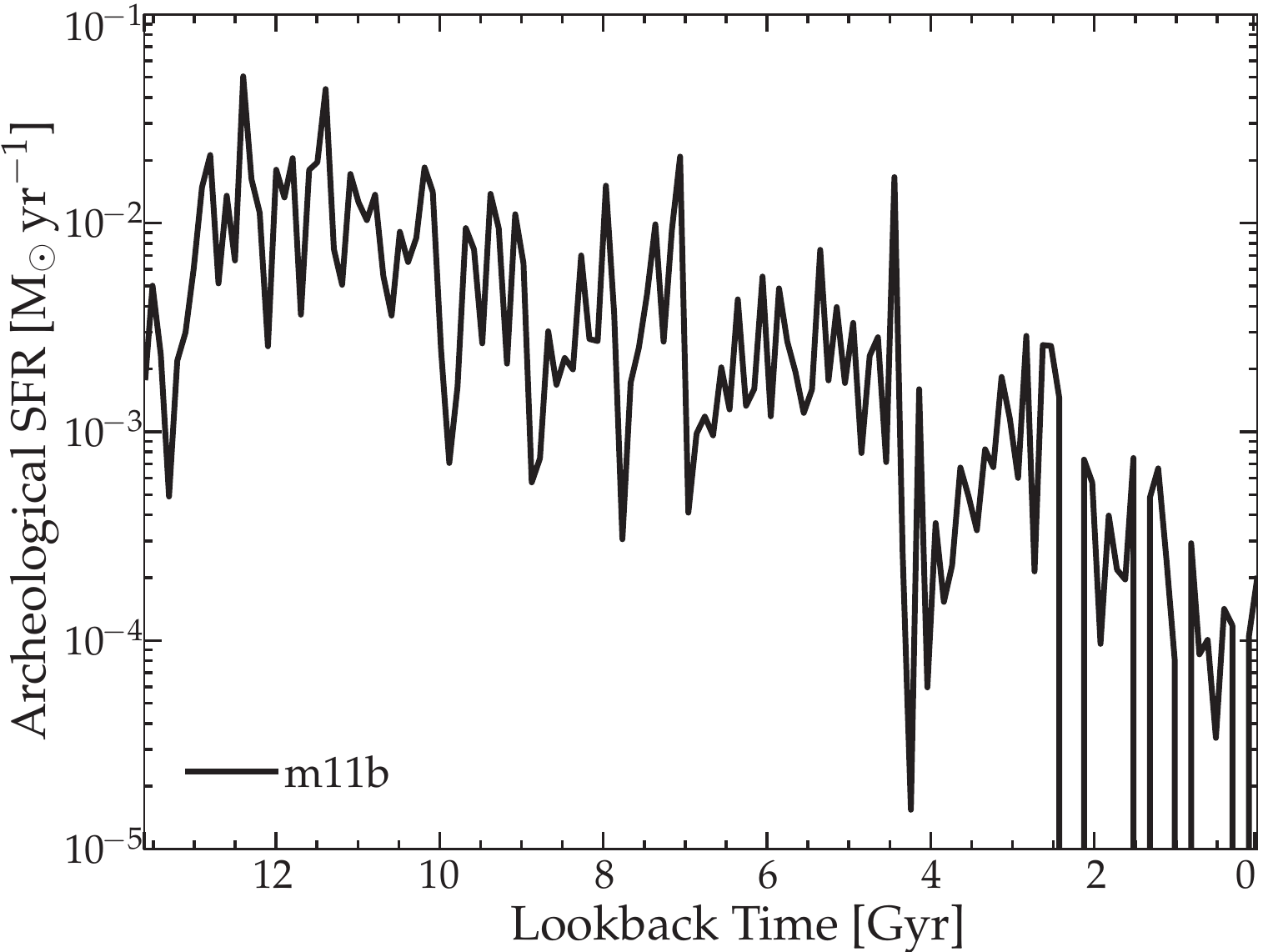}
	\vspace{-0.1cm}
	\caption{Case study of the FIRE-2 default {\bf m11b} (\S~\ref{sec:m11b}), which provides an analog to observed dwarf ($M_{\ast}=6\times10^{7}\,M_{\odot}$) LSB (optical surface brightness $\sim 24-25\,{\rm mag\,arcsec^{-2}}$) galaxies with rigid-body rotation curves and extended disks ($r_{\rm 90,\,HI} \sim 12\,{\rm kpc}$, $\sim 20$ times the optical disk scale-length). 
	{\em Top:} Images of gas and stars (as Fig.~\ref{fig:images}) face-on and edge-on at $z=0$. Surface brightness stretch is a factor of $\sim 100$ in starlight.
	{\em Middle:} Radial profiles at different redshifts of gas surface density $\Sigma_{\rm gas}$, vertical scale-height $H/R$, vertical velocity dispersion $\sigma_{z}/V_{\rm c}$, and Toomre $Q$ of the gas. 
	{\em Bottom:} Circular velocity curve (thick is total, thin is contribution from baryons alone; alternate line for $z=1$ shows just the rotation curve from the primary excluding the region with a merging secondary); angular momentum distribution; and archeological SFR. 
	This illustrates clearly that our ``centrally concentrated'' criterion (which {\bf m11b} meets) still admits large ``cores'' and rising $V_{\rm c}$ profiles, and that systems can become ``less concentrated'' after a disk is established. It also demonstrates an example of a thin gas+stellar disk with bursty SF.
	\label{fig:m11b.demo}\vspace{-0.3cm}}
\end{figure*}

We noted {\bf m11b} -- the lowest-mass system in our ``default'' FIRE suite which forms as disk -- as an exceptional case above, so we discuss it in slightly more detail with this theoretical understanding in mind. Fig.~\ref{fig:m11b.demo} illustrates several key properties of the simulation and its time evolution. This is a dwarf low surface-brightness (LSB) dark-matter dominated galaxy, and at $z=0$ there is a clear disk in both gas and stars. It has remarkably narrow gas $j/j_{\rm c}$ -- i.e.\ it is ``razor thin'' kinematically speaking (in gas). The morphological gas scale height $H/R$ is also extremely small ($\le 0.1$) for a dwarf at this mass scale, as thin or thinner than what are usually called ``superthin'' dwarf disks (e.g.\ compare UGC4278/IC2233, often called ``superthin'' with a de-convolved/intrinsic gas $H/R \approx 0.15$; \citealt{yu:HI.rotation.curve.analysis}). This thickness reflects the ``floor'' from the thermally-supported warm neutral medium (WNM; neutral gas at $\sim8000$\,K, which gives $H/R\approx 0.1$ for this $V_{\rm c}$), with sub-sonic turbulence. The SFH is quite bursty, demonstrating that not only can a disk exist with bursty SF (not just in our idealized tests, but even in default FIRE), but a thin/cold disk can have bursty SF as well. Perhaps most notably, we see $V_{\rm c}(r)$ at $z=0$ rises with $r$ in quasi rigid-body fashion (indicating a ``cored'' mass profile) out to $\sim 10\,$kpc, which at first glance might appear surprising given our concentration criteria, although we see it meets all of our quantitative criteria above (and others below).

Following the galaxy history provides some further insight. The halo initially forms somewhat concentrated, then gets even more ``baryonically concentrated'' from $z \gg 2$ to $z\sim 2$, from some early SF with low angular momentum (we see much of the SF is at early times). When the gas disk starts forming ($z\sim1.7$) the mass profile is more concentrated than it will be at $z=0$, with $V_{\rm c}$ peaking at $r_{\rm Vmax} \sim 4-5$\,kpc. Around $z=1$, a relatively large minor merger comes in: in our spherically-averaged $V_{\rm c}(r)$ as plotted this makes it appear briefly as if the potential is shallow, but this is because of the secondary contamination: if we exclude the merging satellite we see the mass profile of the primary is still relatively concentrated. This merger brings in a substantial gas mass with high impact parameter (circularization radius $\sim 15\,$kpc), which forms much of the extended disk. After this, by $z=0.7$ the secondary is disrupted and the potential is largely relaxed. By $z=0$, the combination of slow accretion onto the halo and continuing bursty SF in the inner disk has made the potential more shallow, pushing $r_{\rm Vmax}$ gradually out to $\sim 10$\,kpc. 

So we clearly see that the potential can continue to become more shallow after disk formation. This also means that while {\bf m11b} still meets our concentrated criterion at $z=0$ (see Figures above), it meets it ``moreso'' when the disk actually starts to form. In contrast, a system like {\bf m11e}, which is somewhat less concentrated at $z=0$ but has no disk, is actually significantly less concentrated  and had even less gas angular momentum at high redshifts, so it was ``further away'' from meeting the centrally-concentrated criterion when the gas was actually accreted (or more formally when the gas ``first attempts to circularize'' but this is more difficult to define rigorously and appears to be close in time to the accretion event). This also clearly emphasizes why our ``sufficiently concentrated'' criterion is a {\em relative} criterion, as it depends on concentration relative to the gas circularization radius.

After the accretion event, the gas surface density in {\bf m11b} steadily grows via accretion while $H/R$ and $\sigma/V_{\rm c}$ drop initially at small/intermediate radii, and then out to larger radii as $z\rightarrow 0$; so the disk is not maintaining $Q\sim1$. Instead, we see that outside of $\sim1\,$kpc, the disk generally has $Q\gg1$ and sufficiently low $\Sigma_{\rm gas}$ to avoid self-shielding to form dense molecular gas (i.e.\ $\Sigma_{\rm gas} \ll 10\,{\rm M_{\odot}\,pc^{-2}}\,(Z_{\odot}/Z)$) so it sits quasi-stably at WNM temperatures while the turbulence damps over $\sim\,$Gyr. Key therefore to the disk being simultaneously very thin while having highly bursty SF is that the SF is concentrated at $\lesssim 1-2\,$kpc, while the disk extends to much larger radii where it can be thermally supported with $H/R \ll 1$.

\subsection{Can More Extended Potentials Suppress Disk Formation?}

\begin{figure}
	\includegraphics[width=0.96\columnwidth]{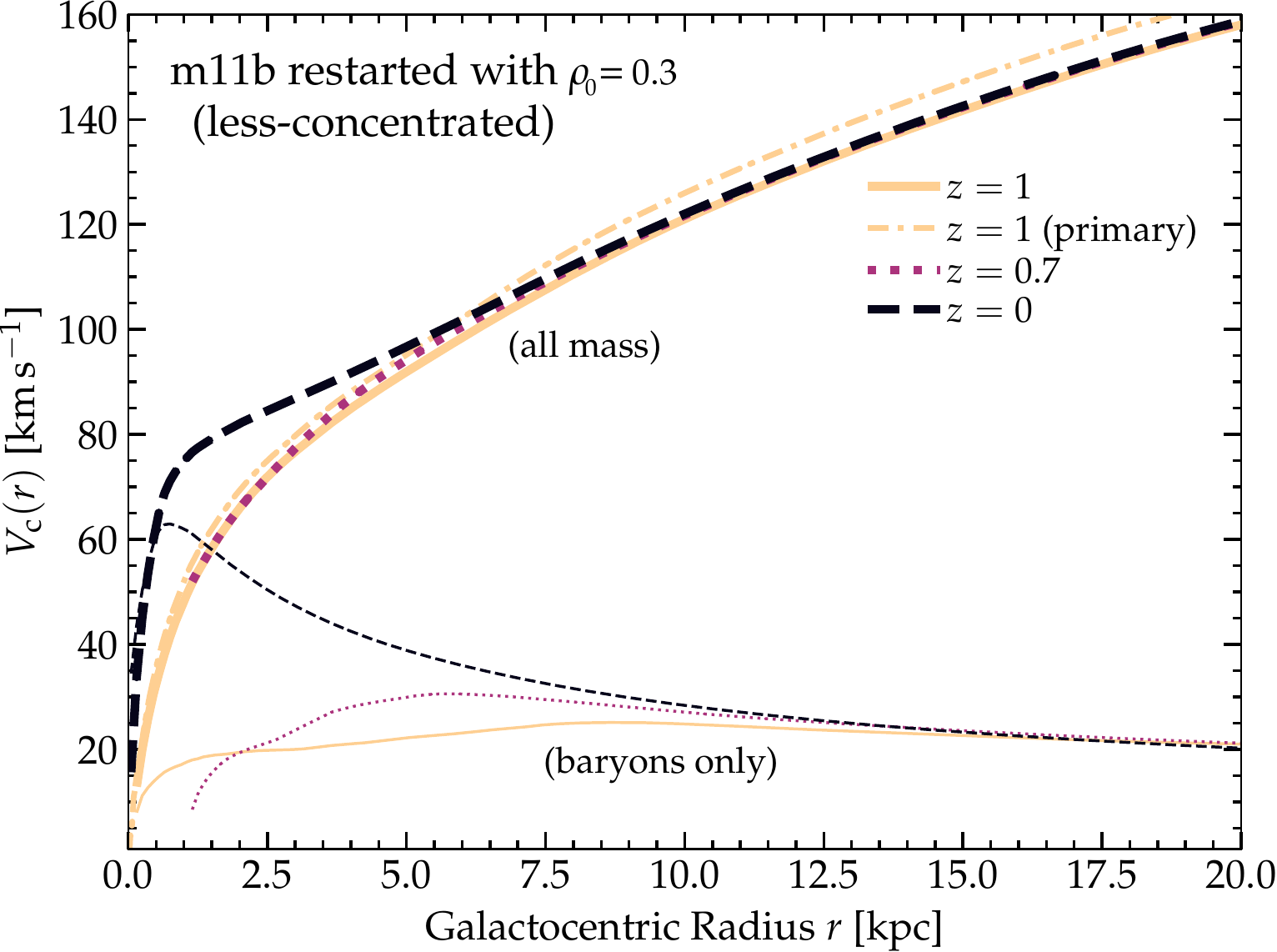}
	\hspace{0.5cm}\includegraphics[width=0.92\columnwidth]{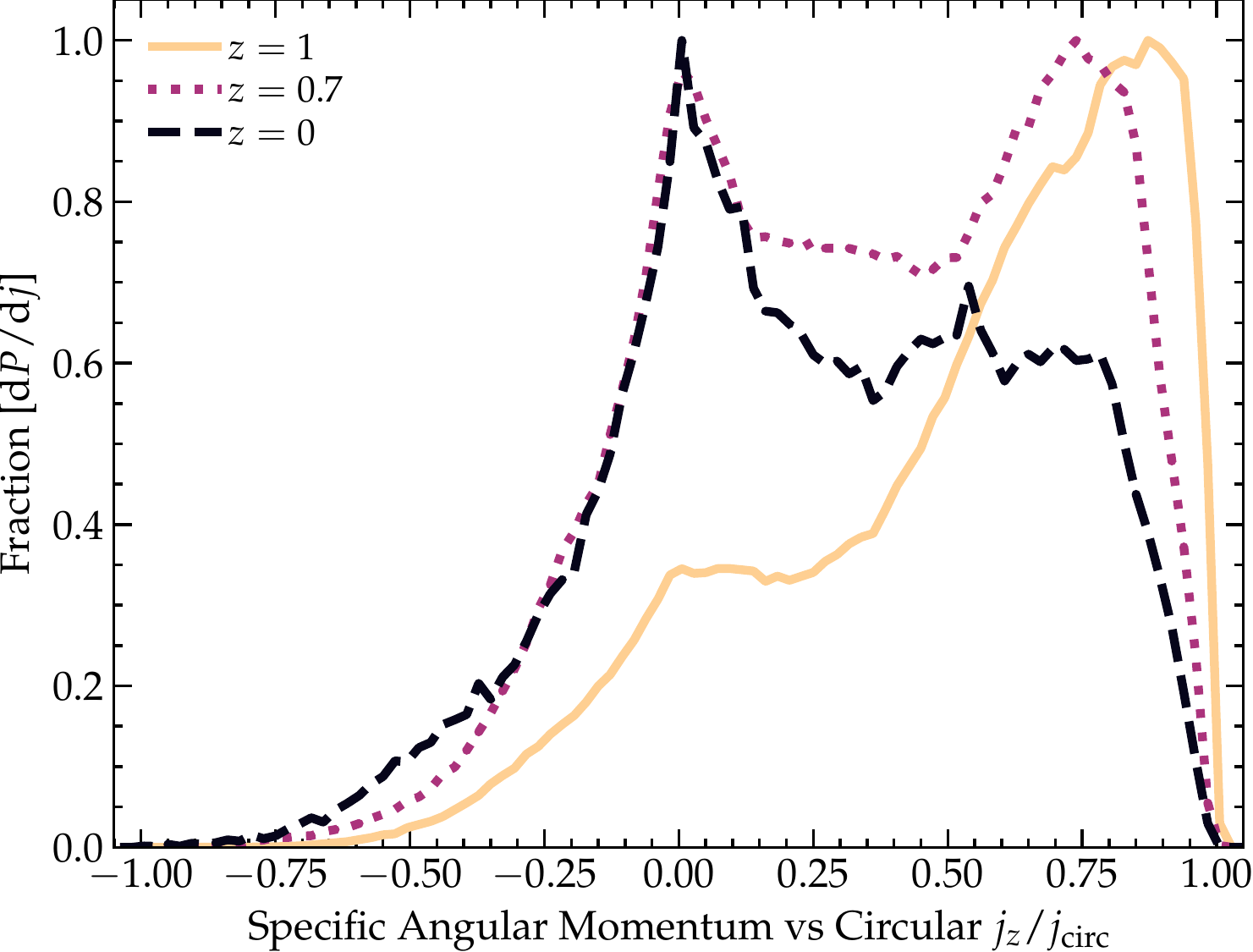}
	\vspace{-0.1cm}
	\caption{Circular velocity profile ({\em top}) and angular momentum distribution ({\em bottom}) of an alternative {\bf m11b} experiment where we restart at $z\approx 1.7$ and add an additional constant-$\rho$ potential with $\rho_{0}=0.3$, making the potential deeper but the effective mass profile less concentrated. While the merger still brings in some angular momentum in gas, it is less coherent to begin and quickly falls into sloshing/breathing modes that lead to essentially no disk at $z=0$. We see that more extended potentials can suppress disk formation where it would otherwise occur in our ``default'' run in Fig.~\ref{fig:m11b.demo}. 
	\label{fig:m11b.restart}}
\end{figure}

Another way to test our hypothesis is to ask if we can {\em prevent} disk formation in a system which would otherwise form a disk by making the mass profile {\em less} concentrated. A natural candidate for this is {\bf m11b}, discussed above: given its history, we re-start at $z=1.7$, before the merger event which builds the disk, and run to $z=0$, adding a constant-density potential with $\rho_{0} = 0.3$. Note that this is strictly additive, so the potential is necessarily deeper and the acceleration and $V_{\rm c}$ are everywhere larger, and the profile still somewhat concentrated from the pre-existing baryons, but less so than in the ``default'' run.

Fig.~\ref{fig:m11b.restart} shows the results. In this run, the merger still brings in material with significant angular momentum forming a transient disky structure, but even at this time, it is never as disky as the more-concentrated ``default'' run. Moreover, after the merger, the gas becomes trapped in a combination of ``sloshing'' and ``breathing'' modes, which ultimately destroy any coherent disk\footnote{Close to $z=0$ we see the galaxy attempt to transiently ``re-form'' a disk as it builds up a central baryonic mass concentration which produces a flat ``shelf'' in $V_{\rm c}$ at small radii, and some small gas disk with effective radius $\sim 1-2$\,kpc can be seen in a couple of snapshots, before it is disrupted by these modes.} as opposed to the default run, where the disk grows in mass and becomes thinner and kinematically more well-ordered.

\begin{figure*}
	\includegraphics[width=0.95\textwidth]{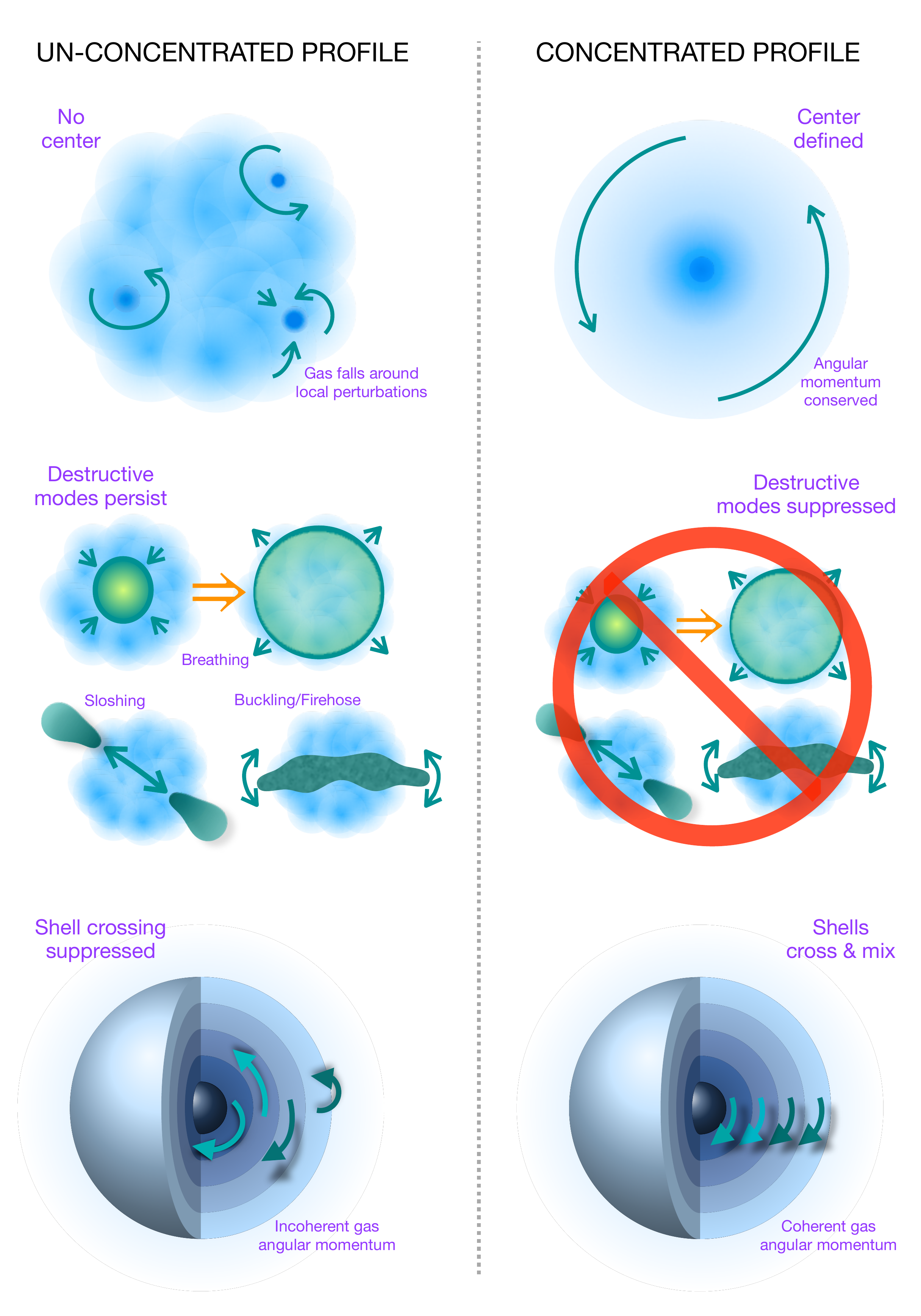}
	\vspace{-0.5cm}
	\caption{Cartoon illustrating the physics of initial gas disk formation, per \S~\ref{sec:diskform}, particularly some of the key reasons why sufficiently centrally-concentrated mass profiles promote disk formation (while un-concentrated profiles inhibit it) from \S~\ref{sec:physics.concentrated}. Green arrows represent trajectories of gas parcels in some mass distribution represented by the blue shading.
	\label{fig:cartoon.concentration}}
\end{figure*}

\begin{figure*}
	\includegraphics[width=0.33\textwidth]{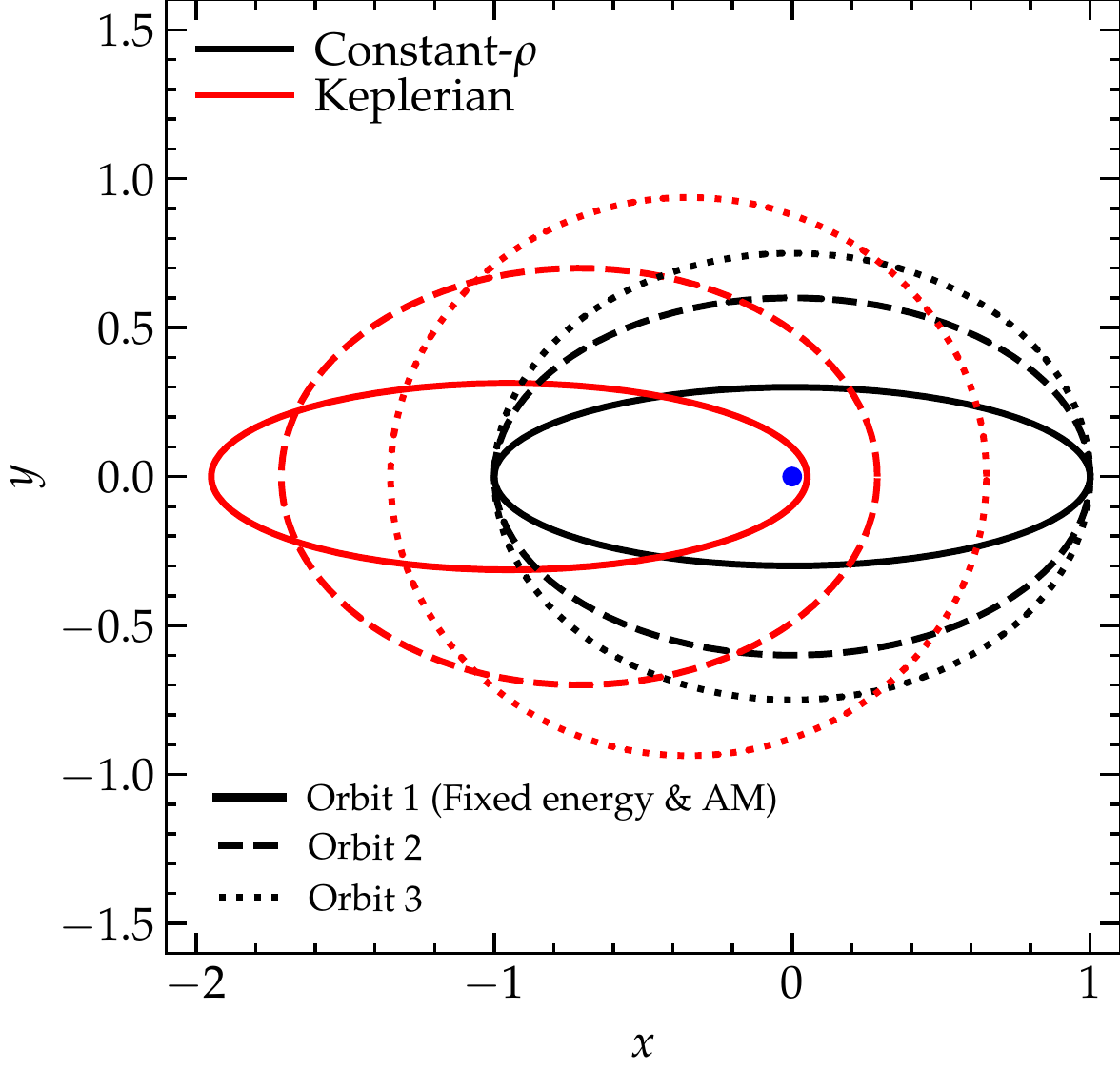}
	\includegraphics[width=0.33\textwidth]{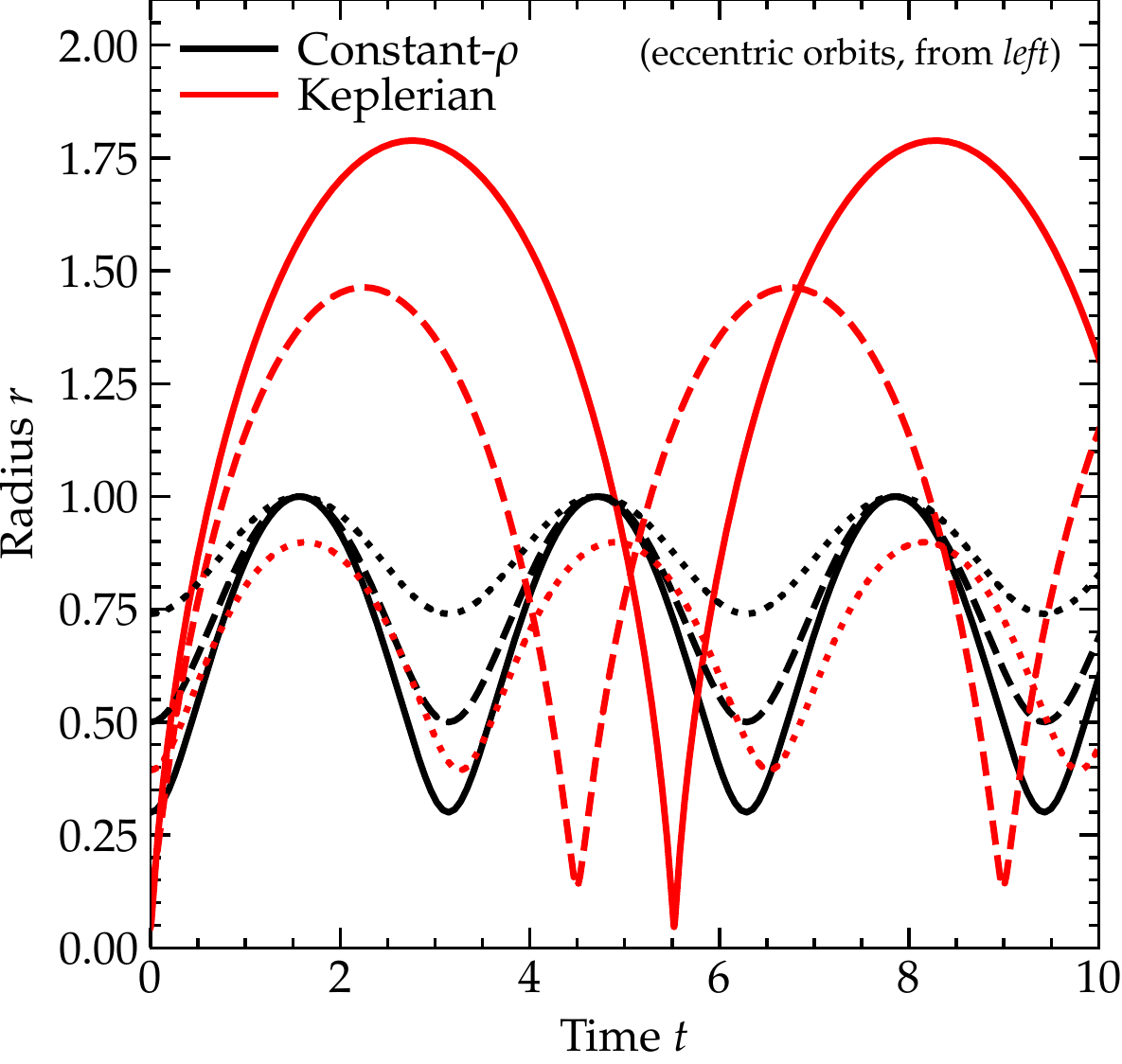}
	\includegraphics[width=0.315\textwidth]{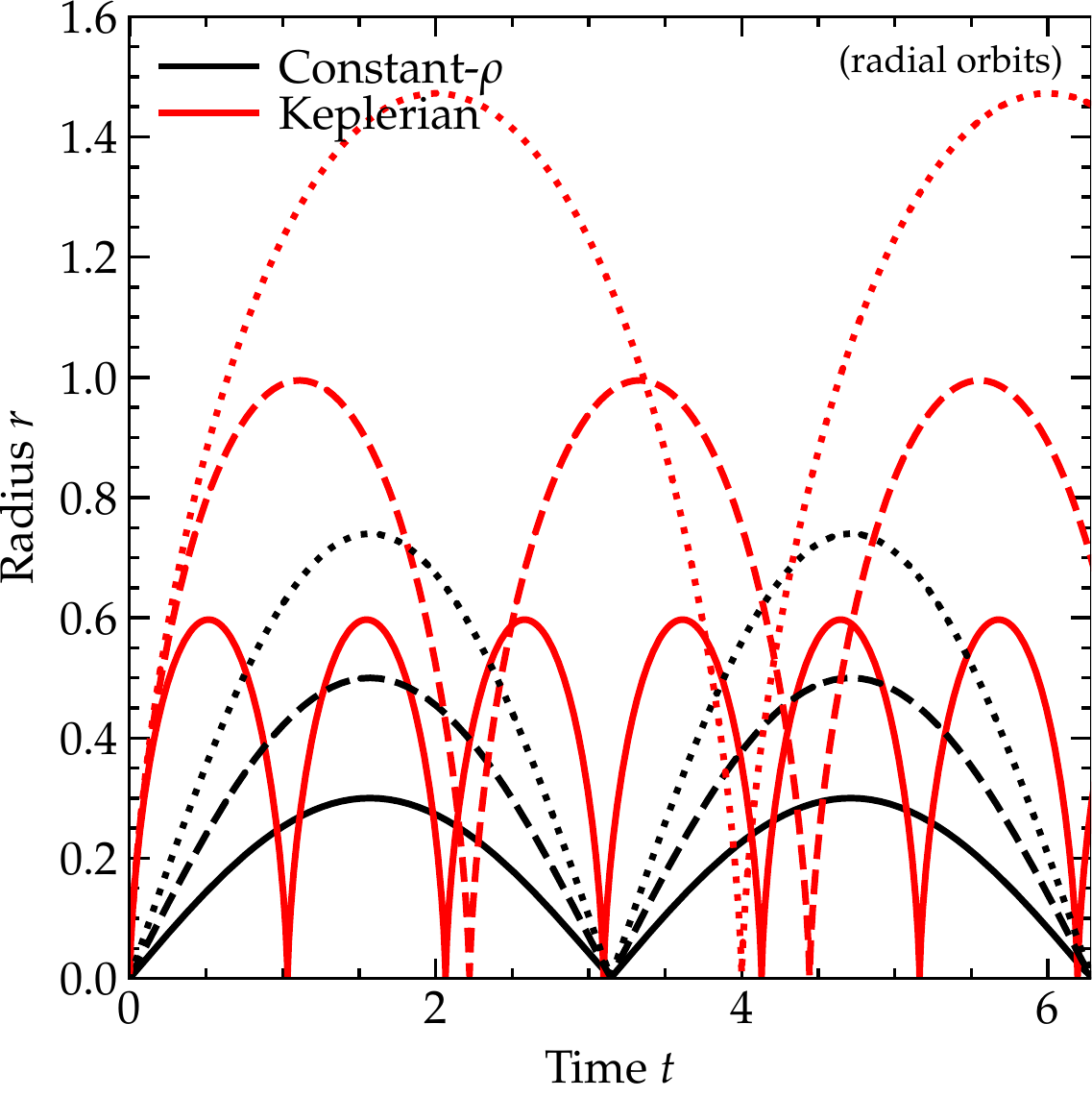}
	\vspace{-0.2cm}
	\caption{Illustration of differences in the types of orbits between centrally-concentrated (here, Keplerian) and extended (here constant-density) mass profiles (see \S~\ref{sec:physics.concentrated}). 
	{\em Left:} Example orbits (in the orbital plane): dot ({\em blue}) shows the potential center (assuming it can be defined). Lines show orbits. For a given pair (e.g.\ {\em solid} lines) the Keplerian and constant-$\rho$ lines correspond to orbits with the same energy and angular momentum assuming the same enclosed mass inside the semi-major axis of the orbit. Constant-$\rho$ potentials admit ``sloshing'' modes across the potential, and highly non-circular/eccentric orbits/streams can co-exist without any orbit crossing. 
	{\em Middle:} Behavior of the same orbits in time. The Keplerian orbits intersect in time, the constant-$\rho$ orbits do not.
	{\em Right:} As {\em middle} but for purely-radial orbits with different energies or apocentric radii. The constant-$\rho$ potential supports coherent global ``breathing modes'' where gas coherently expands and contracts without orbit-crossing. These modes are impossible in a Keplerian potential: the phases decohere and produce mixing, promoting disk formation.
	\label{fig:orbits}\vspace{-0.3cm}}
\end{figure*}

\subsection{Physics: Why Do Centrally-Concentrated Mass Profiles Support Disk Formation?}
\label{sec:physics.concentrated}

It is relatively easy to understand why a sufficiently centrally-concentrated mass profile would support disk formation: we outline several arguments here, which are illustrated heuristically in Fig.~\ref{fig:cartoon.concentration}.

First, the $M_{0}$ and $V_{0}$ potentials correspond to density profiles which rise steeply towards small radii (at least as steep as an isothermal sphere) -- i.e.\ there {\em exists} a well-defined ``center.'' This is very much non-trivial: for sufficiently high-redshift massive galaxies or dwarfs in the ``clumpy'' or bursty mode, there is typically no well-defined galaxy center \citep[see e.g.][and references therein]{ma:fire.reionization.epoch.galaxies.clumpiness.luminosities}. Famously, attempts to observationally define the center of galaxies like the SMC and other Local Group Irregulars have noted that various definitions of kinematic and light-weighted and mass-weighted ``centers'' do not agree \citep{deleo:2020.smc.no.defined.center,2022MNRAS.512.4798C}. This means that in such systems there is no well-defined axisymmetry in the first place, which means angular momentum is not conserved. Related to this, if the central potential is sufficiently ``flat,'' as in e.g.\ our constant-density ($\rho_{0}$) potentials, then any {\em local} maximum in the density (say, from a single gaseous clump or star cluster) is a local extremum and will act like the ``center'' for neighboring orbits, preventing any coherent rotation. Without a clear center, processes like dynamical friction cannot migrate nuclear ``clumps'' or clusters to merge into proto-bulges or nuclear star clusters as well \citep{tremaineweinberg:resonant.dynfric,ma:2021.seed.sink.inefficient.fire,banik:2021.core.stalling.dynfrict}.

Second, it has been known for decades that centrally-concentrated potentials stabilize disks. Specifically, strong non-axisymmetric modes such as spiral arms, eccentric/lopsided $m=1$ modes, and bars are stabilized by the presence of a central mass concentration (often in historical models coming from a dense bulge or star cluster or extremely massive SMBH; see e.g.\ \citealt{bardeen:1975.disk.instability.crit,toomre:spiral.structure.review,goldreichtremaine:spiral.excitement,athanassoula:bar.orbit.morphology,athanassoula:bar.vs.concentration,athanassoula:bar.slowdown,athanassoula:bar.vs.cmc,hernquistweinberg92,barneshernquist96,shen:bar.vs.cmc,bournaud:bar.lifetimes,holley:bar.halo.interaction,dubinski:bar.evol.sim.tests}). The strongly-nonlinear versions of these modes can destroy a nascent disk relatively efficiently by transporting angular momentum on dynamical times \citep{noguchi:merger.induced.bars.dissipationless,elmegreen:bars.vs.companions,shlosman:1993.clumpy.disk.instab.sims,weinberg:bar.halo.res.initial,hopkins:inflow.analytics,debattista:fire.bar.galaxy.m12m}. And it is essentially an identical criterion to the division we see here: systems with a sufficiently peaked central mass density profile (i.e.\ where $V_{\rm c}(r)$ is ``sufficiently shallow''/close-to-constant, or rising to small-$r$) are stabilized against these modes, while systems with rigid-body/core-like (i.e.\ constant-density, or too-rapidly-rising $V_{\rm c}(r)$) mass profiles towards large-$r$ are destabilized. Moreover, certain orbit families (such as box orbits) are not supported if the central potential is sufficiently concentrated (see references above and \citealt{roberts:gas.dynamics.in.bars,schwarz:disk-bar,athanassoula:bar.orbits,hasan:1990.bar.cmc.chaos}), moving the galaxy towards more angular-momentum supported orbits. 

Third, in a disk where $c_{s} \sim 10\,{\rm km\,s^{-1}}$ is not a negligible source of pressure support as noted above, $H/R \sim c_{s}/V_{\rm c}$ can remain below unity (i.e.\ the geometry can be disky) at small $r$ if $V_{\rm c}(r)$ is flat or rising to small $r$, but necessarily approaches spheroidal if $V_{\rm c}(r)$ drops rapidly at small $r$. It is similarly ``easier'' to maintain a disk with $Q\sim 1$ at small $r$.

Fourth, and perhaps most important, the global orbital structure of a centrally-concentrated mass profile is fundamentally different in a couple of crucial ways. For simplicity, let us contrast two idealized cases, which are illustrated in Fig.~\ref{fig:orbits}: a purely Keplerian potential (akin to our large-$M_{0}$ models), and a constant-density sphere potential (akin to our large-$\rho_{0}$ models). Both happen to feature closed, elliptical orbits, but there are major {\em qualitative} differences between the orbital behaviors which have major implications for disk formation. The constant-density potential is just that of a three-dimensional simple harmonic oscillator (SHO), with a universal constant orbital period $T = (3\pi/G\,\rho_{0})^{1/2}$. This means that certain orbit families are not only allowed but are indefinitely stable and can continue, in principle, forever \citep[see also][]{2014ApJ...793...46O}. For example, ``sloshing'' modes, where a ``blob'' of gas oscillates just as one would expect from a simple harmonic oscillator, from one side of the galaxy to the other, with apocentric distances equal on ``both sides'' (i.e.\ from $+x_{0}$ to $-x_{0}$). But even more problematic is the global ``breathing'' mode, where gas at {\em all} radii coherently moves in and out. This is again a natural and completely stable configuration in an SHO potential -- indeed, they are a fundamental feature of such potentials. As a result, these global breathing modes for e.g.\ adiabatic gas will simply continue forever in the absence of external perturbations or dissipation, without orbit crossing or mixing, if they are excited. These are exactly the modes which are identified in many studies as particularly inhibiting to disk formation and destructive of any pre-existing disks in many simulation studies \citep[see e.g.][]{teyssier:2013.cuspcore.outflow,elbadry.2015:core.transformation.stellar.kinematics.gradients.in.dwarfs,el.badry:jeans.modeling.dwarf.coherent.oscillations.biases.mass}. It is commonly stated that these oscillations are driven by feedback, but in fact they do not have to be. Once excited initially (which may come from a burst of star formation and feedback), if the potential resembles constant-density, they will continue stably. Moreover, they are stable both for adiabatic gas (the slow cooling limit) and for isothermal gas (rapid cooling limit) in such a potential. Worse, those studies above and \citet{2005MNRAS.356..107R,pontzen:2011.cusp.flattening.by.sne,governato:2012.dwarf.form.sims,zolotov:2012.baryon.fx.tidal.struct.dwarfs,shen:2014.seven.dwarfs,2017MNRAS.466L...1D,chan:fire.dwarf.cusps,chan:fire.udgs,2019MNRAS.488.2387B} have shown that these breathing modes actually drive the potential towards a ``cored'' (i.e.\ constant-density) mass profile in galaxy centers: meaning that they are {\em self-reinforcing}: the constant-density ($\rho_{0}$) potential becomes a sort of quasi-stable attractor solution.

On the other hand, in a Keplerian potential (or a constant-$V_{\rm c}$ potential or other more centrally-concentrated mass profiles), such orbit families simply do not exist. The ``sloshing'' mode becomes a standard eccentric orbit which (with many coherently excited neighbors) simply forms an eccentric disk. No analog of the global ``breathing'' mode exists\footnote{Formally there is a well-known ``vertical breathing mode'' of a thin disk in a spherical potential, but this arises precisely because the effective potential for small vertical oscillations $|\Delta z| \ll r$ in a thin disk of nearly-circular orbits in a dominant background spherical potential is again that of a harmonic oscillator \citep[see e.g.][]{tothostriker:disk.heating,quinn93.minor.mergers,sellwood:disk.heating,velazquezwhite:disk.heating,hopkins:disk.heating,moster:2010.thin.disk.heating.vs.gas}. These, by definition, require a disk and do not disrupt it, so are not central to our discussion, though they may play a role in the ``vertical disk settling'' process.} -- if gas is spread over a continuum of radii and one attempts to excite such a mode, differential rotation (more formally the increase of orbital period with apocentric radius, which only requires some centrally-rising mass density profile) means that orbit crossings, hence collisions and mixing, are impossible to avoid, even for purely radial orbits. In other words, the radial phases will necessarily de-synchronize on a single outer period, promoting mixing and re-distributing angular momentum between ``shells,'' a crucial part of building a disk \citep[see e.g.][and references therein]{hafen:2022.criterion.for.thin.disk.related.to.halo.ang.mom,gurvich:2022.disk.settling.fire}, and making coherent breathing modes impossible. This also means the action of such modes further ``flattening'' the potential is inhibited. So there is a sort of hysteresis between two quasi-stable solution branches: a centrally-concentrated mass profile which promotes orbit mixing and the formation of a disk with coherent angular momentum, and a constant-density profile which promotes global breathing modes.

Note that the cessation of these global breathing modes in a centrally-concentrated mass profile (which we clearly see in the disky simulations), given the fact that star formation is still bursty, also means that such modes are not strictly necessary for bursty star formation (even if they are often correlated, in practice). One can have bursty star formation without global inflow/outflow breathing modes.

\subsection{Towards a Simple Quantitative Criterion}
\label{sec:disky.quant}

\begin{figure*}
	\includegraphics[width=0.49\textwidth]{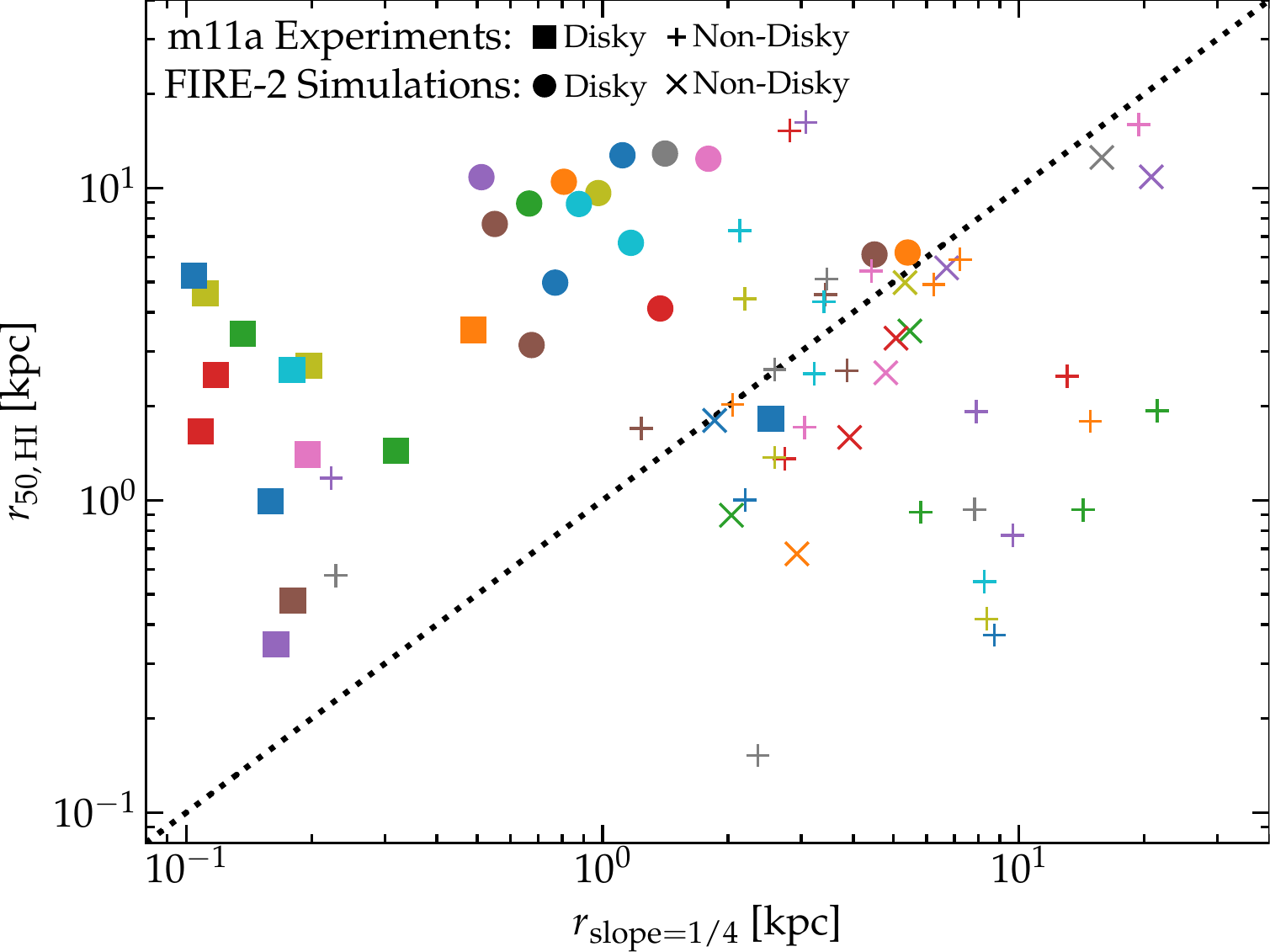}
	\includegraphics[width=0.49\textwidth]{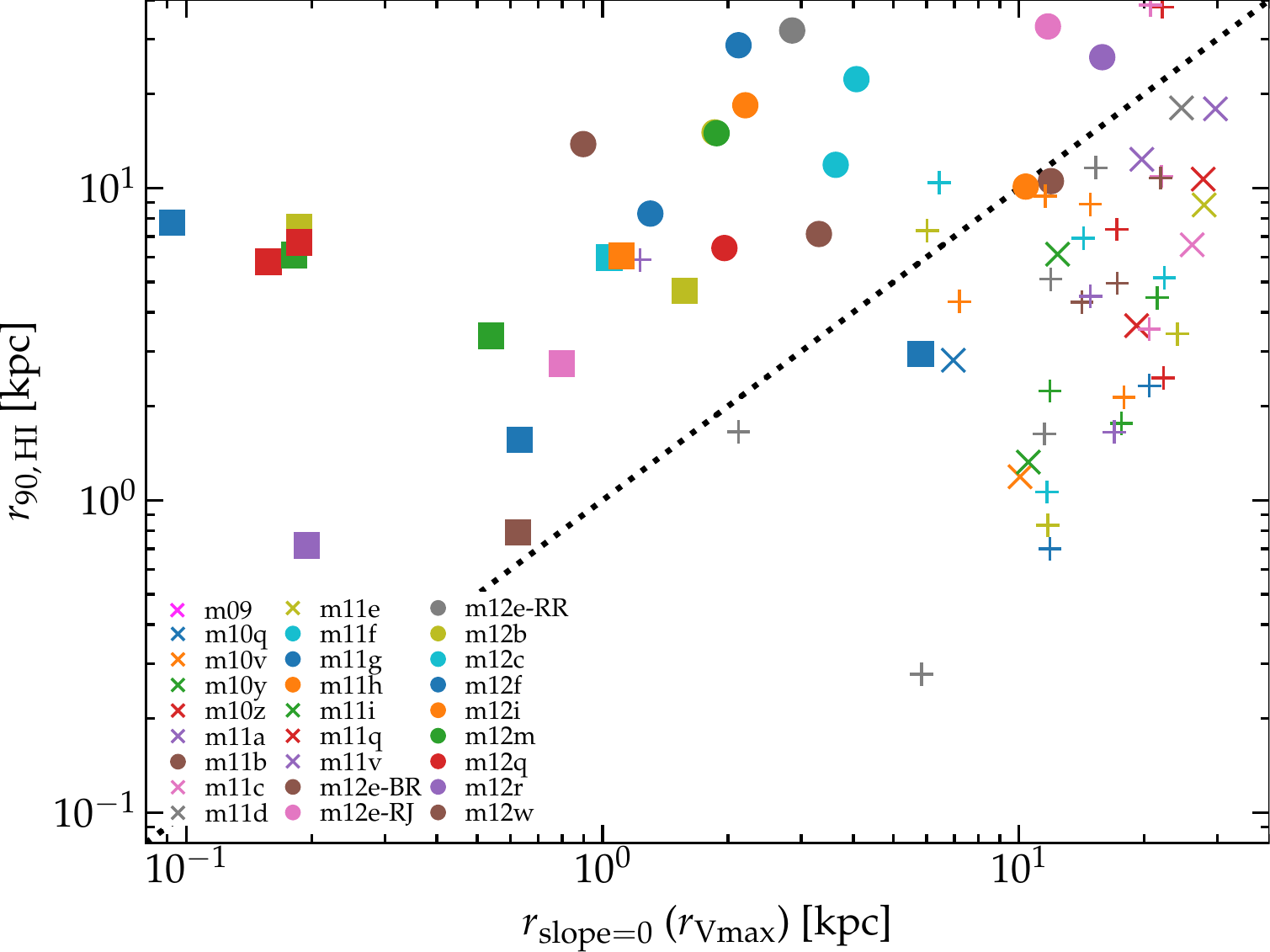} \\
	\includegraphics[width=0.49\textwidth]{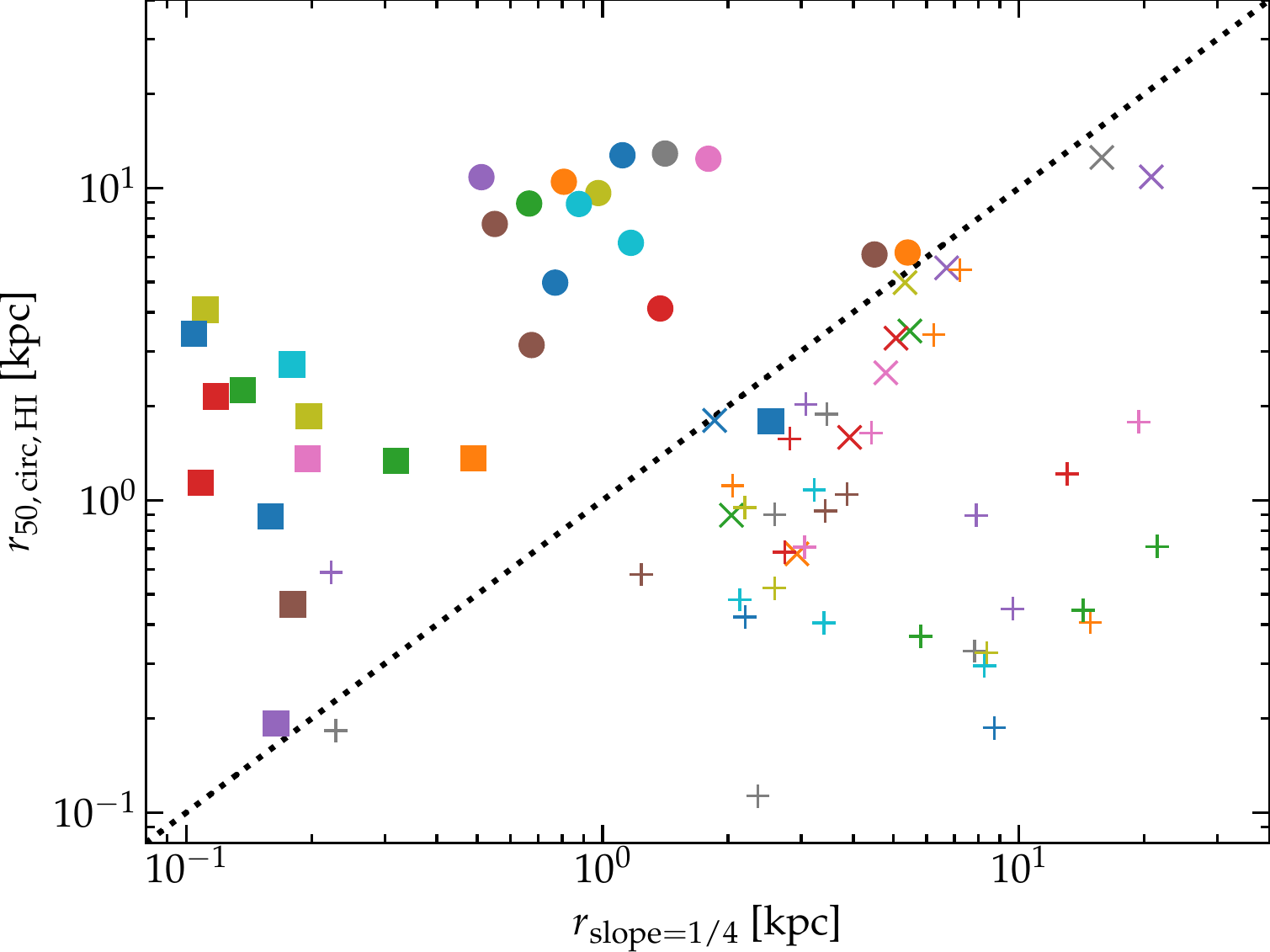}
	\includegraphics[width=0.49\textwidth]{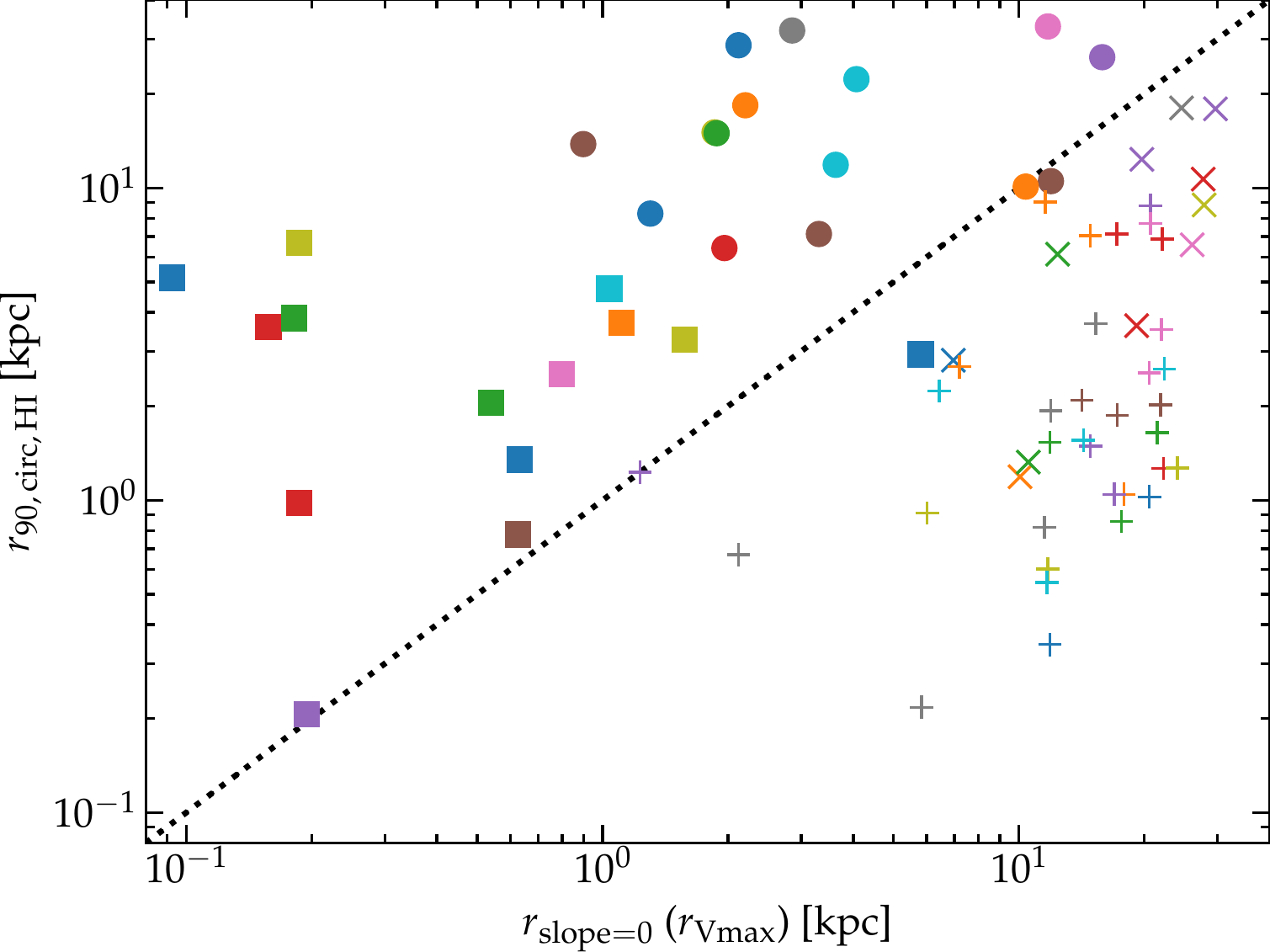} 
	\vspace{-0.2cm}
	\caption{Different proxies for concentration and disk formation (\S~\ref{sec:disky.quant}). In each panel, we denote ``disky'' (filled points) and ``non-disky'' ($+$/$\times$) simulations as labeled and show all FIRE-2 simulations and {\bf m11a} experiments. Dotted lines show equality. 
	{\em Top:} We plot a proxy of disk size: either $r_{\rm 50,\,HI}$ or $r_{\rm 90,\,HI}$ (the $50\%$ or $90\%$ mass inclusion radius of the HI), versus a proxy for the concentration of the mass profile: either $r_{{\rm slope}=1/4}$ or $r_{{\rm slope}=0}=r_{\rm Vmax}$ (the radius where the slope of the mass profile falls below $\partial \ln{V_{\rm c}} / \partial \ln{r} = 1/4$ or $0$). 
	{\em Bottom:} Same, but replacing $r_{\rm 50,\,HI}$ and $r_{\rm 90,\,HI}$ with $r_{\rm 50,\,circ,\,HI}$ and $r_{\rm 90,\,circ,\,HI}$, where each gas cell is assigned a radius $r_{\rm circ}$ of a circular orbit with the same coherent specific angular momentum.
	Our ``concentrated mass profile'' criterion is roughly equivalent to $r_{\rm 50,\,circ\,HI} \gtrsim r_{1/4}$ or $r_{\rm 90,\,circ\,HI} \gtrsim r_{\rm Vmax}$. Using $r_{\rm HI}$ instead of $r_{\rm circ,\,HI}$ we see a broadly similar trend but more scatter: e.g.\ several non-disky {\bf m11a} runs with $r_{\rm 50,\,HI} \lesssim r_{{\rm slope}=1/4}$, but almost all of these are systems where $r_{\rm 50,\,HI}$ appears to be large because the gas is in outflow.
	\label{fig:correlation.reff.rmax}\vspace{-0.3cm}}
\end{figure*}

It is helpful to consider simple scalar proxies for ``sufficiently concentrated mass profiles.'' From our analysis above, this appears to be something like a ``sufficiently close-to-flat or rising-to-small radius'' rotation curve inside of some characteristic radius of the gas: this suggests some simple proxies we might consider. For the rotation curve, a characteristic scale where the rotation curve ceases to be rapidly rising with radius (i.e.\ where the mass profile becomes concentrated) can be defined by $r_{{\rm slope}=x}$, defined as the innermost\footnote{Formally we take the innermost radius outside of the minimum of the central $0.1\,$kpc or $<10\%$ of the gas effective radius, to avoid spurious numerical noise from poor sampling of the potential (or physical but very small-scale structure) at very small $r$.} radius outside of which the logarithmic slope of the rotation curve falls below $\partial \ln{V_{\rm c}} / \partial {\ln{r}}  \le x$. For $x=0$ this defines a rotation curve maximum, so (barring the pathological case of a multiply-peaked rotation curve) this is just $r_{{\rm slope}=0}=r_{\rm Vmax} = r(V_{\rm c} = V_{\rm max})$. Our experiments above and those in \citet{shen:2021.dissipative.dm.dwarfs.fire} which motivated much of our study, show that disk formation is still efficiently promoted for mass profiles $\rho \propto r^{-3/2}$ or steeper (this is the characteristic dSIDM cusp slope in \citealt{shen:2021.dissipative.dm.dwarfs.fire}, and in-between our $a_{0}$ and $V_{0}$ experiments), corresponding to $x=1/4$. But we can of course consider any value of $x$. For the disk, we can define $r_{y,\,{\rm HI}}$, i.e.\ the spherical radius enclosing $y\%$ of the total HI mass, so $r_{50,\,{\rm HI}}$ is the usual effective radius of the gas, while $r_{90,\,{\rm HI}}$ is a typical ``outer'' radius (it makes little difference using this or some typical observational surface-brightness-limit contour as in e.g.\ \citealt{2020A&A...633L...3C}, in practice). We also find it makes relatively little difference for our purposes here if we define these in 3D or face-on 2D projection. 

Comparing a variety of these proxies, we see a clear separation between the disky and non-disky systems in both the default FIRE runs and our idealized tests in Fig.~\ref{fig:correlation.reff.rmax}. The two most useful proxies of those we have considered appear to be: $r_{\rm Vmax}$ and $r_{90,\,{\rm HI}}$ or $r_{{\rm slope}=1/4}$ and $r_{50,\,{\rm HI}}$. If we consider a simple linear scaling, the ``dividing line'' between disky and non-disky is remarkably similar to $r_{\rm Vmax} \lesssim r_{90,\,{\rm HI}}$ or $r_{{\rm slope}=1/4} \lesssim r_{50,\,{\rm HI}}$.\footnote{If we actually fit a power-law bisector we find we can more evenly divide the disky and non-disky simulations with $r_{\rm Vmax}/{\rm kpc} \lesssim (r_{90,\,{\rm HI}}/2.6\,{\rm kpc})^{1.5}$ and $r_{{\rm slope}=1/4}/ {\rm kpc} \lesssim (r_{50,\,{\rm HI}}/0.54\,{\rm kpc})^{0.8}$, but given the small sample and limitations of this approach, we do not consider the deviations from a linear slope here to be particularly physically meaningful.} We stress though that there is nothing particularly special about these exact slope choices: we see a similar quality separation in the data plotting $r_{{\rm slope}=1/4}$ versus $r_{90,{\rm HI}}$ or $r_{75,\,{\rm HI}}$ or $r_{50,\,{\rm HI}}$, and likewise similar separation using $r_{{\rm slope}=0.1}$ instead of $r_{{\rm slope}=1/4}$ or $r_{{\rm slope}=0}$, although the linear ``dividing line'' between disky and non-disky happens to have a coefficient not quite as close to unity (e.g.\ $r_{50,\,{\rm HI}} \gtrsim 0.6\,r_{{\rm slope}=0.1}$).

At a glance plotting all our simulations and all numerical tests here, there are a couple apparent outliers which are not disky. However, upon visual inspection, it is clear almost all of the significant outliers actually have large $r_{y,\,{\rm HI}}$ entirely owing to outflows -- i.e.\ not only is there no gas disk, but almost all the HI is in an outflow from the galaxy. Of course, an outflow can have an arbitrarily large $r$ and would not be meaningful for our criterion. But this does suggest that an even better quantitative criterion would replace $r_{y,\,{\rm HI}}$ with something like the {\em circularization} radius $r_{{\rm circ},\,y,\,{\rm HI}}$ of the HI. We attempt to do so at $z=0$ with a simple proxy: using the value of specific angular momentum $j_{z}$ about the common angular momentum axis already defined (e.g.\ Fig.~\ref{fig:jc}) for each gas resolution element, we define an equivalent circular radius $r_{\rm circ}$ such that $j_{z} = r_{\rm circ}\,V_{\rm c}(r_{\rm circ})$. This of course is not quite the same as a ``circularization radius,'' which requires following the gas dynamics over cosmic time, but it at least captures the leading-order correction, correctly moving systems where $r_{y,\,{\rm HI}}$ is apparently large because of pure outflow or hydrostatic equilibrium gas to lower $r_{\rm circ}$.\footnote{After correcting to $r_{\rm circ}$, while there are a few ``marginal'' galaxies close to the boundary between disky and non-disky systems, there are only two significant outliers in our entire FIRE-2 + {\bf m11a} experiments suite, both variants of {\bf m11a}: run $\eta_{\rm fb}=0.3$ which is labeled non-disky but has $r_{\rm 50,\,circ,\,HI} \sim 3\,r_{{\rm slope}=1/4}$ (and $r_{\rm 90,\,circ,\,HI} \approx r_{{\rm slope}=0}$), and run $a_{0}=10000$, which is has a very thin disk but $r_{\rm 50,\,circ,\,HI} \sim 0.7\,r_{{\rm slope}=1/4}$ ($r_{\rm 90,\,circ,\,HI} \sim 0.5\,r_{{\rm slope}=0}$). The former appears to be at the very early stages of ``beginning'' to form some gaseous disk as the stellar mass has increased to very large masses (making the potential more concentrated and giving rise to these parameters; see Table~\ref{table:sims}). The latter is an extreme and unusual case, where the extremely small disk owes in part to the exceptionally deep and instantaneously-imposed added potential.}

From our case study of {\bf m11b}, it is also clear that it would be better still to replace both $r_{{\rm slope}=x}$ and $r_{{\rm circ},\,y,\,{\rm HI}}$ with their values at the time of accretion or circularization, as opposed to at $z=0$ (as the $z=0$ value alone makes simulations like {\bf m11b} and {\bf m11d/e} appear very similar to one another, when their values at the time of accretion were more discrepant along the lines we predict). We have, in fact, examined all the halos here designated as ``disky'' at times closer to their ``initial'' disk formation times, and find they meet these criteria at those times as well.\footnote{For the massive $\sim 10^{12}\,M_{\odot}$ halos, the central $V_{\rm c}$ is generally less very-sharply-peaked at much earlier times when the disk first begins to form, since by $z=0$ this central peak is dominated by late-time star formation in the galaxy center (see \citealt{hopkins:fire2.methods}). But all of these halos still meet the criteria we define above when their disks form, with some (e.g.\ Romeo) closely analogous to our case study of {\bf m11b} at these times.}

Together these experiments, the transition between behaviors going from our $a_{0}$ to $V_{0}$ models, and the scaling with $r_{{\rm slope}=1/4}$ all suggest that a ``sufficiently centrally concentrated'' mass profile is one in which the inwards radial acceleration $-a_{r}(r) \sim G\,M_{\rm enc}/r^{2} \sim V_{\rm c}^{2}/r$ is a {decreasing} function of radius $r$, at some radii $r$ interior to the gas circularization radii $r_{\rm circ}$ when that gas is able to potentially circularize.

\section{A Critical Escape Velocity/Potential Scale Promotes Smooth Star Formation}
\label{sec:bursty}

The question of ``what promotes smooth star formation'' is somewhat more ambiguous, compared to the disk formation question discussed in the previous section. It is clear, however, that the models with more extended potentials, e.g.\ $q_{0} \gtrsim 5$, $\rho_{0} \gtrsim 1.5$, $a_{0} \gtrsim 1000$, tend to produce smooth SF, even if they generally do not produce disks. Meanwhile, the models with the deepest concentrated potentials ($M_{0} \gtrsim 5$, $V_{0} \gtrsim 100$), or the models with very weak feedback, can produce smooth SF late in their evolution as the central stellar mass and potential build up in a concentrated, very compact disk.

\subsection{What Is Not the Proximate Cause of Smooth SF?}
\label{sec:bursty.ruled.out}

Before going further, we review what our experiments rule {\em out} as the cause of smooth SF, some illustrations of which are summarized in Figs.~\ref{fig:sfhs.weak.bursty.fx}, \ref{fig:profiles.weak.bursty.fx}, \&\ \ref{fig:more.smooth.sfh.tests.cooling} (some additional tests are in \S~\ref{sec:additional.params}, e.g.\ Fig.~\ref{fig:smooth.bursty.tests.nopressure}).

\begin{itemize}

\item{Strong/Weak Feedback:} Per \S~\ref{sec:sffb.fx.weak} (and Fig.~\ref{fig:sfhs.weak.bursty.fx}), changing the strength, form, or rates of stellar feedback have little effect here. Only in the most extreme weak-feedback cases do we see a late-time {\em non-linear} effect via the gravitational potential.\footnote{We have also run simulations where we have varied the numerical form of the feedback coupling between FIRE-1, FIRE-2, and FIRE-3 formulations \citep{hopkins:fire3.methods}; we have varied the terminal momentum and kinetic/thermal energy ratio of feedback as described below; and varied the number of individual SNe by a factor of $10$ in either direction while changing the energy per SNe (so the total specific feedback rate is constant). None of these is able to alter the bursty/non-bursty distinction.}

\item{Star Formation Criteria/Clustering:} Likewise, the SF criteria (Fig.~\ref{fig:sfhs.weak.bursty.fx}) do not appear to directly influence this (\S~\ref{sec:sffb.fx.weak}).\footnote{In addition to the variations described in \S~\ref{sec:sffb.fx.weak}, we have varied the SF criteria by turning off/on a molecular gas requirement, virial threshold, converging flow criterion, varied the density threshold for SF from $0.1-10^{3}\,{\rm cm^{-3}}$, and even considered a model with only a density threshold for SF at $n=0.1\,{\rm cm^{-3}}$ and a constant {\em depletion time} for SF for all particles (independent of density) so the SFR was strictly proportional only to the mass above this threshold density, not any density substructure in the medium. All of these tests still produced bursty SF.}

\begin{figure*}
	\includegraphics[width=0.48\textwidth]{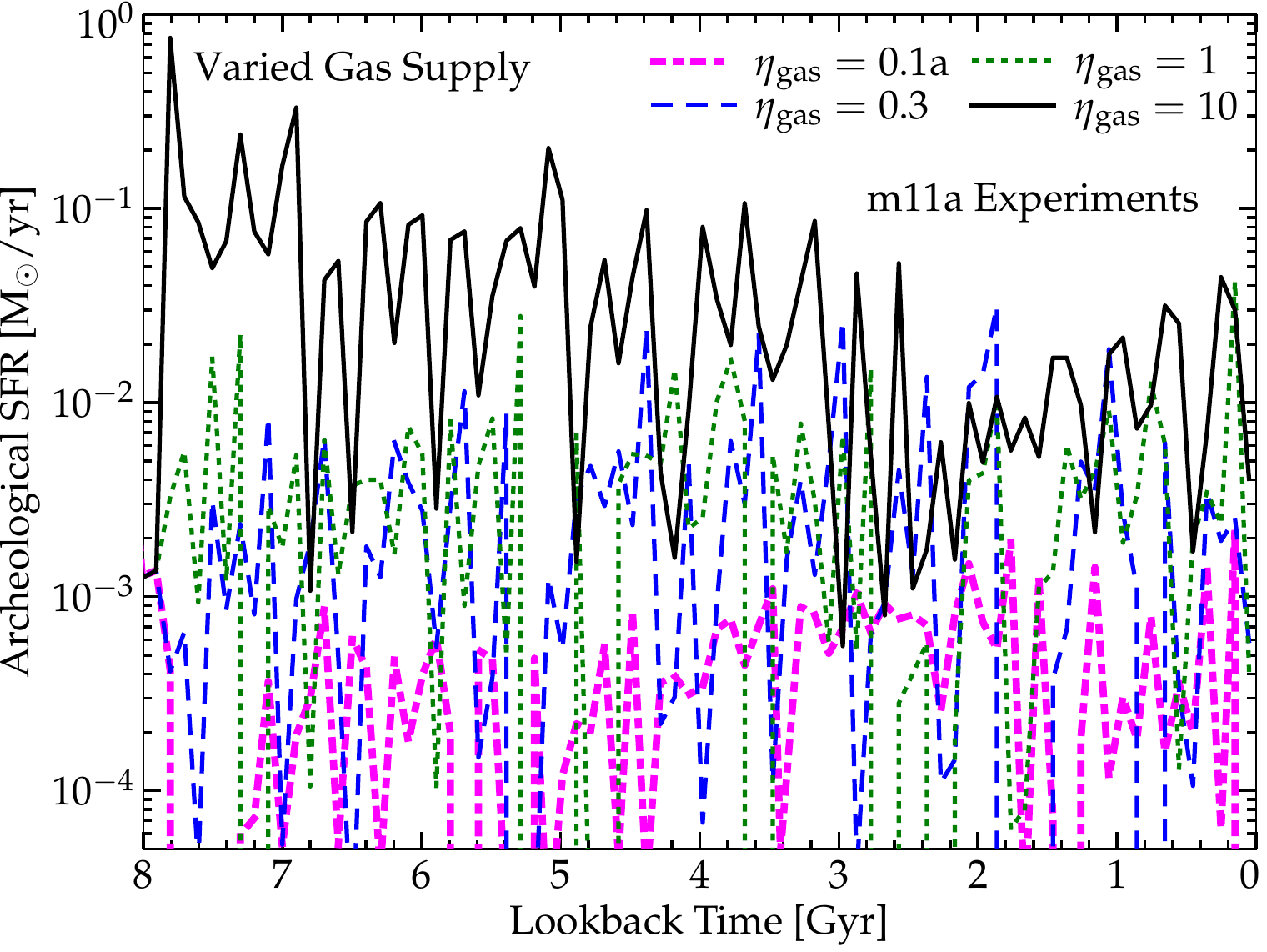}
	\includegraphics[width=0.48\textwidth]{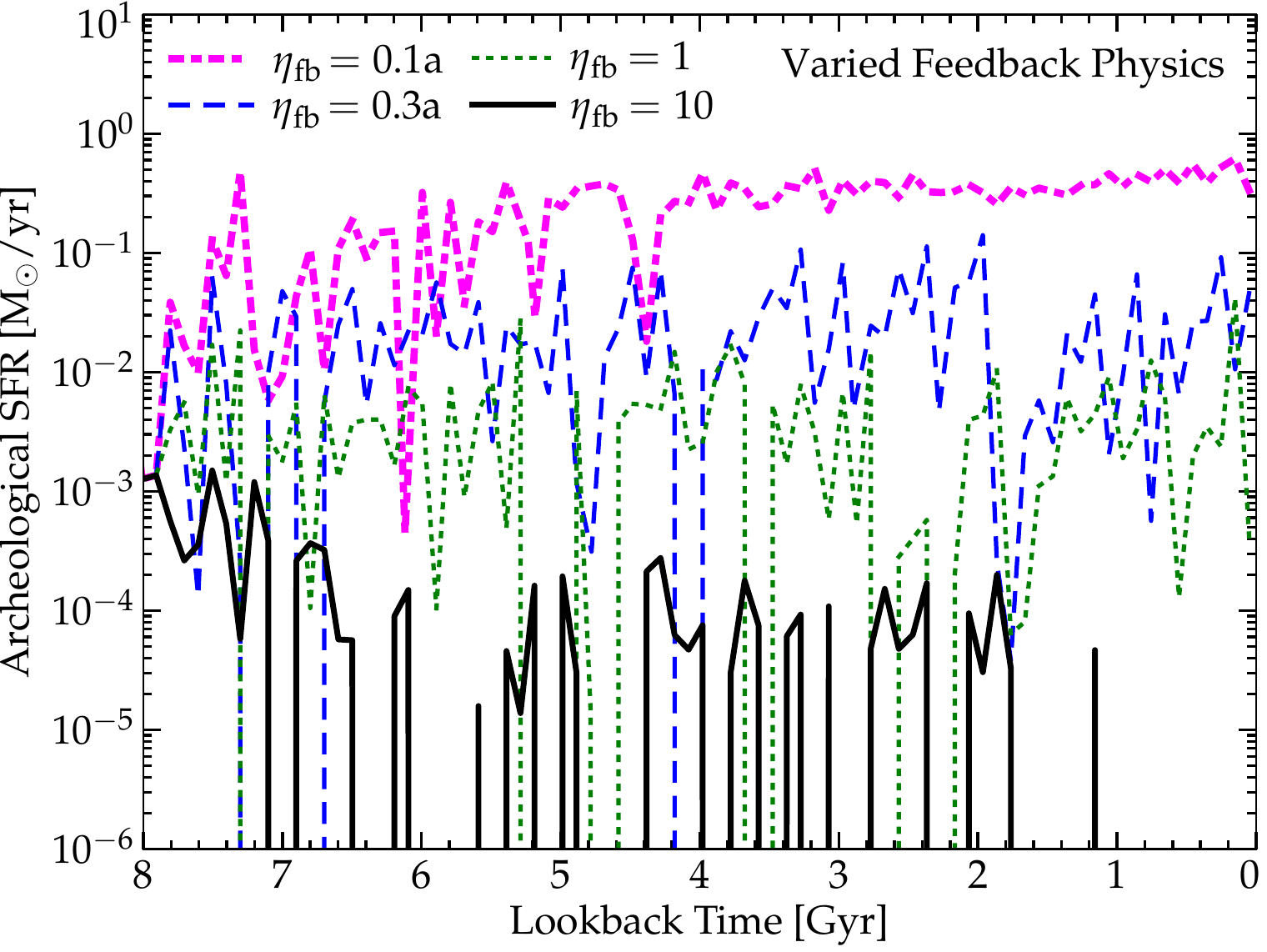} \\
	\includegraphics[width=0.48\textwidth]{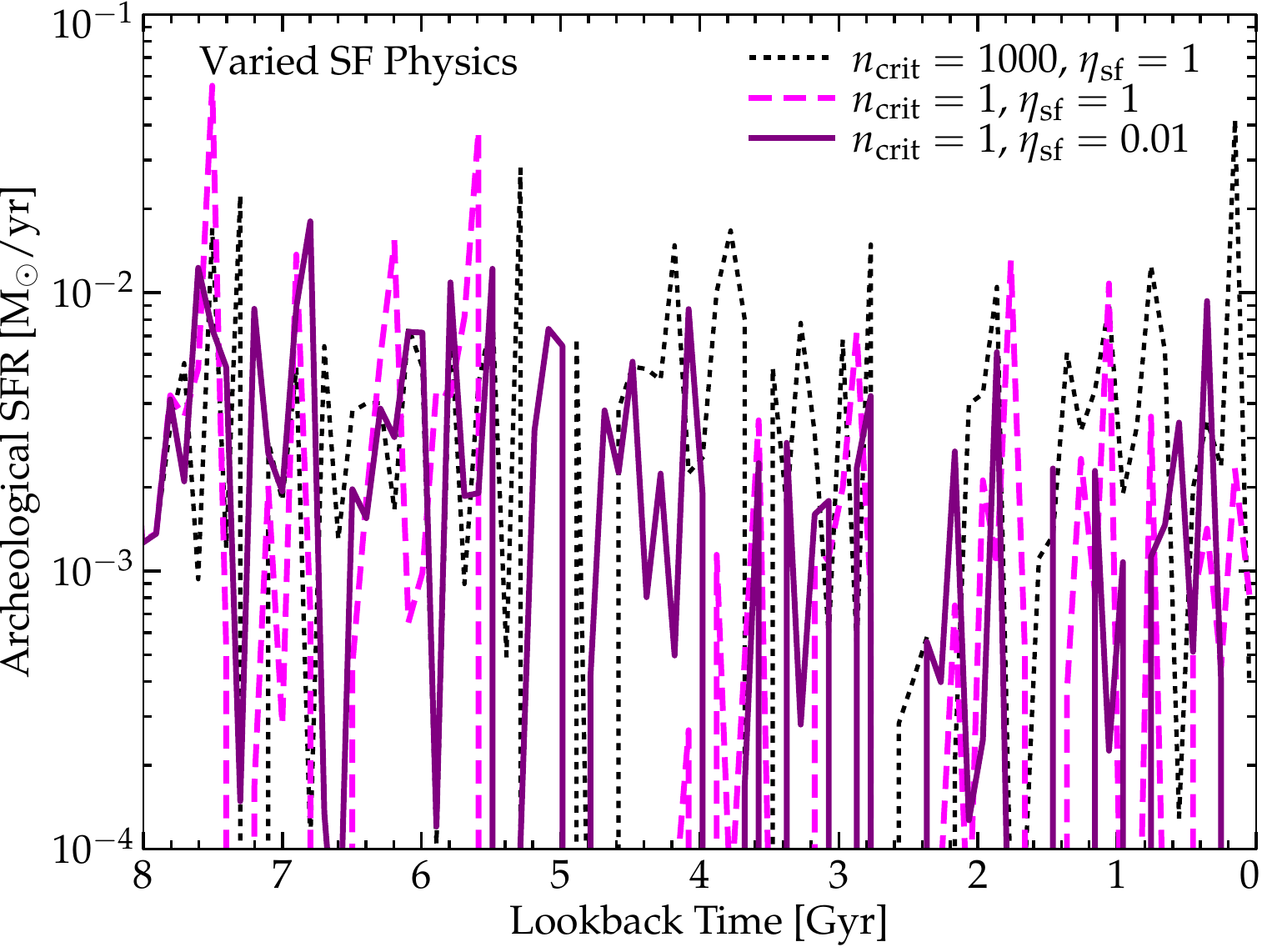}
	\includegraphics[width=0.48\textwidth]{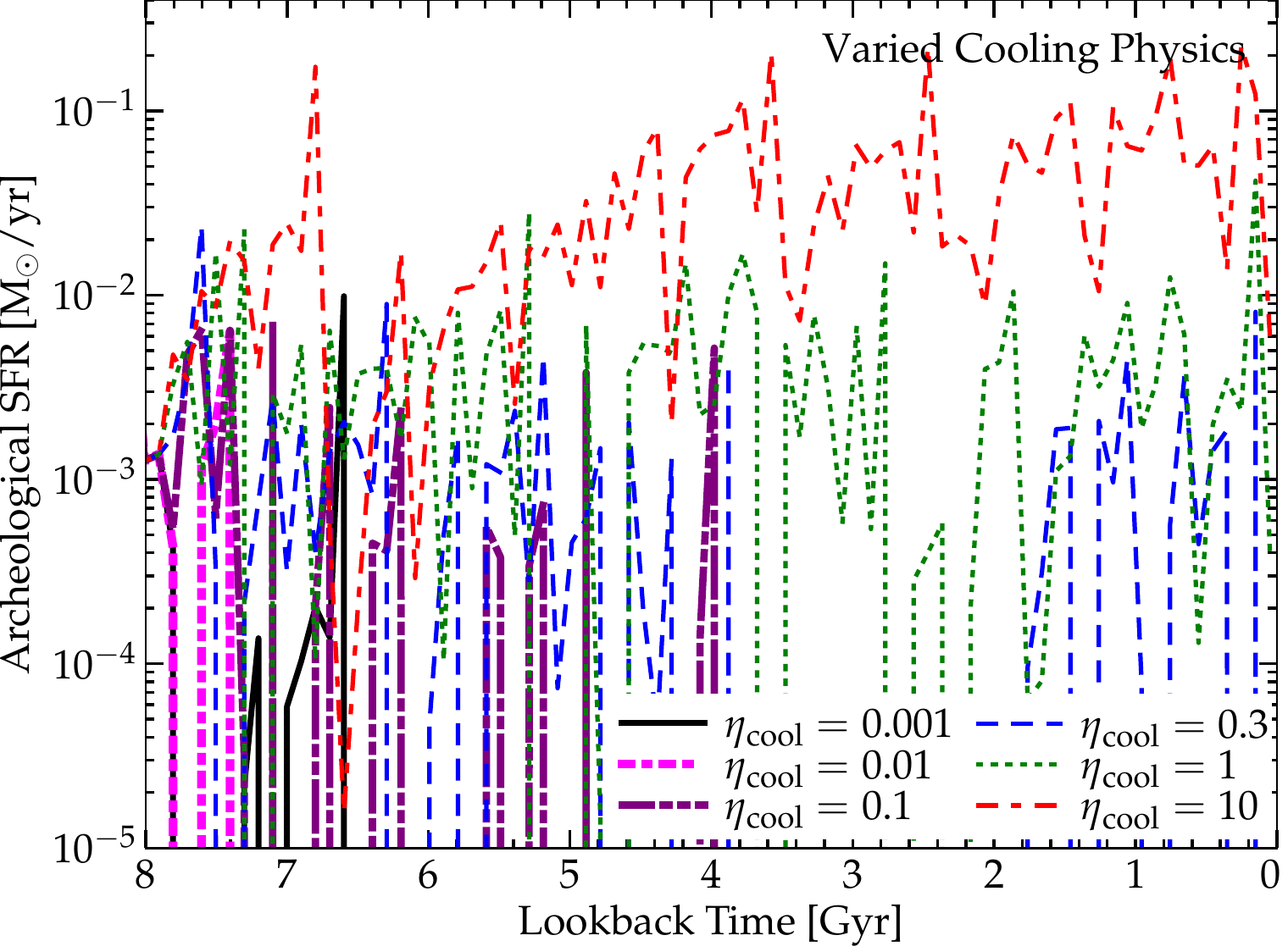}
	\vspace{-0.1cm}
	\caption{Example SFHs (as Fig.~\ref{fig:sfr}) versus physics variations from Table~\ref{table:sims} for {\bf m11a} (see \S~\ref{sec:bursty.ruled.out}). Changing the gas supply/content, star formation model or rate per freefall time, feedback strength, or gas thermodynamics/cooling rates does not cause ``smooth'' SF. The only cases which appear to transition to smooth SF are the weakest feedback examples, which do so after producing a stellar mass far larger than observed and building up deep potential wells (discussed below).
	\label{fig:sfhs.weak.bursty.fx}\vspace{-0.3cm}}
\end{figure*}

\begin{figure*}
	\includegraphics[width=0.32\textwidth]{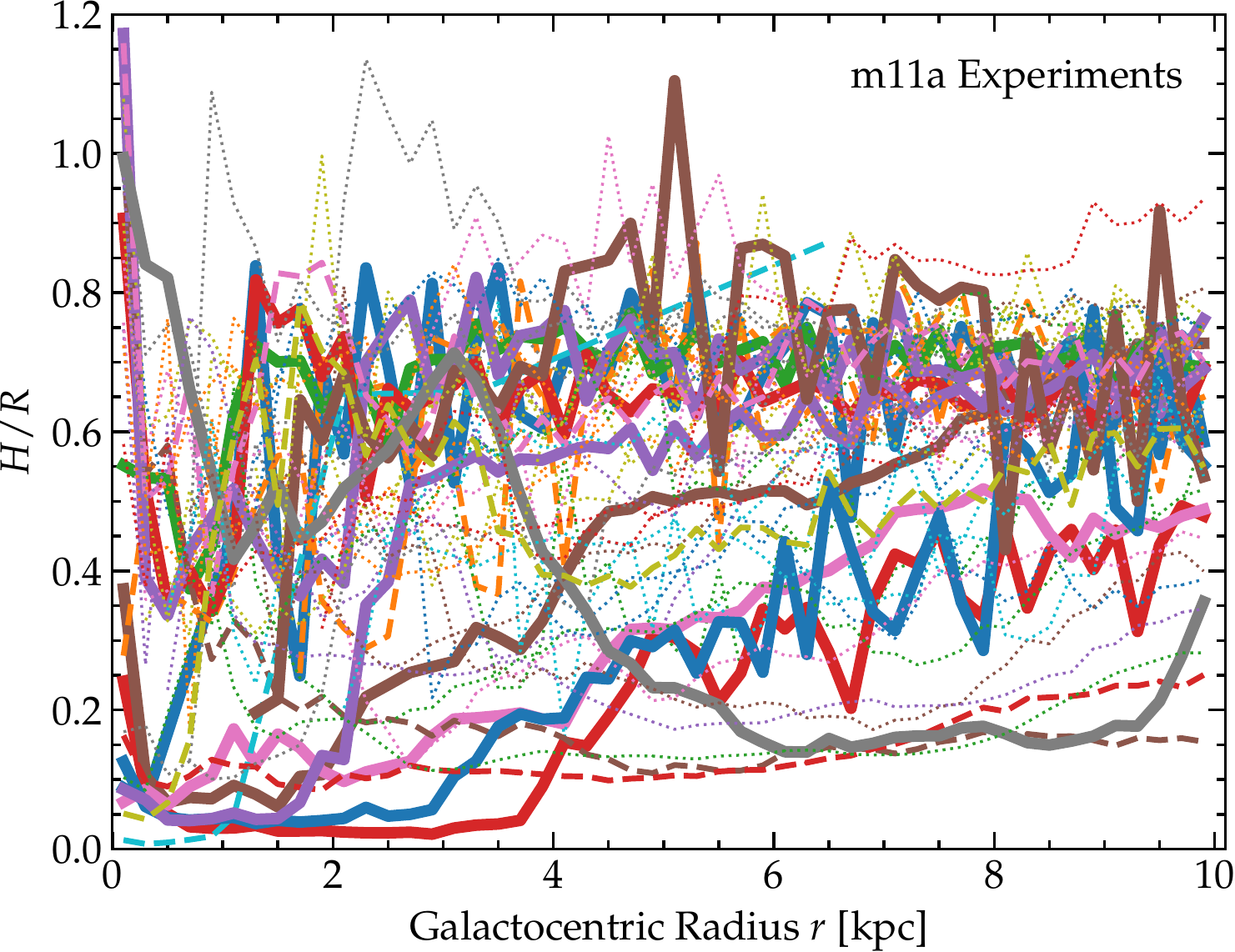}
	\includegraphics[width=0.32\textwidth]{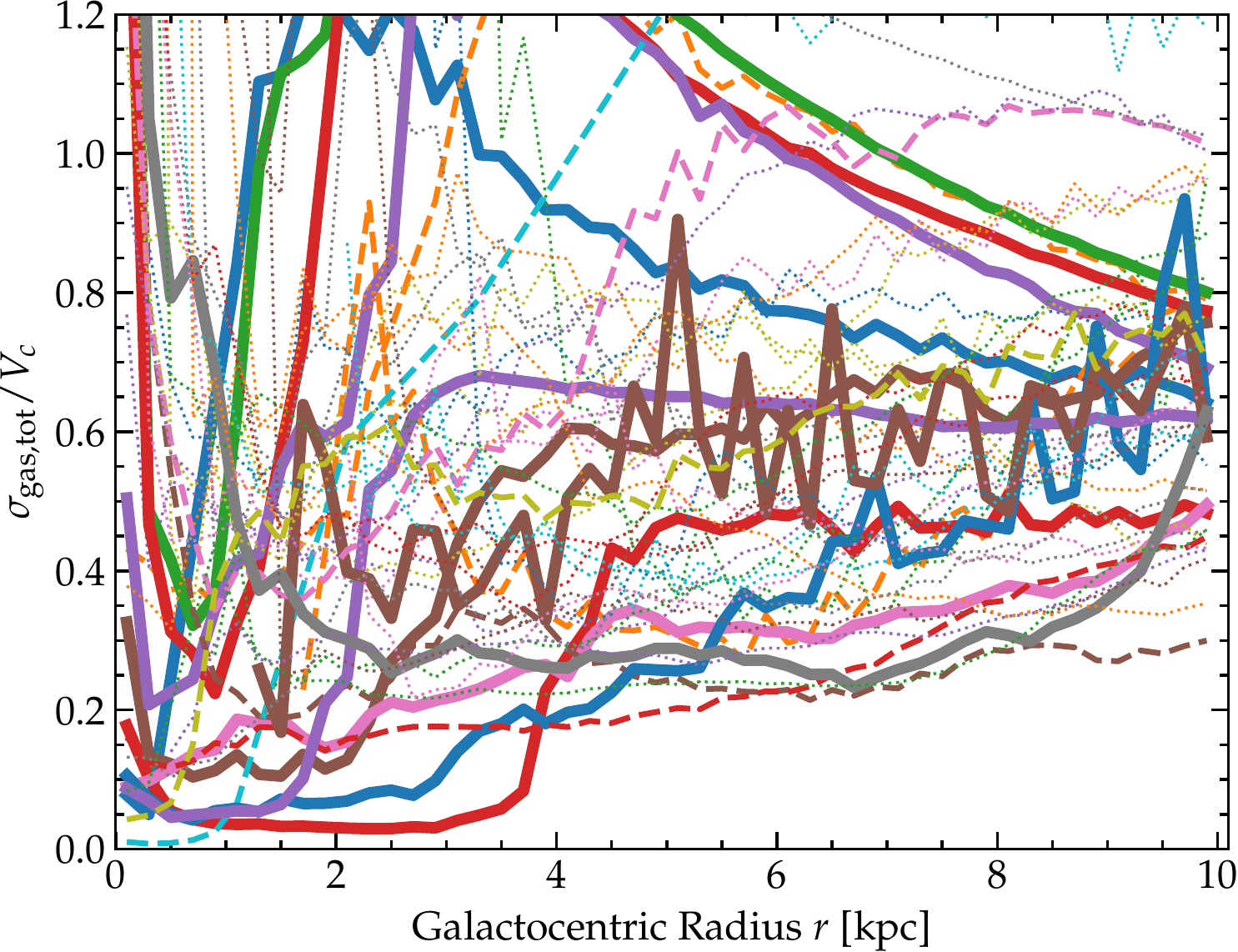} 
	\includegraphics[width=0.32\textwidth]{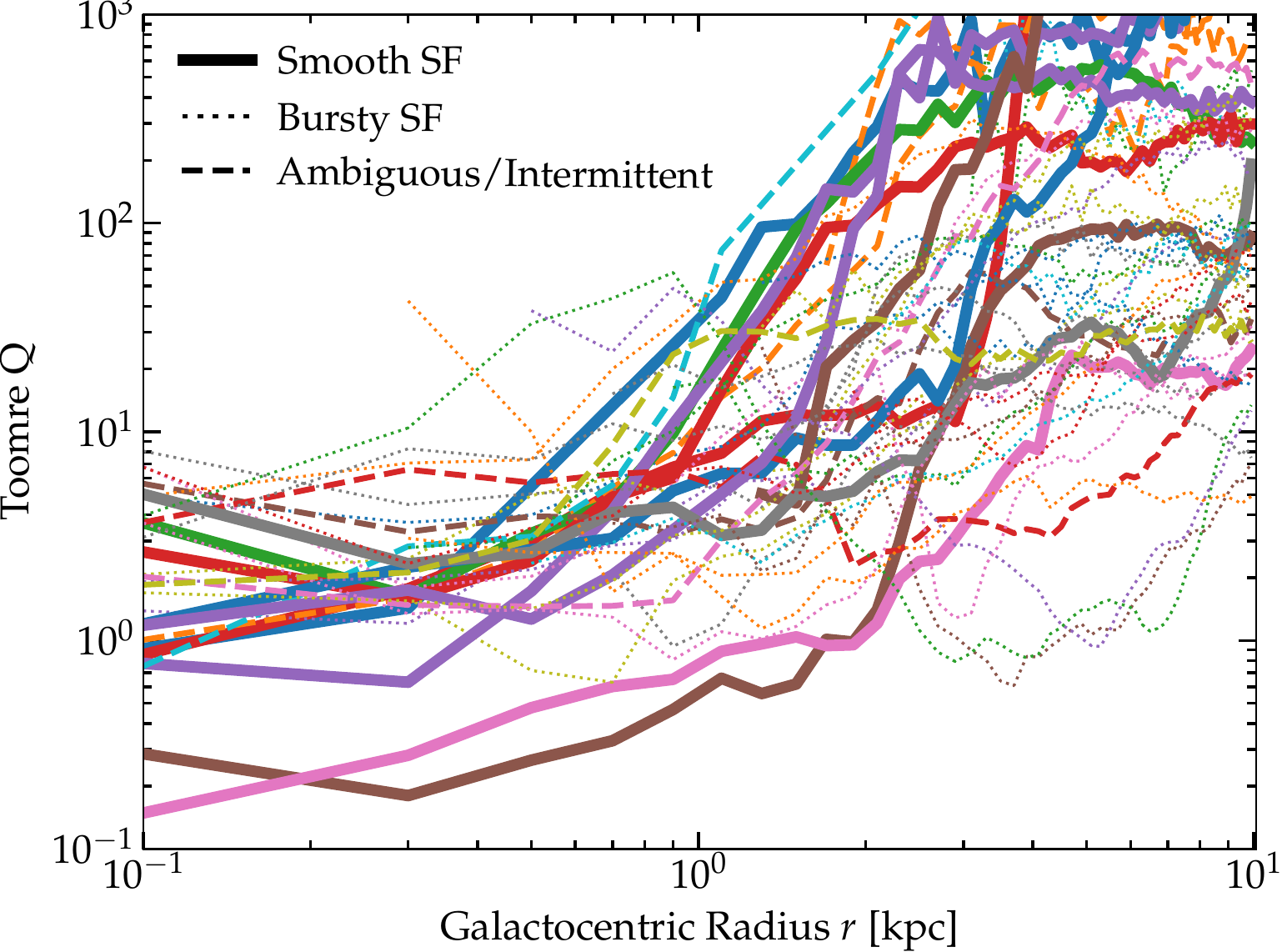} \\
	\includegraphics[width=0.32\textwidth]{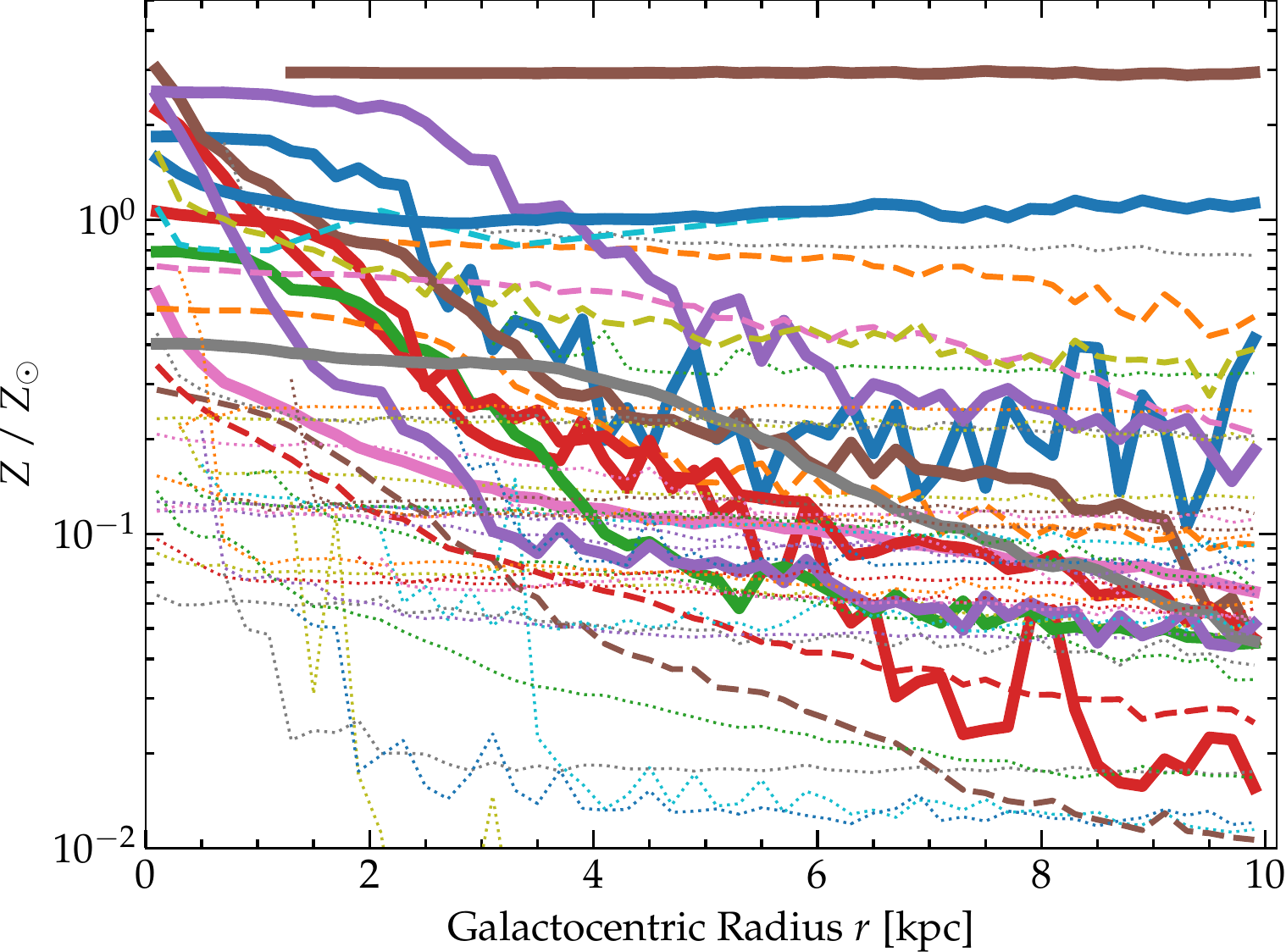} 
	\includegraphics[width=0.32\textwidth]{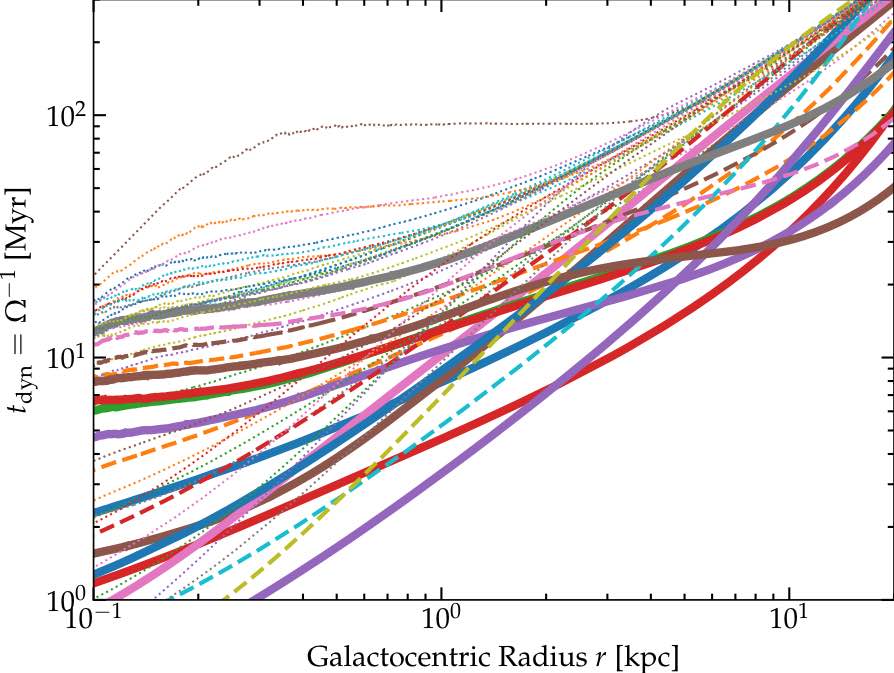}
	\includegraphics[width=0.32\textwidth]{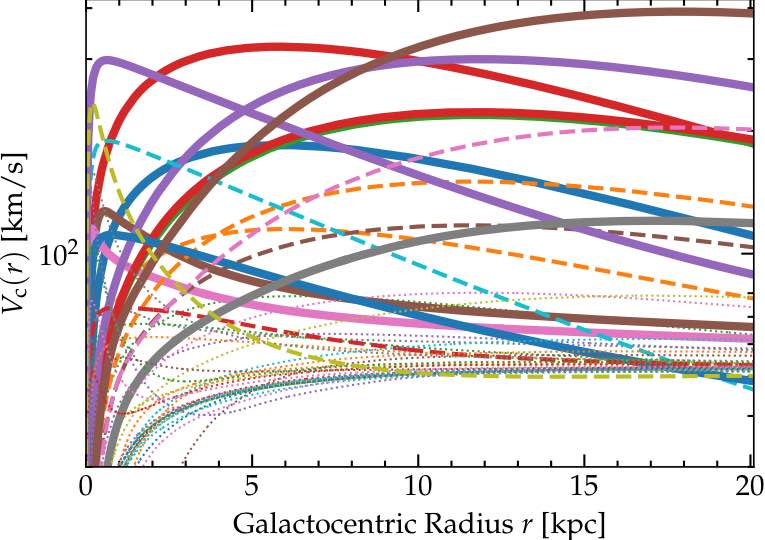}
	\vspace{-0.1cm}
	\caption{Radial ($z=0$) profiles of various quantities for all Table~\ref{table:sims} experiments for {\bf m11a} (as Fig.~\ref{fig:other.disky.profiles}), labeled by whether the SFR is clearly smooth ({\em thick solid}), clearly bursty ({\em thin dotted}) or ambiguous/intermittent ({\em medium dashed}). 
	{\em Top Left:} Vertical scale-height $H/R$.
	{\em Top Right:} Turbulent+thermal dispersion $\sigma/V_{\rm c}$.
	{\em Middle Left:} Toomre $Q$.
	{\em Middle Right:} Gas-phase metallicity $Z$.
	{\em Bottom Left:} Disk dynamical time $t_{\rm dyn}\equiv 1/\Omega$.
	{\em Bottom Right:} Circular velocity $V_{\rm c}$.
	Most of these show no separation at all between smooth and bursty systems. There is a hint in $V_{\rm c}$ where higher {\em maximum} $V_{\rm c}$ may correlate (but no trend with the shape of $V_{\rm c}$), and some enhancement of $Z$ in smooth systems: we discuss these more clearly below.
	\label{fig:profiles.weak.bursty.fx}\vspace{-0.3cm}}
\end{figure*}

\begin{figure}
	\includegraphics[width=0.97\columnwidth]{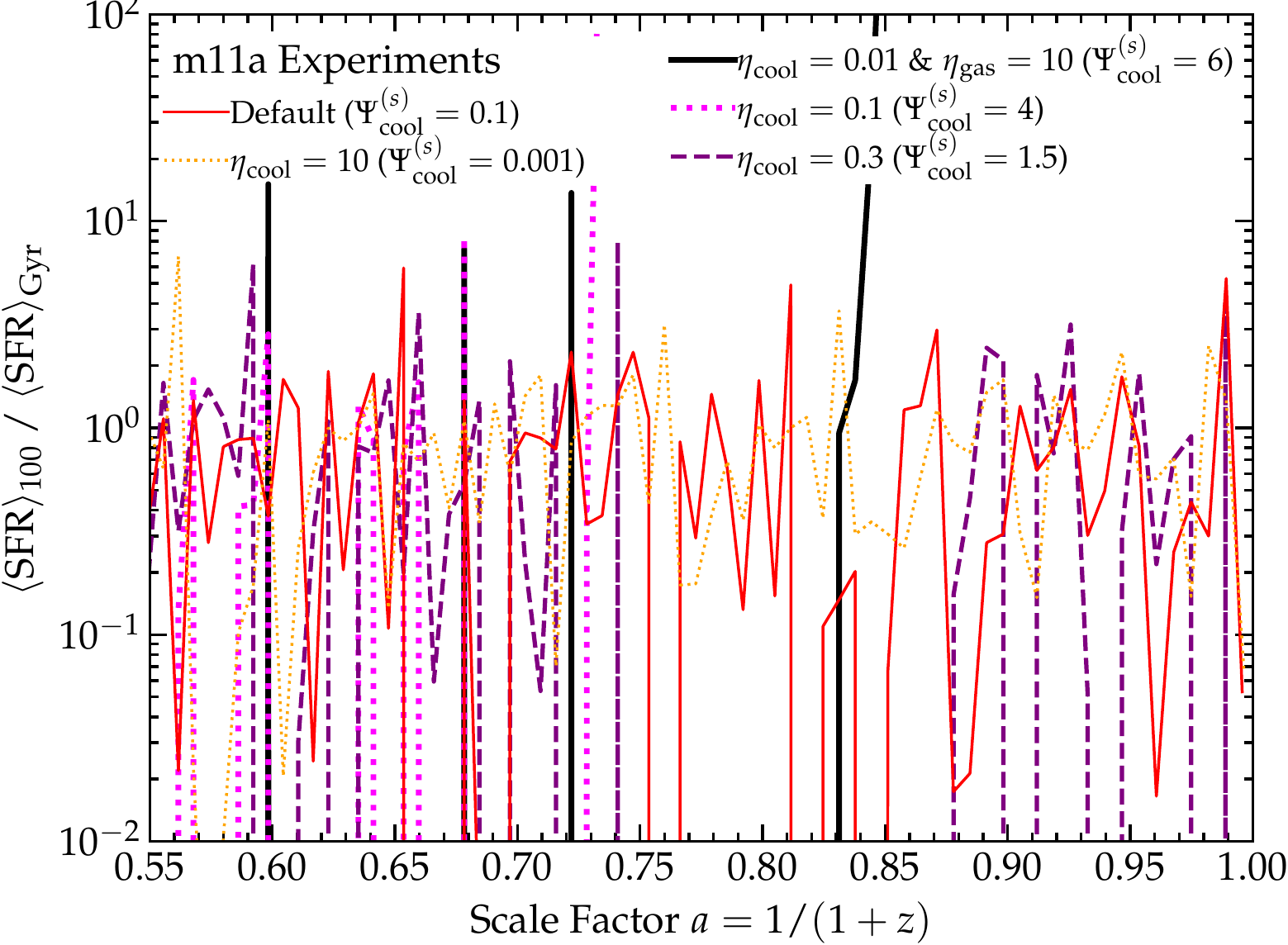}
	\includegraphics[width=0.97\columnwidth]{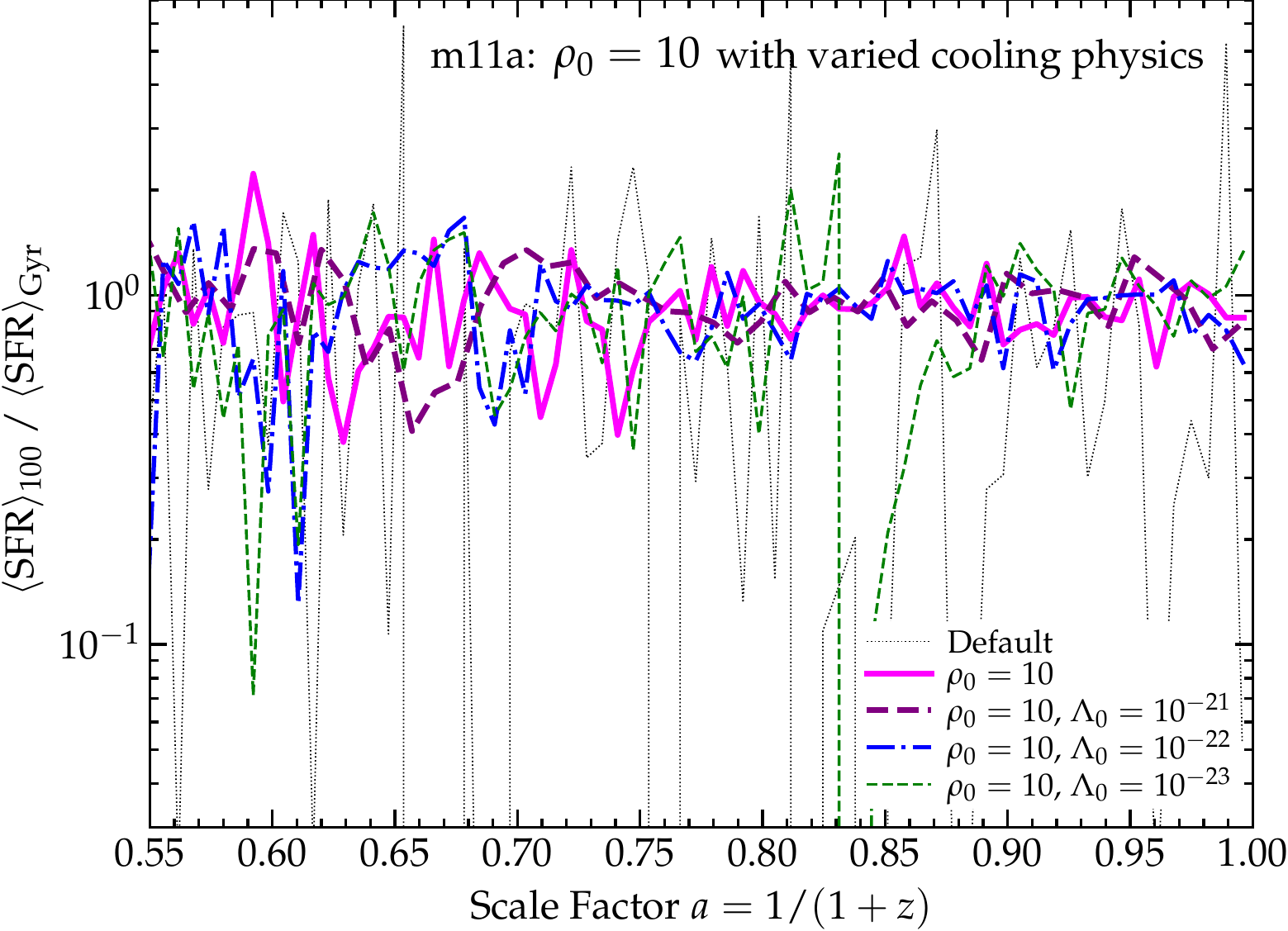}
	\includegraphics[width=0.95\columnwidth]{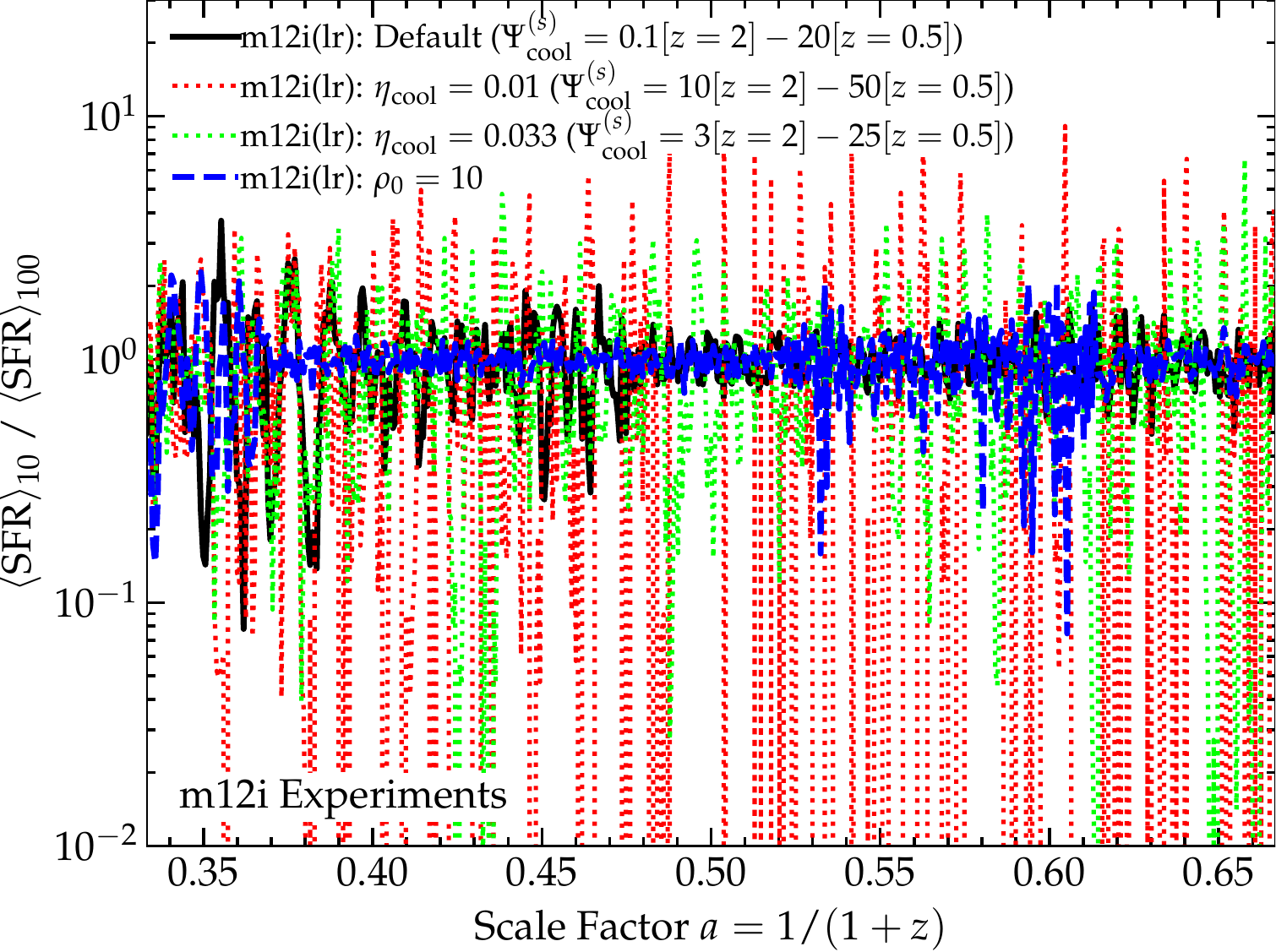}
	\vspace{-0.1cm}
	\caption{Normalized SFHs as Figs.~\ref{fig:sfr.vesc.time.weak.fb}-\ref{fig:sfr.smooth.rhomodel}, demonstrating the weak effects of cooling physics.
	{\em Top:} {\bf m11a} experiments (as Fig.~\ref{fig:sfhs.weak.bursty.fx}), with the value of  $\Psi_{\rm cool}^{(s)}(0.1\,R_{\rm vir}) \equiv t_{\rm cool}^{(s)}/t_{\rm dyn}$ (measured from the profiles at $z=0$, following \citealt{stern21_ICV}). We include a variant with gas supply increased ($\eta_{\rm gas}=10$) to make up for less cooling ($\eta_{\rm cool}=0.01$). which forms $9\times10^{7}\,M_{\odot}$ worth of stars (clearly does not quench) and can reach even higher $\Psi_{\rm cool}^{(s)}$. There is no transition to smooth SF above $\Psi_{\rm cool}^{(s)} > 1$. 
	{\em Middle:} {\bf m11a} experiments with $\rho_{0}=10$ (a much deeper potential), which produces {\em smooth} SF, varying the cooling function. Increasing the cooling rate does not ``re-introduce'' burstiness.
	{\em Bottom:} Restarts of {\bf m12i} (a Milky Way-mass halo) run from $z=2$ to $0.5$, with the values of $\Psi_{\rm cool}^{(s)}$ for the cooling physics variants quoted. The default run the SF becomes smooth by the \citet{yu:2021.fire.bursty.sf.thick.disk.origin} definition at $a\sim 0.5$ (coincidentally when $\Psi_{\rm cool}^{(s)}$ exceeds $\sim1$), but the runs with lower $\eta_{\rm cool}$ ($\Psi_{\rm cool}^{(s)} \gg 1$ at all times) do not become smooth. Alternatively a run with the added potential $\rho_{0}=10$ shows an immediate decrease in burstiness.
	\label{fig:more.smooth.sfh.tests.cooling}}
\end{figure}

\item{High/Low Gas Fractions and the Gas Clump/Toomre Scale:} Per \S~\ref{sec:thermo} and Figs.~\ref{fig:sfhs.weak.bursty.fx} \&\ \ref{fig:profiles.weak.bursty.fx}, changing the gas supply either directly or via modified cooling does not appear to be able in any of our experiments to induce smooth SF.\footnote{We also considered experiments where we varied the gas mass $\eta_{\rm gas}$ only in the ISM or CGM (inside/outside $10\,$kpc), and experiments where we simultaneously adjusted $\eta_{\rm cool}$ and $\eta_{\rm gas}$ so as to maintain the same SFR in the ISM of the galaxy as our ``default'' run. None of these changed our conclusions.} Appendix~\ref{sec:additional.params} and Fig.~\ref{fig:orr.fgas.criterion} explicitly demonstrate some additional related variables such as gas surface densities, optical depths through the disk, and the \citet{orr:2021.bubble.breakout.model} ``critical gas fraction'' based on bubble breakout models do not appear to correlate with the bursty/smooth SF distinction.

\item{Cooling Times Long or Short Compared to Freefall Times:} Likewise, arbitrarily changing the cooling time, including both variations with expected median $t_{\rm cool} \gg t_{\rm freefall}$ or $\ll t_{\rm freefall}$ in the halo, does not influence burstiness (\S~\ref{sec:thermo}). As shown in Fig.~\ref{fig:more.smooth.sfh.tests.cooling} and discussed in detail below, we have also considered variations of $\eta_{\rm cool}$ at different halo mass scales, in experiments which are already in the ``smooth'' SF regime, and simultaneous variations of $\eta_{\rm gas}$ and $\eta_{\rm cool}$ or $\eta_{\rm fb}$ and $\eta_{\rm cool}$ designed to keep the SFR and/or feedback strength identical to the ``default'' runs. We have also varied the shape of the cooling function, and modified the cooling function only in the ISM and/or CGM (or even arbitrarily removing thermal pressure from the CGM or forcing an isothermal ``cold'' equation-of-state; Appendix~\ref{sec:additional.params}). All of these tests support our conclusion that $t_{\rm cool}$ has little or no direct effect on bursty SF. Likewise (Fig.~\ref{fig:profiles.weak.bursty.fx}) comparing different metallicities appears to have no systematic effect.\footnote{We re-ran simulations arbitrarily multiplying all metallicities by a factor $\eta_{\rm Z}=(0.01,\,1,\,10)$ and found no effect on our conclusions here.}

\item{Dynamical Times Long or Short Compared to Massive Stellar Evolution Times:} ``Steady-state'' star formation models generally assume that feedback is proportional to the SFR, which in turn assumes the dynamical time is long compared to the massive star main-sequence lifetime ($t_{\ast} \sim 10-30\,$Myr), so some have suggested the transition between ``bursty'' and ``smooth'' SF may correspond to when the dynamical time $t_{\rm dyn} = 1/\Omega = r/V_{\rm c}$ at the effective-star forming radius drops below $t_{\ast}$ \citep{torrey.2016:fire.galactic.nuclei.star.formation.instability,cafg:bursty.sf.toymodel}. In Fig.~\ref{fig:profiles.weak.bursty.fx} we see no trend with $t_{\rm dyn}/t_{\ast}$ in burstiness: we have bursty simulations with $t_{\rm dyn}(r_{\rm eff}) \gg t_{\ast}$ and $t_{\rm dyn} \ll t_{\ast}$. Moreover our cases with smooth SF in deep potentials correspond to cases where the added potential specifically made $t_{\rm dyn}$ much {\em smaller}, moving it from $t_{\rm dyn} > t_{\ast}$ to $t_{\rm dyn} \ll t_{\ast}$.

\item{Some Absolute Value of Halo Mass, Stellar Mass, or SFR:} We clearly see that at a given stellar or halo mass (the latter similar in all our {\bf m11a} re-runs) or SFR, we can have bursty or non-bursty SF. Moreover, the value of SFR (hence ``absolute'' feedback strength) at which our experiments which do show smooth SF transition to such ranges widely, from $\sim 10^{-3} - 1\,{\rm M_{\odot}\,yr^{-1}}$ (seen directly in Fig.~\ref{fig:sfhs.weak.bursty.fx}, and reflected in the factor $\gtrsim 100$ difference in the stellar masses formed at late times in Table~\ref{table:sims}). Below, we also repeat some of our key experiments at different halo mass scales from $10^{10}-10^{12}\,M_{\odot}$ (Fig.~\ref{fig:more.smooth.sfh.tests.cooling}), and we obtain the same conclusions.

\item{Presence/Absence of a Disk:} Per \S~\ref{sec:decoupling}, we clearly see that the formation of a disk does not necessarily imply smooth SF, and vice versa. Though a very thin disk may be correlated with smooth SF, a point we discuss further below, it is not always so (see \S~\ref{sec:m11b}).

\item{{\em Shape} of the Potential:} Per Fig.~\ref{fig:profiles.weak.bursty.fx}, we see no obvious correlation with the {\em shape} of the potential or effective mass profile: while more extended potentials can induce smooth SF without forming a disk, we also see smooth SF in some of the most dramatically compact potentials (e.g.\ $V_{0}=300$, $\eta_{\rm fb}=0.1$). 

\item{$V_{\rm c}$ or Acceleration scale {\em at the effective radius of star formation} (or gas, or stars):} Considering all the circular velocity curves in e.g.\ Fig.~\ref{fig:profiles.weak.bursty.fx}, we see a suggestion that smooth SF may be related to the maximum of the $V_{\rm c}$ ($\gg 100\,{\rm km\,s^{-1}}$) or acceleration scale ($\gg ({\rm a\ few}) \times 10^{-9}\,{\rm cm^{2}\,s^{-1}}$) at any radius beyond some limit. But (a) this requires we neglect the $M_{0}$ models or sufficiently small radii where $V_{\rm c}$ diverges; and (b) the radius where this maximum occurs varies widely and, in the more extended potential models ($\rho_{0}$, $q_{0}$) is set by our imposed cutoff, with no clear correlation with any galaxy property. Plotting the curves scaled by the effective radius of the stars, gas, or star formation we see these vary widely and do not correspond to where the system is actually forming stars. For many of the extended-potential models, the actual SF and gas are confined to  significantly smaller radii, where the acceleration or $V_{\rm c}$ scale is much smaller as well. But some others like the high-$V_{\rm c}$ models behave in the opposite manner. 

\end{itemize}

\subsection{The Escape Velocity Scale and Behavior of ``Blowouts'': The Role of Confinement}
\label{sec:vesc.bursty}

What does, in fact, appear to link all of the simulations with smooth SF is the behavior of gas after it is acted upon by stellar feedback.

\begin{figure}
	\includegraphics[width=0.95\columnwidth]{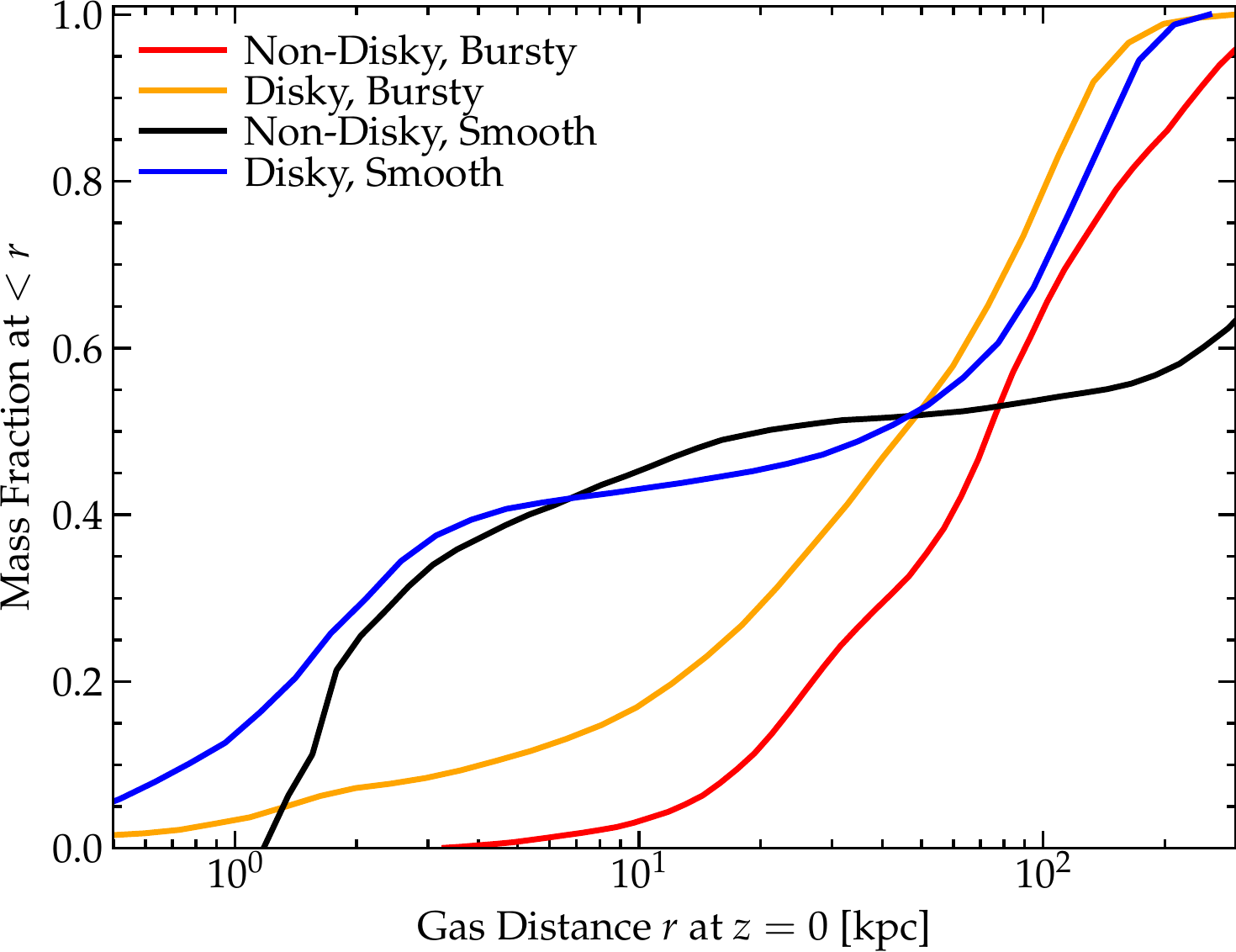}
	\vspace{-0.1cm}
	\caption{Cumulative distribution of distances at $z=0$, of gas which was ``in the galaxy'' (at $<5\,$kpc from the center) at $z=0.5$, in the runs from Fig.~\ref{fig:images}. In the ``bursty'' experiments, almost all the gas in the galaxy at this earlier time has escaped far from it by $z=0$. In the ``smooth'' runs, tens of percent (up to half) the gas (not already consumed by star formation) remains ``within the galaxy'' by the same definition. We also see this reflected in retained metals (Fig.~\ref{fig:profiles.weak.bursty.fx}). There is some (order-unity) outflow/escape to large radii, but the smooth systems are ``confined'' (\S~\ref{sec:vesc.bursty}).
	\label{fig:distance.blown.out}}
\end{figure}

\subsubsection{Confinement in Practice}

In Fig.~\ref{fig:distance.blown.out}, we illustrate this by comparing the positions of the same gas at two different, well-separated times in our {\bf m11a} experiments. Specifically we select gas first at $z=0.5$ which is ``inside the galaxy'' (chosen ad hoc to be within a radial distance $r_{0.5} \le 5\,$kpc at $z=0.5$), and then we plot the distribution of galacto-centric distances of the same gas cells at $z=0$. Because our code is Lagrangian, we can follow individual parcels over time in this manner \citep[see][for more detailed studies]{angles.alcazar:particle.tracking.fire.baryon.cycle.intergalactic.transfer,hafen:2018.cgm.fire.origins,hafen:2019.fire.cgm.fates}. Note that we only select gas which has not turned into stars by $z=0$. Of course the specific initial radius and time chosen are rather arbitrary here, but our qualitative results are not particularly sensitive to the choice, so long as $r_{0.5}$ includes some gas near the galaxy center at the earlier time, and the time spacing is many galaxy dynamical times. We compare a representative simulation from each of four groups: (1) not disky, with bursty SF (our default run); (2) disky, with bursty SF ($M_{0}=1$); (3) not disky, with smooth SF ($q_{0}=120$); disky, with smooth SF ($V_{0}=100$). In the ``bursty'' cases, we find that almost all the gas which was in/near the star-forming region at $z=0.5$ has been expelled by $z=0$, mostly to beyond the virial radius of the halo. In the ``smooth'' cases, we find that while there is still significant outflow, and order-unity (tens of percent) fraction of the gas {\em remains} at small radii (this is even more dramatic if we include the gas which remained at small radii but turned into stars). This appears robust across the different simulations in these sub-categories.

This suggests some sort of ``confinement'' at work. This may also be reflected in the morphology of the star-forming gas. For example, in all of our {\bf m11a} experiments which {\em do not} have a disk but {\em do} exhibit smooth star formation (e.g.\ $q_{0}=120$, $\rho_{0} \gtrsim 5$, $a_{0} \sim (1-3)\times10^{3}$), the star formation is dominated by gas in a compact, dense, quasi-spherical (but internally clumpy/irregular) central ``blob'' (akin to a single extremely-massive GMC complex), surrounded by a highly-depleted (low-density) gaseous ``halo.'' In the systems which are {both} disky and exhibit clearly smooth SF (e.g.\ $V_{0} \gtrsim 100$, $a_{0}\gtrsim 10^{4}$, $\eta_{\rm fb}\lesssim 0.1$), the star-forming gas disk tends to be either radially compact (extent $\lesssim 1-2\,$kpc) or vertically compact (i.e.\ very thin).

\begin{figure*}
	\includegraphics[width=0.48\textwidth]{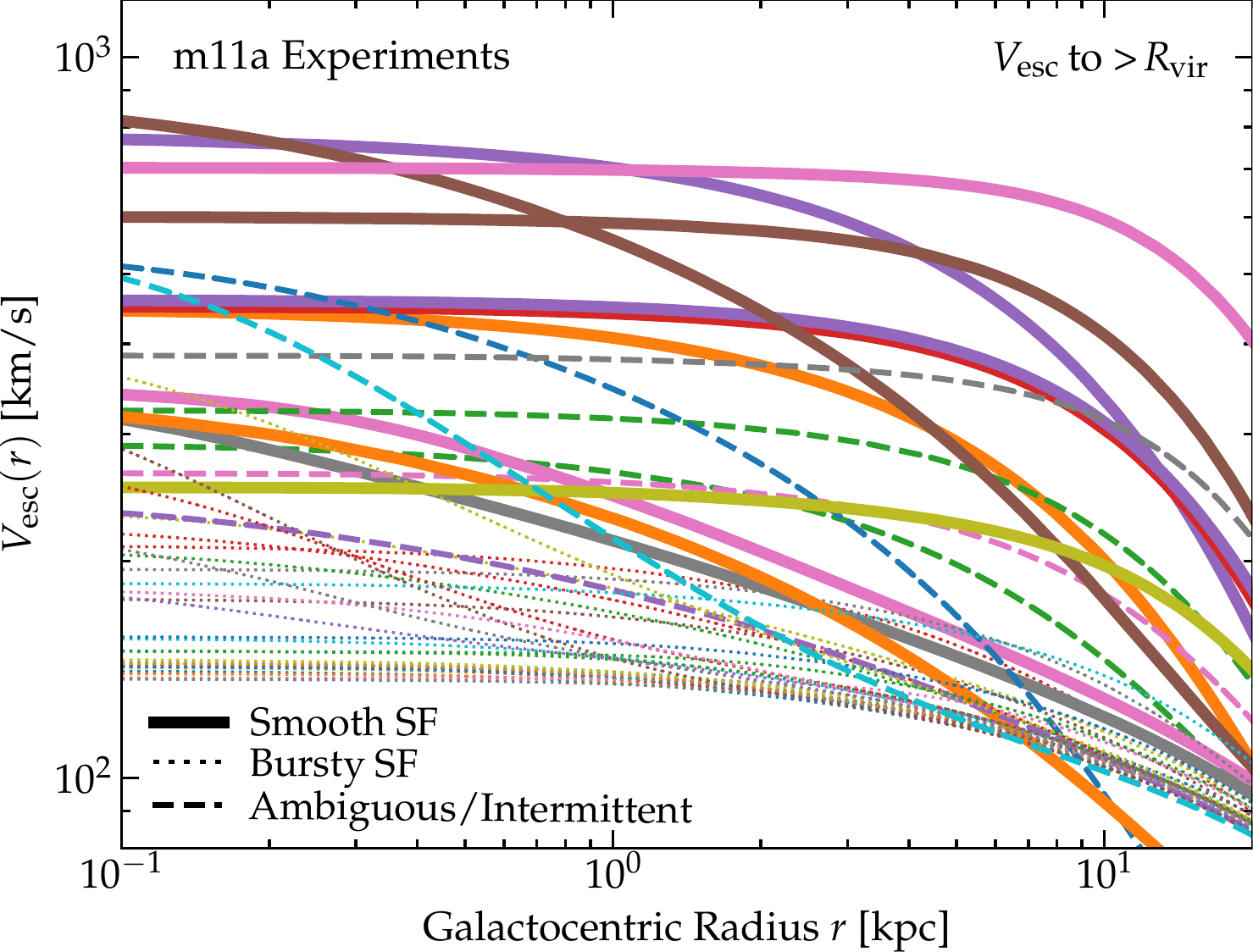}
	\includegraphics[width=0.48\textwidth]{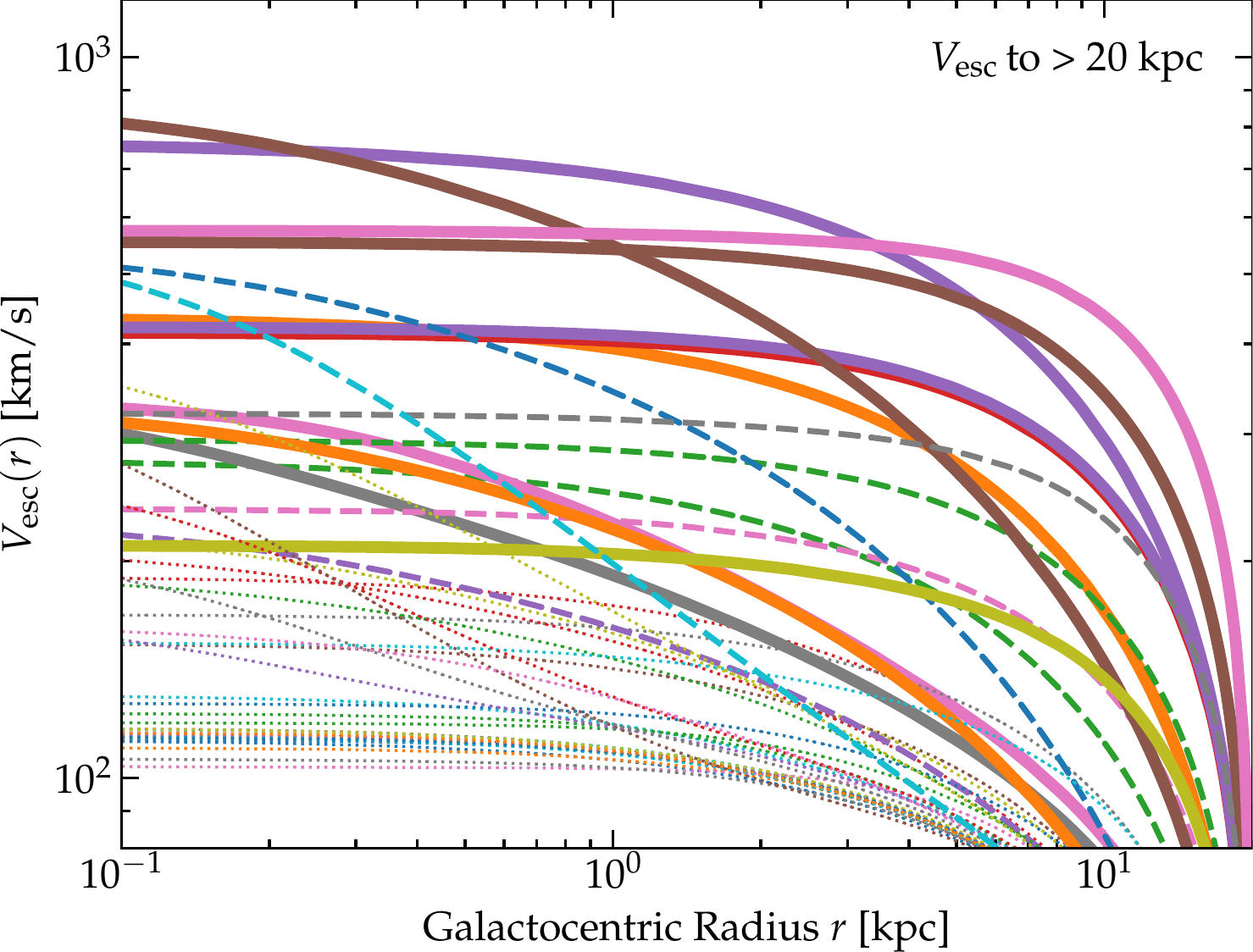}
	\vspace{-0.1cm}
	\caption{Profiles of the escape velocity $V_{\rm esc}$ from radius $r$ to $>R_{\rm vir}$ ({\em left}) or to $>20\,$kpc ({\em right}), for the smooth ({\em thick solid}), intermediate/ambiguous ({\em medium dashed}), and bursty ({\em thin dotted}) systems in our Table~\ref{table:sims} {\bf m11a} experiments. There is a clear separation here, with all galaxies at $V_{\rm esc}(r\sim r_{\rm eff,\,sfr} \sim 1\,{\rm kpc}) \gtrsim 200\,{\rm km\,s^{-1}}$ being at least somewhat ``smooth,'' and all systems at lower $V_{\rm esc}$ at these radii being bursty. The central escape velocity or potential well depth appears to be a much better predictor of ``smooth'' star formation (vs.\ other variables in Figs.~\ref{fig:sfhs.weak.bursty.fx}-\ref{fig:profiles.weak.bursty.fx} or Appendix~\ref{sec:additional.params}), and naturally explains the confinement effect in Fig.~\ref{fig:distance.blown.out} (see \S~\ref{sec:vesc.bursty}).
	\label{fig:Vesc}\vspace{-0.3cm}}
\end{figure*}

\subsubsection{Relation to the Escape Velocity Scale in our Idealized Tests}

Fig.~\ref{fig:Vesc} compares the escape velocity ($V_{\rm esc}$) scale of the {\bf m11a} simulation experiments with smooth or bursty SF. We compute $V_{\rm esc}$ from a given radius $r$ assuming the potential is spherically symmetric, as: 
\begin{align}
\label{eqn:vesc} V_{\rm esc}^{2}(r) = \int_{r}^{\infty} 2\,a_{\rm spherical}\,dr
\end{align}
(corrections for non-spherical terms are small). Here, we see a much more striking trend than in $V_{\rm c}(r)$, $a(r)$, $M_{\rm enc}(r)$, or ${\rho}(r)$ (quantities which only depend on the enclosed mass inside some radius $r$, as opposed to on the exterior $r^{\prime}>r$ distribution outside $r$), especially evaluated at the effective radius of SF $r_{\rm eff}$ in each galaxy. All of the simulations which feature smooth SF have an escape velocity at $r_{\rm eff}$\footnote{For the {\bf m11a} experiments, in Fig.~\ref{fig:Vesc}, it makes little difference if we plot $V_{\rm esc}(r)$ versus absolute physical $r$ or $r$ scaled to some effective radius $r_{\rm eff}$ or $r_{i}$. The result is also qualitatively similar to the $V_{\rm esc}$ to $>20\,$kpc plot for any outer radius in the range $\sim10-30\,$kpc.} of $V_{\rm esc}(r_{\rm eff}) \gtrsim 220\,{\rm km\,s^{-1}}$. Also, every simulation with $V_{\rm esc} \gtrsim 220\,{\rm km\,s^{-1}}$ (from initial $r\sim 1\,$kpc) features somewhat smooth SF, and everything with $V_{\rm esc}\gtrsim 400\,{\rm km\,s^{-1}}$ (again from $r\sim 1\,$kpc) falls into the un-ambiguous very-smooth-SF category (see also Table~\ref{table:sims}). Interestingly the slightly lower $V_{\rm esc}$ ($\sim 200-300\,{\rm km\,s^{-1}}$) but still ``very smooth'' systems are also disks, perhaps suggesting a slightly-weaker $V_{\rm esc}$ criterion for disks, though we do not see an obvious strong trend.\footnote{We note that we see a quite similar result (with perhaps a very slightly cleaner separation between smooth and non-smooth models) if we plot instead a Bernoulli-like parameter $\sqrt{V_{\rm esc}^{2} - V_{\rm c}^{2}}$ (the effective escape velocity for a vertical outflow from a circular thin disk), instead of just $V_{\rm esc}$. The threshold at $\sim 400\,{\rm km\,s^{-1}}$ is almost unchanged by this given the significantly smaller $V_{\rm c}$ at those radii, though the lower threshold at $\sim 220\,{\rm km\,s^{-1}}$ moves to $\sim 200\,{\rm km\,s^{-1}}$.}

This escape velocity scale of $\sim 220\,{\rm km\,s^{-1}}$ corresponds to a central gravitational potential scale $-\Phi = (1/2)\,V_{\rm esc}^{2} \sim  2.5\times10^{14}\,{\rm erg\,g^{-1}}$, and a virial temperature scale $T_{\rm vir} = -\Phi\,\mu\,m_{p}/3\,k_{\rm B}\sim 0.6\times10^{6}$\,K (i.e.\ ``escape temperature'' $\gtrsim 10^{6}\,$K). This combination (given the halo mass and virial temperature of {\bf m11a}) is suggestive of a picture where the escape velocity and/or potential scale becomes sufficiently large that a significant amount of gas can no longer be expelled from the galaxy via feedback (although some clearly still escapes). 

This immediately explains how the experiments with more extended potentials can produce smooth SF, even without a disk: their lack of a centrally-concentrated profile inhibits disk formation (\S~\ref{sec:physics.concentrated}), but even though $V_{\rm c}$ at small radii can be quite small, $V_{\rm esc}$ can be quite large. For e.g.\ our constant-$\rho_{0}$ models, including just the potential of the added $a_{\ast}(r)$ term, we have $V_{\rm c} \propto r \rightarrow 0$ as $r\rightarrow 0$, but $V_{\rm esc}(r\rightarrow 0) = (4\pi\,\rho_{0}/3)^{1/2}\,r_{0} \sim 300\,{\rm km\,s^{-1}}\,(\rho_{0}/5)^{1/2}$ (for $r_{0}$ defined in Eq.~\ref{eqn:a.ext.full}) -- since $V_{\rm esc}$ depends on the integral, the matter at large radii (well beyond the star-forming galaxy) contributes importantly to ``confinement.''

\subsubsection{The Escape Velocity to What Distance?}

When we refer to $V_{\rm esc}(r)$ in terms of the initial radius $r=r_{i}$, it is important to define to which outer or final radius $r_{f}$ we define the escape speed. In idealized problems, one often implicitly refers to the escape velocity to infinity, $V_{\rm esc}(r_{i}=r,\,r_{f}\rightarrow \infty)$. But in a cosmological simulation, this is not well-defined. Above we used $V_{\rm esc}$ to the virial radius, but this is also somewhat pathological: for an NFW halo, there is a large contribution to $V_{\rm esc}$ to $r_{\rm vir}$ from mass near $r_{\rm vir}$ itself (owing to the relatively shallow mass profile) -- but the gas does not need to travel anywhere nearly so far as $r_{\rm vir}$ to ``blow out'' the entire mass of the ISM and totally suppress star formation, and impose a very long recycling time to re-accrete. Intuitively, a better ``final'' radius would be sometime like a few times the effective radius of the star-forming gas, or the radius where the density is suppressed by a critical factor or the turnaround time reaches some critical value. In future work, we will explore in greater detail the Lagrangian trajectories of the material ejected in bursts following \citet{hafen:2018.cgm.fire.origins,hafen:2019.fire.cgm.fates}, but lacking more detailed models, we can estimate the effects of this by simply considering $V_{\rm esc}$ to some representative smaller radius of $\sim 10-20\,$kpc. 

Therefore in Fig.~\ref{fig:Vesc} we repeat the exercise described above for an outer radius of $\sim20\,$kpc, but we have also experimented with radii from $\sim10-30\,$kpc in physical units or set to some multiple of the star-forming effective radius of the galaxy. We see similar separation in each case: if anything, the separation is slightly ``cleaner'' using $V_{\rm esc}$ to something like $\sim 10-20\,$kpc.

\begin{figure}
	\includegraphics[width=0.95\columnwidth]{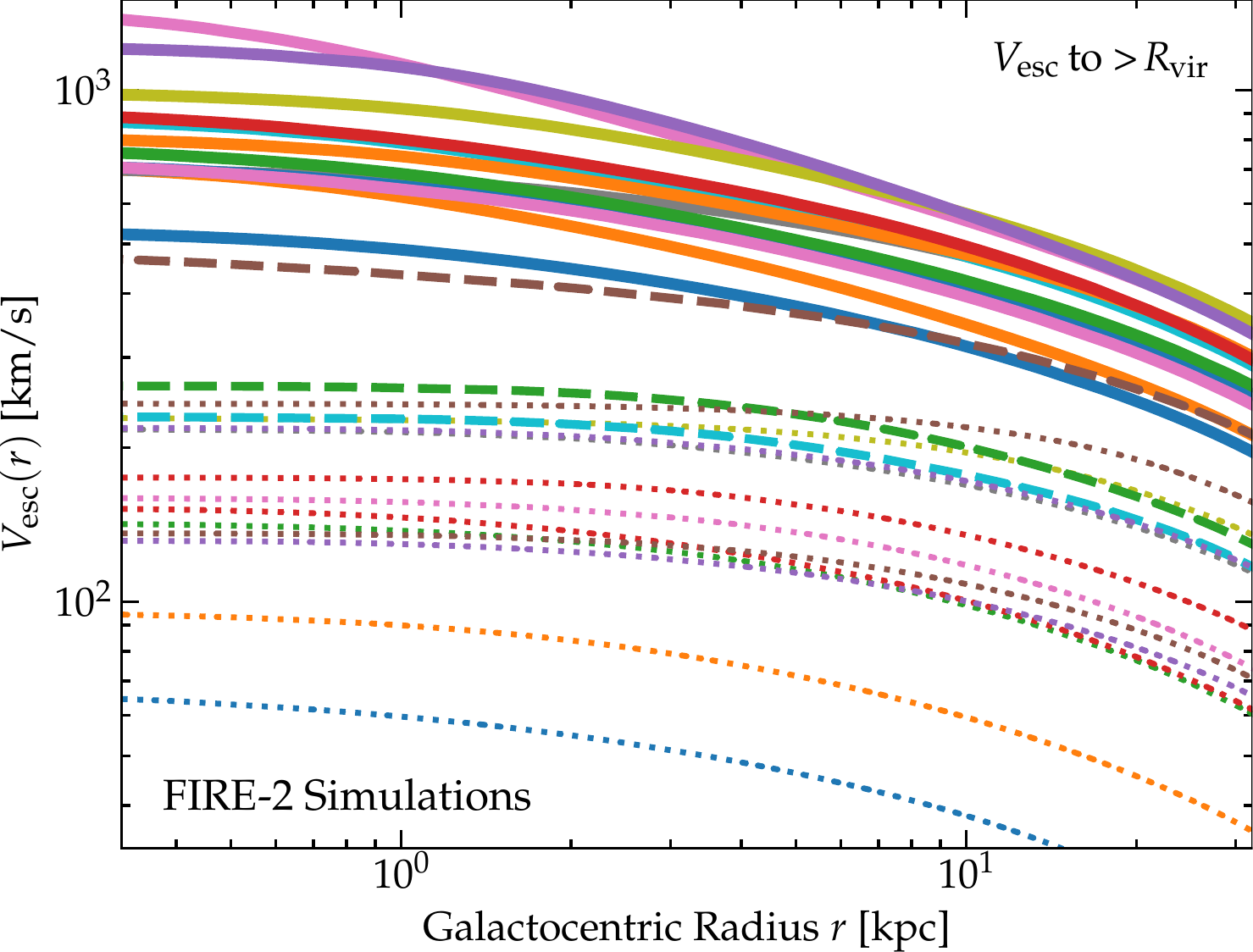}
	\includegraphics[width=0.95\columnwidth]{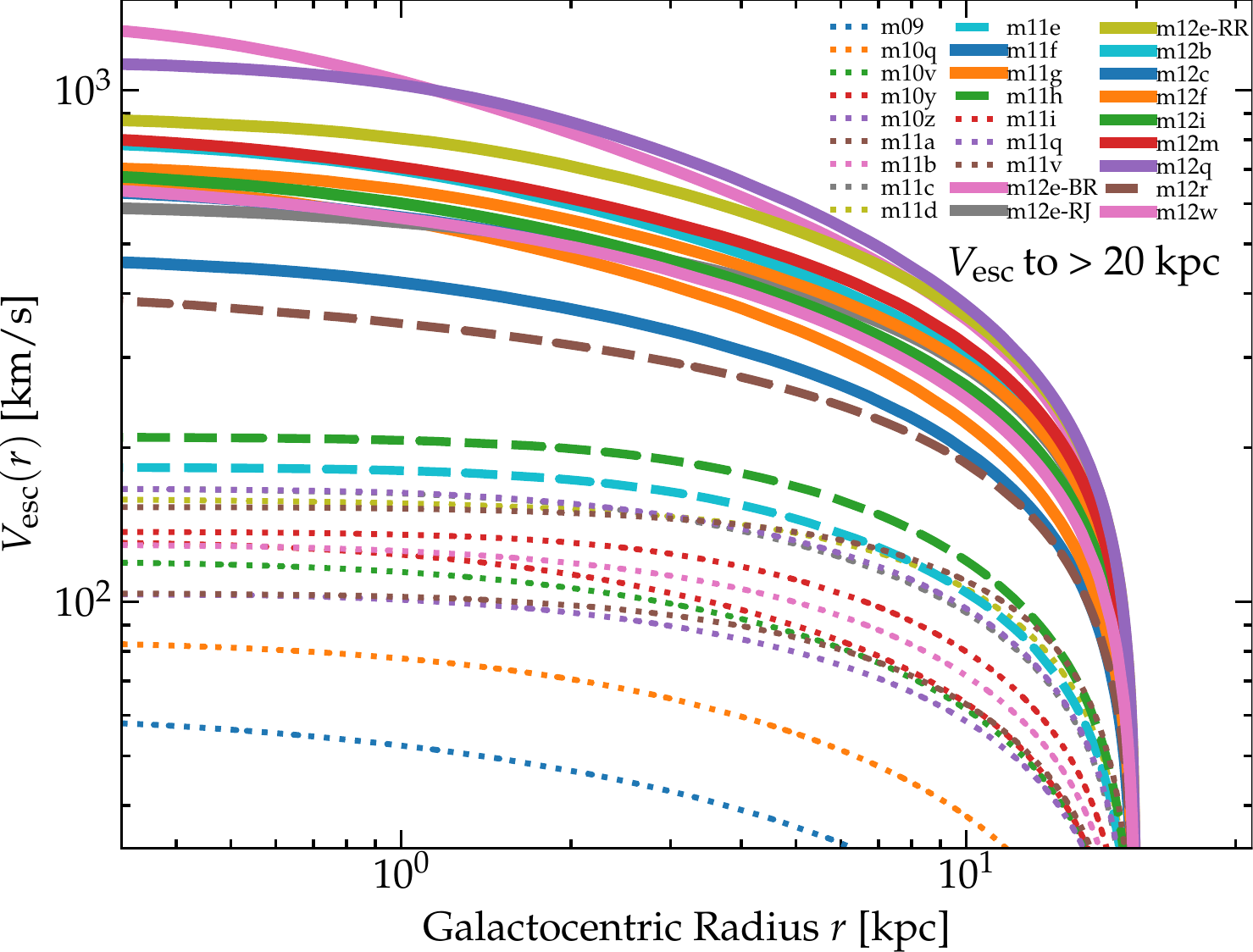}
	\includegraphics[width=0.97\columnwidth]{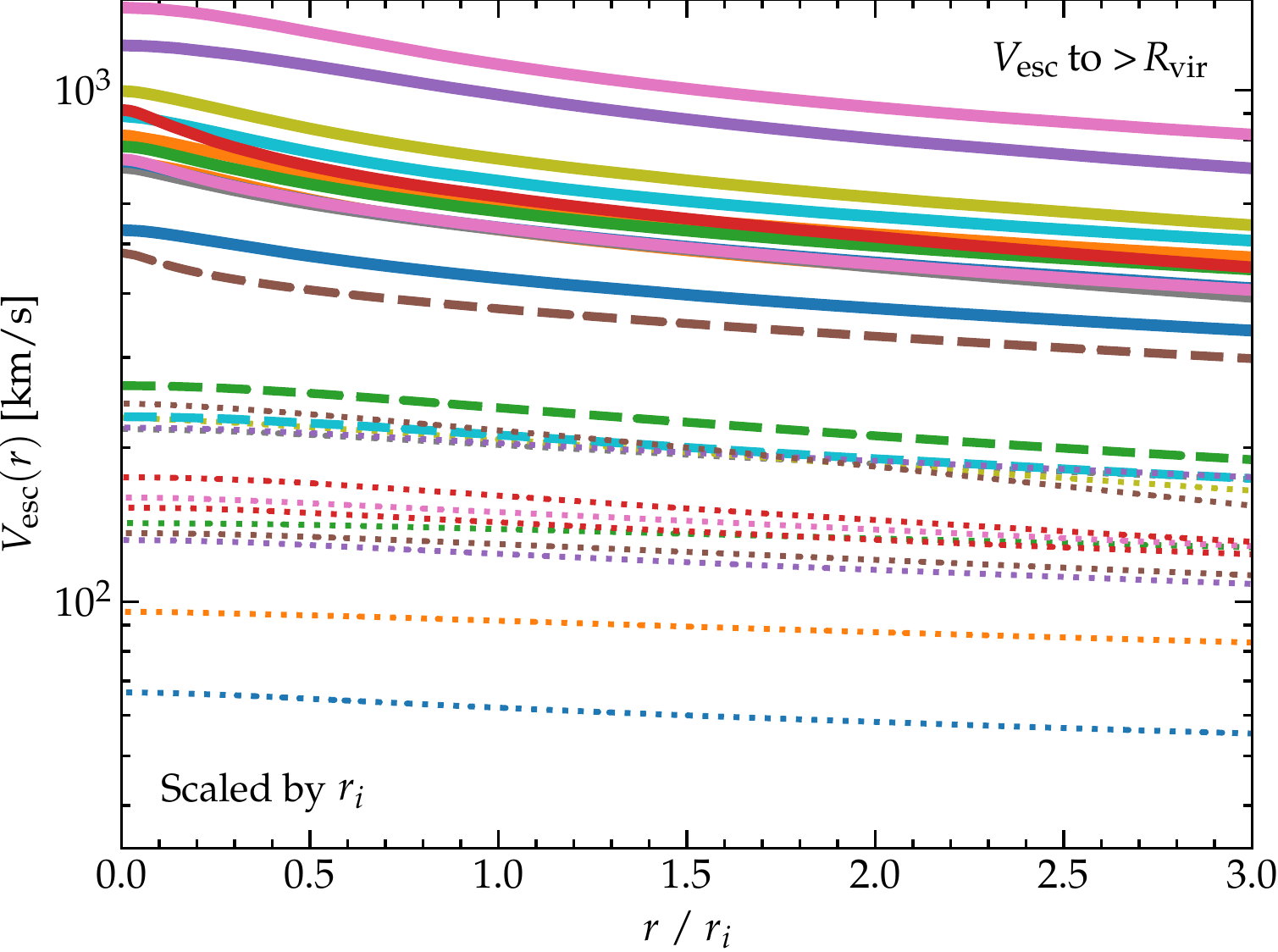} 
	\vspace{-0.1cm}
	\caption{Escape velocity curves as Fig.~\ref{fig:Vesc} but for all the ``default physics'' FIRE-2 simulations in \citet{hopkins:cr.mhd.fire2} at $z=0$, labeled by whether they exhibit clearly smooth SF ({\em solid}), intermediate ({\em dashed}), or bursty ({\em dotted}) SF. We compare $V_{\rm esc}$ to $>R_{\rm vir}$ ({\em top}) or $>20\,$kpc ({\em middle}), and versus absolute radius or scaled to the outer gas radius $r_{i}$ as in Fig.~\ref{fig:Vc.fire.disks} ({\em bottom}). see a similar behavior as for the {\bf m11a} experiments in Fig.~\ref{fig:Vesc}. 
	\label{fig:Vesc.fire.smooth}}
\end{figure}

\subsubsection{Validation in the FIRE Simulation Suite}

Fig.~\ref{fig:Vesc.fire.smooth} repeats the comparison of $V_{\rm esc}$ as Fig.~\ref{fig:Vesc}, but for our entire ``default'' FIRE suite. 
We see the same conclusion appears: $V_{\rm esc}$ appears to be an excellent proxy for the ``bursty'' or ``non-bursty'' status of the simulation SFH, moreso than any of the other parameters we study. Again, considering $V_{\rm esc}$ to more modest radii appears to provide slightly cleaner separation, compared to $V_{\rm esc}$ measured to the virial radius. Examining a broad range of snapshots and cosmic times (not shown here, for brevity), we find this difference is especially helpful for separating the more massive halo progenitors at high redshift, where $V_{\rm esc}$ to $\sim 10-20\,$kpc can be relatively small, and the systems are indeed bursty, but owing to the contribution of extended dark matter at large radii, $V_{\rm esc}$ at the virial radius or even larger can be significant (even though this has almost no influence on the gas dynamics, since gas expelled in the bursts does not travel so far).

\begin{figure}
	\includegraphics[width=0.95\columnwidth]{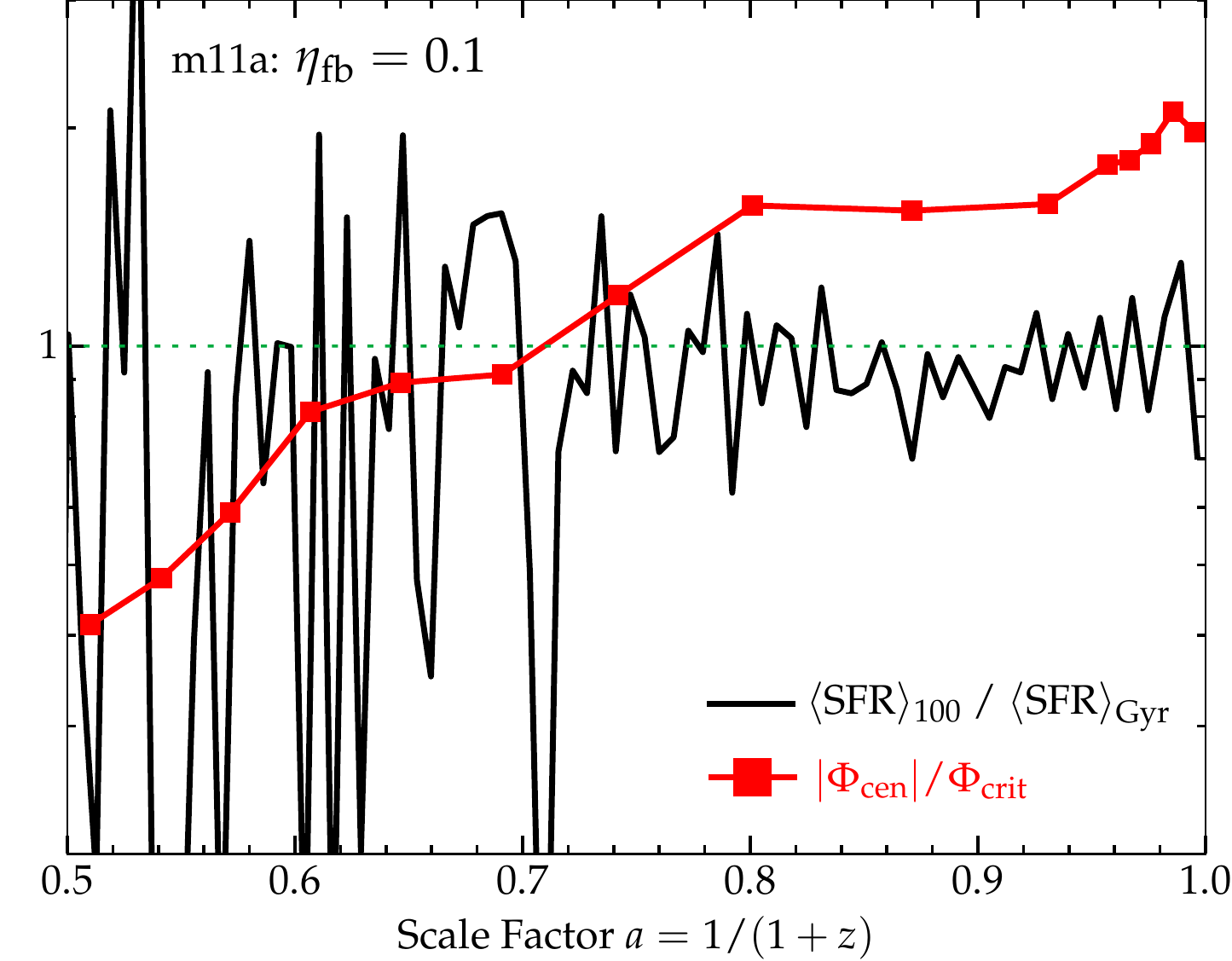}
	\vspace{-0.1cm}
	\caption{Evolution of the bursty-to-smooth star formation transition of a representative very-weak-feedback case in our {\bf m11a} experiments, $\eta_{\rm fb}=0.1$ (see Fig.~\ref{fig:weak.fb}). 
	We show star formation history (SFH; {\em black}), plotting the ratio of the SFR averaged in $\sim 100$\,Myr intervals to that in $\sim1\,$Gyr intervals as a way to show short-timescale fluctuations, versus scale factor $a$ for times after the experiment has begun ($z=1$, $a=0.5$). And we compare the central potential, $|\Phi_{\rm cen}| \equiv \langle V_{\rm esc}(r<1\,{\rm kpc})^{2} \rangle/2$ (averaged in the central $\sim 1\,$kpc) at each snapshot. As the galaxy builds up mass the potential increases rapidly. $\Phi_{\rm cen}$ crosses the critical value of $\Phi_{\rm crit} \sim (220\,{\rm km\,s^{-1}})^{2}/2$ (\S~\ref{sec:vesc.bursty}) around $a \sim 0.75$  and the SFR becomes visibly smoother.
	\label{fig:sfr.vesc.time.weak.fb}}
\end{figure}

\begin{figure}
	\includegraphics[width=0.99\columnwidth]{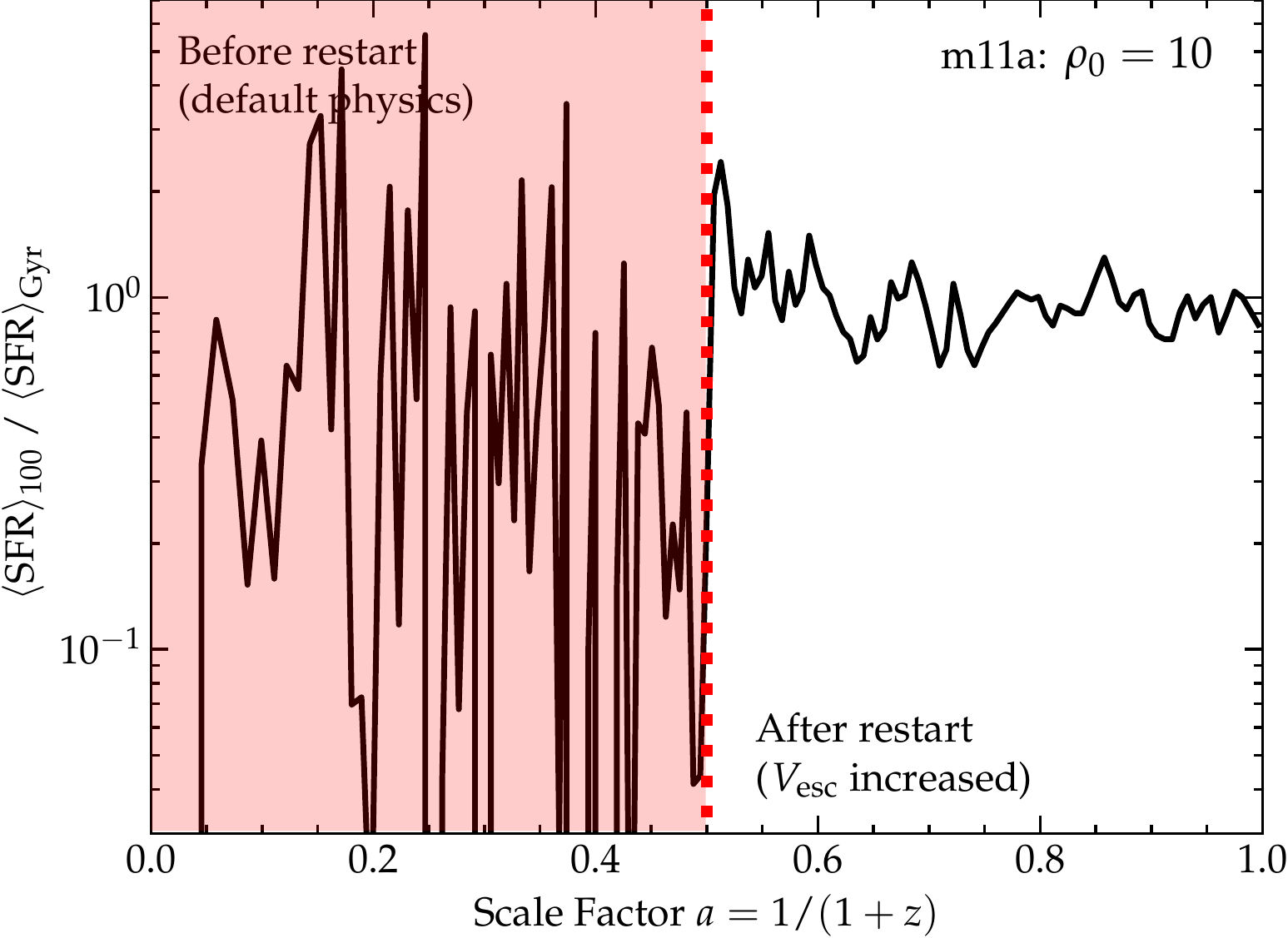}
	\vspace{-0.1cm}
	\caption{Normalized star formation history (rolling $100\,$Myr average divided by $1\,$Gyr average, as Fig.~\ref{fig:sfr.vesc.time.weak.fb}) versus scale factor $a$, showing {\em all} times before and after the time when we modify the physics of our simulation ($a=0.5$; {\em red dotted}) by imposing an additional constant-density central potential here with $\rho_{0}=10$ (in our {\bf m11a} experiments). This increases the central $V_{\rm esc}$ from $\sim 100\,{\rm km\,s^{-1}}$ to $\sim 400\,{\rm km\,s^{-1}}$: we see an almost immediate transition to smooth SF (in contrast to the long time required for the potential to build up and burstiness decrease in the weak-feedback case of Fig.~\ref{fig:sfr.vesc.time.weak.fb}).
	\label{fig:sfr.smooth.rhomodel}}
\end{figure}

\subsubsection{Example Time Evolution: Non-Linearly Suppressing Bursty SF Via Weak Feedback}

As with our disk tests in Fig.~\ref{fig:weak.fb}, we noted above that our weakest-feedback simulations eventually do transition to ``smooth'' SF at later times, owing to the nonlinear effects of star formation changing the potential. We test this in Fig.~\ref{fig:sfr.vesc.time.weak.fb}, where we plot the SFH in normalized scale (making it easier to see the ``bursty'' versus ``smooth'' behavior) as a function of time for our weakest-feedback $\eta_{\rm fb}=0.1$ case, alongside the central potential or escape velocity (mass-weighted average for gas in the central $\sim1$\,kpc, roughly the effective radius of star formation). We see that the system does not transition immediately to smooth SF, as we would expect if this were purely a function of feedback efficiency: instead, bursty SF persists, but the SFR is very high in absolute units as we see in Fig.~\ref{fig:sfhs.weak.bursty.fx} (as expected given the weak feedback), so the central potential rapidly builds up as the baryon density (mostly stars) accumulates, and the SFH becomes smooth once the potential crosses the critical threshold we anticipate.

Contrast Fig.~\ref{fig:sfr.smooth.rhomodel}, which shows the same for our $\rho_{0}=10$ model, where because we impose the modified potential, the potential and escape velocity go from well below the ``critical'' value to well above the critical value almost instantaneously upon our restart. As a result, we see a corresponding almost instantaneous change in the SFH from bursty to smooth.

\subsubsection{Tests at Other Mass Scales}
\label{sec:bursty.tests.m12.m10}

We can further test this at different mass scales: as with our disk tests in \S~\ref{sec:disk.test.vs.mass}, we have repeated our experiments re-starting simulations {\bf m10q} (a dwarf with halo mass $<10^{10}\,M_{\odot}$) and {\bf m12i} (a Milky Way-mass system). We have re-simulated both adding a constant-density potential with $\rho_{0}=10$, and varying the cooling rates.

Galaxy {\bf m10q}, being another dwarf with naturally ``bursty'' SF in our default runs, shows very similar behavior to {\bf m11a} in basically every experiment, and adding a large-$\rho_{0}$ potential produces smooth SF akin to our {\bf m11a} tests, while changing the cooling time has no effect on burstiness.

The more interesting case is {\bf m12i}, as it has a much higher SFR, gas density, virial temperature, and is closer to ``naturally'' crossing the burst-smooth threshold on its own (in our ``default'' runs). In fact Fig.~\ref{fig:more.smooth.sfh.tests.cooling} shows it does transition from bursty to smooth, as defined by \citet{yu:2021.fire.bursty.sf.thick.disk.origin} or \citet{gurvich:2022.disk.settling.fire}, at this resolution at around $z\approx 1$.\footnote{We choose to use a different averaging timescale for this run because it makes the differences between runs more clear, but the results are robust to whatever timescales we average the SFR over.} This corresponds very closely to when the escape velocity $V_{\rm esc}$ to $10$ or $20\,$kpc crosses the critical value (see Fig.~\ref{fig:Vesc.fire.smooth}). Adding the potential with $\rho_{0}=10$ introduces this transition almost immediately, even at $z=2$, when the default run has not yet transitioned. For this run, in our simulations varying $\eta_{\rm cool}$ and related parameters, we explicitly quote the value of 
 $\Psi^{(s)}_{\rm cool} \equiv t_{\rm cool}^{(s),\,0.1\,R_{\rm vir}}/t_{\rm ff}$ from \citet{stern21_ICV} -- the ratio of the cooling time to free-fall time of CGM gas at $0.1\,R_{\rm vir}$ as it ``would be'' if the gas were in spherical hydrostatic pressure equilibrium assuming the exact simulation mass and metallicity and gas density profile ``as-is.'' This tends to systematically rise with time as the potential gets deeper, so we show the range of values over which the SFH is plotted (from $z=2$ at restart to $z=0$). We see in the default run that the transition from ``bursty'' to ``smooth'' SF happens to coincide with $\Psi^{(s)}_{\rm cool}$ rising above a value $\sim1-2$; but by reducing $\eta_{\rm cool}$, we clearly see that even if we make $\Psi^{(s)}_{\rm cool}$ very large, we do not automatically introduce a transition to smooth SF. In fact, the simulations with small $\eta_{\rm cool}$ (the largest $\Psi^{(s)}_{\rm cool}$), because they have lower SFRs, do not form many stars or deepen the potential so they remain below the critical $V_{\rm esc}$ for the entire duration of the re-simulation: thus they end up non-linearly being {\em more bursty} than the ``default'' run.
 
For completeness, we also repeated the experiment with $\eta_{\rm cool}=0.01$ in {\bf m12i}, but also multiplying the feedback rates by $\eta_{\rm fb}=0.2$, which in combination gives approximately the same SFR as our ``default'' simulation. This gives results similar to that ``default'' simulation: $\eta_{\rm cool}$ does not cause the SFR to become smooth, though it does eventually become so at the same time as our default run once sufficient stars form to deepen the potential and exceed the critical $V_{\rm esc}$.

\begin{figure*}
	\includegraphics[width=0.93\textwidth]{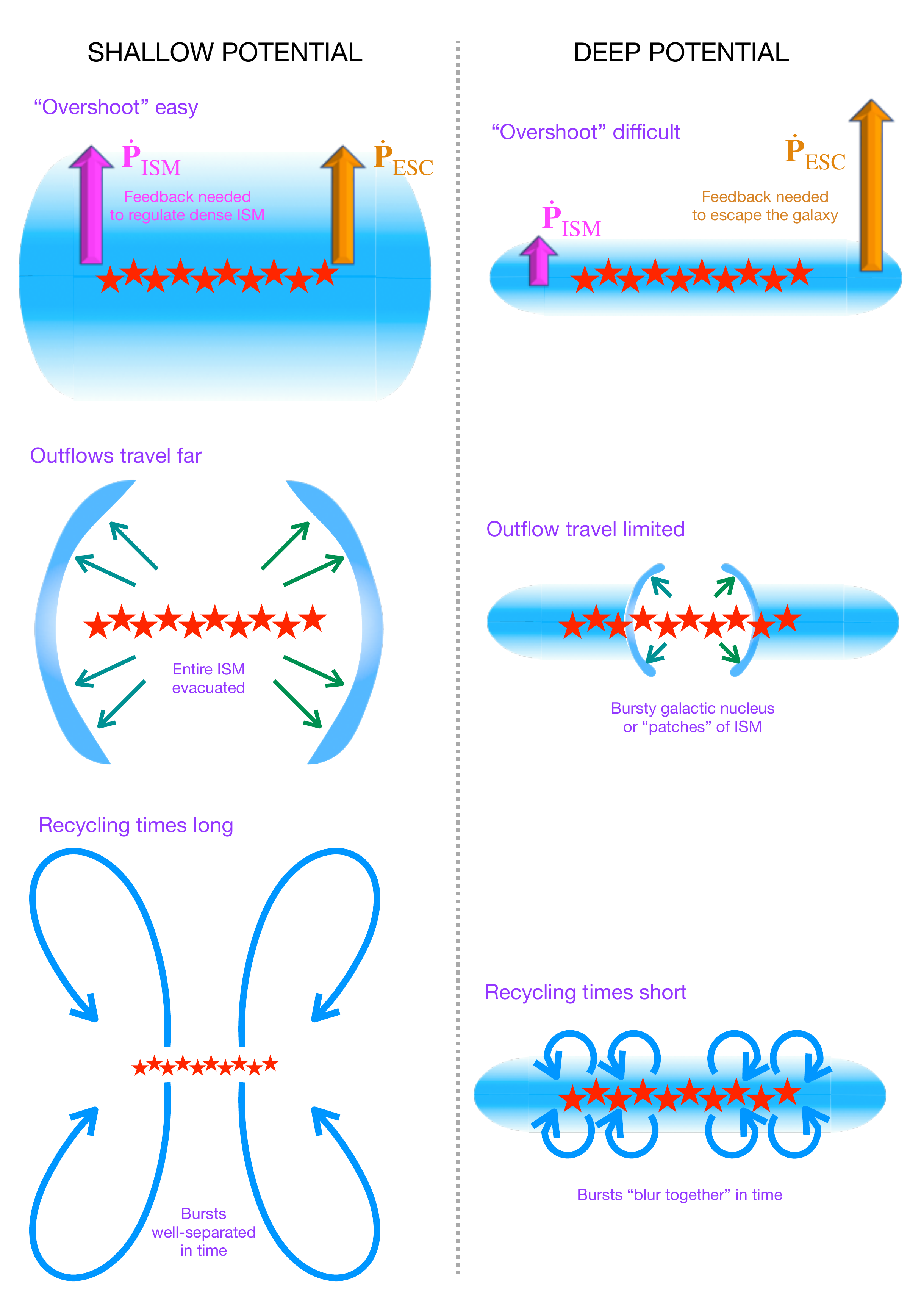}
	\vspace{-0.2cm}
	\caption{Cartoon illustrating the physics of smooth versus bursty star formation, per \S~\ref{sec:bursty}, illustrating some of the key reasons why sufficiently deep potential wells suppress bursty star formation by preventing ``overshoot'' (\S~\ref{sec:overshoot.tests}) and ejection of the ISM with long recycling times. Blue shaded material and arrows represent gas under the influence of gravity plus feedback from young, massive stars (represented by the red stars).
	\label{fig:cartoon.potential}}
\end{figure*}

\subsection{How Does the Potential Scale Suppress Burstiness?}
\label{sec:bursty.physics}

It is not surprising, in principle, that a deeper potential would produce less bursty star formation by suppressing the ability of the galaxy to eject most of its gas supply (as illustrated in Fig.~\ref{fig:cartoon.potential}), and indeed qualitatively similar effects have been seen in many other simulations going back more than a decade: see e.g.\ the experiments in \citet{stinson:2007.dwarf.gal.sims.burst.sfh}, who found weaker bursts in deeper potentials. Material accelerated to some velocity below $V_{\rm esc}$ will turn around and recycle in galactic fountains instead of escaping to large radii, again illustrated in  Fig.~\ref{fig:cartoon.potential}. And in toy models for continuous energy or momentum-conserving winds, the outflow energy or momentum-flux requirement scales as $\dot{E}_{\rm out} \sim \dot{M}_{\rm out}\,V^{2}_{\rm esc}(r_{\rm launch})/2$, $\dot{P}_{\rm out} \sim \dot{M}_{\rm out}\,V_{\rm esc}(r_{\rm launch})$, respectively. Moreover, there are related effects which can further enhance confinement: if the CGM is in quasi-virial equilibrium (as we see), then the pressure rises as $P \sim \rho\,V_{\rm esc}^{2}$ and the work required to launch an outflow increases as $\sim M_{\rm gas,\,cgm}\,V_{\rm esc}^{2}$, which can further confine gas within or near the galaxy \citep[][]{muratov:2015.fire.winds,muratov:2016.fire.metal.outflow.loading,hopkins:2020.cr.outflows.to.mpc.scales}. 

What is less obvious is why this would be the dominant criterion we see, why (upon further reflection) these effects could not simply be offset with ``more feedback'' (i.e.\ bursty SF with a different normalization), and why there should be any kind of threshold behavior (instead of a smooth continuous dependence). 

One idea, which seems plausible, stems from the fact that the critical $V_{\rm esc}$ we find is suspiciously close to the ``cooling velocity'' $v_{\rm cool}$ of a standard energy-conserving bubble (shell speed at which e.g.\ an expanding SNe remnant will cool). This suggests a picture where once $V_{\rm esc} \gtrsim v_{\rm cool}$, hot gas (which has not entrained sufficient mass to decelerate and cool) can still ``vent'' out of the galaxy smoothly (explaining the non-negligible remaining fast outflows in e.g.\ Fig.~\ref{fig:distance.blown.out}), but cold gas (which has cooled/decelerated below $v_{\rm cool}$ and therefore $V_{\rm esc}$) cannot. So venting will ``leak'' feedback, but most of the star-forming gas mass will stay in the galaxy. This is similar to the picture suggested in \citealt{muratov:2015.fire.winds,sparre.2015:bursty.star.formation.main.sequence.fire} for MW-mass halos after they have ``settled.''\footnote{Nominally this might appear sensitive to thermodynamics or feedback physics, but for any Sedov-Taylor-like expansion, if we solve for $v_{\rm cool}$ as the velocity of the energy-conserving flow where $t_{\rm cool} = t_{\rm expansion}$ (here $t_{\rm cool} \propto k_{\rm B}\,T / n,\Lambda$; $t_{\rm expansion} \propto r / v$; with $E \propto M_{\rm swept}\,v^{2}\sim$\,constant), for any $\Lambda$ which is not an extremely-strong function of temperature (i.e.\ any reasonably physical $\Lambda$), the resulting $v_{\rm cool}$ is an extremely weak function of the ambient properties, injection energy, and thermodynamics. For example, if $\Lambda$ were a constant, we get $v_{\rm cool} \propto E_{51}^{1/11}\,n_{1}^{2/11}\,\Lambda^{3/11}$ (in terms of the injection energy $E_{0}$ and ambient density $n$). For a more realistic cooling curve shape around the energies/temperatures of interest here, we get an even weaker dependence: inserting numbers gives $v_{\rm cool} \sim 200\,{\rm km\,s^{-1}}\,E_{51}^{0.06}\,n_{1}^{0.1}\,\Lambda_{-22}^{0.2}$ \citep[see e.g.][]{iffrig:sne.momentum.magnetic.no.effects,martizzi:sne.momentum.sims,walch.naab:sne.momentum,gentry:sne.momentum.boost}. So even order-of-magnitude variations to $\Lambda$ produce modest variations at most in $v_{\rm cool}$. In fact, almost all our simulations with different $\eta_{\rm cool}$ modifying $\Lambda$ (or $\eta_{\rm gas}$ modifying $n$, or $\eta_{\rm fb}$ modifying $E_{51}$) would be insufficiently-large variations to shift the threshold $V_{\rm esc}$ into the values we measure for otherwise ``default'' {\bf m11a} properties, potentially explaining why they had weak effects there.}

\begin{figure}
	\includegraphics[width=0.97\columnwidth]{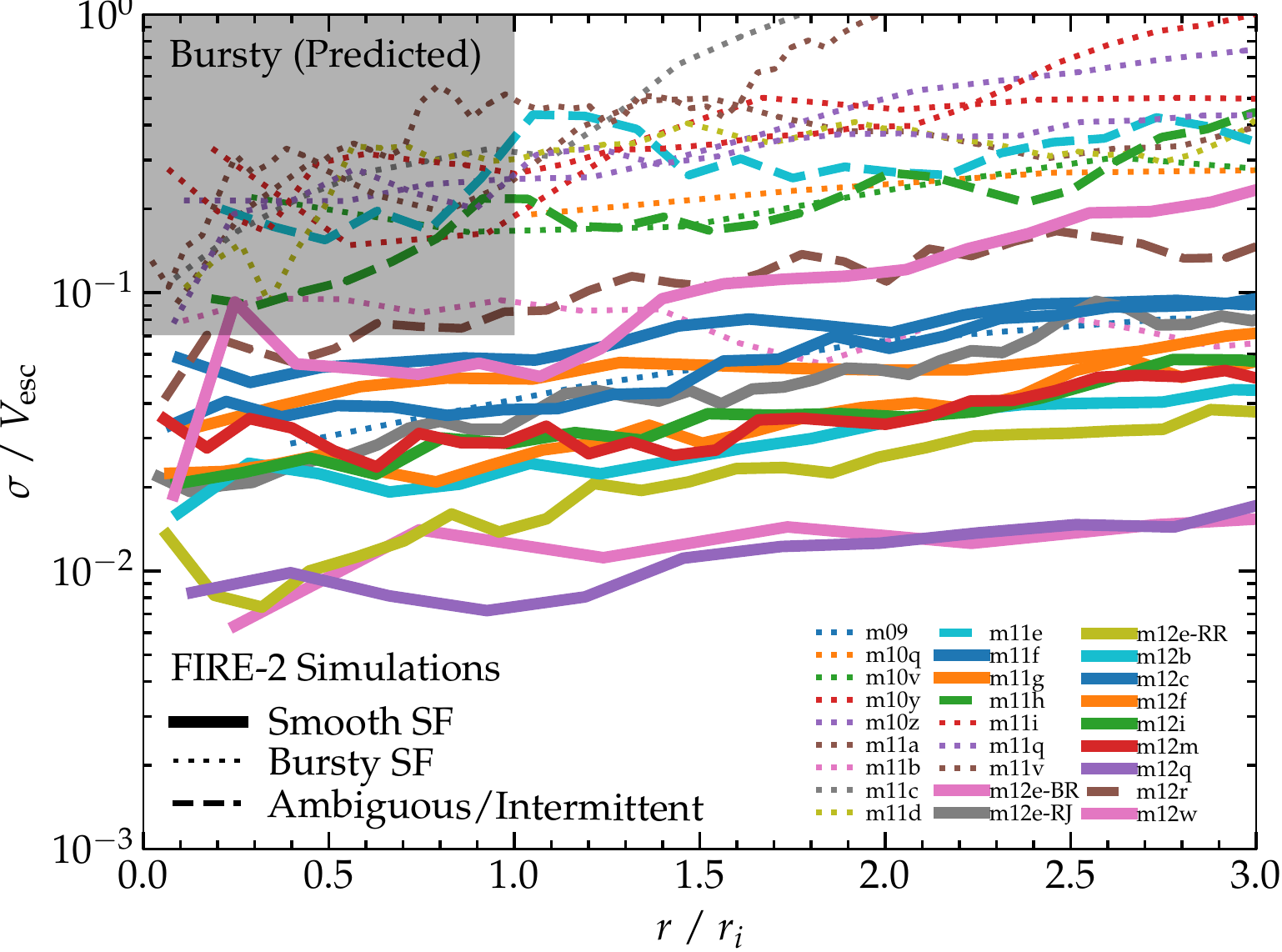} 
	\vspace{-0.1cm}
	\caption{The ratio of $\sigma/V_{\rm esc}$, for all the  ``default physics'' FIRE-2 simulations labeled by their smooth/bursty SFHs (as Fig.~\ref{fig:Vesc.fire.smooth}). The ``overshoot'' model of \citet{hayward.2015:stellar.feedback.analytic.model.winds} (\S~\ref{sec:bursty.physics}) predicts this parameter (not necessarily the absolute value of $V_{\rm esc}$ alone) should discriminate between bursty and smooth SF, where systems in the shaded range ($\sigma/V_{\rm esc}>0.07$ in the star-forming ISM, defined here as $r \lesssim r_{i}$) would be bursty. The model separation is not quite as clean as in Fig.~\ref{fig:Vesc.fire.smooth} and is primarily driven by $V_{\rm esc}$, but the value predicted does appear to match the separation in the simulations.
	\label{fig:sigma.Vesc.fire.smooth}}
\end{figure}

A second idea comes from \citet{hayward.2015:stellar.feedback.analytic.model.winds}, and is tested in Fig.~\ref{fig:sigma.Vesc.fire.smooth} and illustrated heuristically in Fig.~\ref{fig:cartoon.potential}. They argue that if stellar feedback drives gas motions which produce density fluctuations (e.g.\ turbulence, shocks, with some dispersion $\sigma$) with momentum input rate $\dot{P} \sim \langle p / m_{\ast} \rangle\,\dot{M}_{\ast}$ (where $\langle p/m_{\ast} \rangle \sim 3000\,{\rm km\,s^{-1}}$ is the SSP-lifetime-integrated momentum per unit mass; see \citealt{ostriker.shetty:2011.turb.disk.selfreg.ks}) balancing dissipation, then those density fluctuations naturally produce under and over-dense ``patches'' of the ISM, the less-dense of which will be ejected by said momentum flux (in a momentum-driven outflow, i.e.\ momentum $\sim M_{\rm gas,\,patch}\,V_{\rm esc}$). While the expressions derived therein for e.g.\ star formation rates and outflow rates/mass loadings are rather complicated, the authors argue that the fraction of the ISM mass which will be ejected in a dynamical time (their ``$f_{\rm out}$'' parameter) follows a threshold behavior with the simple parameter $\sigma/V_{\rm esc}$.\footnote{Note that in \citet{hayward.2015:stellar.feedback.analytic.model.winds}, the authors explicitly state that it is the escape velocity which enters their scalings used here. However for analytic simplicity they then simply assumed $V_{\rm esc} \sim \sqrt{2}\,V_{\rm c}$ for some characteristic $V_{\rm c}$. Since the distinction is important here, we explicitly use $V_{\rm esc}$ where their derivation did indeed involve the escape velocity.} Specifically, for $\sigma/V_{\rm esc}$ larger than some threshold value, $f_{\rm out} \sim 1$, corresponding to a ``blowout'' or total ejection, while (because of the lognormal-like scaling of compressive density fluctuations) for $\sigma_{z}/V_{\rm esc}$ below a threshold value $\sim 0.07$\footnote{This comes from taking their quoted value $\sqrt{2}\,\sigma_{\rm 3D}/(V_{\rm esc}/\sqrt{2}) \sim 0.3$ and converting to our $V_{\rm esc}$ and 1D $\sigma=\sigma_{z}$.} this is exponentially suppressed. We can also immediately extend their argument to an energy-driven outflow (as they discuss, though they argue this is likely less relevant), which multiplies their criterion by $(\langle \epsilon/m_{\ast} \rangle / \langle p/m_{\ast} \rangle) /  V_{\rm esc} \sim v_{\rm cool} / V_{\rm esc} $ (where $ \langle \epsilon/m_{\ast} \rangle \sim (10^{51}\,{\rm erg}/(100\,M_{\odot})) \sim v_{\rm cool}\, \langle p/m_{\ast} \rangle$ essentially by definition, since $v_{\rm cool}$ defines the efficacy of conversion of energy into momentum). Importantly, in these models, because it is the {\em same} feedback driving the SFR and outflows, the properties of the feedback itself factor out entirely from the relevant scalings.

We plot this in Fig.~\ref{fig:sigma.Vesc.fire.smooth} and find it gives a similar separation to $V_{\rm esc}$ between models, and it is worth noting that the bursty-smooth transition actually does appear to occur around the analytically predicted critical value. But we caution that most of the separation comes from $V_{\rm esc}$ itself (there is not a clear trend in $\sigma$), so we cannot say that our numerical experiments above favor this model over a simple absolute threshold in $V_{\rm esc}$. But we explore it further below.

\subsubsection{Discriminating Between these Possibilities: The Absolute Velocity Scale of Feedback and Cooling Velocity Do Not Appear Critical}

In order to discriminate between these two hypotheses, we have therefore considered a number of additional experiments. 

First, to explore the role of the ``cooling velocity'' $v_{\rm cool}$, we note that especially in our low-resolution massive halos (e.g.\ {\bf m12i}, re-started from $z=2$ at low resolution $\sim 6\times10^{4}\,M_{\odot}$ as we did in \S~\ref{sec:bursty.tests.m12.m10} above) that the cooling radii of SNe are rarely resolved. Instead, $v_{\rm cool}$ appears largely {\em implicitly} in our SNe prescription as it determines the ``terminal momentum'' which is adopted analytically to calculate the thermal-to-kinetic energy ratio of coupled SNe momentum in the limit where this is locally unresolved \citep[see][for details]{hopkins:sne.methods}. We therefore consider re-starts of both {\bf m11a} and {\bf m12i} varying the ``terminal momentum'' by an order of magnitude in either direction (equivalent to assuming $v_{\rm cool}$ varies by in order of magnitude in the opposite direction). We have also considered experiments where we explicitly disable any thermal energy deposition in SNe (making it purely ``cold,'' i.e.\ kinetic feedback). 

More radically, we also considered a model where we discretize the local SNe into fixed-velocity probabilistic ``kicks'' with a fixed characteristic velocity $\Delta v_{\rm target}$, in order to test whether there was a clear transition from bursty to smooth SF as we lowered  $\Delta v_{\rm target}$ below some critical value. Specifically in these tests when we would otherwise couple some momentum $\Delta {\bf p}_{j}$ to a gas cell $j$, we instead probabilistically couple either nothing or a discrete momentum $\Delta {\bf p}_{j}^{\rm target} = m_{j}\,\Delta v_{\rm target}\,\Delta \hat{\bf p}_{j}$ in the same direction with a probability $p \equiv |\Delta {\bf p}_{j}|/|\Delta {\bf p}_{j}^{\rm target}|$ (ensuring the same momentum deposition per unit time in an average sense), where we vary $\Delta v_{\rm target}$ between $\sim 10-1000\,{\rm km\,s^{-1}}$.

Although these experiments can change the absolute efficacy of feedback (hence the SFR), we find that {\em none} of them qualitatively changes the system from bursty to smooth SF in {\bf m11a} (or otherwise appreciably influences ``how bursty'' the SF is). This appears to contradict the hypothesis that there is an absolute critical velocity scale of $V_{\rm esc} \sim 200\,{\rm km\,s^{-1}}$ set either by the cooling velocity or some other physics.

\begin{figure}
	\includegraphics[width=0.95\columnwidth]{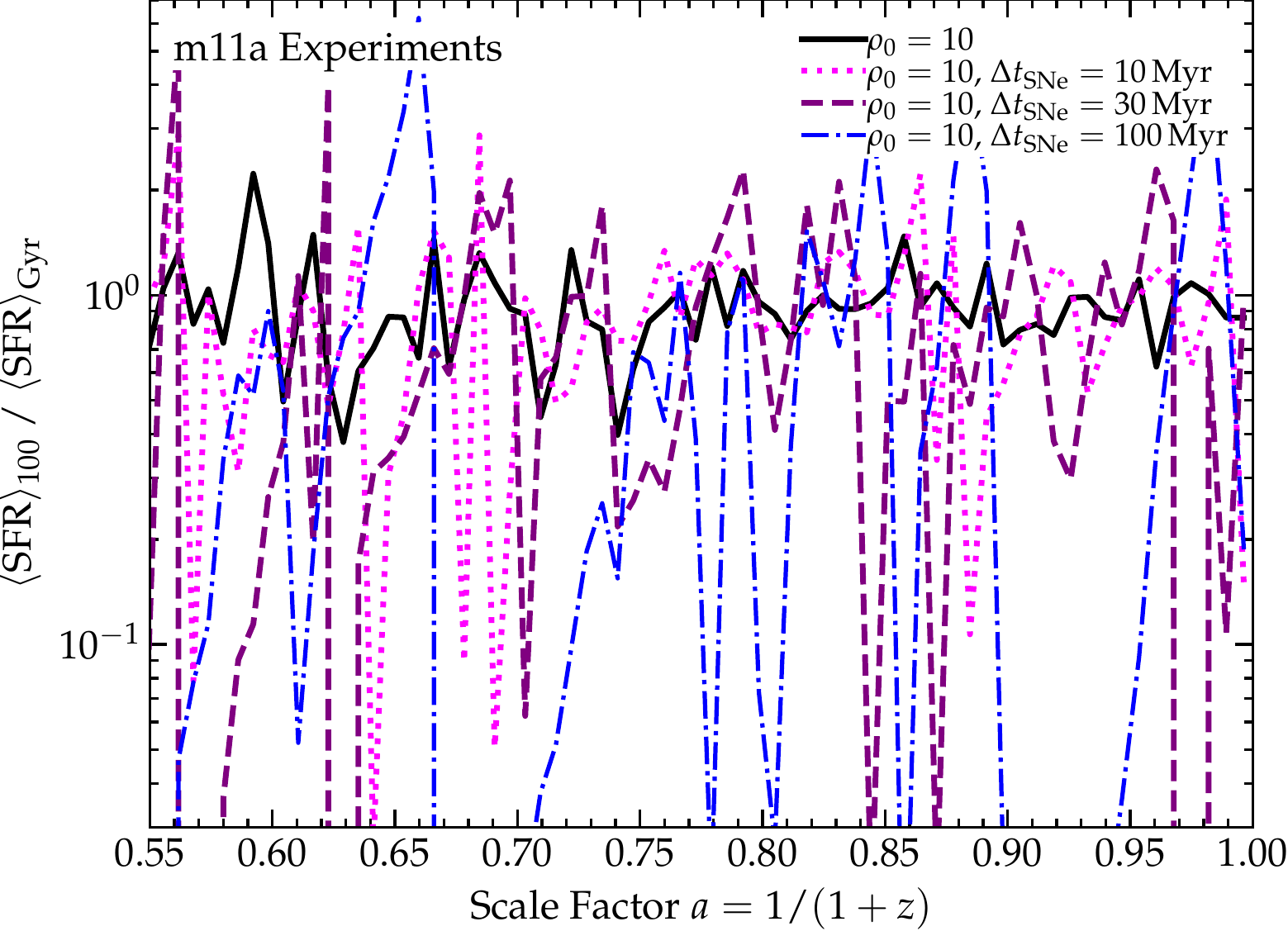}
	\includegraphics[width=0.95\columnwidth]{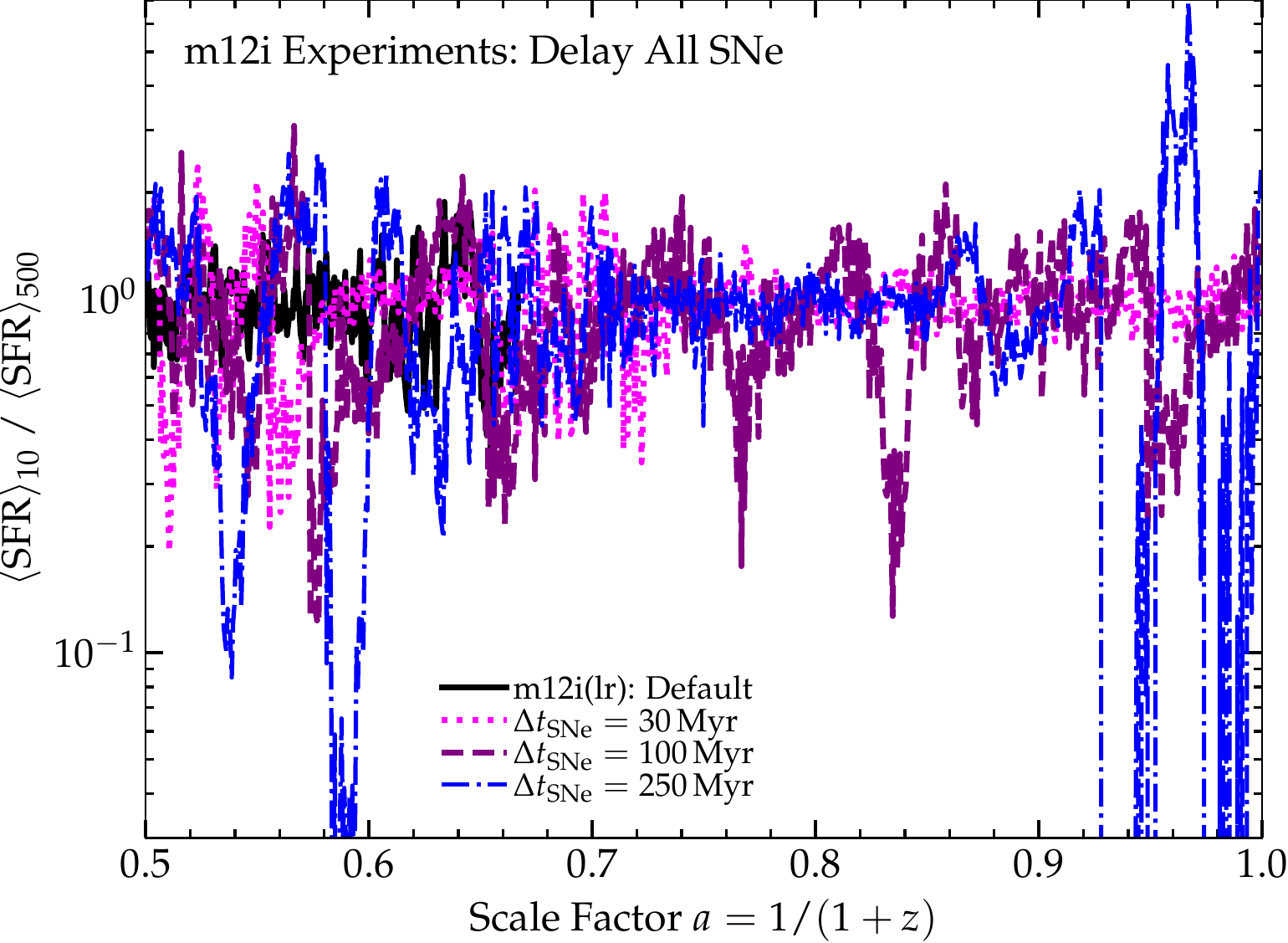}
	\vspace{-0.1cm}
	\caption{Experiments (\S~\ref{sec:overshoot.tests}) which demonstrate how feedback produces bursty SF via ``overshoot'' wherein SF in the galaxy begins to self-regulate or shut down but has some probability of ejecting material completely, producing a quench cycle. We test this by considering runs which would otherwise give {\em smooth} SF, but shift the SNe delay time distribution by a uniform delay $\Delta t_{\rm SNe}$, which (if large enough) prevents small-scale self-regulation.	
	{\em Top:} {\bf m11a} tests with the added potential $\rho_{0}=10$, but (optionally) some $\Delta t_{\rm SNe}$: the ``normal'' version of this run produces smooth SF (unlike the default {\bf m11a} which is bursty), but we can make the SF bursty once more by also adding a significant $\Delta t_{\rm SNe}$.
	{\em Bottom:} {\bf m12i} tests. The default version and those with sufficiently small $\Delta t_{\rm SNe} \ll 30\,$Myr produce smooth SF over this range of times, but adding larger $\Delta t_{\rm SNe}$ makes SF bursty.
	\label{fig:sfh.vs.delay}}
\end{figure}

\subsubsection{Discriminating Between these Possibilities: The  \citet{hayward.2015:stellar.feedback.analytic.model.winds} ``Overshoot Probability'' Appears Critical}
\label{sec:overshoot.tests}

But none of these tests contradict the \citet{hayward.2015:stellar.feedback.analytic.model.winds} hypothesis, because these properties of feedback ``factor out'' of that calculation. This is just one example of a broad category of ``overshoot'' models which, in their simplest form, argue that feedback will {\em locally} cause SF to cease at some level in the ISM (with, of course, this ``self-regulated'' SFR or young stellar mass inversely proportional to the feedback efficiency, for given ISM properties),\footnote{Here we refer to the feedback on $\sim$\,kpc ISM scales, still largely dominated by SNe, and distinct in principle from e.g.\ feedback regulating collapse in very dense gas such as HII regions.} but this feedback has some probability of ``overshooting'' and providing enough momentum to eject most of the ISM into an outflow at $V_{\rm esc}$ out to some radius a few times larger than its present location, leading to a dramatic suppression of SF (with the ``required'' amount of SF or young stars proportional to $V_{\rm esc}$ and inversely proportional to the feedback efficiency). Crudely, one can think of the ``overshoot probability'' as scaling with the ratio of the ``feedback required to locally regulate ISM SF'' (the ``numerator'') to the ``feedback required to eject'' (the ``denominator''), as illustrated in  Fig.~\ref{fig:cartoon.potential}. Since it is the ``same'' feedback doing both, the feedback properties themselves factor out. In the \citet{hayward.2015:stellar.feedback.analytic.model.winds} model, they assume that since it is specifically the dense, cool, {\em star-forming} gas of interest for the SFR variability, the feedback of interest is momentum-conserving, and that something like $\sigma_{\rm eff}/\Omega\sim (c_{\rm s}^{2} + \sigma_{\rm turb}^{2})^{1/2}/\Omega$ traces the pressure gradient scale length of the ISM, which gives a ratio or overshoot probability scaling proportional to $\sim \sigma_{\rm eff}/V_{\rm esc}$. 

Our previous tests are consistent with this, but primarily varied the ``denominator'' ($V_{\rm esc}$) by adding extended potentials. To test this class of models further, since simply varying the overall feedback strength/momentum/energy or rates will not change this overshoot probability, we desire a way to divorce the strength of feedback driving galactic outflows (i.e.\ SNe) from local self-regulation/star formation. 

We therefore consider the following experiment: we set the SNe rate ``by hand'' (instead of allowing it to self-regulate) by turning off SNe from stars which form after the {\bf m11a} simulation restart (at $z<1$), and assigning a fixed specific SNe rate $\dot{N}_{\rm SNe}/m_{\ast}$ to the {\em old} stars formed before $z>1$.\footnote{We still allow the young stars to act via radiative feedback and O/B or AGB mass-loss (this does not change our conclusions if we disable it as well but allows the SF to self-regulate in very dense clumps, preventing some pathological cases).} We do this with an arbitrary duty cycle by turning the SNe on for some duration $\Delta t_{\rm on}$ then off for $\Delta t_{\rm off}$, etc. For an ``always on'' model or during the ``on'' phases with some duty cycle, we see a clear threshold behavior: above a critical SNe rate, feedback injection exceeds gravity and the entire ISM is expelled explosively, quenching the galaxy. Below the critical rate, or during ``off'' cycles (provided the earlier feedback is not so strong that it blows everything out of $R_{\rm vir}$ and prevents any re-accretion), even at just a factor $\sim 2$ lower injection rate, the injection is weaker than gravity, gas collapses, and star formation occurs in a dense ($\lesssim 200\,$pc) nuclear gas concentration, at a rate limited by the cooling of new gas onto the galaxy ($\sim 0.3-1\,{\rm M_{\odot}\,yr^{-1}}$, comparable to the rate seen in our weakest-feedback $\eta_{\rm fb} \lesssim 0.1$ runs). Importantly, however, unlike even the lowest $\eta_{\rm fb}$ runs (in which the SFR remained bursty until sufficient stellar mass builds up, as we saw in Fig.~\ref{fig:sfr.vesc.time.weak.fb}), the SFR almost immediately transitions in these ``weak'' cases to smooth SF. These results may seem obvious, but they confirm several important assumptions: (1) the ``bursty'' SF cycles are fundamentally {\em feedback-driven}, not the direct causal result of variations in accretion rates or cooling or other galaxy (thermo)dynamics; (2) that stellar feedback driving the ``quenching'' in burst-quench cycles is {\em dominated by SNe}, not ``early'' (radiative+O/B mass-loss) feedback; (3) the ``critical'' SNe rate corresponds closely to the {\em time-averaged} value in our ``default'' simulation (here $1\,$SNe per $\sim 10^{4}\,{\rm yr}$), as expected if the default model is self-regulating; and (4) there is a {\em sharp threshold behavior}, i.e.\ at least in dwarfs which would naturally be ``bursty'' like this particular {\bf m11a} experiment, there is almost no ``middle ground'' between ``insufficient SF to self-regulate'' and ``expelling the ISM'' (so the ``overshoot probability'' must be $\mathcal{O}(1)$ in the ``default'' runs). 

Another way to test the ``overshoot'' argument is to de-synchronize the feedback timescales, shown in Fig.~\ref{fig:sfh.vs.delay}. We run a series of experiments where we shift the SNe delay time distribution (DTD) by a fixed $\Delta t_{\rm delay} = (10,\,30,\,100,\,300)\,{\rm Myr}$ (so for e.g.\ a shift of $100\,$Myr, the first SNe will not occur until $\sim 103\,$Myr after a star particle forms). In the context of an ``overshoot'' model, the longer delays would mean CC SNe could not locally regulate collapse in the ISM on shorter timescales, so should push the system to more bursty SF. We therefore implement this in setups which would otherwise have {\em smooth} SF: we re-run our low-resolution {\bf m12i} at $z<1$ (where the default run has already transitioned to smooth SF) and an {\bf m11a} re-start with the added extended potential $\rho_{0}=10$ (which again made the SFR smooth, with the ``default'' SNe DTD). We see in both cases that this {\em ``de-synchronization'' is able to re-introduce bursty SF}. The {\bf m11a} runs look bursty for all of the delay values tested; for even a $\sim30\,$Myr delay the {\bf m12i} run goes from having a ``smooth'' SFR to a ``sawtooth'' pattern where each burst causes a rapid drop in the SFR on a dynamical time, then the SFR rises over a longer timescale (regulated by the cooling or recycling time) during which the SFR is smooth because the lagging SNe cannot ``catch up'' until an overshoot leads to a quench cycle. Longer $\Delta t_{\rm delay}$ produces higher-amplitude burst/quench cycles (more severe ``overshoot'').  

This appears to confirm something like the \citet{hayward.2015:stellar.feedback.analytic.model.winds} ``overshoot'' hypothesis. Exactly what the best proxy for the overshoot probability actually remains an open question: we test some other possible parameterizations below, and find none of them provide a more accurate predictor than the simple scaling from \citet{hayward.2015:stellar.feedback.analytic.model.winds}, but exploring this further is an important subject for future work.

\begin{figure}
	\includegraphics[width=0.95\columnwidth]{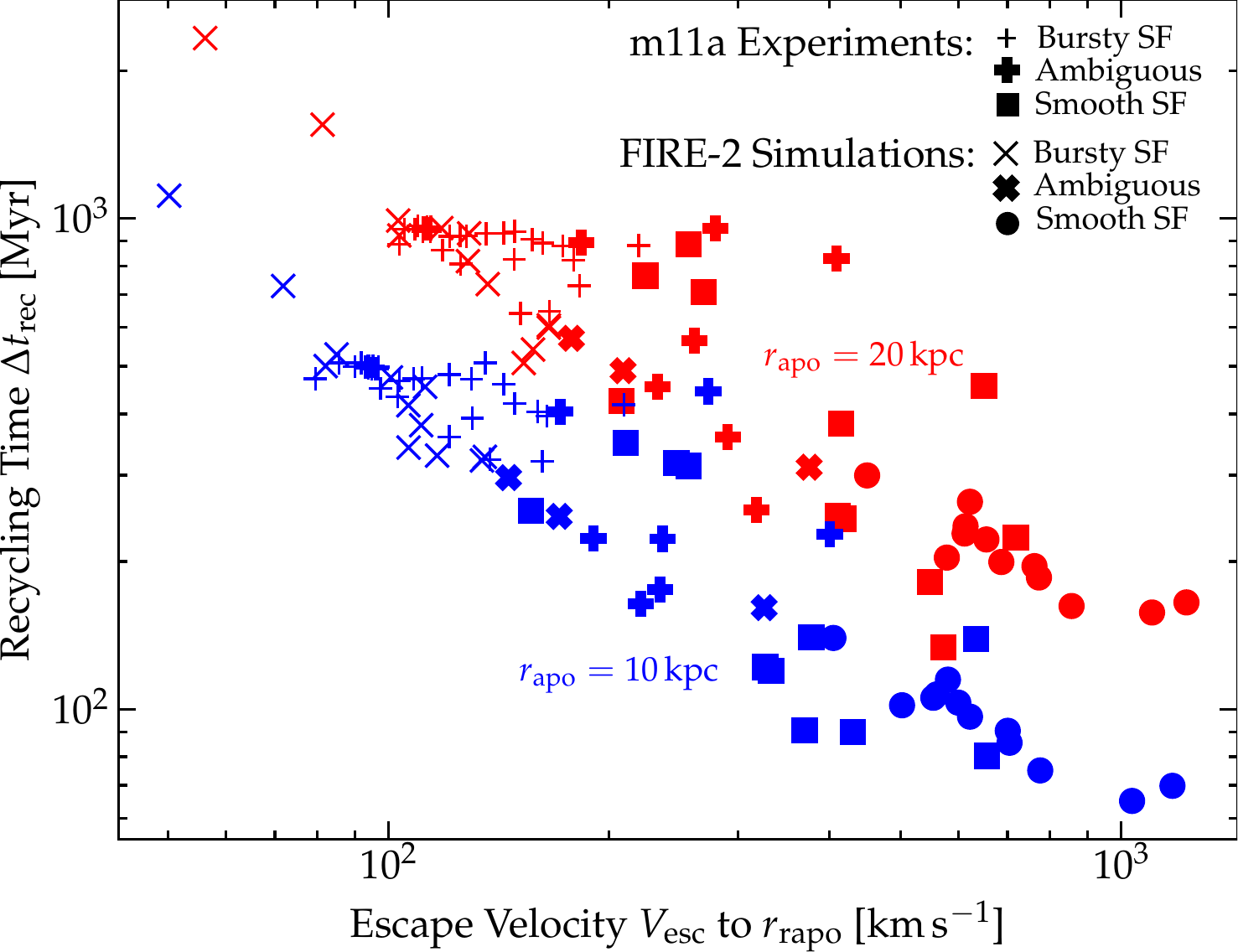}
	\includegraphics[width=0.95\columnwidth]{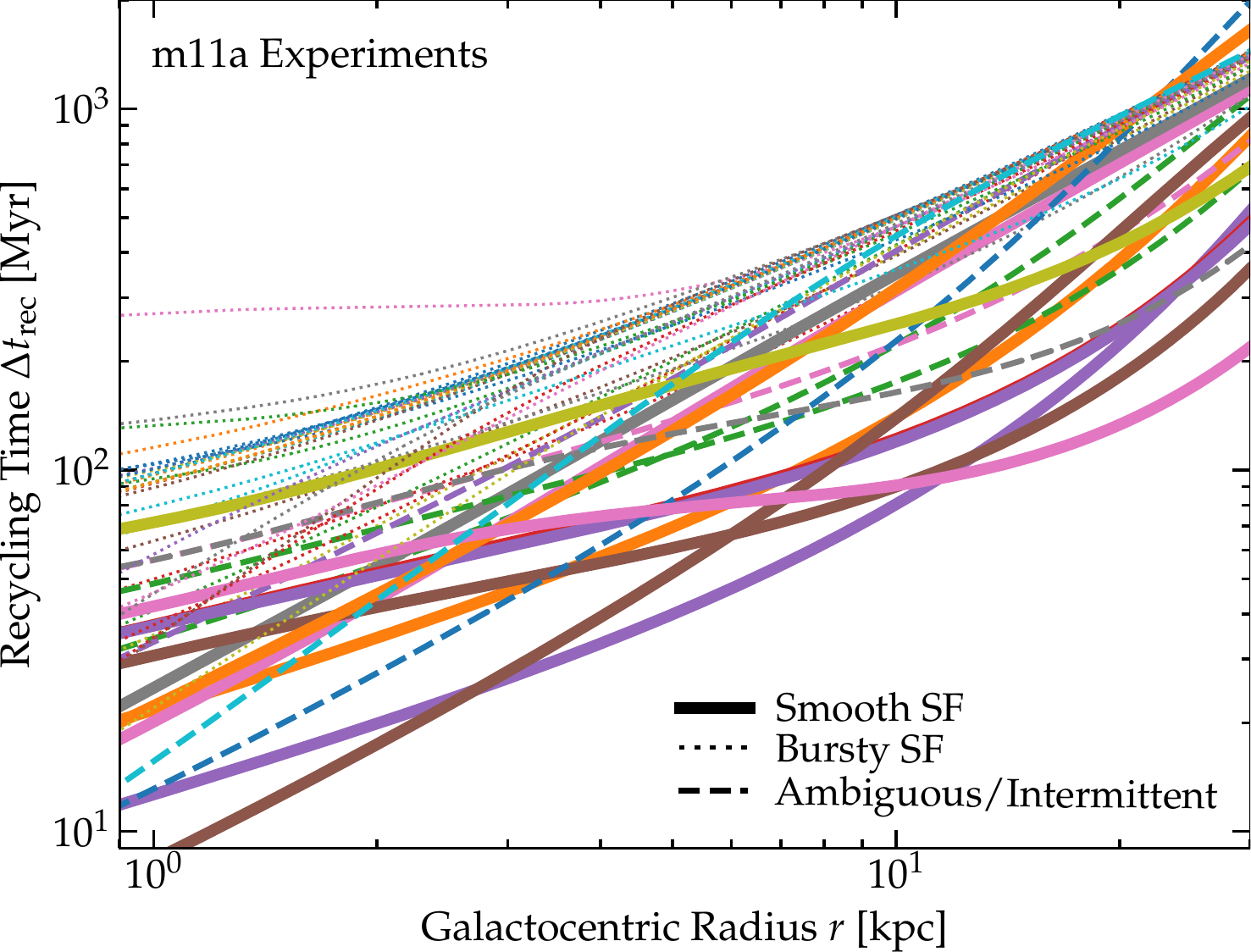}
	\includegraphics[width=0.95\columnwidth]{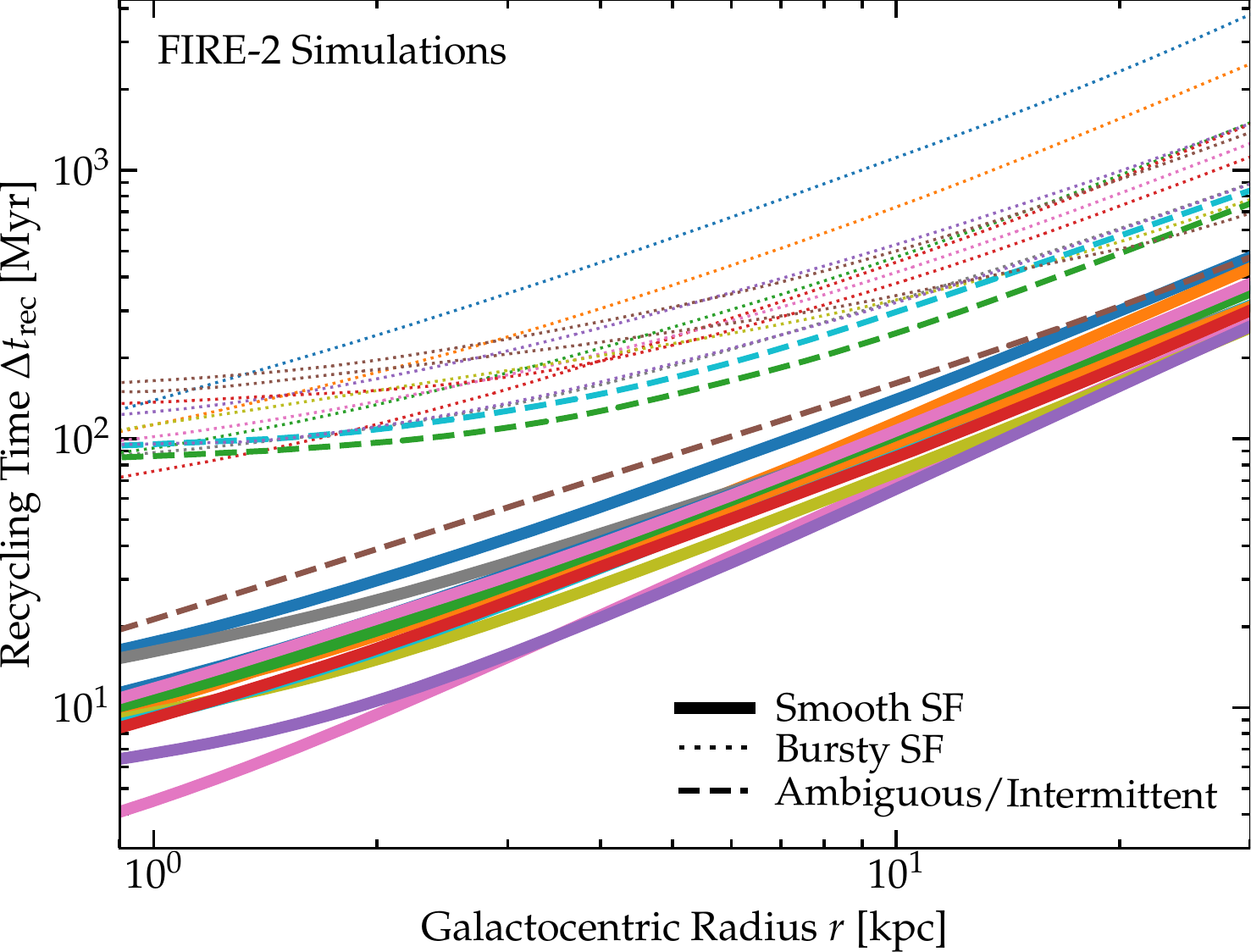}
	\vspace{-0.1cm}
	\caption{Comparison of ``recycling times'' $\Delta t_{\rm rec}$ for ballistically-ejected gas on radial orbits from small radii to reach $r = r_{\rm apo}$ and return (\S~\ref{sec:recycling.times}).
	{\em Top:} Values of $\Delta t_{\rm esc}$ for different $r_{\rm apo}=10$ or $20\,{\rm kpc}$ versus the escape velocity $V_{\rm esc}$ to the same radii, for all of our {\bf m11a} experiments (coded by whether they have ``smooth'' SF, as Fig.~\ref{fig:Vesc}) and the same for all our FIRE-2 default simulations (as Fig.~\ref{fig:Vesc.fire.smooth}). 	{\em Middle:} Profile of $\Delta t_{\rm esc}$ versus $r_{\rm apo}$ for the {\bf m11a} runs, coded by SF status as Fig.~\ref{fig:Vesc}.
	{\em Bottom:} Same for the FIRE-2 runs. 	
	For a fixed $r_{\rm apo}$, there is a strong inverse correlation between $V_{\rm esc}$ and $\Delta t_{\rm esc}$; and we see separation in $\Delta t_{\rm rec}$ as well as $V_{\rm esc}$. But there appears to be a sharper division in ``bursty'' versus ``smooth'' SF with $V_{\rm esc}$, and the precise values of $V_{\rm esc}$ which divide these regimes show a much weaker dependence on the precise value of $r_{\rm apo}$. 
	Still, this means deeper potentials lead not only to less efficient gas ejection, but faster recycling of that gas, further suppressing bursty SF.	\label{fig:recycling.time.vs.vesc}}
\end{figure}

\subsubsection{Relation to the ``Recycling Time'' of Ejected Gas}
\label{sec:recycling.times}

Related to the ``overshoot'' question, it is worth noting that there is a close physical relationship between the escape velocity from small $r_{i}$ to some outer radius $r_{f}$ ($V_{\rm esc}(r_{i},\,r_{f})$) and the ``recycling time'' for material ejected to $r_{f}$ (but not unbound) to return to $r_{i}$. The latter depends on details of the orbit and interactions between gas and other non-thermal forces \citep[see e.g.][]{angles.alcazar:particle.tracking.fire.baryon.cycle.intergalactic.transfer,hafen:2018.cgm.fire.origins,hafen:2019.fire.cgm.fates}, but for the sake of a simple quantitative single valued-comparison, we can define a characteristic ``return'' or ``recycling'' time $\Delta t_{\rm rec}(r=r_{f})$ as the time required for a ballistic trajectory of a test-particle kicked initially from $r_{i} = 0$ on a purely-radial orbit with apocentric radius $r_{f}$ to return to $r_{i}$. This is determined by the shape of $V_{\rm c}(r)$, but for a wide range of profile shapes corresponds roughly to $\sim 3/\Omega(r_{f}) \sim 3\,r_{f}/V_{\rm c}(r_{f})$. Fig.~\ref{fig:recycling.time.vs.vesc} shows this explicitly. We plot the profiles of $\Delta t_{\rm rec}(r=r_{f})$, coded by the smooth/bursty SF status, for both our {\bf m11a} and FIRE-2 runs, which show clear separation as might be expected based on the arguments above, where the simulations with deeper potentials and smooth SF show smaller/faster $\Delta t_{\rm rec}(r=r_{f})$. We plot the correlation between $\Delta t_{\rm rec}(r=r_{f})$ and $V_{\rm esc}$ directly, measured at two different apocentric/escape radii ($10$ and $20$\,kpc). As expected, there is a strong anticorrelation between the two, though it is not perfect. For a sufficiently extended potential so that $V_{\rm c}$ rises with $r$, $V_{\rm esc}(r_{i} \sim 0,\,r_{f} = r_{\rm apo})$ scales as $\sim \sqrt{2}\,V_{\rm c}(r_{\rm apo})$, so for fixed $r_{\rm apo}$ the correlation with $\Delta t_{\rm rec}(r=r_{f})$ is natural, but for more concentrated mass profiles $V_{\rm esc}$ scales more closely with $V_{\rm c}(r_{i})$. We see in Fig.~\ref{fig:recycling.time.vs.vesc} that there is a sharper separation in $V_{\rm esc}$ compared to $\Delta t_{\rm rec}(r=r_{f})$; moreover as $\Delta t_{\rm rec}(r=r_{f})$ scales linearly or super-linearly with $r_{\rm apo}$ while $V_{\rm esc}$ scales much more weakly, the ``critical'' value of $V_{\rm esc}$ is much more weakly dependent on the choice of $r_{\rm apo}$ compared to some critical $\Delta t_{\rm rec}(r=r_{f})$. All of this suggests that the transition to smooth SF is not quite as simple as a pure threshold in $\Delta t_{\rm rec}(r=r_{f})$. Nonetheless, it also clearly shows that as $V_{\rm esc}$ increases, $\Delta t_{\rm rec}(r=r_{f})$ also decreases, so it not only becomes more difficult to ``overshoot,'' but material ejected in such an event will recycle more rapidly, becoming more akin to a galactic fountain, and less like extended burst-quench cycles. This will further suppress bursty SF as the potential becomes deeper, as summarized in  Fig.~\ref{fig:cartoon.potential}.

\subsection{Some Alternative Scalings (Which Do Not Appear to Explain the Experiments)}
\label{sec:alt.scaling.failures}

Briefly we can note, and test in Appendix~\ref{sec:additional.params}, some alternative scalings broadly akin to those above, which do not appear to explain our simulation results. 

\citet{hayward.2015:stellar.feedback.analytic.model.winds} note an alternative critical threshold criterion for $f_{\rm out}$ or the overshoot probability, directly motivated by the models in \citet{ostriker:2010.molecular.reg.sf,2018ApJ...859...68K}, which would be applicable if the galactic gas were entirely pressure-supported via photo-ionization and/or photo-electric heating feedback (e.g.\ in the WNM or WIM maintained by stars within the galaxy, with negligible turbulence), giving a threshold when $Z_{\rm gas}\,\Sigma_{\rm gas} / V_{\rm esc} \lesssim 5\times10^{-3}\,{\rm M_{\odot}\,pc^{-2}\,km^{-1}\,s}$. 

We can also assume a global model of the galaxy as a single-phase slab or sphere where feedback balances gravity, with the SFR in either (a) in the $t_{\rm dyn} \ll t_{\ast}$ limit, in which case the stellar mass formed ``per episode'' in idealized simulations of gas clouds/patches is $M_{\ast} \sim (a/\langle \dot{p}/m_{\ast} \rangle)\,M_{\rm gas}$ \citep[at larger $M_{\ast}$, the gas is blown out by ``early'' feedback with a momentum flux per unit stellar mass $\langle \dot{p}/m_{\ast} \rangle \sim 10^{-7}\,{\rm cm^{2}\,s^{-1}}$; see][]{fall:2010.sf.eff.vs.surfacedensity,grudic:sfe.cluster.form.surface.density,grudic:max.surface.density,grudic:mond.accel.scale.from.stellar.fb,2018ApJ...859...68K,hopkins:2021.bhs.bulges.from.sigma.sfr}, or (b) in the $t_{\rm dyn} \gg t_{\ast}$ limit, with momentum injection from feedback proportional to the SFR balancing gravity so $\dot{M}_{\ast} \sim (a/\langle p/m_{\ast} \rangle)\,M_{\rm gas}$ \citep{murray:momentum.winds,thompson:rad.pressure,hopkins:rad.pressure.sf.fb}. Next assume the ``later'' feedback or a $\mathcal{O}(1)$ fraction of either the integrated (mostly from SNe) (A) energy ($E_{\rm SNe}$, with SSP-integrated energy per unit mass deposited $=\langle \epsilon/m_{\ast} \rangle\,M_{\ast}$) or (B) momentum ($P_{\rm SNe} = \langle p / m_{\ast} \rangle$) drives an energy ($E_{\rm out} \sim (1/2)\,M_{\rm out}\,V_{\rm esc}^{2}$) or momentum-conserving ($P_{\rm out} \sim M_{\rm out}\,V_{\rm esc}$) outflow, respectively, with the total mass $M_{\rm out}$ integrated over the ``burst'' in case (a) or a dynamical time in case (b). This gives the criterion for blowing out all the gas from the galaxy as, for each case a dimensionless number exceeding unity: (aA) $(2\,\langle \epsilon/m_{\ast} \rangle/\langle \dot{p}/m_{\ast} \rangle)\,(a/V_{\rm esc}^{2})$, (aB) $(\langle p/m_{\ast} \rangle/\langle \dot{p}/m_{\ast} \rangle)\,(a/V_{\rm esc})$, (bA) $(2\,\langle \epsilon/m_{\ast} \rangle/\langle p/m_{\ast} \rangle)\,(V_{\rm c}/V_{\rm esc}^{2})$, (bB) $V_{\rm c}/V_{\rm esc}$. We can also modify each of these to a ``thin disk'' version of the same model, where instead of using the acceleration scale $a$ appropriate for a sphere, if we assume vertical outflow from a thin disk being centrifugally supported, the effective acceleration scale becomes $a \rightarrow (\sigma/V_{\rm c})\,a$. 

We have tested each of these in turn (plotting each dimensionless parameter versus radius for all our simulations). We find no evidence for separation between the ``bursty'' and ``smooth'' simulations in criteria (aA) or (aB) or (bB), and only weak separation -- much less clear than the separation between bursty and smooth in $V_{\rm esc}$ alone -- in criteria (bA). The same is true for the ``thin disk'' versions of (aA) and (aB). Note that the ``thin-disk'' versions of (bA) and (bB) become dimensionally akin to the \citet{hayward.2015:stellar.feedback.analytic.model.winds} models already discussed (so these do, of course, show separation, as described above). Moreover, in idealized experiments where we modify the ratio of different feedback processes to modify the pre-factors here, we do not see a clear burst-smooth transition, so we conclude that none of these more complicated scalings represents an improvement over a threshold at some absolute value of $V_{\rm esc}$ or the \citet{hayward.2015:stellar.feedback.analytic.model.winds} criterion alone.

In Appendix~\ref{sec:additional.params} we further consider models such as those in \citet{orr:2021.bubble.breakout.model} for a critical gas fraction based on assumptions about SNe super-bubble breakout, or those in \citet{krumholz:2010.instab.turb.in.disks,forbes:2011.thick.disk.torque.evol} based on the gas surface density and/or optical depth/self-shielding of the disk, and find these show no separation (in either our {\bf m11a} experiments or FIRE suite) between bursty and non-bursty systems. These models are also inconsistent with many of our tests including those varying $\eta_{\rm gas}$, or $\eta_{\rm cool}$, or $Z$, or the specific feedback energy, hence our not considering them in more detail.

\subsection{Relation to ``Inner CGM Virialization''}
\label{sec:icv}

It is worth discussing how this escape velocity-based behavior can provides a natural explanation for the {\em apparent} tight correlation between the cessation of ``bursty'' star formation and the transition in CGM thermodynamics identified in \citet{stern21_ICV}, where gas at $\sim 0.1\,R_{\rm vir}$ becomes quasi-hydrostatic (at roughly the virial temperature), a transition they termed ``inner CGM virialization'' (ICV). \citet{stern21_ICV} showed that ICV occurs when the ratio of cooling to freefall time exceeds a critical value $t_{\rm cool}^{(s)}/t_{\rm ff} \gtrsim 2$, for the cooling time $t_{\rm cool}^{s}$ defined as that which gas would have in a virialized hydrostatic halo (given the {\em actual} galaxy mass profile, gas mass, etc.). We have verified that this relation between ICV and the ratio $t_{\rm cool}^{(s)}/t_{\rm ff}$ holds in the experiments here, {\em provided} we use the appropriately modified potential, cooling functions, metallicity, gas mass profile, and other parameters matched correctly to each individual experiment. 

Now recall from \S~\ref{sec:thermo} that if ICV were directly the cause of smooth SF, then it should be directly evident in our experiments modifying the cooling function $\eta_{\rm cool}$ (as this allows us to freely vary $t_{\rm cool}^{(s)}/t_{\rm ff}$ well above and below this limit).\footnote{Specifically using the definition of $\Psi^{(s)}_{\rm cool} \equiv t_{\rm cool}^{(s),\,0.1\,R_{\rm vir}}/t_{\rm ff}$ from \citet{stern21_ICV} measured at $0.1\,R_{\rm vir}$, our simulations with $\eta_{\rm cool}=(0.001,\,0.01,\,0.1,\,0.3,\,1,\,10)$ give $\Psi^{(s)}_{\rm cool}=(300,\,15,\,4,\,0.5,\,0.1,\,0.001)$ at $z=0$. Our $\eta_{\rm cool}=0.01$ + $\eta_{\rm gas}=10$ run gives a value of $\Psi^{(s)}_{\rm cool} \approx 6$. In Fig.~\ref{fig:more.smooth.sfh.tests.cooling}, we showed that we can retain bursty SF in {\bf m12i} experiments with $\Psi^{(s)}_{\rm cool}$ up to $\sim 50$.} It should also be the case that our experiments modifying $\eta_{\rm gas}$ should result in ICV according to the criterion in \citet{stern21_ICV}, as $t_{\rm cool}^{s} \propto 1/\eta_{\rm gas}$.\footnote{Our $\eta_{\rm gas}=0.1$a and $0.3$a runs give $\Psi^{(s)}_{\rm cool}\approx 1$, while $\eta_{\rm gas}=(0.1,\,0.3)$ give $\Psi^{(s)}_{\rm cool}\approx 0.2$, and $\eta_{\rm gas}=10$ gives $\Psi^{(s)}_{\rm cool}=0.05$.} Fig.~\ref{fig:more.smooth.sfh.tests.cooling} summarizes some representative examples of these experiments, plotting some of the {\bf m11a} experiments with varied $\eta_{\rm cool}$ and $\eta_{\rm gas}$ and $\Psi^{(s)}_{\rm cool} \ll 1$, $\Psi^{(s)}_{\rm cool}\sim 1$, and $\Psi^{(s)}_{\rm cool} \gg 1$. We have also experimented with changing the initial metallicity and find it has no effect (\S~\ref{sec:other.physics}), and recall in Fig.~\ref{fig:profiles.weak.bursty.fx} we see some trend towards higher metallicities in the experiments with smooth star formation (opposite the prediction if larger $t_{\rm cool}$ resulted in less-bursty SF). As noted above, we can simultaneously vary $\eta_{\rm cool}$ and $\eta_{\rm fb}$ to keep the SFR and/or feedback strength fixed, and this does not change our conclusions regarding burstiness. We can also test the converse in Fig.~\ref{fig:more.smooth.sfh.tests.cooling}: for a simulation which otherwise produces smooth SF, the ICV hypothesis would predict that {\em increasing} the cooling rate above some critical value would produce bursty SF. If anything, we see the opposite. Fig.~\ref{fig:smooth.bursty.tests.nopressure} presents some even more extreme tests, in which we explicitly re-start simulations otherwise in the ``smooth'' SF regime (either default {\bf m12i} or {\bf m11a} with $\rho_{0}=10$) at different times enforcing an absolute temperature maximum of $\le 10^{4}\,$K everywhere, or removing all thermal gas pressure in the hydrodynamic Riemann problem in the code for gas in the CGM -- in either case designed to ensure that it is impossible for the CGM to actually be in thermal virial equilibrium -- and see that these remain ``smooth.''

However, in the criterion/conditions for ICV to occur, just like with the canonical scaling for the cooling rate at $R_{\rm vir}$ which separates the traditional ``cold mode'' and ``hot mode'' of accretion \citep[see][]{silk:1977.galaxy.cooling.fragmentation,rees:1977.tcool.tdyn.vs.mhalo,binney:1977.weak.vir.shocks}, the most important variable by far is the virial temperature $T_{\rm vir} \propto \Phi_{\rm c} \propto V_{\rm esc}^{2}$, hence the potential or escape velocity. Indeed, if we take the approximate scaling for $t_{\rm cool}^{(s)}/t_{\rm ff}$ (using a simple approximation for the cooling function) given in  \citet{stern21_ICV} Eqs.~17-18, the dependence on escape velocity is very strong, $t_{\rm cool}^{(s)}/t_{\rm ff} \propto V_{\rm esc}^{4.4}$. Meanwhile at fixed $V_{\rm esc}$ for standard CDM halos all other terms appear with a very weak dependence: there is almost no redshift or halo mass dependence and only linear dependence on metallicity and gas mass ($Z^{-1}$, $n_{\rm gas}^{-1}$). In fact, if we insert the same canonical values for other parameters used in \citet{stern21_ICV} for FIRE halos at $z=0$ in their Eq.~18, we obtain $t_{\rm cool}^{(s)}/t_{\rm ff}(r=0.1\,R_{\rm vir}) \sim 2\,(V_{\rm esc}[<0.1\,R_{\rm vir}]/220\,{\rm km\,s^{-1}})^{4.4}$ -- i.e.\ their critical value of $t_{\rm cool}^{(s)}/t_{\rm ff}(r=0.1\,R_{\rm vir}) \sim 2$ corresponds almost exactly with the critical escape velocity threshold we identify here. 

There thus seems to be a coincidence that the criterion for ICV happens to match the critical escape velocity threshold identified here, and this explains the frequent coincidence in timing between ICV and the end of bursty star formation in our ``default'' FIRE-2 simulations, even if they are not directly causally connected. The similarity of the two thresholds may not, actually, be entirely coincidental: in the ICV argument, again akin to the traditional cold/hot halo argument, this strong dependence and threshold reflect relatively inefficient gas cooling at $T \gtrsim 10^{6}\,$K, akin to the arguments cited above regarding $v_{\rm cool}$, but also meaning that ``confinement'' of the outflows in an overshoot-type model will be further aided by the pressure of the virialized CGM (by definition, for a halo which has undergone ICV, the ``PdV'' work required to expel the ISM beyond the ICV radius must be order-of-magnitude comparable to the energetic cost required to overcome the escape velocity to the same radius). But crucially, we show that it is not really the virialization of the gas alone which confines the outflows and suppresses bursty star formation. Rather, ICV as defined in \citet{stern21_ICV} will typically occur as a consequence of the inner halo and galaxy crossing the same critical $V_{\rm esc}$/overshoot threshold at which bursty star formation is suppressed.

\section{Relation to Vertical Disk ``Settling'' \&\ Very Thin Disks}
\label{sec:settling}

\begin{figure}
	\includegraphics[width=0.99\columnwidth]{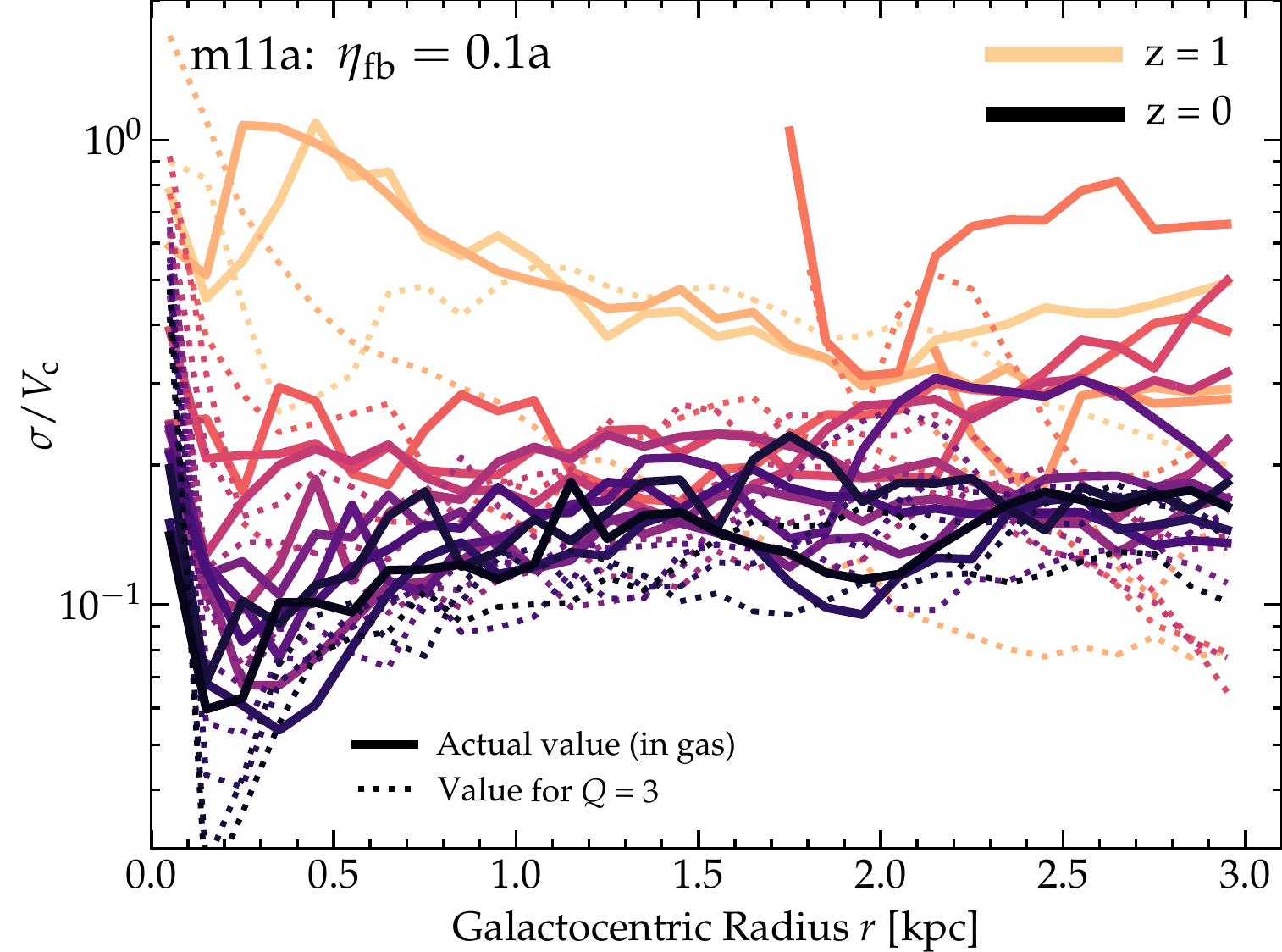}
	\vspace{-0.1cm}
	\caption{Evolution of the vertical disk structure in our $\eta_{\rm fb}=0.1$a {\bf m11a} experiment, with different colors encoding different redshifts from $z=1$ ({\em light}) to $z=0$ ({\em dark}) as Fig.~\ref{fig:weak.fb}. We show the radial profile of $\sigma/V_{\rm c}$ ({\bf solid}; $H/R$ is similar) and the value needed to maintain $Q\approx 3$ or the ``effective gas/thin disk fraction'', i.e.\ $\sigma/V_{c} \sim 3 \times \pi\,G\,R\,\Sigma_{\rm eff}/V_{\rm c}^{2} = 3\times (\pi\,r^{2}\,\Sigma_{\rm thin}[r])/M_{\rm enc}[<r]$. Here $\Sigma_{\rm thin}$ is simply defined as the surface density of gas+stars interior to the gas disk scale height. The $H/R \sim \sigma/V_{\rm c}$ in the disk crudely follows the expectation for constant-$Q$ evolution, once a disk is actually formed, as described in \S~\ref{sec:settling}.
	\label{fig:vertical.settling.time}}
\end{figure}

The generally-accepted proximate physical cause of ``vertical disk settling'' -- which we define for this discussion specifically as the decrease in disk thickness $H/R$ at a given galacto-centric radius -- is more straightforward. For some dwarf galaxy disks (e.g.\ {\bf m11b} discussed in \S~\ref{sec:m11b}) and the outskirts of extended (non self-shielding) HI disks extending well beyond the star-forming radii in galaxies, the gas can ``settle'' (turbulence can damp) until it is thermally-supported at ``warm'' temperatures ($c_{\rm s} \sim 10\,{\rm km\,s^{-1}}$) and so its scale height $H/R \sim c_{\rm s}/V_{\rm c}$ is largely determined by $V_{\rm c}$ alone. But provided that the potential is sufficiently deep and/or gas surface density is sufficiently large, the disk cannot maintain $Q\gtrsim 1$ with isothermal WNM/WIM-phase gas. In our experiments here (see Fig.~\ref{fig:profiles.weak.bursty.fx}) and an enormous array of previous simulation work making many different assumptions about thermodynamics, feedback, and galaxy properties, disks in this limit robustly self-regulate to maintain a turbulent Toomre $Q\sim 1$ (at the order-of-magnitude level) in the {\em star forming} gas disk \citep[see e.g.][]{tasker:2009.gmc.form.evol.gravalone,kim:2011.sf.selfreg.disks,kim:disk.self.reg,hopkins:rad.pressure.sf.fb,hopkins:fb.ism.prop,cacciato:2011.analytic.disk.instab.cosmo.evol,ceverino:2013.rad.fb,gatto:2017.silcc.self.regulation,orr:non.eqm.sf.model}. This is also ubiquitously observed in both local and high-redshift galaxies in both $H/R$ and $\sigma/V_{\rm c}$ \citep{leroy:2008.sfe.vs.gal.prop,walter:2008.things,swinbank:clumps,swinbank:2012.population.disks.q1.highz.clumps,kennicutt.evans:2012.sf.review,genzel:2015.molecular.sf.law.highz.q1,wisnioski:2015.disk.settling.consistent.with.Q1}. In that limit, $H/R \approx \sigma / V_{\rm c} \approx M_{\rm gas,\,disk}(<r) / M_{\rm enc}(<r)$ (see \S~\ref{sec:thermo}). Disks in this limit therefore become morphologically thinner and dynamically colder as the ``gas fraction'' $M_{\rm gas,\,disk}(<r) / M_{\rm enc}(<r)$ declines. Obviously at {\em a given radius} this can occur either by $M_{\rm gas,\,disk}$ decreasing (e.g.\ gas depletion, star formation, outflows, starvation; see also \citealt{hafen:2022.criterion.for.thin.disk.related.to.halo.ang.mom}) or $M_{\rm enc}$ (i.e.\ $V_{\rm c}$ at that $r$) increasing (buildup of a deeper potential, formation of stars in a compact core). And if evaluated at e.g.\ the disk effective radius, this can change simply by moving $r$.

Our experiments clearly show that simply modifying $M_{\rm gas}(<r) / M_{\rm enc}(<r)$ alone, before a disk forms at all, does {\em not} necessarily produce disks, let alone vertical disk settling (aka thin disks). The key, as discussed in \S~\ref{sec:thermo}, is that the gas must {\em already} be in a stable, self-regulated disk structure, which requires different conditions. Thus there are effectively three conditions: (1) a gas disk must have already formed (the mass profile was sufficiently centrally-concentrated at some point); (2) the circular velocity $V_{\rm c}(r) \gg 10\,{\rm km\,s^{-1}}$ (more like $V_{\rm c} \gtrsim 100\,{\rm km\,s^{-1}}$), so that the thermal warm-medium support cannot alone maintain a very thick disk (for direct observational examples, see \citealt{bigiel:2010.outer.disk.qlarge.nosf}, and for more theoretical discussion see \citealt{hafen:2022.criterion.for.thin.disk.related.to.halo.ang.mom}); (3) the effective gas fraction $M_{\rm gas}(<r) / M_{\rm enc}(<r) $ has decreased to values $\ll 1$ (for observations, see references above).

This immediately explains why almost all of our ``very thin'' star-forming disks appear to exhibit smooth SF. A very thin disk, requires, by definition, $\sigma/V_{\rm c} \ll 1$, and given realistic thermal physics if the disk is sufficiently dense to self-shield and cool and form stars, this requires $V_{\rm c} \gtrsim 100\,{\rm km\,s^{-1}}$. And of course for any realistic potential $V_{\rm esc}(r) \gtrsim \sqrt{2}\,V_{\rm c}(r)$. So it is almost automatically true that any system which can form a very thin disk {\em which is also star forming} (as distinct from a thermally-supported, non-self-shielding, non-star forming extended disk, akin to that in e.g.\ {\bf m11b} discussed previously) must have met our criteria for the cessation of bursty star formation, discussed above.

Fig.~\ref{fig:vertical.settling.time} shows one example of this, in our $\eta_{\rm fb}=0.1$ run (a weak-feedback simulation which, owing to rapid SF building up the potential and making it more concentrated, forms a ``very thin'' disk by $z=0$). First we note that indeed we do see $H/R \sim \sigma/V_{\rm c}$, with some fluctuations at any given instant owing to non-trivial out-of-equilibrium behaviors \citep[see][for a detailed discussion of this]{gurvich:2022.disk.settling.fire,hafen:2022.criterion.for.thin.disk.related.to.halo.ang.mom}. Second, we do see a trend between $\sigma/V_{\rm c}$ both as a function of radius and time and the value expected for a constant $Q\approx 2-3$, though it is not exactly one-to-one. But for the latter, it is important to carefully account for the two-component (stellar+gas) effective $Q$ \citep[e.g.][]{romeo:1992.two.component.dispersion}: we see significantly worse correlation if we assume the ``gas-only'' $Q_{\rm gas} \equiv \sigma\,\kappa / (\pi\,G\,\Sigma_{\rm gas})$ were constant.

\section{Cosmological Origins of these Conditions \&\ Relation to Observations}
\label{sec:obs}

\begin{landscape}
\begin{figure}
%\begin{sidewaysfigure*}
    \hspace{0.25cm}\includegraphics[width=0.97\linewidth]{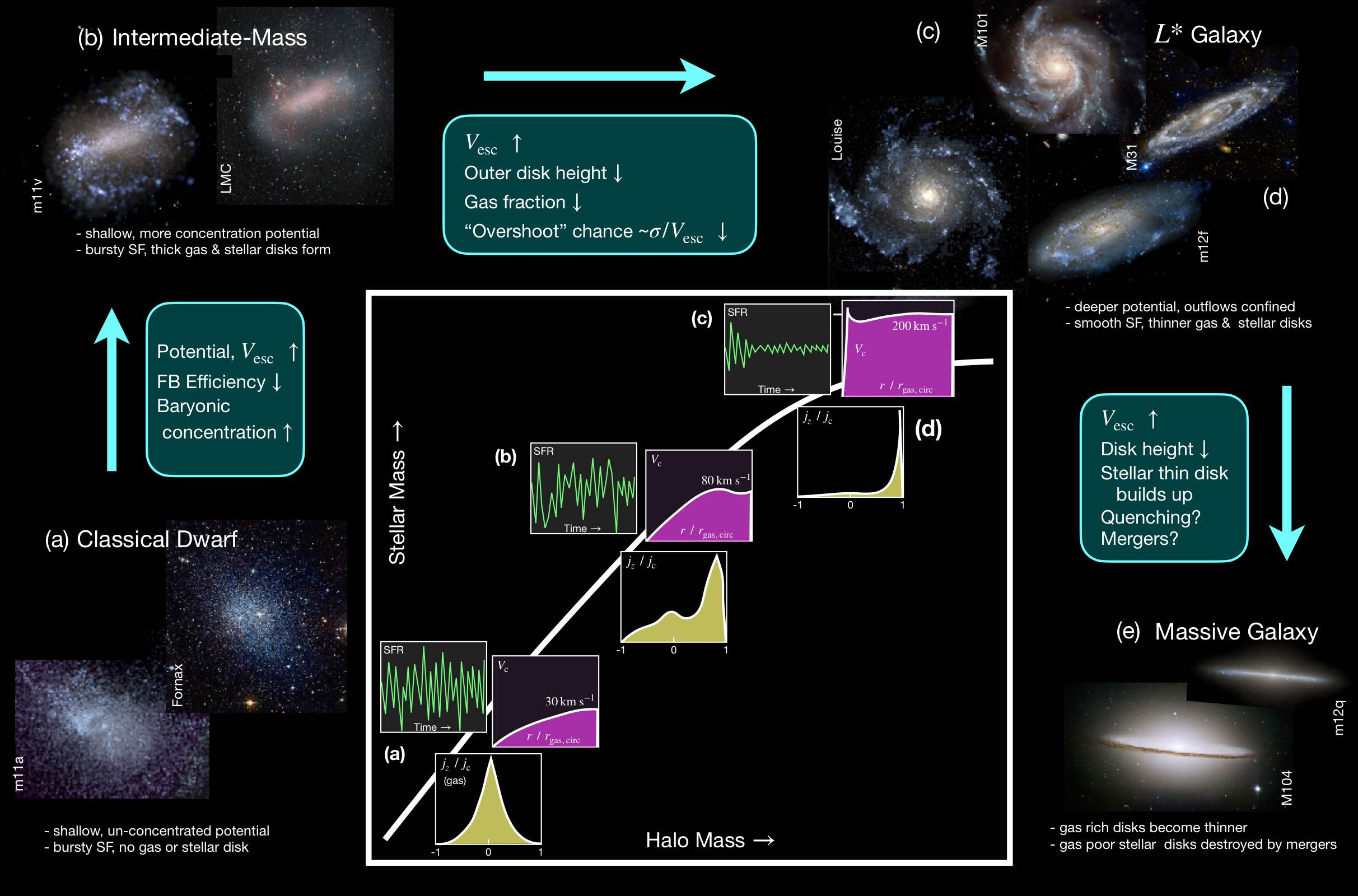}
    \vspace{-0.25cm}
    \caption{Cartoon illustrating a common (but by no means unique) ``track'' taken by our default FIRE-2 simulation galaxies (per \S~\ref{sec:obs}) through the space from a bursty, spheroidal system {\bf (a)}, bursty system with a thick gas disk {\bf (b)}, through developing a smooth SFR {\bf (c)} and thin disk {\bf (d)}, and ultimately evolving into a massive galaxy {\bf (e)} whose subsequent disk survival depends on physics of quenching, mergers and other evolution beyond the scope of this paper. The sequence in stellar and halo mass ({\em inset}) from {\bf (a)}$\rightarrow${\bf (e)} can be thought of as a mass sequence at $z\sim 0$, or (heuristically) as a time sequence in the evolution of a massive system today. At each stage we illustrate an example FIRE-2 galaxy and observed system ({\em images}), with an example of the SFR versus time, circular velocity $V_{\rm c}$ as a function of radius $r$ (relative to the gas circularization radius $r_{\rm circ}$), and gas-phase angular momentum distribution $j_{z}/j_{\rm c}$.
    {\em Image Credit:} (a) ESO/Digitized Sky Survey 2; (b) ESO/VMC Survey; 
    (c) ESA/NASA/CFHT/NOAO/K.~Kuntz, F.~Bresolin, J.~Trauger, J.~Mould, Y.-H.~Chu, D.~Martin; 
    (d) NASA/JPL-Caltech; (e) NASA/ESA and The Hubble Heritage Team (STScI/AURA).
    \label{fig:cartoon.time.evol}}
% a: https://esahubble.org/images/heic1425h/
% b: https://www.eso.org/public/images/eso1914a/
% c: https://esahubble.org/images/heic0602a/
% d: https://www.nasa.gov/mission_pages/galex/pia15416.html
% e: https://esahubble.org/images/opo0328a/
    \vspace{-0.5cm}
%\end{sidewaysfigure*}
\end{figure}
\end{landscape}

A heuristic illustration of a typical (though not necessarily unique) mass or time sequence in our default FIRE-2 simulations, as it evolves from a galaxy with a shallow potential and un-concentrated mass profile (with no disk and bursty star formation) to a system with a concentrated mass profile in a deep potential well (with a disk and smooth star formation), is shown in Fig.~\ref{fig:cartoon.time.evol}. We discuss this and other physical pathways in cosmological settings, and their connection to observations, below.

\subsection{Physical Mechanisms for Achieving Centrally-Concentrated Mass Profiles}

By far the simplest (in ``default'' FIRE simulations across different masses) way to realize a centrally-concentrated mass profile is via some condensation of baryons (whether or not they turn into stars) in the halo center, as the baryons are dissipative. Baryons will naturally fall into the halo center as they cool, and can further lose angular momentum and energy and become denser under the action of gravitational torques and instabilities. In our experiments here, we see that for our {\bf m11a} system, if the central structure is sufficiently concentrated (sub-kpc sizes), then it only requires $\sim 10^{9}\,M_{\odot}$ worth of mass to significantly re-structure the potential ($\sim 2\%$ of the halo mass or $\sim 10\%$ of the baryonic mass), and we stress that this is beginning from an initial condition which is strongly cored in the center (so this is a ``worst-case'' scenario in many ways, compared to a system which begins from a steeper central profile). In fact, \citet{elbadry:fire.morph.momentum} found that the best correlator (of properties they surveyed including mass and spin) with gas ``diskiness'' in the FIRE simulations is the degree of ``baryonic concentration,'' exactly as this scenario would predict. The precise form of such a concentration (a compact spheroid, nuclear star cluster, bulge, dense gas clouds, hyper-massive black hole, etc.) will of course depend on the baryonic physics, but is not particularly important for our purposes. 

Achieving such a concentration will generically get ``easier'' as galaxies increase in mass from dwarf to $\sim L_{\ast}$ systems as shown in Fig.~\ref{fig:cartoon.time.evol}, as feedback becomes less able to unbind the baryons and the observed central baryonic masses increase \citep[e.g.][and references therein]{behroozi:2019.sham.update}. In FIRE for example, while there are lower-mass exceptions, this most commonly occurs around a stellar mass scale $M_{\ast}\gtrsim 10^{10}\,M_{\odot}$, as in the high-redshift progenitor galaxy many dense star clusters which form rapidly and efficiently in those conditions \citep{grudic:balls.of.fire.starcluster.from.galaxy.sims} merge together within the central $\lesssim 1\,$kpc to form a massive, dense proto-bulge (\citealt{ma:2020.globular.form.highz.sims}; X.\ Ma et al., in prep.), around which a disk begins to form. In dwarf galaxies even smaller than our {\bf m11a} case here, nuclear star clusters are common and have masses and sizes which might be sufficient for them to play a similar role in many cases \citep[see][]{boker02:nuclei.ids,boker04:nuclei.scalings,milosavljevic:diss.nuclei.formation,seth:nuclear.star.clusters,mccrady:m82.sscs}. If so, this would be challenging to capture in simulations since, although the {\em initial formation} of star clusters can be resolved in state-of-the-art simulations, their long-term survival and evolution (critical if they are to play a role in disk formation) are not (see \citealt{kim:gc.form.FIRE,ma:2020.globular.form.highz.sims}).

But is it possible to achieve this in an entirely dark matter-dominated system? Yes: we clearly see this in some FIRE examples (\S~\ref{sec:m11b}). For a standard (non-cored) NFW halo, $\rho \propto (r/R_{\rm s})^{-1}\,(1+r/R_{\rm s})^{-2}$ with $R_{\rm s} \equiv R_{\rm vir}/c_{\rm vir}$ the scale-length defined in terms of the concentration $c_{\rm vir}$. This gives $a(r) \rightarrow 0.34\,G\,M_{\rm vir}/R_{s}^{2}\sim$\,constant (akin to our $a_{0}$ models) as $r\rightarrow 0$, i.e.\ $V_{\rm c} \propto r^{1/2}$ increasing modestly with $r$. But the density profile steepens at larger radii: the critical $r_{{\rm slope}=1/4}$ from \S~\ref{sec:disky.quant} (Fig.~\ref{fig:correlation.reff.rmax}) is $\sim 0.55\,R_{\rm s} \sim 12\,{\rm kpc}\,(M_{\rm vir}/10^{12}\,M_{\odot})^{0.43}\,(c_{\rm vir} / \langle c_{\rm vir}(M_{\rm vir},\,z) \rangle)^{-1}$,\footnote{We obtain this scaling by inserting the virial definitions from \citealt{bryan.norman:1998.mvir.definition} and the mean concentration-mass-redshift relation $\langle c_{\rm vir}(M_{\rm vir},\,z) \rangle$ from \citealt{dutton.maccio:2014.concentration.mass.relation}.} and the ``turnover radius'' $r_{\rm Vmax}=r_{{\rm slope}=0}$ (outside of which $V_{\rm c}$ decreases) is $\sim2.2\,R_{\rm s}$. For comparison, a typical HI disk size\footnote{We estimate this by combining the typical ratios of HI $r_{\rm 50,\,HI}$ and $r_{\rm 90,\,HI}$ to optical stellar disk sizes from \citet{2021AJ....161...71H}, with the optical size-virial mass relation estimated in \citet{kravtsov:2013.size.vs.rvir.relation}.} $r_{\rm 50,\,HI} \sim 7\,{\rm kpc}\,(M_{\rm vir}/10^{12}\,M_{\odot})^{0.33}$ (or $r_{\rm 90,\,HI} \sim 2-3\,r_{\rm 50,\,HI}$), a factor $\sim 2$ smaller than $r_{{\rm slope}=1/4}$. So (as we see in our simulations), the potential would not be sufficiently concentrated from dark matter alone in most cases. Since the scatter in $c_{\rm vir}$ is comparably small ($\sim0.1\,$dex), variations in $c_{\rm vir}$ will not generally be able to bring $r_{{\rm slope}=1/4} \lesssim r_{\rm 50,\,HI}$. But the scatter in HI disk size is quite large ($\sim 0.3-0.5\,$dex), and since we really care about some circularization radius at an earlier time it could be even larger. So it is reasonable to expect that in some cases, a sufficiently-concentrated NFW profile with a comparably large gas circularization radius could meet our disk formation criterion. 

As we explored earlier in detail, this is likely playing a role in the case of the default-FIRE {\bf m11b} run discussed above (the lowest-mass default-FIRE run analyzed here which forms a disk, similar in mass to {\bf m11a} studied here). As discussed in \citet{elbadry:fire.morph.momentum}, that galaxy does not have an especially unusual halo concentration or spin. However as we showed, it does initially form somewhat early and it accumulates a more centrally-concentrated early-forming baryonic mass (so $r_{\rm Vmax}$ is more like $\sim 4-5\,$kpc, rather than the expected $\sim 12\,$kpc given its mass). Later, an accretion event brings in some new gas with a high impact parameter (much larger than $\sim \lambda_{\rm spin}\,R_{\rm vir} \sim 5\,$kpc), which can circularize outside $r_{\rm Vmax}$ producing an unusually prominent and extended ($r_{\rm 50,\,HI} \gtrsim 12\,{\rm kpc}$) disk for a halo of this mass.

Of course, if the dark matter profile is ``cored'' out to reasonably large radii, then it will be even less concentrated, pushing $r_{{\rm slope}=1/4}$ and $r_{\rm Vmax}$ further out and potentially inhibiting initial disk formation if there is no baryonic concentration.

Finally, a host of exotic mechanisms could produce more centrally-concentrated profiles. Extremely massive BHs in dwarf centers would act like our $M_{0}$ models, but we stress that the masses required (e.g.\ $\sim 10^{9}\,M_{\odot}$ for {\bf m11a}, larger than the galaxy stellar mass) are far in excess of observed SMBHs in dwarf galaxies \citep[e.g.][]{reines:2020.off.nuclear.agn.in.dwarfs}. While collisionless cold dark matter models (let alone models of warm or elastic-scattering self-interacting dark matter [SIDM]; see \citealt{robles:sidm.on.fire,fitts:2019.fire.dwarfs.sidm.wdm.cdm}) tend to predict shallow or cored central dark matter profiles, a broad class of dark matter models produce more-dense centers, including many scalar field or ``fuzzy'' dark matter models \citep{robles:2019.fdm.solitons.more.dense.dwarfs} and the incredibly broad category of models which allow dissipation in the dark sector (dSIDM; \citealt{shen:2021.dissipative.dm.dwarfs.fire,xiao:2021.smbh.seeds.dissipative.dm}). In experiments with dSIDM models, for example, \citet{shen:2021.dissipative.dm.dwarfs.fire} find that there is a one-to-one correspondence between models which produce steeper dark ``cusps'' in their centers (often in the form of density profiles which have $\rho \propto r^{-3/2}$ in their center, so $r_{{\rm slope}=1/4}\rightarrow 0$ and $r_{\rm Vmax}$ is reduced by a factor of $\sim 2-3$ at masses $M_{\rm vir} \sim 10^{10}-10^{11}\,M_{\odot}$) and those which produce disks in the dSIDM models where there were none in CDM.

\subsection{Physical Mechanisms for Achieving High Escape Velocity Scales (Low Overshoot Probability)}

Achieving a large escape velocity or potential scale is in many ways more straightforward: if one needs to exceed our critical $V_{\rm esc}$ (or fall below the critical $\sigma_{\rm eff}/V_{\rm esc}$) on scales where star formation occurs in real galaxies, then for realistic mass profiles (where density is higher at small radii, and $\sigma_{\rm eff} \gtrsim 10\,{\rm km\,s^{-1}}$) this will typically involve building up something like a total mass of $\gtrsim 3\times10^{10}\,M_{\odot}$ inside of the central few kpc (again illustrated in Fig.~\ref{fig:cartoon.time.evol}).

For an NFW-like halo alone (using the scalings above and assuming a cutoff outside the splashback radius), $V_{\rm esc}(r\rightarrow 0) \approx 450\,{\rm km\,s^{-1}}\,[(c/\langle c\rangle)\,M_{12}]^{0.3}$, so this threshold potential or $V_{\rm esc}$ is crossed at a mass scale of $M_{\rm vir} \approx (1-1.5)\times 10^{11}\,M_{\odot}\,(\langle c(M_{\rm vir},\,z) \rangle/c)$ (depending on exactly which radius must exceed the critical $V_{\rm esc}$), independent of redshift. If the halo center is strongly cored, this moves up in mass but just to $\sim (2-3)\times10^{11}\,M_{\odot}$. For low halo masses ($M_{\rm vir} \lesssim 10^{11}\,M_{\odot}$), we do not expect the baryons to significantly influence this, except perhaps to introduce cores which slightly decrease $V_{\rm esc}$ (although given the contribution to $V_{\rm esc}$ from large radii, we find our strongest-cored galaxies only decrease $V_{\rm esc}$ by $\sim 25\%$), given the observed low baryon-to-halo mass ratios. But at higher masses, where the central baryon fractions and star formation efficiencies are high, the baryons can add significantly to $V_{\rm esc}$, making the transition more rapid -- this is part of why we see a ``jump'' in the central $V_{\rm esc}$ in the FIRE systems (which sample halos at factor $\sim 3$ increments) in Fig.~\ref{fig:Vesc.fire.smooth}.

Given the above, we expect at $z=0$ that halos of larger mass have, on average, crossed some $V_{\rm esc}$ or potential threshold earlier in their progenitor history. And at a given mass, more concentrated (earlier-forming) halos will again have done so at earlier times.

Of course, the same exotic mechanisms discussed above (e.g.\ different dark matter physics) could similarly increase or decrease the halo escape velocity, shifting when the critical potential threshold is reached. 

In \S~\ref{sec:overshoot.tests} we showed that sufficiently large changes to the structure of the ISM and feedback rates (e.g.\ producing very different clustering of star formation in time, or large changes to the SNe delay time distribution) can also shift the threshold for transitioning to smooth SF at fixed $V_{\rm esc}$. However, barring radical revisions to our understanding of star formation and stellar evolution, these are generally more subtle or second-order effects (their expected variation is generally much weaker than the quite-large variation in $V_{\rm esc}$ across galaxy populations).

\subsection{Correlation with Vertical Disk ``Settling''}

As discussed above in \S~\ref{sec:settling}, vertical disk settling requires (1) a disk, and (2) $\sigma/V_{\rm c} \ll 1$. Given the empirical thermal floor in $\sigma$ of $\sim 10\,{\rm km\,s^{-1}}$, and that $V_{\rm esc} \gtrsim \sqrt{2}\,V_{\rm c}$, this means that in general, settling to very thin $H/R \ll 1$ in an actively star-forming disk (as opposed to more extended non-star-forming gas disk) will require a system meeting both our ``disk formation'' and ``smooth SF'' criteria already, so the physical requirements discussed above will also apply. This automatically explains, as discussed above, why there appears to be a correlation in cosmological simulations between the cessation of bursty SF and the onset of disk settling. However, the disky+smooth SF criteria we outline are not strictly sufficient to immediately produce a ``fully settled'' (i.e.\ very thin) disk: one could have $\sigma/V_{\rm c}$ still large even with sufficiently large $V_{\rm esc}$ in a disk owing either to (a) a potential with $V_{\rm c} \ll V_{\rm esc}$ or (b) a large gas fraction in a disk self-regulating to maintain $Q\sim 1$ (so $\sigma(r)/V_{\rm c}(r) \sim M_{\rm gas,\,disk}(<r) / M_{\rm enc}(<r)$) or some other mechanism maintaining large $H/R$. In such a case, we would expect $M_{\rm gas,\,disk}(<r) / M_{\rm enc}(<r)$ and therefore $\sigma/V_{\rm c}$ to gradually decline over time for one of several reasons: the disk could be ``starved'' and deplete gas via inner halo virialization (less efficient cooling) or ``mass quenching'' \citep{dave:2011.mf.vs.z.winds,2022MNRAS.512.3806V,gurvich:2022.disk.settling.fire}, or a rapid increase with growing mass in the efficiency of star formation and/or outflows \citep{dekel:2009.clumpy.disk.evolution.toymodel,ceverino:2013.rad.fb,hayward.2015:stellar.feedback.analytic.model.winds,barro:2017.critical.sigma.1.ridgeline.for.quenching}, or simply (at fixed gas supply) gradual buildup of the potential ($V_{\rm c}$ and $M_{\rm enc}(r)$) as the galaxy grows. And, of course, a prominent thin stellar disk requires first forming a star-forming thin gas disk, followed by sufficient time to form enough stars (without strong heating/disruption) in that disk to be appreciable. These processes mean that settling can, in principle, be a more gradual, {continuous} process of decreasing $H/R$ over time \citep[e.g.][]{guedes:2011.cosmo.disk.sim.merger.survival,ma:2016.disk.structure}, and so can continue on for well after the ``initial'' disk formation or transition from bursty to smooth SF, though they are often associated in time (as shown in Fig.~\ref{fig:cartoon.time.evol}).

\subsection{Does One Have to Happen Before the Other?}

In principle, we show that disk formation could occur before or after the transition to ``smooth'' star formation. One could imagine building up a deep (large absolute value) potential in a relatively ``flat'' (e.g.\ constant-$\rho$) density profile, transitioning to smooth SF by prohibiting cold gas escape, which might lead to more efficient star formation and baryonic concentration, until a steeper mass profile is built up and a disk can be stabilized. But most often in cosmological simulations (see Fig.~\ref{fig:cartoon.time.evol}), the order is the opposite: a disk forms which goes through a relatively brief intermediate phase where SF is still bursty (and the disk may be disrupted by particularly strong feedback events, mergers, etc.), then the galaxy transitions to smooth SF, and the disk gradually settles over the remainder of the Hubble time \citep{ma:radial.gradients,ma:2016.disk.structure,stern21_ICV,hafen:2022.criterion.for.thin.disk.related.to.halo.ang.mom,gurvich:2022.disk.settling.fire}. From the arguments above, this is natural: the initial disk formation criterion can be met at any mass scale, while the smooth SF criterion generally requires crossing a relatively large threshold. Still, from the arguments above, the expected halo masses where both will occur are not so different, and in practice simulations see they are often relatively closely associated in time \citep{yu:2021.fire.bursty.sf.thick.disk.origin,gurvich:2022.disk.settling.fire}. Meanwhile, vertical settling requires (by definition) $\sigma_{\rm eff}/V_{\rm c} \ll 1$, where $\sigma_{\rm eff}$ (which should include thermal+magnetic support) is almost never below $\sim 10\,{\rm km\,s^{-1}}$ in the {\em star-forming} parts of disks, so given $V_{\rm esc} \gtrsim \sqrt{2}\,V_{\rm c}$, this essentially requires the smooth SF criterion is already met; that plus the fact that settling occurs over $\sim$\,Gyr timescales means that it -- insofar again as it applies to the star-forming and subsequent stellar disk (see \S~\ref{sec:m11b}) -- will smoothly follow this transition.

\subsection{Distinction Between ``Formation'' or ``Transition'' and Galaxies Today -- Can You ``Switch Back''?}

It is generically quite hard to imagine physical processes which significantly {\em decrease} the central escape velocity/potential well -- as even processes which ``inflate'' the dark matter or baryonic matter distribution via relaxation tend to very weakly modify the specific binding energy \citep[see][]{barnes:disk.halo.mergers,boylankolchin:mergers.fp,hopkins:cores}. So once a galaxy is well into the ``smooth'' star formation phase, it seems unlikely it would revert to bursty behavior unless either (a) it was very close to the boundary/critical potential, (b) some strong external perturbation (e.g.\ a merger) drives a starburst,  or (c) the galaxy expands so that the inner region (with high $V_{\rm esc}$) is depleted, and the later star formation occurs at some significantly larger radius $r$ where the potential well is more shallow. 

On the other hand, it is very easy to imagine any of a large number of well-known mechanisms which can destroy or kinematically heat disks, after they form. Of course, galaxy mergers are one obvious example \citep{toomre77,schweizer83:review,barnes:disk.halo.mergers,hernquist:review93}, with an efficiency which is known to strongly depend on e.g.\ the gas-richness (and therefore prior feedback/star formation history) of the galaxies \citep{walker:disk.fragility.minor.merger,robertson:disk.formation,hopkins:disk.survival.cosmo,hopkins:disk.survival,hopkins:disk.heating,hammer:obs.disk.rebuilding,governato:disk.rebuilding,moreno:2021.galaxy.merger.sims}. Generations of inflows with different angular momentum can produce misaligned disks which generate strong mutual torques and low net angular momentum \citep{sales:2012.coherence.of.accretion.key.for.disk.am.in.final.galaxy,vandevoort:misaligned.disk,2017ApJ...835..289S,kretschmer:thin.disk.filament.form.but.really.avoiding.destruction,hafen:2022.criterion.for.thin.disk.related.to.halo.ang.mom}. Disk instabilities (bars, arms, clumps) can redistribute mass within the disk and build pseudobulges \citep{roberts:gas.dynamics.in.bars,schwarz:disk-bar,weinberg:bar.dynfric,combes:pseudobulges,shlosman:1993.clumpy.disk.instab.sims,noguchi:1999.clumpy.disk.bulge.formation,athanassoula:bar.halo.growth,bournaud:gas.bar.renewal.destruction,berentzen:gas.bar.interaction,read:thick.disk.cosmo.sims,debattista:fire.bar.galaxy.m12m}. And there are many other ways to produce non-linearly different morphologies well after some ``initial'' disk formation event. This means that cosmologically time-integrated quantities like the bulge-to-disk ratio, disk mass, size, scale height, etc., will necessarily be complicated functions of the formation history, environment, mass, and other properties of galaxies. All of these can be {strongly} non-linearly influenced by the details of stellar feedback and gas thermodynamics \citep[see e.g.][]{sommerlarsen99:disk.sne.fb,okamoto:feedback.vs.disk.morphology,scannapieco:fb.disk.sims,piontek:feedback.vs.disk.form.sims,moreno:2019.fire.merger.suite,2021MNRAS.501...62N,2022MNRAS.515.3406M}. For example it has been known for decades that, absent feedback entirely, galaxies convert most of their baryons into stars rapidly after first halo collapse at very high redshifts, leading to pure hierarchical assembly at later times via collisionless/gas-poor/``dry'' mergers which efficiently destroy any stellar disks that might have formed at those early times (see references above). This is also illustrated heuristically in Fig.~\ref{fig:cartoon.time.evol}. So we stress that we are {\em not} claiming that all systems today with centrally-concentrated mass profiles must have disks, nor that the ``degree of concentration'' necessarily correlates with the ``diskiness'' of the $z=0$ galaxy. Instead, we are making a much more limited and specific claim: that a necessary condition for the {\em initial} formation of some disky gaseous structure is a (relatively) sufficiently-concentrated mass profile at that time. 
,

\subsection{What Does This Imply About Observed Galaxies \&\ Disky Structure or Bursty Star Formation?}

\subsubsection{Disky Structure}

\begin{figure*}
	\vspace{-0.1cm}
	\hspace{-0.4cm}
	\includegraphics[width=0.34\textwidth]{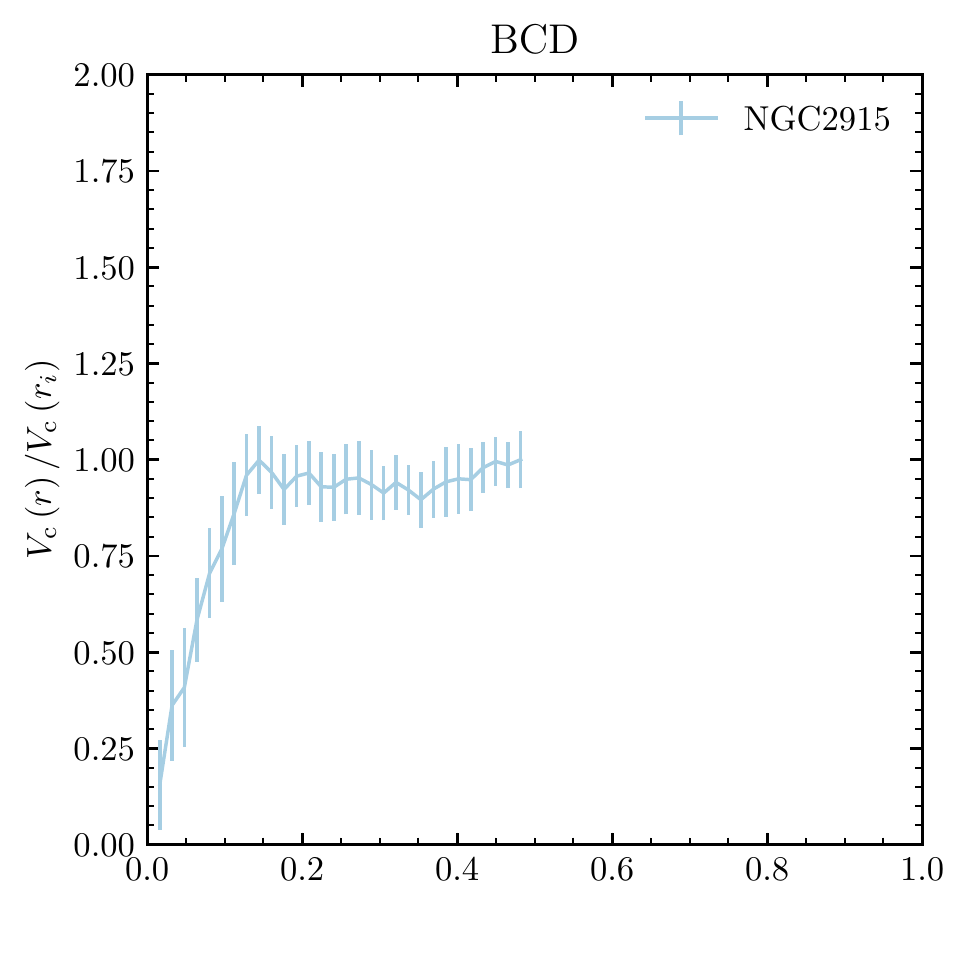}
	\hspace{-0.4cm}
	\includegraphics[width=0.34\textwidth]{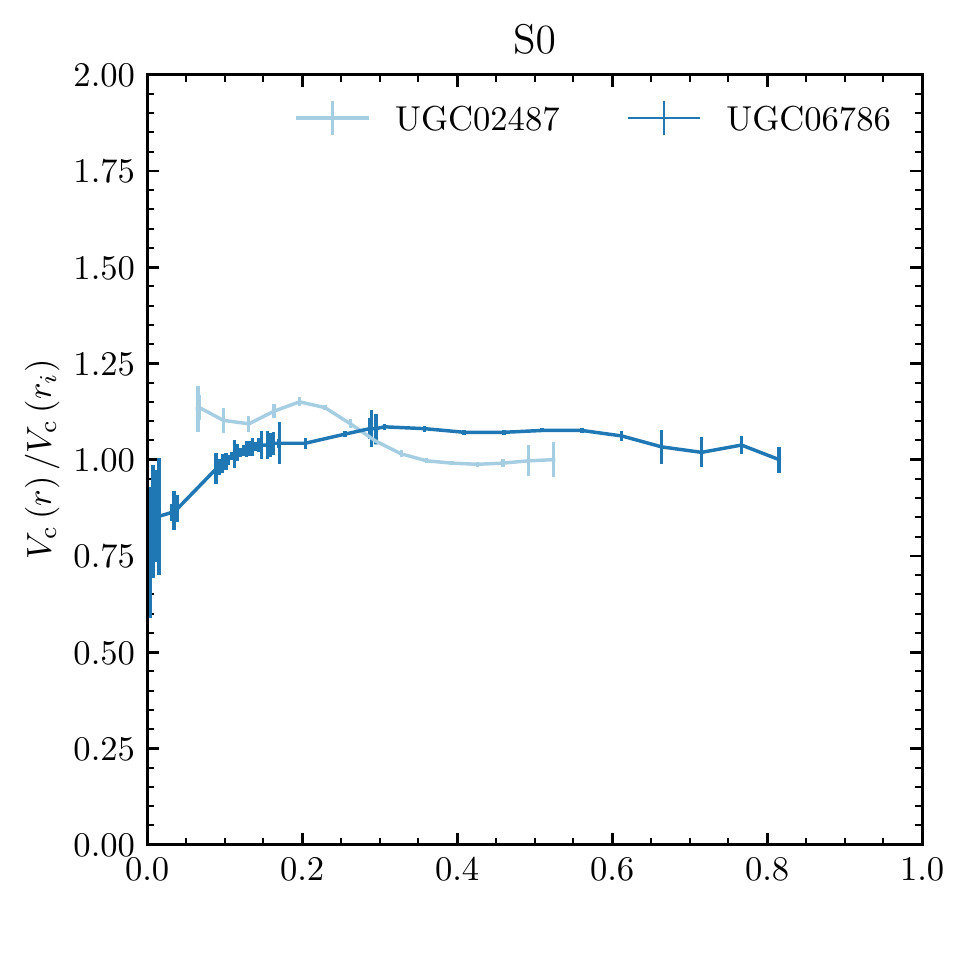}
	\hspace{-0.4cm}
	\includegraphics[width=0.34\textwidth]{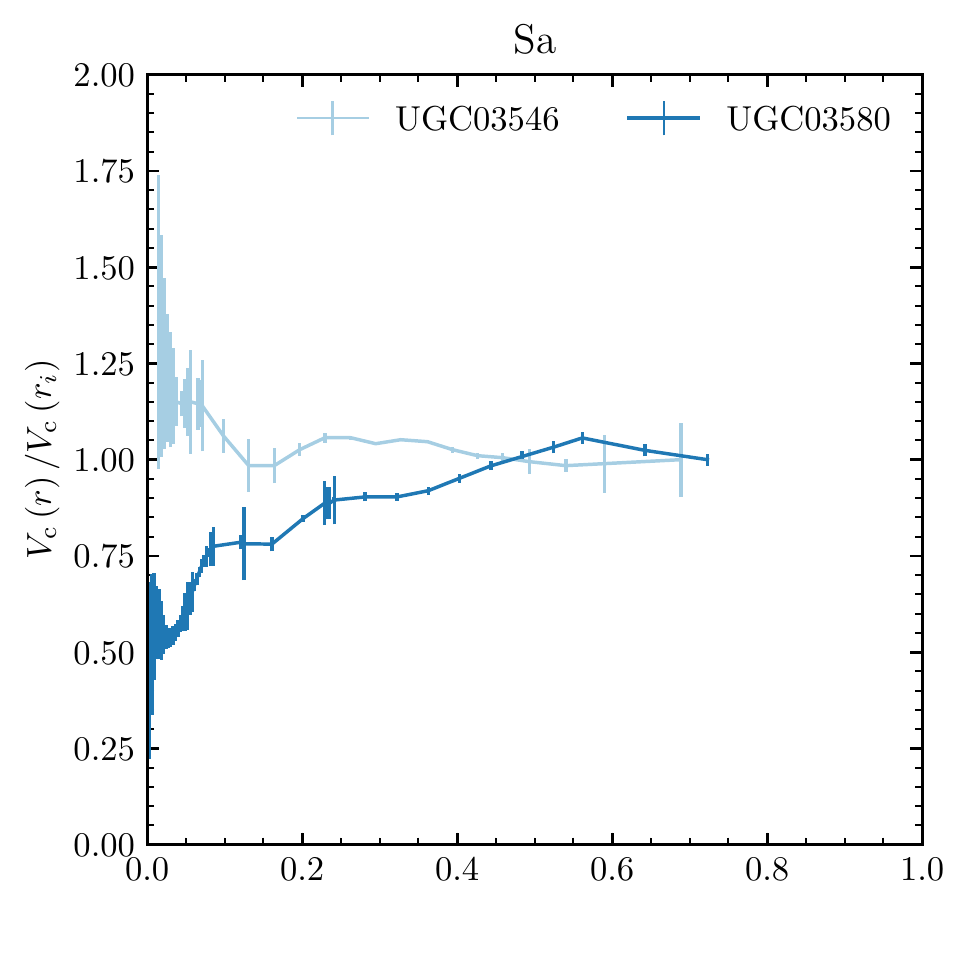}\\
	\vspace{-0.52cm}
	\hspace{-0.4cm}
	\includegraphics[width=0.34\textwidth]{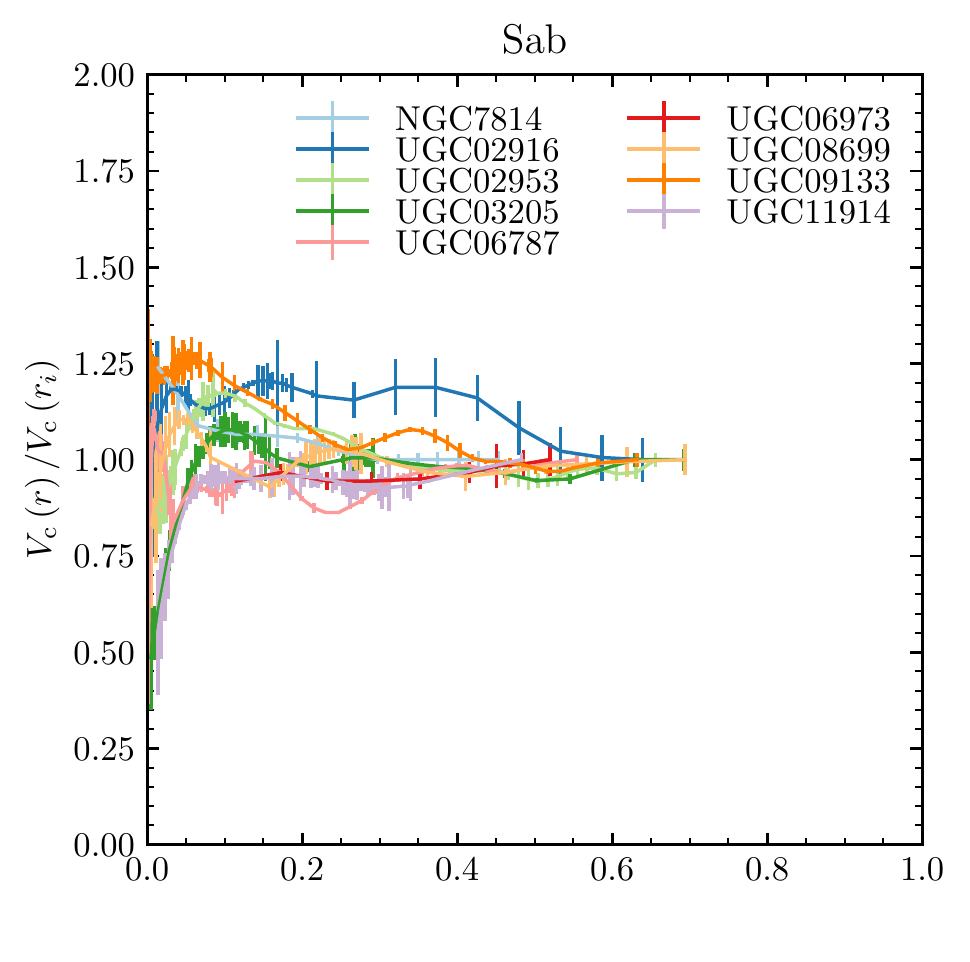}
	\hspace{-0.4cm}
	\includegraphics[width=0.34\textwidth]{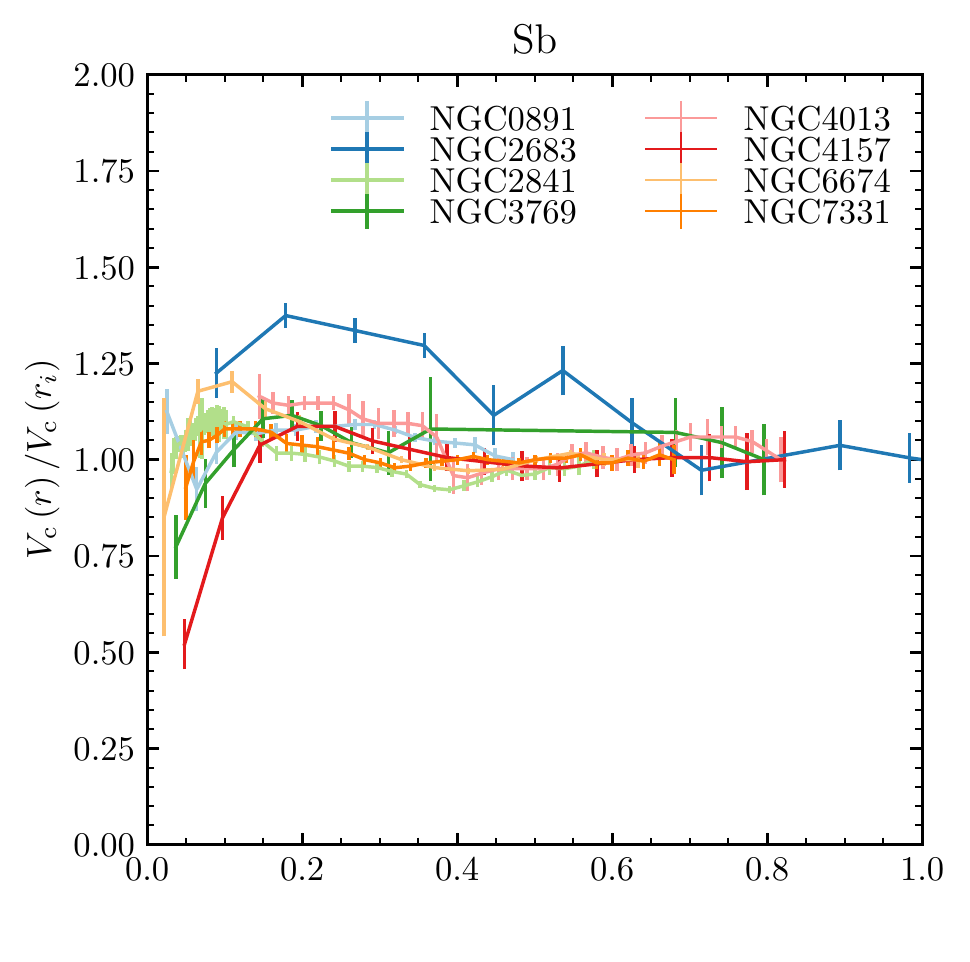}
	\hspace{-0.4cm}
	\includegraphics[width=0.34\textwidth]{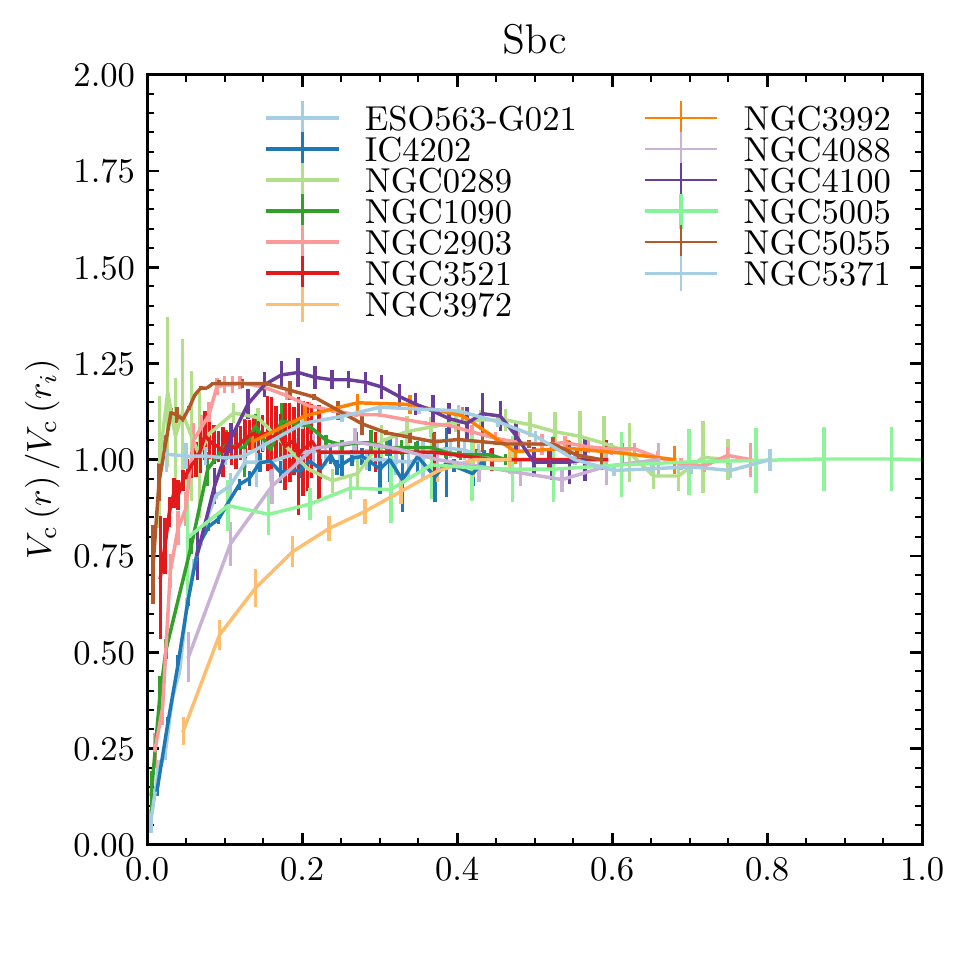}\\
	\vspace{-0.52cm}
	\hspace{-0.4cm}
	\includegraphics[width=0.34\textwidth]{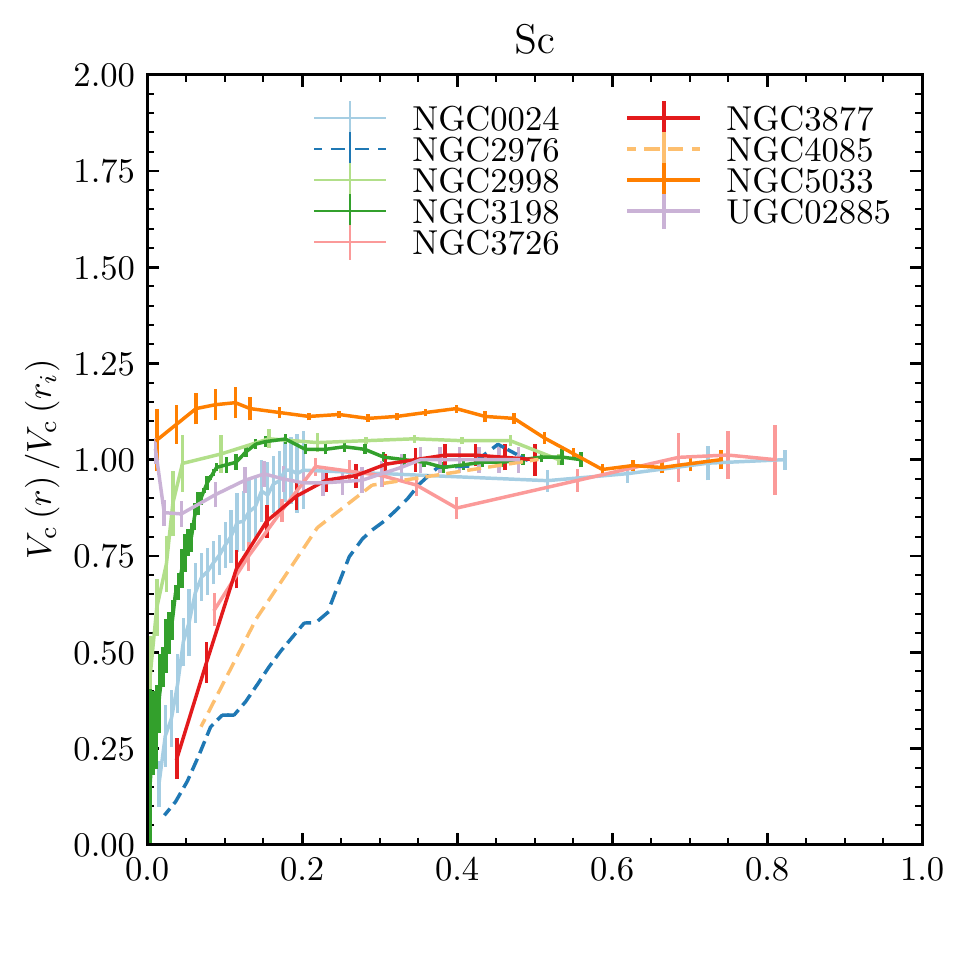}
	\hspace{-0.4cm}
	\includegraphics[width=0.34\textwidth]{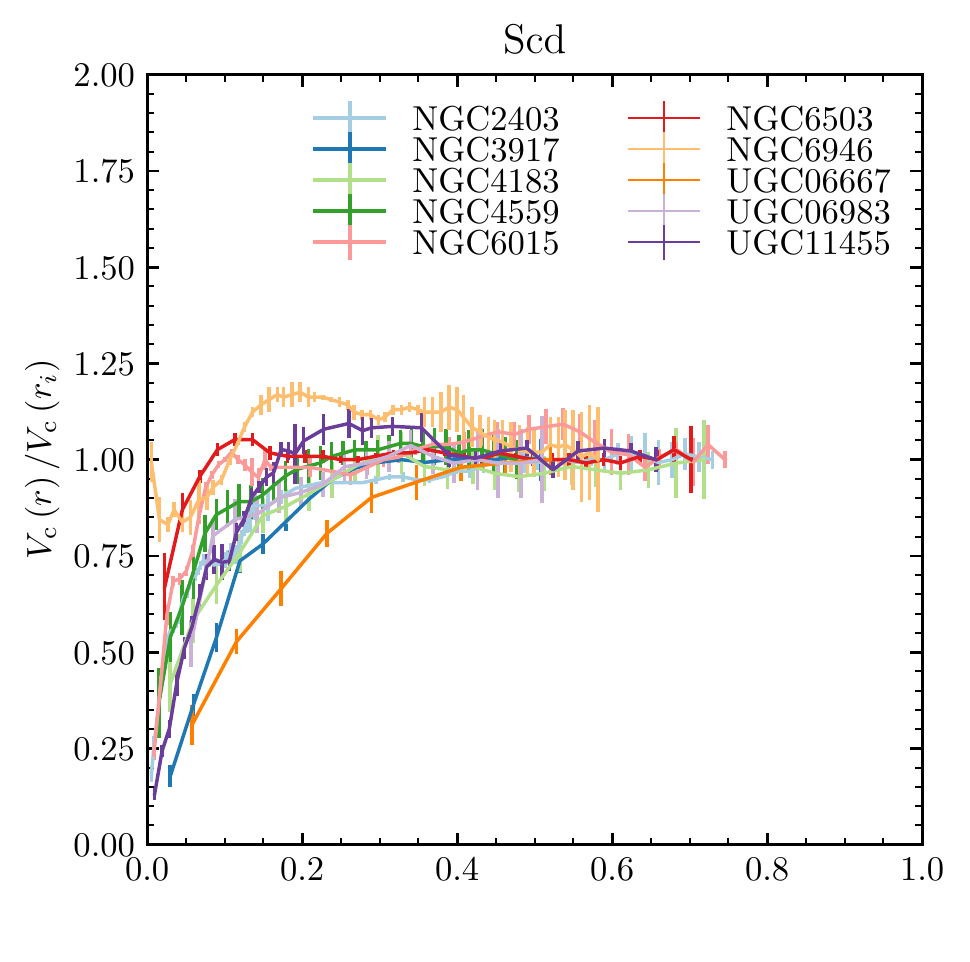}
	\hspace{-0.4cm}
	\includegraphics[width=0.34\textwidth]{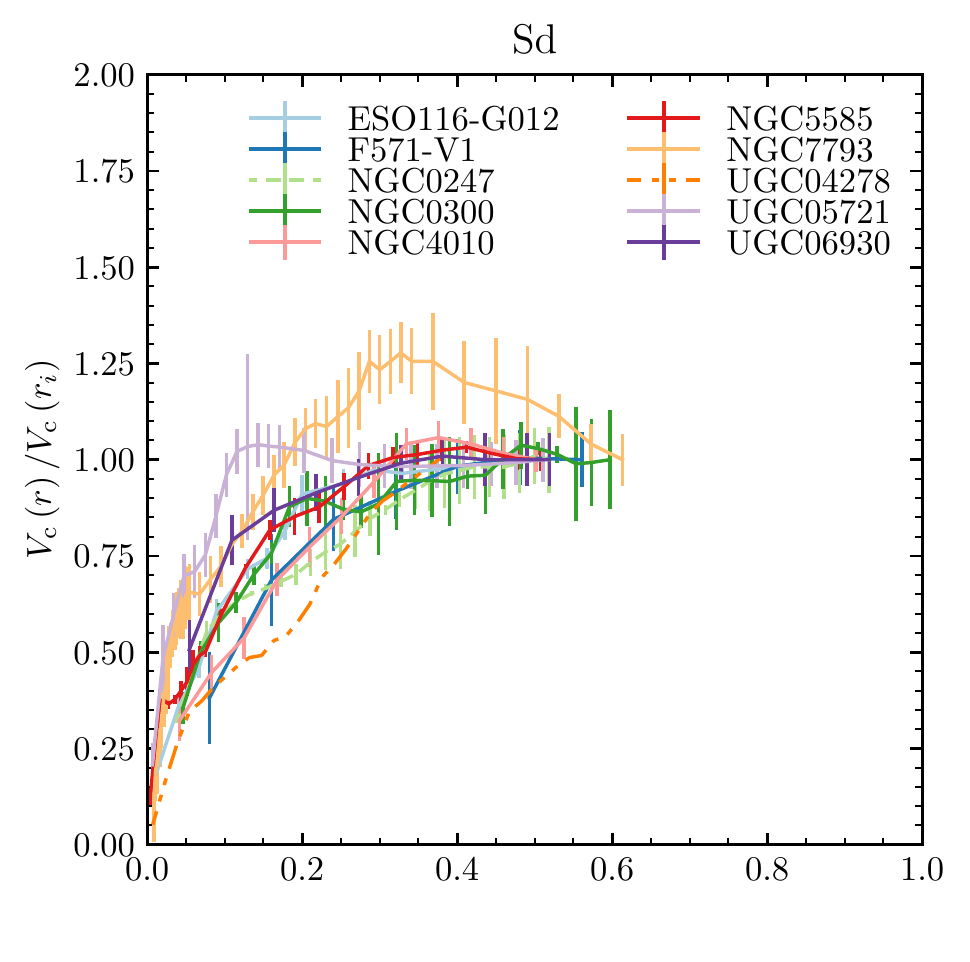}\\
	\vspace{-0.52cm}
	\hspace{-0.4cm}
	\includegraphics[width=0.34\textwidth]{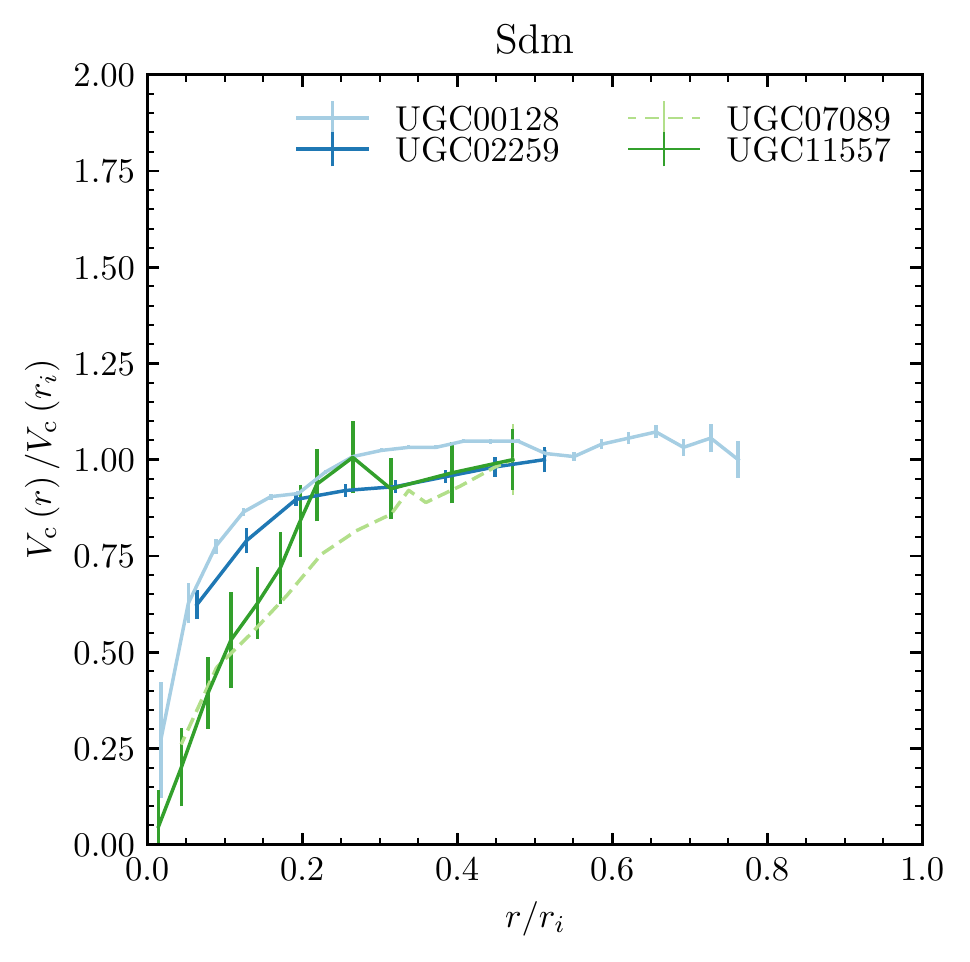}
	\hspace{-0.4cm}
	\includegraphics[width=0.34\textwidth]{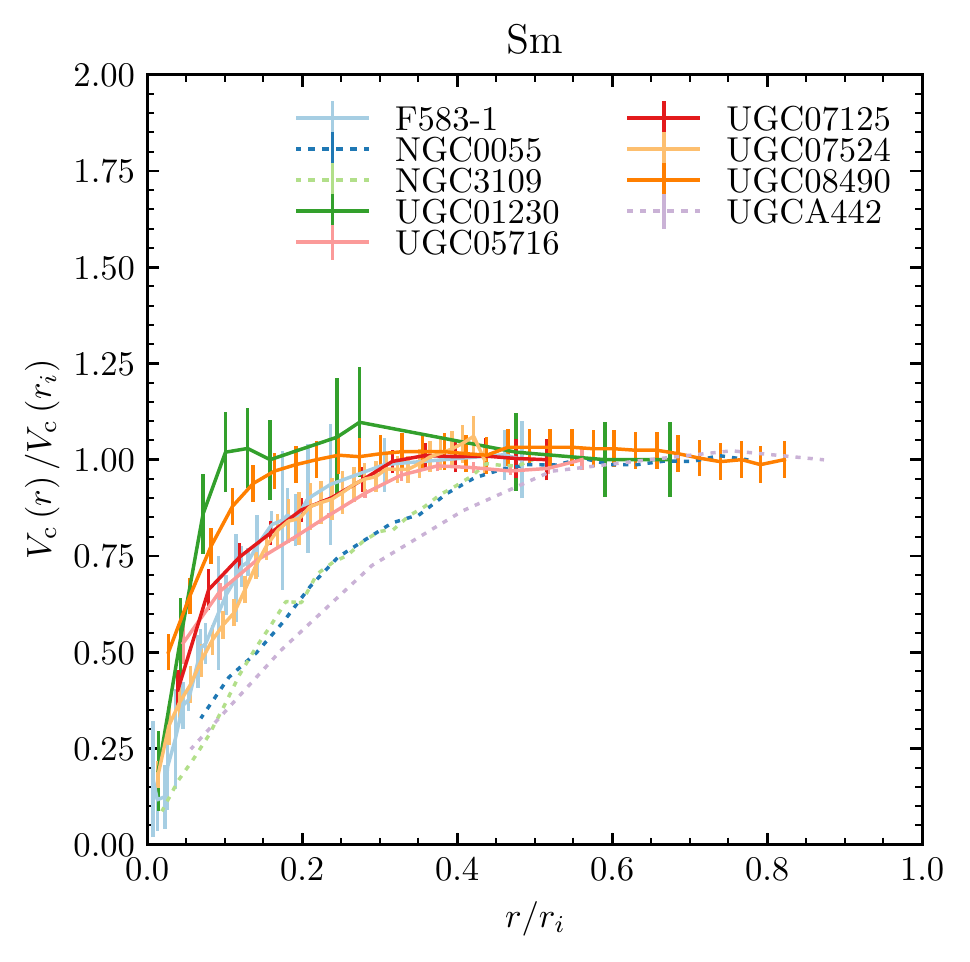}
	\hspace{-0.4cm}
	\includegraphics[width=0.34\textwidth]{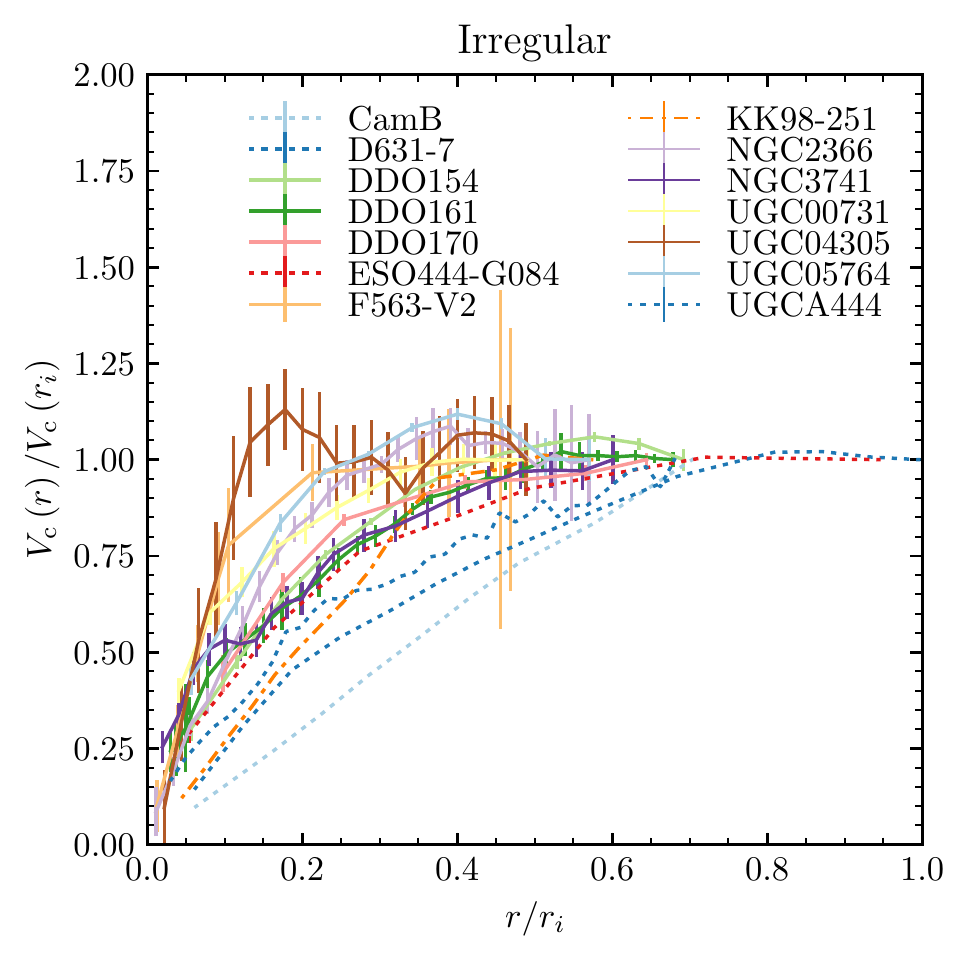}
	\vspace{-0.4cm}
	\caption{HI-inferred circular velocity curves of all SPARC galaxies from \citet{li:2020.sparc.rotation.curve.data} for which sufficient data was available to place them on a plot akin to our Fig.~\ref{fig:Vc.fire.disks} (for which the rotation curve is measured out to at least $\gtrsim 0.4\,r_{i}$ in order to have a meaningful comparison, giving $>90$ galaxies). For galaxies without rotation curves extending to $r=r_{i}$ we normalize to $V_{\rm c}$ measured at the maximum observed radius. We group the galaxies by the morphology given therein. Interacting systems are labeled with dashed lines, and systems which would robustly be classified as ``not a disk'' both morphologically and kinematically by both our criterion here and that in \citet{elbadry:fire.morph.momentum} are dotted. UGC4278 is labeled without error bars as its H$\alpha$-inferred $V_{\rm c}(r)$ disagrees with (is more concentrated than) the HI-inferred $V_{\rm c}(r)$ by a larger margin than the statistical errors. The vast majority of the SPARC sample, including almost all more-ordered systems and those with visible stellar disks (Sd or earlier), lies in the same regime as the FIRE systems and experiments here which form disks (Fig.~\ref{fig:Vc.fire.disks}). The two most potentially-significant outliers, NGC3972 and UGC6667 are near the boundary of our ``disky'' FIRE sample but still show $r_{\rm Vmax} \lesssim 0.5\,r_{i}$ and do not appear any significantly less concentrated than e.g.\ {\bf m11b} or {\bf m11h}.
	\label{fig:SPARC.comparison}}
\end{figure*}

We stress that there is no obvious contradiction between our criterion for disk formation and any observations of disk galaxies themselves, insofar as we are aware. Importantly, the ``centrally concentrated mass profile'' criterion does not contradict observations of disk rotation curves or structure in low-surface brightness and/or dwarf galaxies. First off, as is obvious from Figs.~\ref{fig:Vc.vs.disk}-\ref{fig:Vc.fire.disks}, the rotation curves $V_{\rm c}(r)$ of our disky galaxies can be and often are {\em rising} with radius $r$ at radii up to $\sim 0.5-5\,$kpc (or up to $\gtrsim 2$ times the extended HI disk effective radius) -- in more extreme cases like {\bf m11b} and {\bf m11h} up to $\sim 8-12\,$kpc -- and while they are falling at larger radii they tend to be extremely flat over radii $\sim 1-30\,$kpc (out to $\gtrsim 6$ times the HI effective radii). This means it is absolutely allowed, in principle, for the central $\lesssim {\rm few}\,$kpc of the mass profile to have a ``cored'' halo. This is all quite consistent with observed rotation curves \citep[see e.g.][and references therein]{deblock:2008.core.cusp.halo.low.densities.in.things,lelli:2016.sparc.rotation.curves,lang:2020.phangs.rotation.curves} -- and rigorous examples of this comparing the default FIRE simulations here to observed galaxies in their rotation curve structure, Tully Fisher relation, radial acceleration relation, etc., can be found in \citet{el.badry:jeans.modeling.dwarf.coherent.oscillations.biases.mass,elbadry:fire.morph.momentum,elbadry:HI.obs.gal.kinematics,chan:fire.udgs,garrisonkimmel:fire.morphologies.vs.dm}. 

In Fig.~\ref{fig:SPARC.comparison}, we show this somewhat more quantitatively: we take every galaxy in the published SPARC sample \citep{li:2020.sparc.rotation.curve.data} for which sufficient data is available (compiling data from \citealt{li:2020.sparc.rotation.curve.data,2020MNRAS.495.2867P,2020MNRAS.497.4795I,2021AJ....161...71H,2021AJ....161...93D,2021ApJ...913...53P} and additional references therein\footnote{Classifications and HI profiles used to determine $r_{i}$ are taken from \cite{deblock:1996.lsb.dwarf.HI,1997PhDT........13V,1998A&AS..131...73M,2000A&AS..141..469H,2000AJ....120.3027C,2001AJ....121.3026W,2001AJ....122..825B,2001AJ....122.2381M,2003NewA....8..267B,2003ApJ...596..957S,2004A&A...424..509B,2005AJ....129..698M,2005A&A...442..137N,2008MNRAS.386..138B,2008MNRAS.386.1667B,2008AJ....136.2648D,2008AJ....136.2563W,2010ApJ...716..792F,2011ApJS..192....6L,2011MNRAS.415..687B,2011AJ....142..121H,2012MNRAS.425.2083K,2014NewA...26...40W,2016AAS...22711105P,2016MNRAS.460..689R,2016MNRAS.460.2143W,2016MNRAS.463.4052P,2016AJ....152..157L,2017MNRAS.464.2419S,2018A&A...614A.143M,2018MNRAS.478.1611K,2018MNRAS.479.4136K}.}) to place them on the same Fig.~\ref{fig:Vc.fire.disks} (showing the shape of the rotation curve versus gas disk extent) as our FIRE galaxies. The SPARC systems which might plausibly have gas disks by our criteria (some clearly do not, as they both feature $H/R\gtrsim 0.3$ and a highly clumpy/irregular/asymmetric morphology) all lie in the overlapping parameter space to our FIRE disky systems, i.e.\ have ``sufficiently concentrated'' mass profiles. In fact, by the definition in Fig.~\ref{fig:correlation.reff.rmax}, all of these systems rather easily meet our criterion ($r_{\rm slope=1/4} \lesssim r_{\rm 50,\,HI}$ or $r_{\rm Vmax} \lesssim r_{90,\,{\rm HI}} \sim r_{i}$). We see that this is especially true for the more kinematically and morphologically well-ordered systems (i.e.\ the systems which are not Irr or Sm). 

Crucially, our requirement that the mass profile be centrally concentrated is a relative criterion that only applies at radii similar to or inside the initial circularization radii of extended gas disks. And indeed, almost every galaxy in surveys like THINGS, PHANGS, or SPARC (as representative samples of ordered, thin gas disks at low masses) is known to show $V_{\rm c}(r)$ shallow/close to flat or falling with $r$ outside $\sim 2-4$\,kpc. And in fact, those studies have consistently shown that on average, low-mass thin-gas-disk galaxies show excessive ``baryonic concentration'' as defined in \citet{elbadry:fire.morph.momentum} and discussed above (see e.g.\ Fig.~58 in \citealt{deblock:2008.core.cusp.halo.low.densities.in.things}, showing that the central $\sim 10\,$kpc are often strongly baryon-dominated in such systems). 

Moreover, there is no contradiction between a requirement of central concentration and observations of ``bulge-free'' disks (observationally defined as disks without a high-Sersic-index excess central light profile above the inward extrapolation of an exponential disk; see e.g.\ \citealt{kormendy:2010.bulgeless.galaxies.vs.mergers,kormendy:2012.spheroidals,simmons:2013.bulgeless.disks.without.bhs.but.dont.count.pseudobulges}). 
First, the concentration we discuss could be (and in some of our default FIRE simulations clearly is) entirely dark, owing to dark matter (e.g.\ a cusp, or sufficiently concentrated cored-NFW halo), or SMBHs, or gas, or other components (as discussed above). 
Second, a central light concentration could still be present, in the form of a pseudobulge, inner disk, or barlens/ring, all of which would still be classified as ``bulge-free'' via any of the criteria above \citep[see references above and][]{kuijken:pseudobulges.obs,debattista:pseudobulges.a,fisher:pseudobulge.ns,erwin:2015.pseudo.bulges,athanassoula:2015.bulgeless.bars.face.on}. 
Third, a central luminous mass could be small, less than $\sim 10\%$ of the baryonic mass in our experiments here, which is where it would generally be called ``bulge-free'' in most observational diagnostics \citep[see][]{gao:2017.bulge.disk.decomposition.fitting}.  
Fourth, an early-forming central mass concentration could be physically relaxed or dissolved after a disk forms: a star cluster can be dissolved by N-body relaxation and/or core-collapse, or inflated by mergers, or a collisionless ``cusp'' in stars or dark matter can be flattened into a core later if bursty star formation continues \citep{naab:dry.mergers,pontzen:2011.cusp.flattening.by.sne,teyssier:2013.cuspcore.outflow}. 
Fifth, such a concentration can simply be ``buried'' underneath subsequent star formation: e.g.\ the $\sim 10^{9}\,M_{\odot}$ central masses within a $\sim$\,kpc in our most extremely $M_{0}$ models, even if in stars, would by the time the galaxy grows more massive be easily ``hidden'' underneath the light profile of a standard low-surface-brightness disk  \citep{courteau:disk.scalings}.

There are, of course, outliers expected for any such nonlinear process as galaxy formation. One apparent example, at face value, would be NGC 925 (NGC3972 and UGC6667, noted in Fig.~\ref{fig:SPARC.comparison}, are qualitatively similar but less extreme, so less interesting here), which is a relatively massive $M_{\ast} \sim 3\times10^{9}\,M_{\odot}$ disk with a rising gas rotation speed (not necessarily $V_{\rm c}$) out to $r\sim 10\,$kpc (then flat/falling; see \citealt{deblock:2008.core.cusp.halo.low.densities.in.things,wiegert:2011.spiral.HI.models}). This is marginally larger than most of the low-density disks in our FIRE sample, but we stress that examples like {\bf m11b}, which we study in detail (\S~\ref{sec:m11b}), have similar profiles and still fall within our ``sufficiently centrally concentrated'' parameter space. Moreover it is worth noting that the gas disk in NGC 925 is very extended ($\gtrsim 15\,$kpc) and gas-poor in the center -- so in a comparison of $r_{{\rm slope}=1/4}$ versus $r_{\rm 50,\,HI}$ or $r_{\rm Vmax}$ versus $r_{\rm 90,\,HI}$ (as Fig.~\ref{fig:correlation.reff.rmax}), or plotting $V_{\rm c}$ versus $r/r_{i}$ (as Figs.~\ref{fig:Vc.fire.disks} or \ref{fig:SPARC.comparison}), it appears marginal (like our {\bf m11b} and {\bf m11h} cases), but not an actual outlier. Indeed, even if the dark matter is ``strongly cored'' in this system, the {\em stellar} mass profile appears to trace a stellar density which is concentrated in our sense, consistent with a Mestel disk in 2D or a singular isothermal sphere $\rho_{\ast} \propto r^{-2}$ (or even slightly steeper) in 3D, dominating over the inferred dark matter+gas density inside $\lesssim 4\,$kpc, and many have noted that the total density profile at small radii is difficult to model and could in fact be steeply rising \citep{pineda:2017.rotation.curve.fitting.drawn.to.core.even.for.vc.curves,oman:2019.thick.disks.systematically.underestimate.central.vc.in.tilted.ring.models,roper:2022.model.vc.curves.incorrect.core.inference}. Even if the mass profile were not compact, the morphology is unusual, as NGC 925 appears to be (a) strongly-barred; (b) globally asymmetric (in both the mass distribution and shape of the spiral arms); (c) warped; and (d) very thick (with a cold HI vertical scale height of $\sim 0.5-1\,$kpc, i.e.\ $H/R\sim 0.5-1$ inside $\lesssim 2\,$kpc; \citealt{bacchini:2019.things.galaxy.height.kinematic.profiles,patra:2019.molecular.scale.height.dwarfs,patra:HI.scale.height.dwarfs}). So the gas disk could be transient, as suggested by \citet{heald:2011.halogas.HI.obs} who describe NGC 925 as a ``prototypical minor merger'' with the outer gas disk in fact being an accreted stream of gas from an interacting companion \citep[see also][]{pingel:2018.gas.env.local.dwarf.galaxies}; or the bar could have flattened the profile from an earlier, more-concentrated configuration \citep{weinberg:bar.halo.res.initial,berentzen:bar.evol.in.cuspy.halos}.

We stress that we are not purporting to explain nor predict the structure of disk rotation curves or light profiles here. Whether or not some galaxy model can reproduce the observations above from first principles is sensitive to physics far beyond what we explore. We simply wish to stress that none of these observations appear to {\em prima facia} contradict (indeed, many of them appear to support) our conclusions regarding causal mechanisms here.

When comparing to observations, it is important to note, as shown explicitly in detail in \citet{elbadry:fire.morph.momentum}, that ``disky'' by our criteria is not the same thing as ``showing an ordered velocity shear/gradient viewed edge-on.'' There are many examples of systems which show no disk whatsoever in $j/j_{\rm c}$, or any other detailed 3D morphological or kinematic measurement, but would show a clear velocity shear edge-on (this can arise for many reasons, for example, bi-directional outflows). 
%A systematic study is in \citet{elbadry:fire.morph.momentum} but we also show a number of examples both in our default FIRE simulations and {\bf m11a} experiments in Appendix~\ref{sec:example.velocity}. 
In future work, it will be particularly interesting to compare the $V_{\rm c}$ profiles which would be observationally inferred from these systems in realistic mock observations: given the similarity of the velocity moment and HI surface brightness maps to certain observed systems, it is almost certainly the case that some would be fit (incorrectly) via tilted ring models to ``thin disks,'' which could be important for understanding how these galaxies are used to infer circular velocity curves and other constraints \citep{oman:2019.thick.disks.systematically.underestimate.central.vc.in.tilted.ring.models}.

\subsubsection{Bursty Star Formation}

It is much more challenging to directly compare any criterion for ``bursty'' vs.\ ``smooth'' star formation to individual observed galaxies, since our assessment of this is based on the entire time-history of the galaxy, and (at least at present) it is simply not possible to reconstruct the archeological star formation history of individual galaxies (even the Milky Way, Magellanic Clouds, and Andromeda) at $\lesssim 100\,$Myr time resolution over $\sim\,$several Gyr.\footnote{Importantly, we stress that there is a difference between ``burstiness'' as we define it, and assessing whether or not a given galaxy at the time observed is undergoing or coming out of a {\em single} starburst episode (defined as a star formation event with higher SFR than the past-averaged history of the galaxy), as any single starburst could be (in principle) a one-off event caused by phenomena such as mergers or disk instabilities.} 
Rather, studies trying to assess this have had to rely on {\em populations} of galaxies with semi-direct diagnostics (e.g.\ comparing the dispersion of SFR tracers nominally sensitive to stars of different masses and therefore ages, as in \citealt{sparre.2015:bursty.star.formation.main.sequence.fire,2021MNRAS.501.4812F}), or on much more indirect arguments about correlations between burstiness and stellar/gas kinematics or abundance patterns \citep{elbadry.2015:core.transformation.stellar.kinematics.gradients.in.dwarfs,ma:radial.gradients,emami:2021.fire.testing.bursty.sf.size.fluct.corr,patel:2022.mw.abundance.patterns.vs.bursty.sf.and.mergers}. While there is qualitative agreement across these indicators that dwarf galaxies tend to have ``more bursty'' SFHs, with a transition to smoother SFHs that is qualitatively consistent with the mass scales we predict above, the uncertainties are still large, and this is largely restricted to qualitative statements \citep{2016ApJ...833...37G,kurczynsk:candels.sfr.main.sequence.scatter.evol,emami:2019.bursty.dynamics.similar.to.fire.in.obs.but.fire.too.rapid.in.intermediate,2021MNRAS.506.1346K,atek:2022.sf.bursty.in.dwarfs}. Indeed, {\em quantitative} interpretation of indicators like the H$\alpha$ to UV ratio (and even what timescales these indicators actually probe) remains controversial and uncertain \citep[see references above and][]{2021MNRAS.501.4812F}. And we are not, here, attempting to make any quantitative prediction for the magnitude or timescales of ``burstiness,'' only a prediction for the conditions where this decreases.

Compounding these complications, the escape velocity $V_{\rm esc}$ is perhaps one of the most difficult-to-measure properties of galaxies, as it requires detailed mass models from $\sim 1\,$kpc to $\sim 100\,$kpc. Even in the Milky Way (the one galaxy where reliable estimates exist from multiple techniques), estimates of $V_{\rm esc}$ {\em at the Solar circle} vary from $\sim 400-700\,{\rm km\,s^{-1}}$ (though most favor values closer to $\sim 450\,{\rm km\,s^{-1}}$; see \citealt{necib:2022.mw.escape.velocity.estimate}) and vary at $r\lesssim 1\,$kpc by a factor of two.

\section{Conclusions}
\label{sec:conclusions}

In this paper, we considered an extensive set of numerical experiments in cosmological simulations, as well as validation against the suite of existing FIRE galaxy simulations, in order to explore different proposed mechanisms for the onset of disk formation and/or cessation of ``bursty'' star formation. Our experiments are designed to allow us to separate different physical processes which are normally tightly-correlated in galaxy formation simulations and therefore complicate physical interpretations. We provide more detailed physical understanding of these processes and show that we see clear evidence of the typical proximate physical cause of each. Our major conclusions and their relation to real, observed galaxies are discussed in detail above. In summary, our most important conclusions are:

\begin{enumerate}

\item{The key criterion for initially forming a disk is the development of a sufficiently centrally-concentrated mass profile. A radial acceleration which is (even weakly) increasing towards smaller radii, interior to the gas circularization location, promotes initial disk formation and stabilizes these proto-disks for many reasons (\S~\ref{sec:diskform}). We discuss a variety of mechanisms by which this might occur in nature, and how it relates to observed disk galaxy structure (\S~\ref{sec:obs}). This mass/acceleration profile does not necessarily have to persist after the disk forms, and of course other physics can destroy disks that have previously formed.}

\item{Disk formation and ``smooth'' star formation are not one-to-one connected, even if in practice they tend to occur at about the same time: one can, in principle, form a well-ordered disk which still exhibits ``bursty'' star formation, and conversely one can have a system with no disk whatsoever but ``smooth'' star formation (\S~\ref{sec:decoupling}). However, it does appear that very thin disks, if most of the star formation is inside the thin disk itself, will usually meet the criteria for smooth star formation.}

\item{The key criterion for star formation becoming ``smooth'' appears to be the potential or escape velocity scale at the radii of star formation (to some radii $\gtrsim 10-20\,$kpc) crossing a critical threshold (also lowering the re-accretion timescale for material ejected), given crudely by $\sigma_{\rm eff}/V_{\rm esc} \lesssim 0.05$ or $V_{\rm esc} \gtrsim 200\,{\rm km\,s^{-1}}$. Once this scale is exceeded, a large fraction of material which would otherwise be ejected from the galaxy, in particular cold, mass-loaded outflows of gas which could otherwise form stars, is trapped and/or confined (or more rapidly recycled) within the galaxy and stabilizes the SFR (\S~\ref{sec:bursty}). The similarity of this $V_{\rm esc}$ threshold and the threshold for inner CGM virialization (ICV) also provides a natural explanation for why phenomena like ICV and the onset of disk settling/thin disk formation tend to accompany the bursty-smooth transition.}

\item{The gas supply/fraction to/in the galaxy, cooling rates and physics, criteria for and densities of star formation, and strength/rates/forms of stellar feedback, {\em do not} play a direct causal role in either the initial formation of galaxy disks, nor in the transition from bursty to smooth star formation (\S~\ref{sec:sffb.fx.weak}-\ref{sec:thermo}). Of course, these physics can {\em indirectly} influence disk formation via their non-linear influence on the gravitational potential evolution, and influence the amount of disk material which can be accreted, but they are not directly causally related.}

\item{Disk ``vertical settling'' (becoming morphologically thinner and kinematically colder) is a distinct process not strictly causally related to either disk formation or the cessation of bursty star formation (\S~\ref{sec:settling}). Settling occurs after a disk can initially form, and once the potential is sufficiently deep that gas with thermal $c_{s} \sim 10\,{\rm km\,s^{-1}}$ cannot maintain a very thick system, as disks self-regulate at marginal stability ($Q\sim 1$) so $H/R \sim \sigma/V_{\rm c} \sim M_{\rm gas,\,disk}(<r)/M_{\rm enc}(<r)$ decreases. We show that decreasing this quantity {\em before} the disk forms has no effect on disk formation or the cessation bursty star formation (\S~\ref{sec:thermo}).}

\end{enumerate}

Of course, all of the above do not form a complete cosmological explanation for all galaxy properties. Nor are they always sufficient to automatically cause a transition: for example if there were no accreting gas, or it had no net angular momentum, obviously no gas disk could form, and strong dynamical perturbations (e.g.\ mergers) could induce bursts and prevent disk formation. Alternatively galaxies could occasionally form (often transient) disky structures or briefly appear to undergo episodes of ``smooth'' SF under special conditions without meeting these criteria. Our goal here was instead to understand which of several physical possibilities most directly and generically are required for these processes. The statistics of large populations (and therefore exceptions to the rule) are beyond the scope of our sample. And the origins in a cosmological sense of any one of these phenomena (e.g.\ the formation of a more-centrally-concentrated mass profile) are not explained here (although we do briefly discuss various possibilities in \S~\ref{sec:obs}), and could be myriad. In future work we hope to explore this, as well as more diverse conditions for the processes above.

\acknowledgments{We thank Alyson Brooks, Vadim Semenov, and Charlie Conroy for helpful conversations during the development of this draft. Support for PFH was provided by NSF Research Grants 1911233 \&\ 20009234, NSF CAREER grant 1455342, NASA grants 80NSSC18K0562, HST-AR-15800.001-A. Numerical calculations were run on the Caltech compute cluster ``Wheeler,'' allocations FTA-Hopkins supported by the NSF and TACC, and NASA HEC SMD-16-7592. MBK acknowledges support from NSF CAREER award AST-1752913, NSF grants AST-1910346 and AST-2108962, NASA grant 80NSSC22K0827, and HST-AR-15809, HST-GO-15658, HST-GO-15901, HST-GO-15902, HST-AR-16159, and HST-GO-16226 from the Space Telescope Science Institute, which is operated by AURA, Inc., under NASA contract NAS5-26555. JS was supported by the Israel Science Foundation (grant No. 2584/21).
CAFG was supported by NSF through grants AST-1715216, AST-2108230,  and CAREER award AST-1652522; by NASA through grants 17-ATP17-0067 and 21-ATP21-0036; by STScI through grants HST-AR-16124.001-A and HST-GO-16730.016-A; by CXO through grant TM2-23005X; and by the Research Corporation for Science Advancement through a Cottrell Scholar Award. ZH was supported by a Gary A. McCue postdoctoral fellowship at UC Irvine. The Flatiron Institute is supported by the Simons Foundation. This work made use of the FIRE data repository hosted by the Flatiron Institute.}

\datastatement{The data supporting this article are available on reasonable request to the corresponding author.} 

%\bibliography{/Users/phopkins/Dropbox/Public/ms}
\bibliography{ms_extracted}

\begin{appendix}

\section{Some Additional Parameters}
\label{sec:additional.params}

\begin{figure*}
	\includegraphics[width=0.32\textwidth]{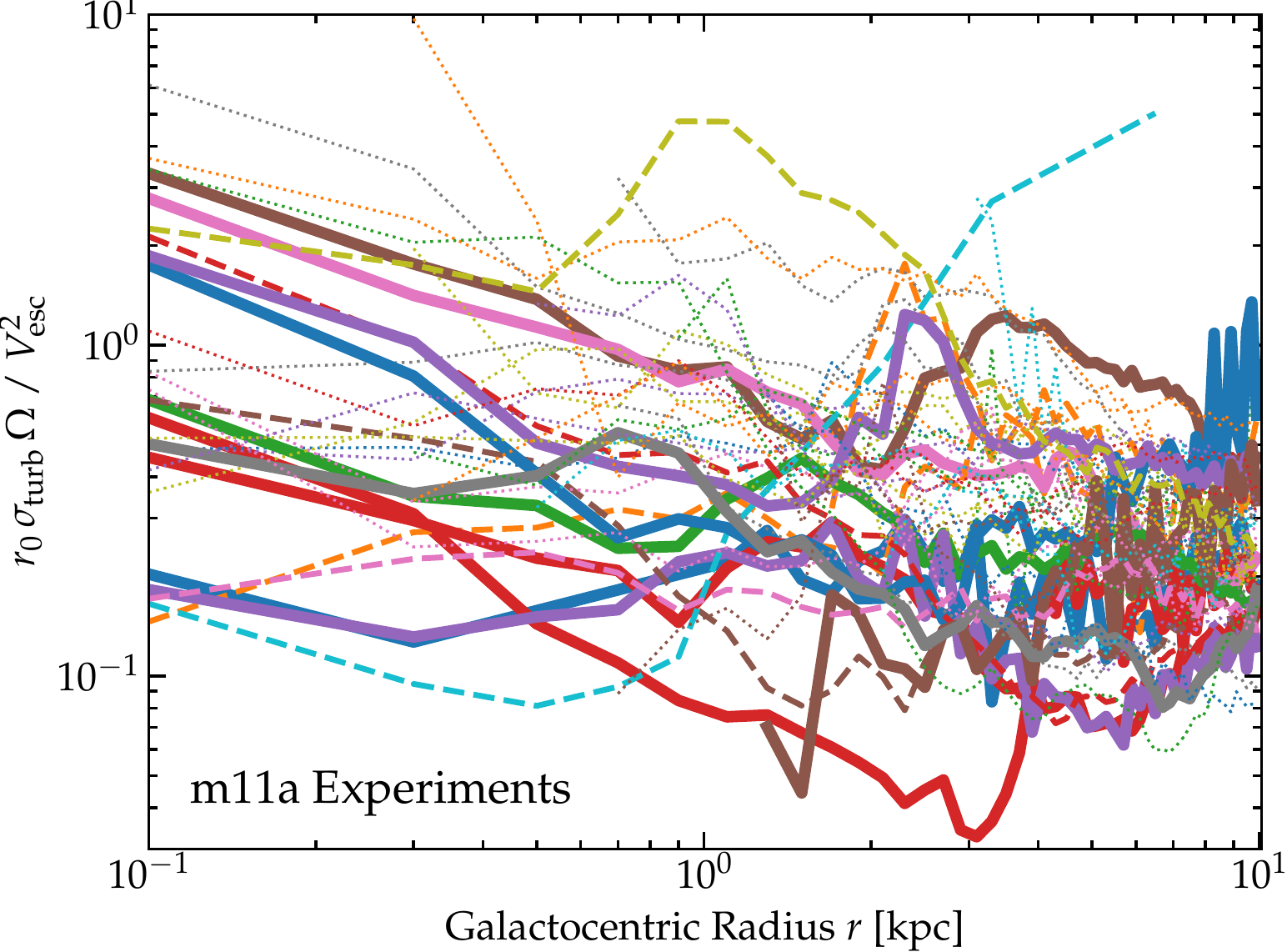} 
	\includegraphics[width=0.32\textwidth]{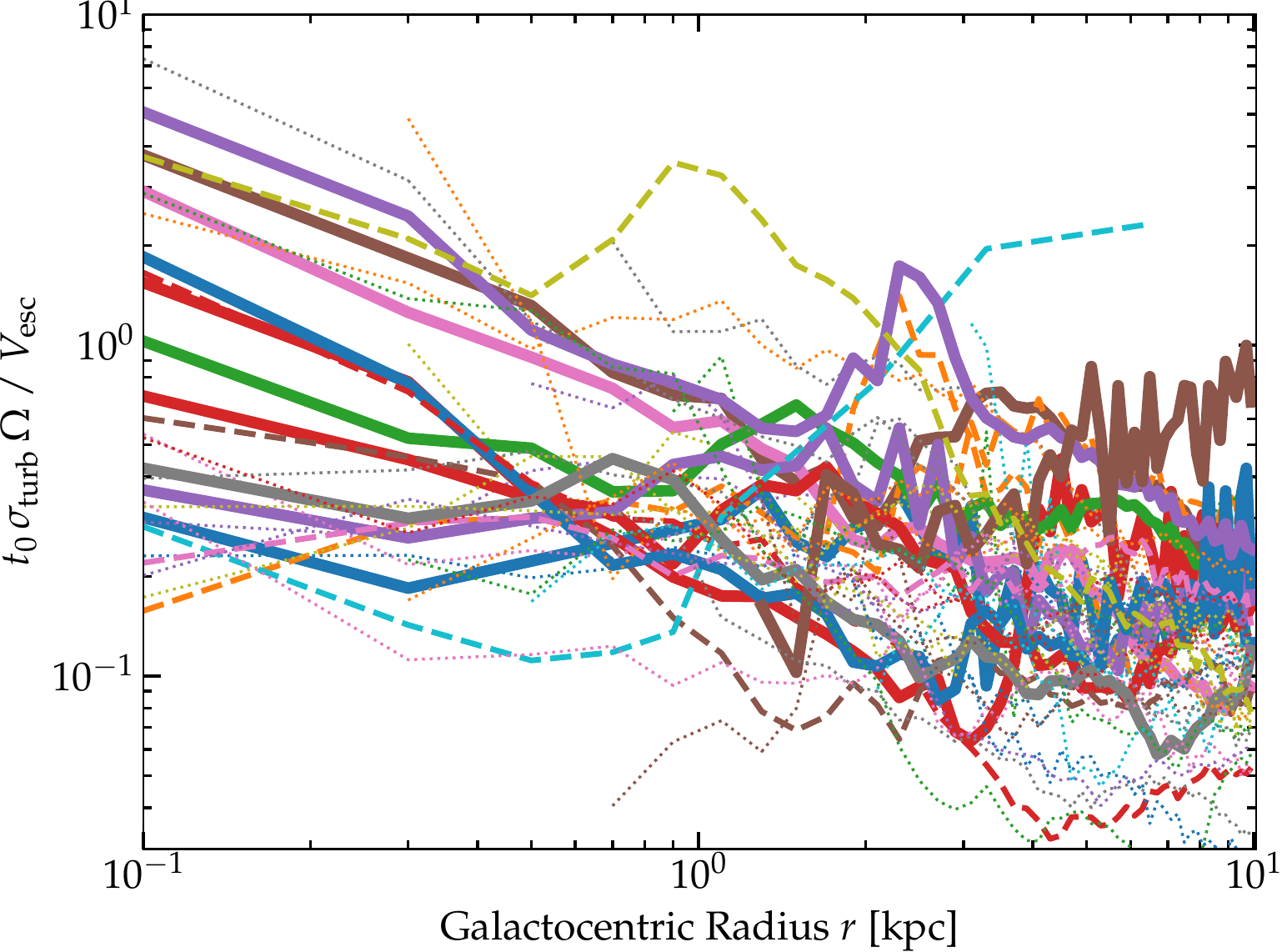}
	\includegraphics[width=0.32\textwidth]{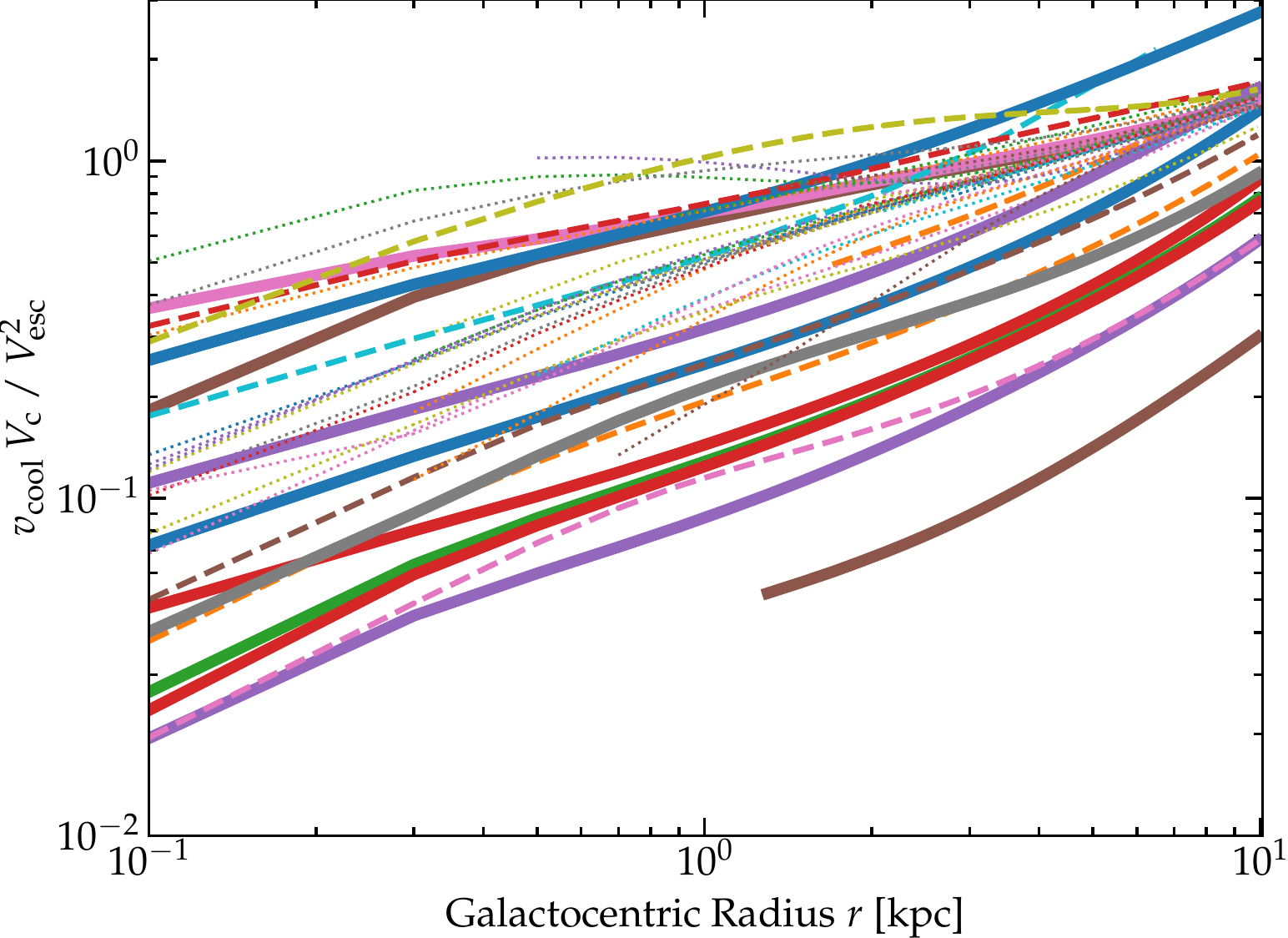} \\
	\includegraphics[width=0.32\textwidth]{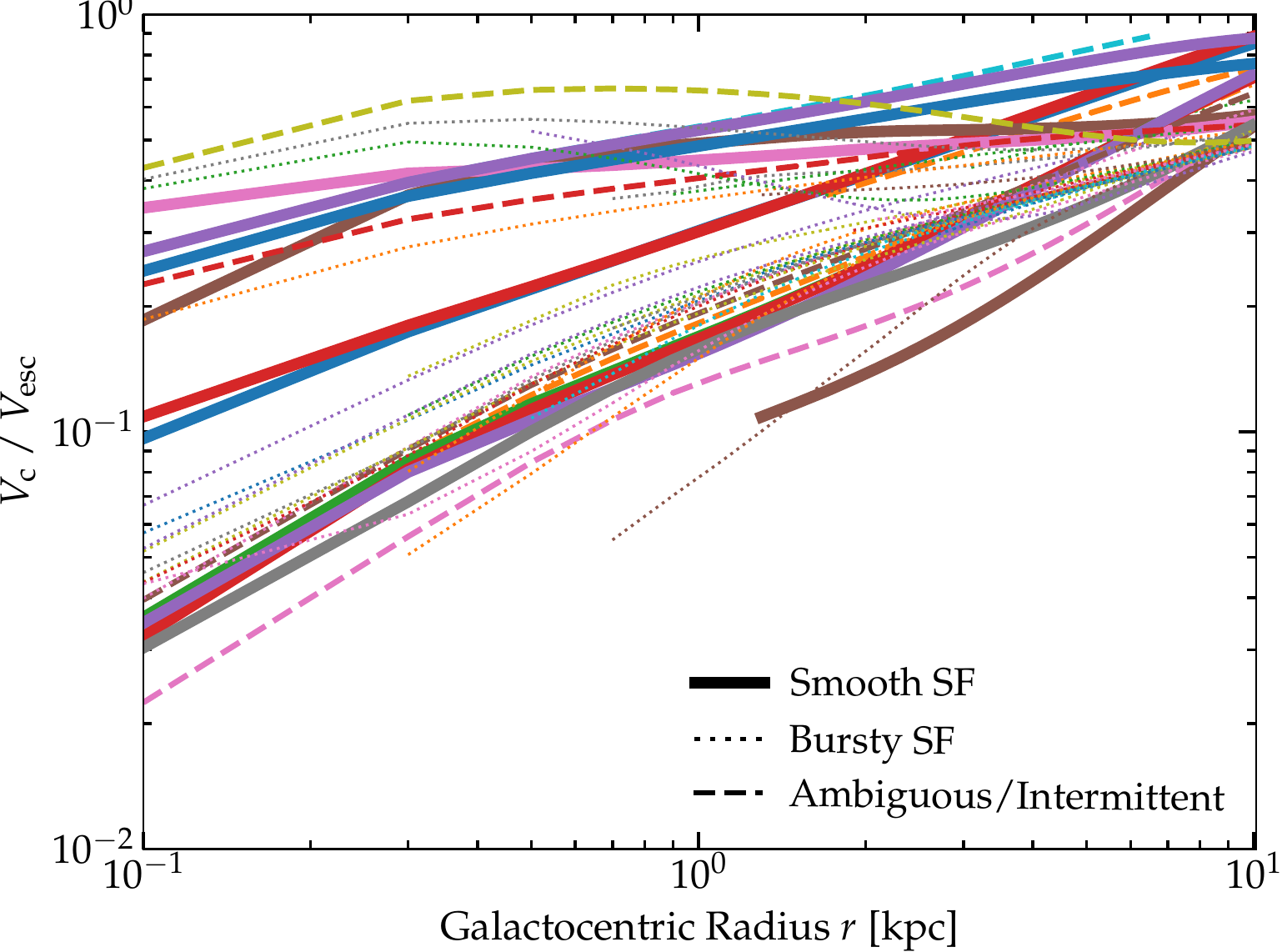} 
	\includegraphics[width=0.32\textwidth]{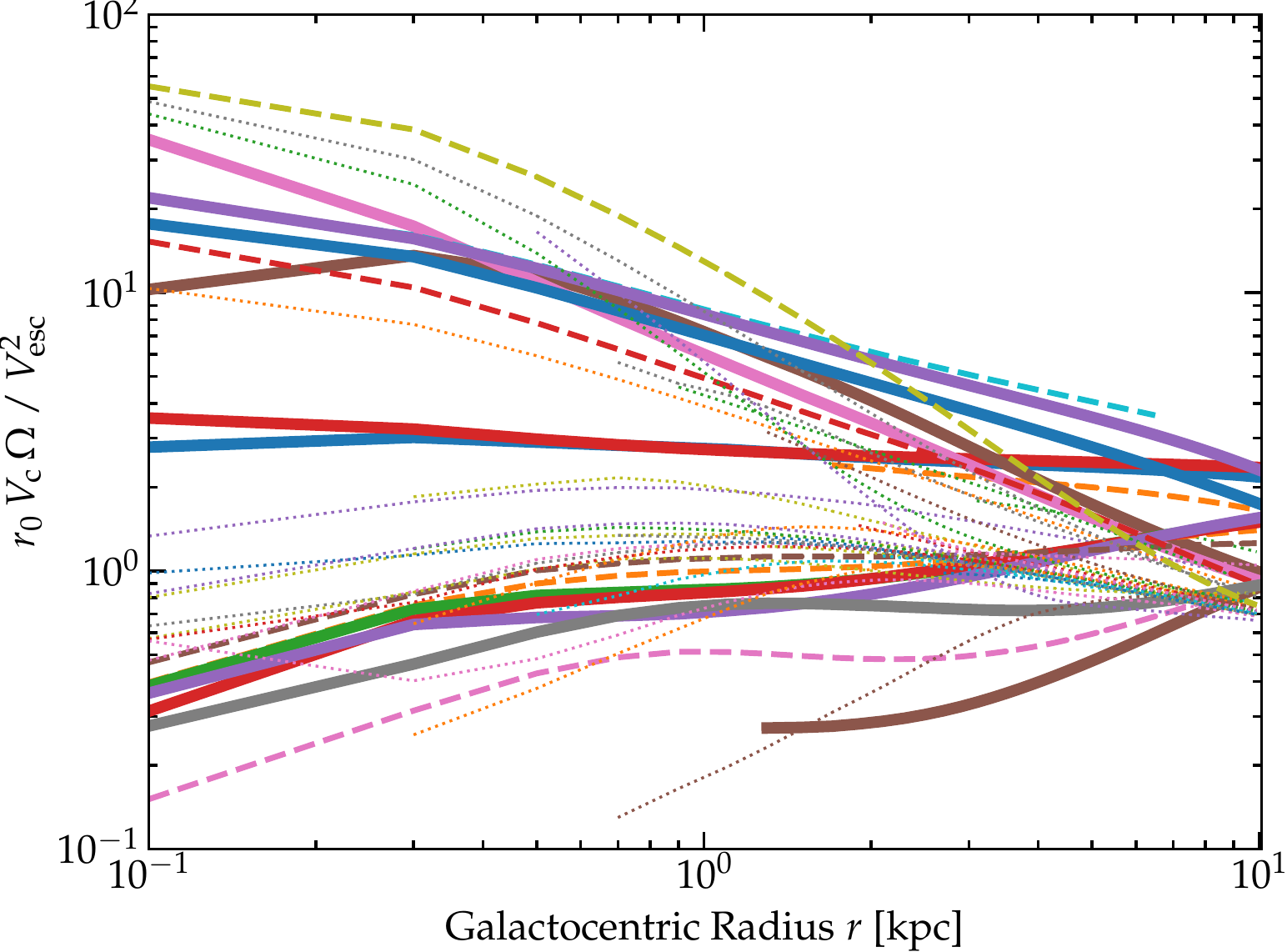}
	\includegraphics[width=0.32\textwidth]{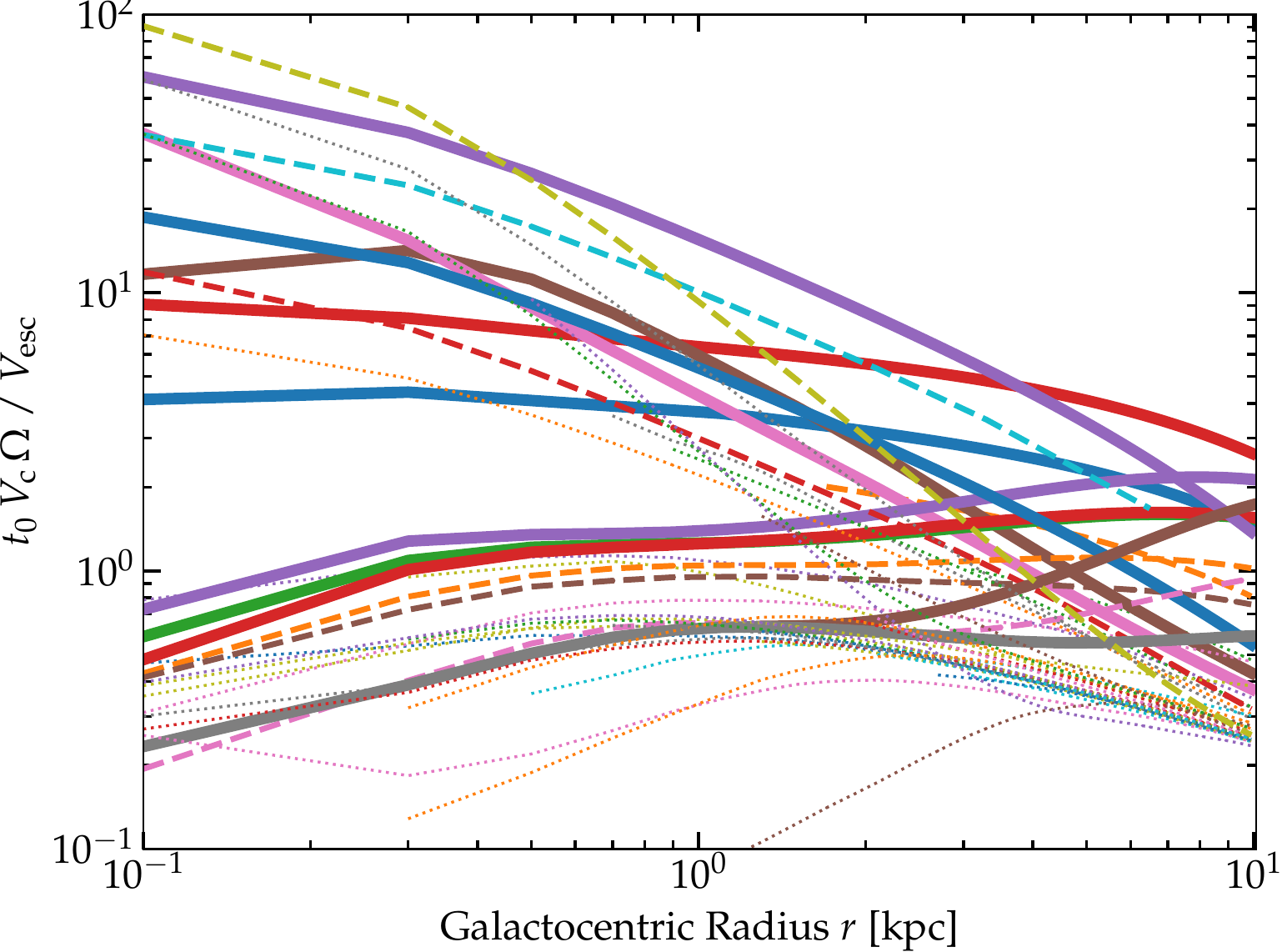} \\
	\includegraphics[width=0.32\textwidth]{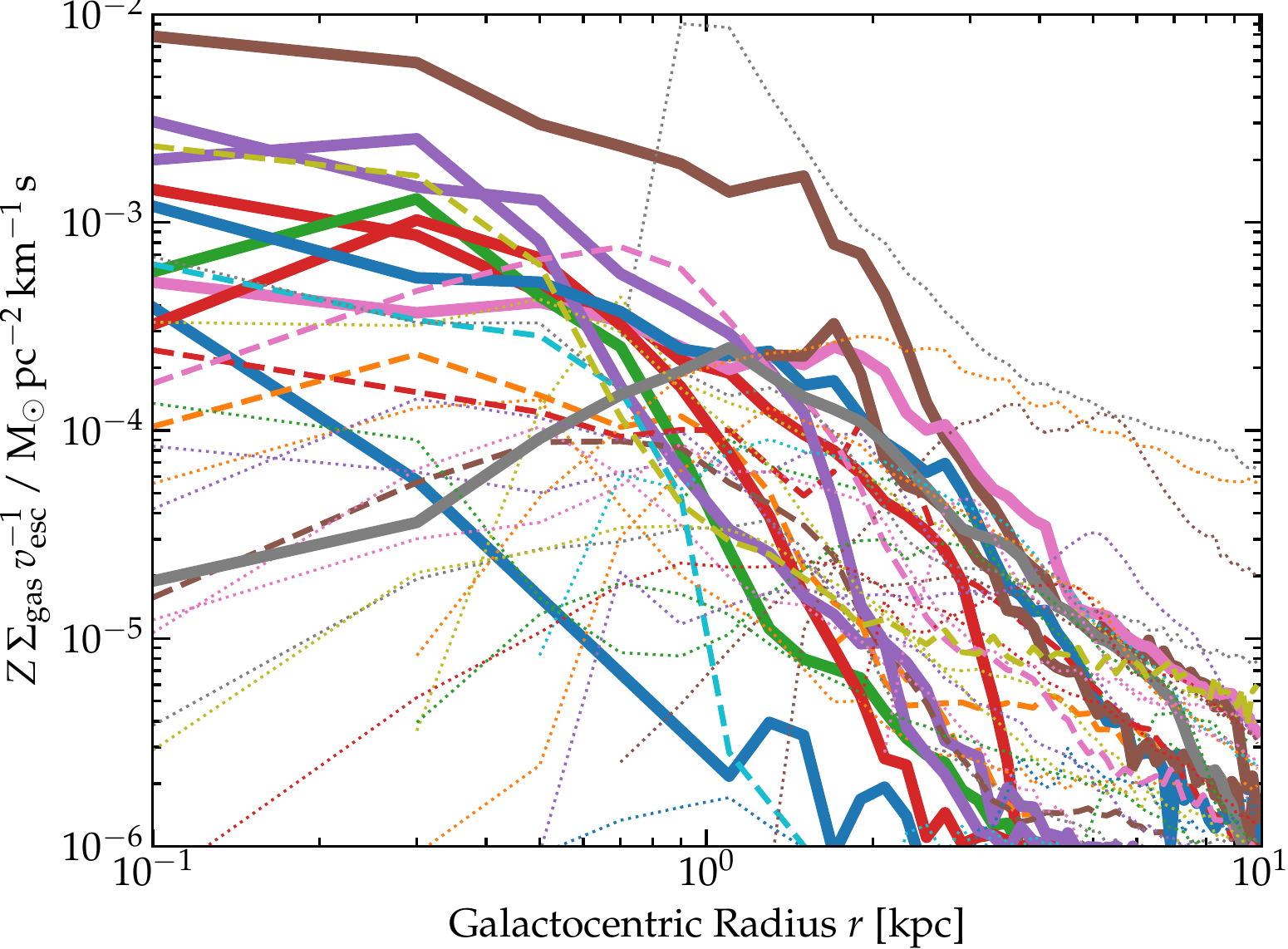}
	\includegraphics[width=0.32\textwidth]{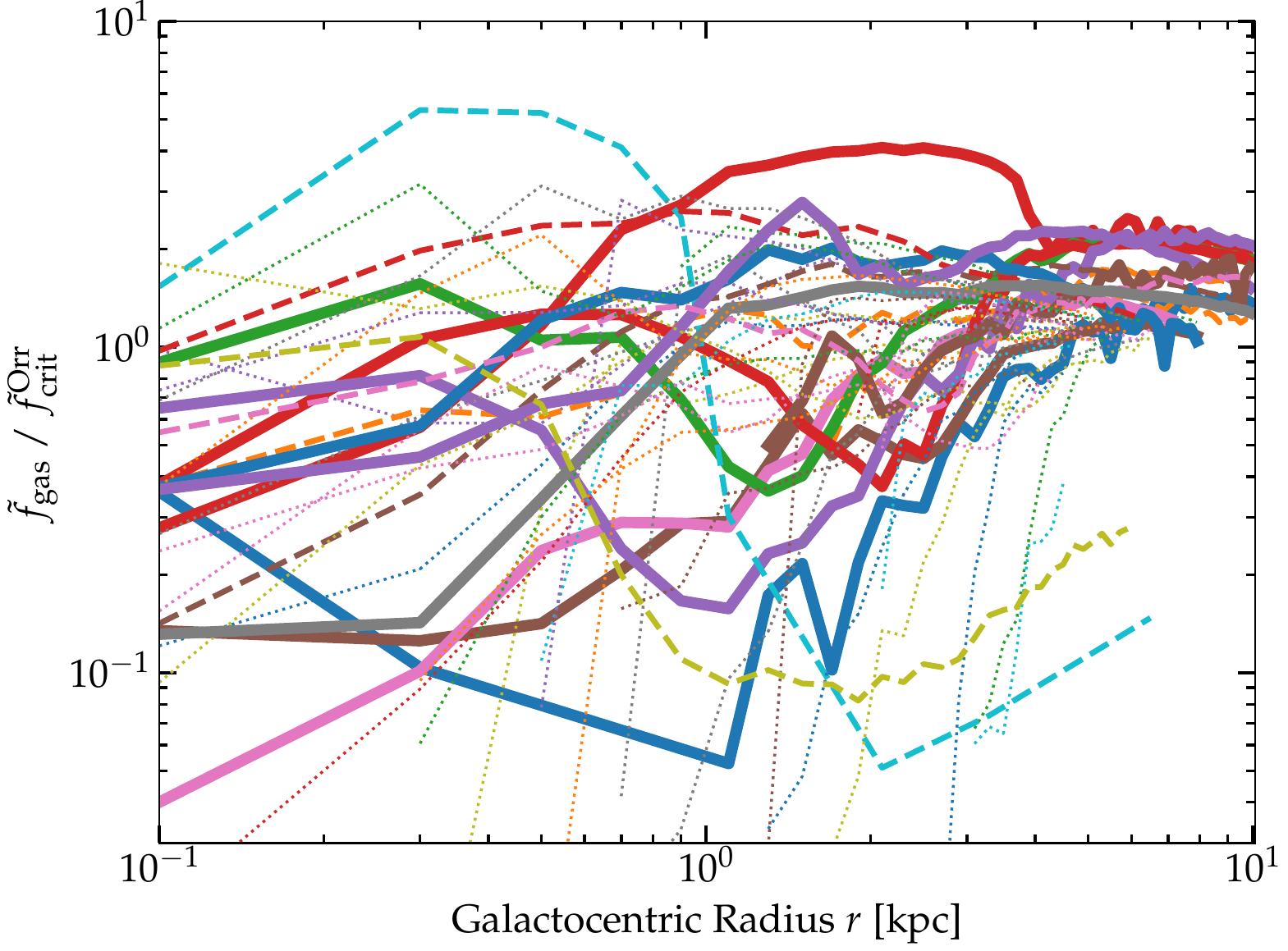}
	\includegraphics[width=0.32\textwidth]{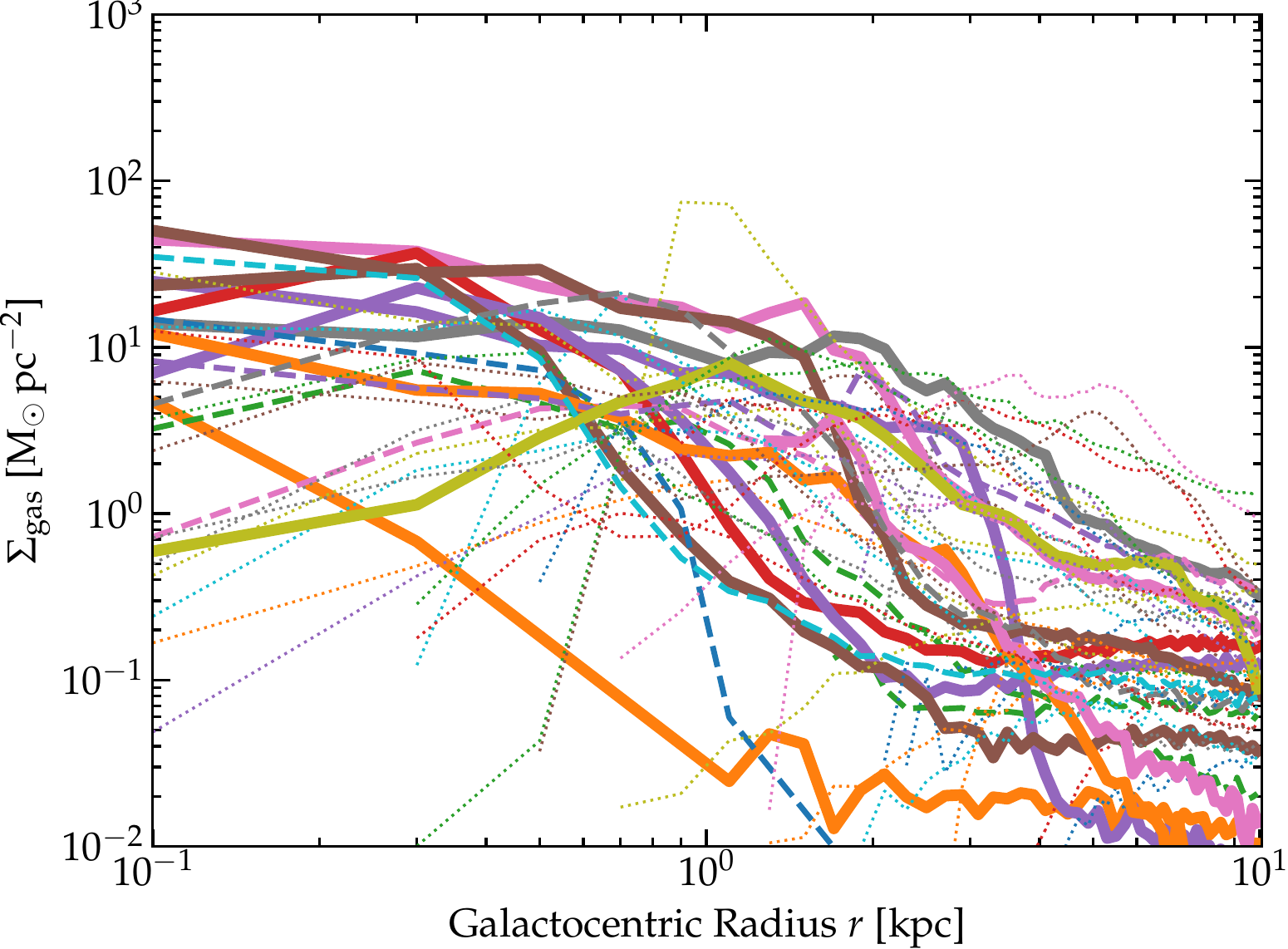} \\
	\includegraphics[width=0.32\textwidth]{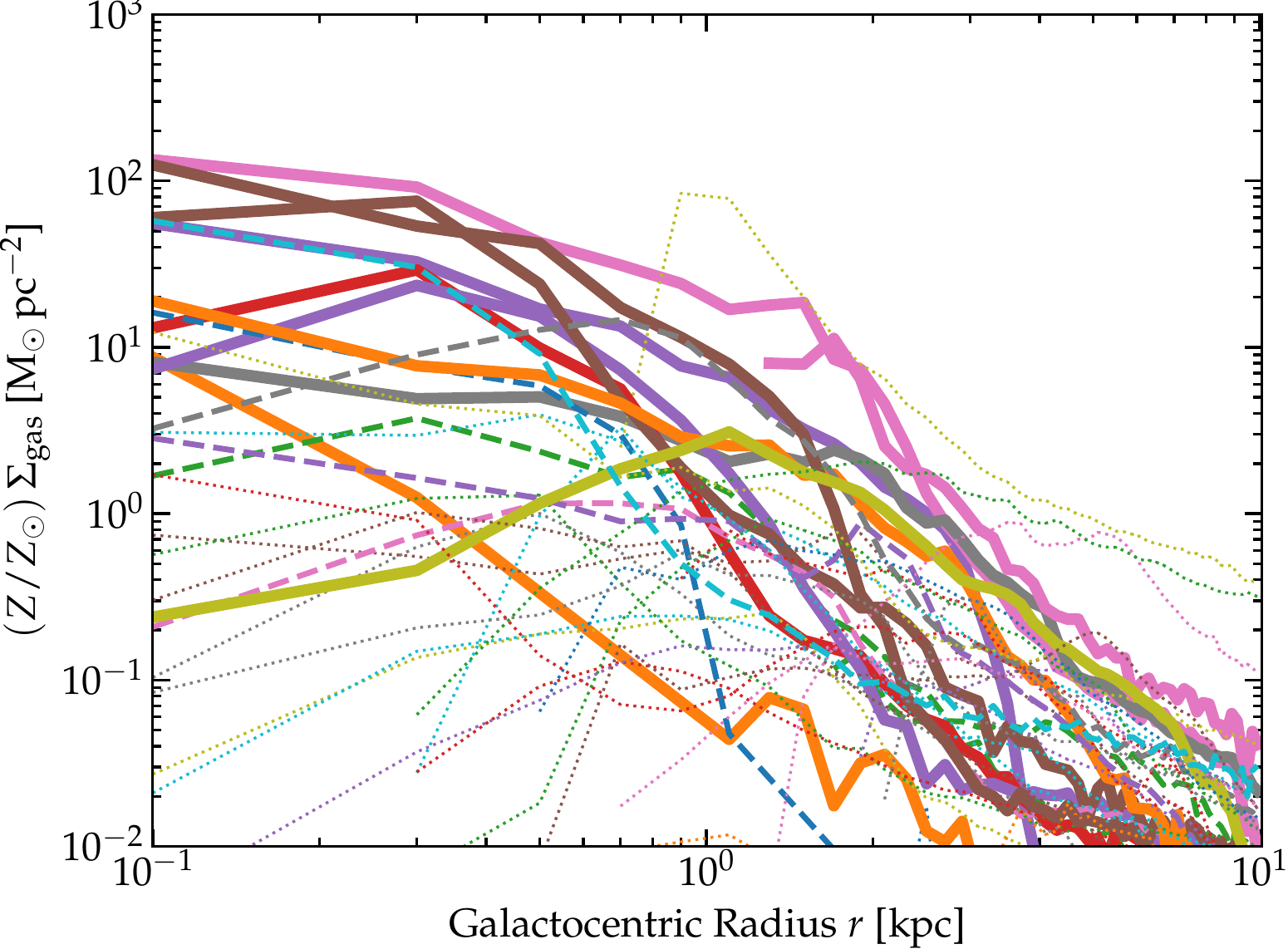}	
	\vspace{-0.1cm}
	\caption{Profiles of various parameters comparing bursty SF ({\em thin dotted}), ambiguous/intermittent ({\em dashed}) and smooth SF ({\em thick solid}) systems in our {\bf m11a} experiments, as Fig.~\ref{fig:Vesc} in the main text. The different parameters here are broadly motivated by different assumptions about feedback and self-regulation (see text, \S~\ref{sec:vesc.bursty}). These include various dimensional combinations of $r$, $\Omega$, $V_{\rm c}$, and $V_{\rm esc}$, as well as the \citet{orr:2021.bubble.breakout.model} criterion of the ratio of disk baryonic gas fraction in an annulus $\tilde{f}_{\rm gas} \equiv \Sigma_{\rm gas}(r) / (\Sigma_{\rm gas}(r) + \Sigma_{\ast}(r))$ versus their critical $\tilde{f}_{\rm crit}^{\rm Orr}$ which is a function of disk properties, and disk surface/column density $\Sigma_{\rm gas}$ and a simple proxy for dust optical depth/shielding $Z\,\Sigma_{\rm gas}$. None of these appears to show a clear separation or correlation with smooth SF.
	\label{fig:bursty.criteria.gallery}}
\end{figure*}

\begin{figure}
	\includegraphics[width=0.95\columnwidth]{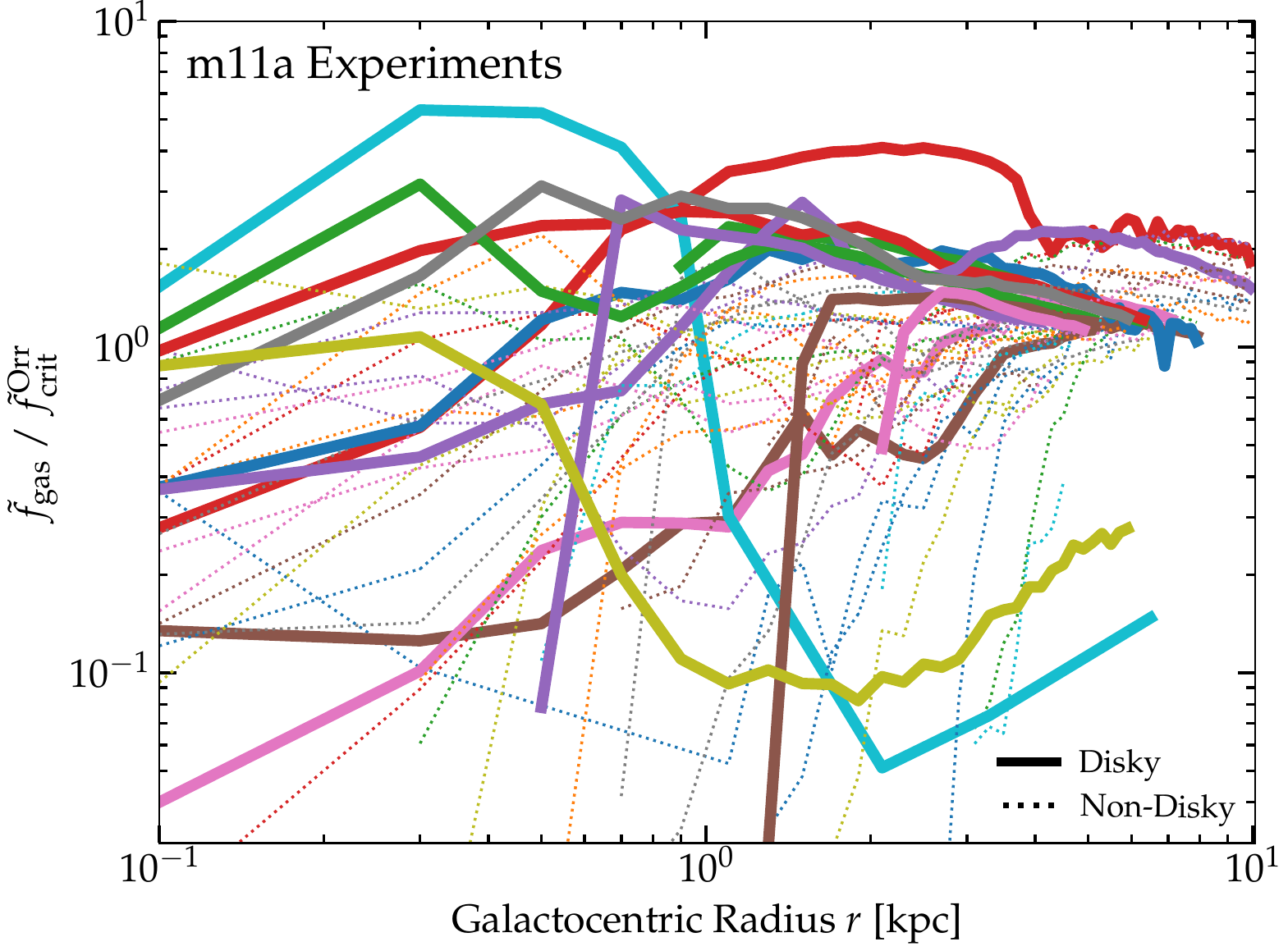}
	\vspace{-0.1cm}
	\caption{\citet{orr:2021.bubble.breakout.model} criterion versus diskiness of the {\bf m11a} experiments. The prediction is that clustered SNe in a purely-momentum-conserving phase break out or are confined at values above/below unity, respectively. There is no strong trend here, and the weak trend seen (with disky systems at slightly higher $\tilde{f}_{\rm gas}/\tilde{f}_{\rm crit}^{\rm Orr}$) is opposite the predicted dependence if confinement were important for disk formation.
	\label{fig:orr.fgas.criterion}}
\end{figure}

\begin{figure*}
	\includegraphics[width=0.45\textwidth]{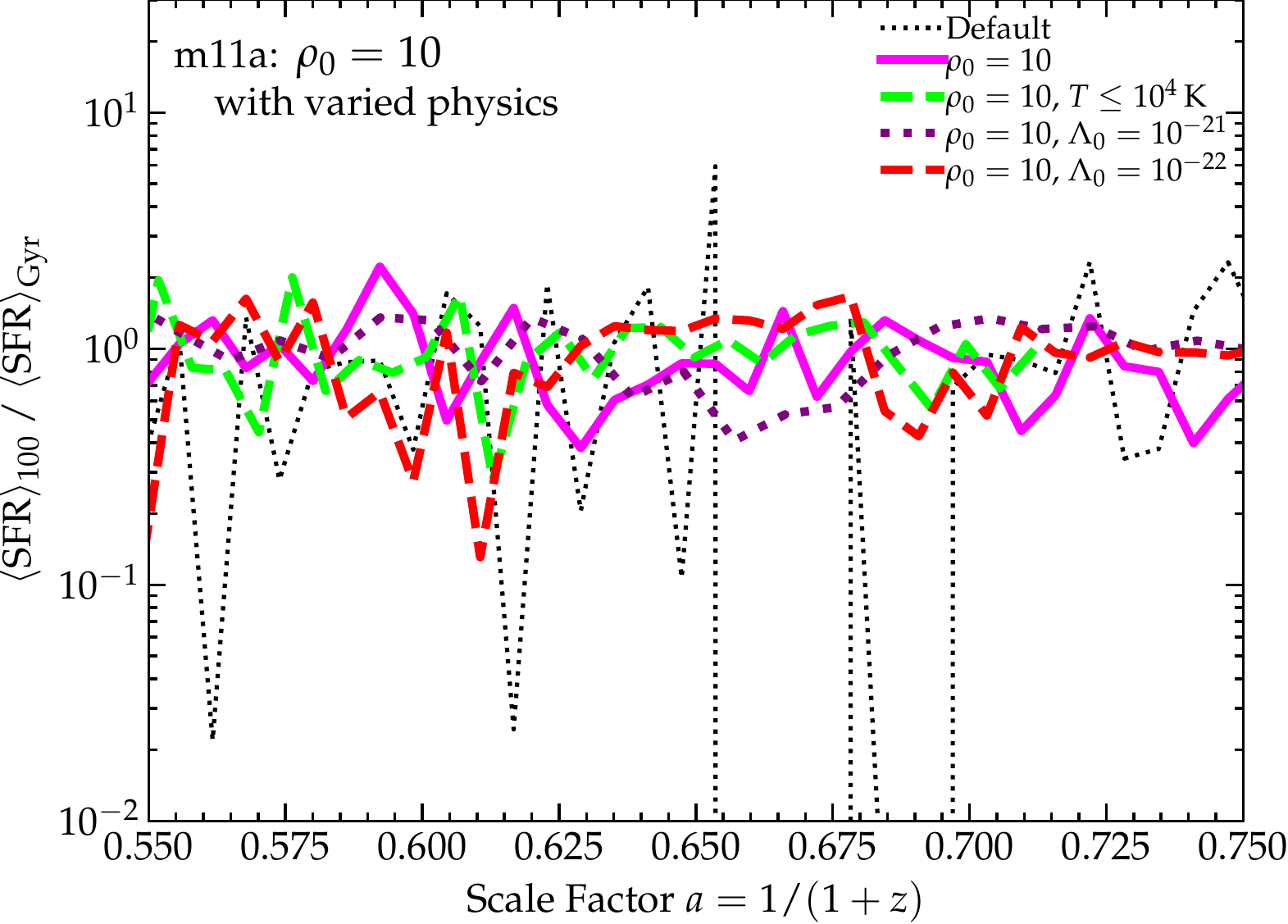} 
	\includegraphics[width=0.45\textwidth]{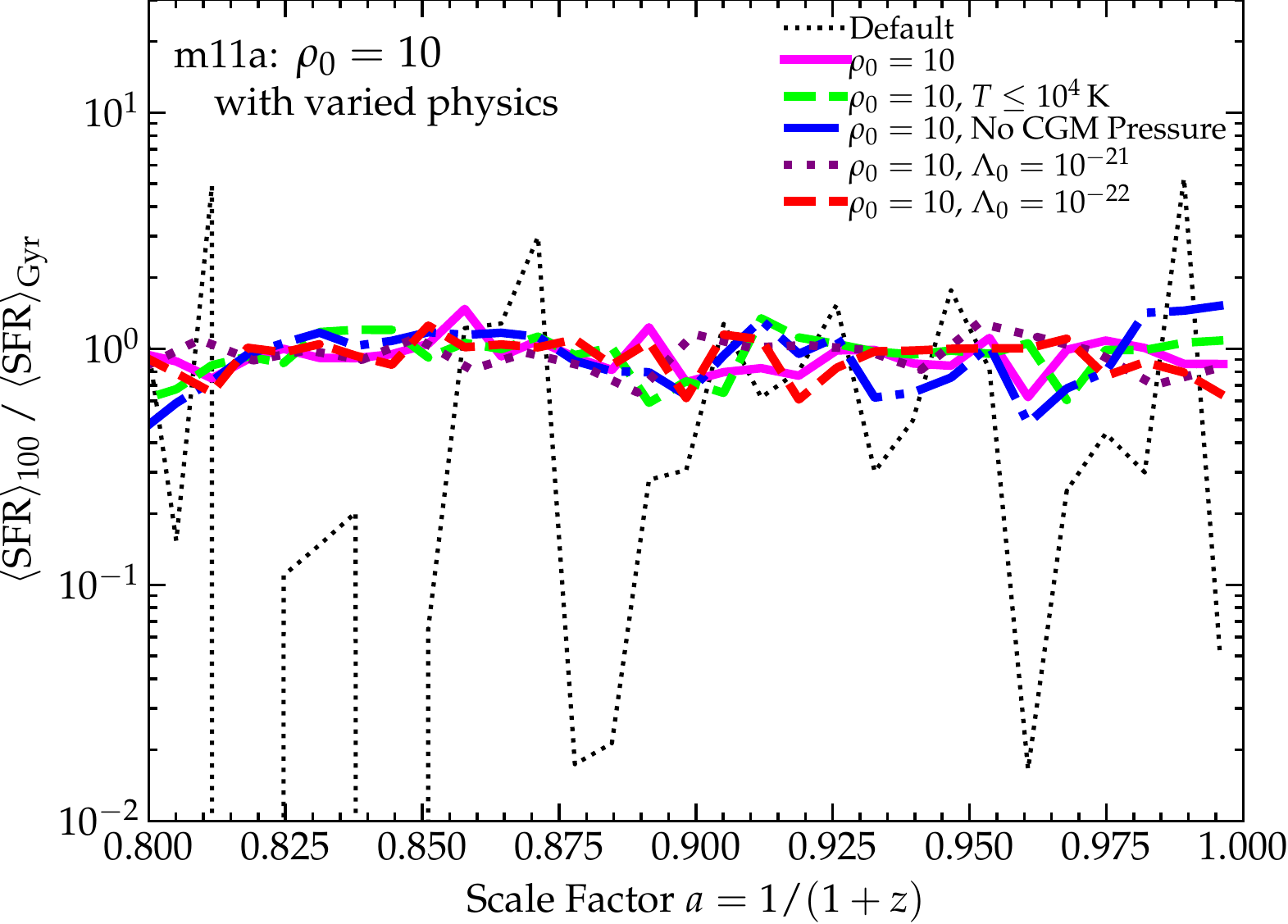} \\ 
	\includegraphics[width=0.45\textwidth]{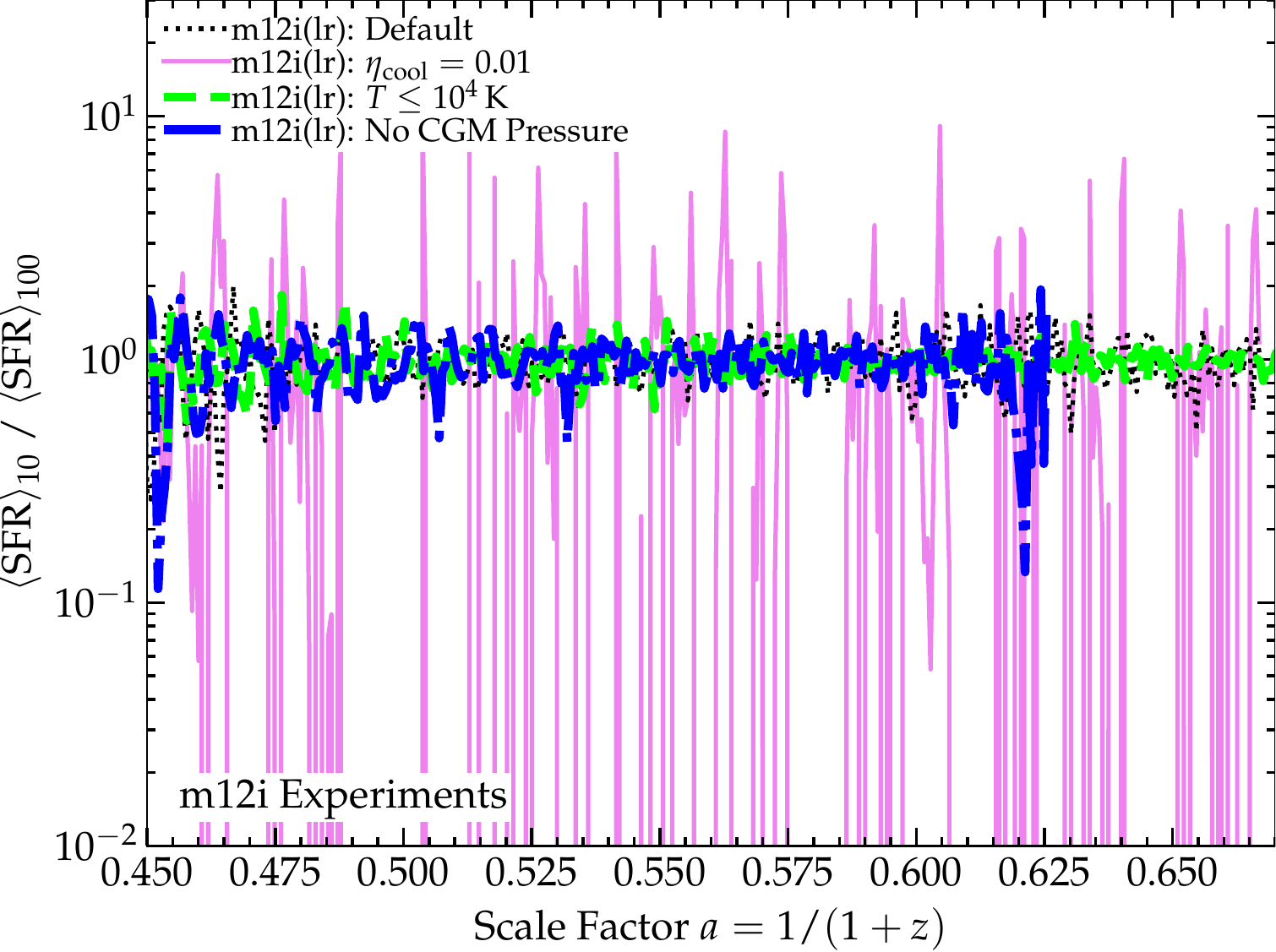} 
	\includegraphics[width=0.45\textwidth]{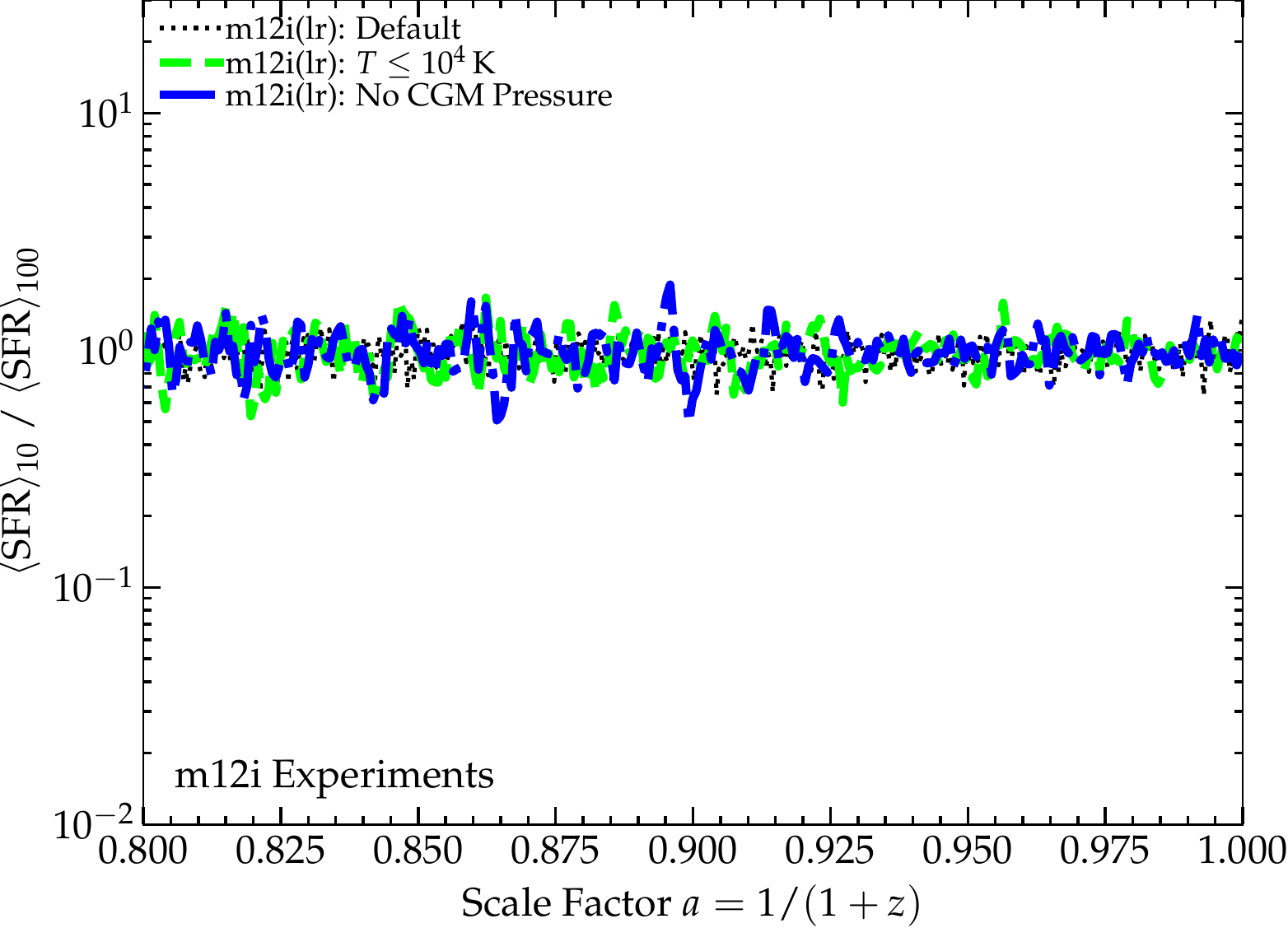} \\ 
	\vspace{-0.1cm}
	\caption{Additional tests of the form in Fig.~\ref{fig:more.smooth.sfh.tests.cooling}, exploring how the thermodynamics of the CGM gas does or does not influence the ``smooth'' versus ``bursty'' SFHs. {\em Top:} {\bf m11a} experiments, comparing the default (bursty) run with variations run with a deeper potential ($\rho_{10}=10$), which produces smooth SF. We re-start and run for a modest time either at early times ({\em left}) or at late times ({\em right}), with the modified cooling functions shown in Fig.~\ref{fig:more.smooth.sfh.tests.cooling} or the much more extreme variations of (a) setting the gas to obey a strictly isothermal equation-of-state above a temperature {\em maximum} at $T=10^{4}$\,K (the equivalent of $\Lambda \rightarrow \infty$ for $T>10^{4}$\,K), or (b) setting $P\rightarrow 0$ the hydrodynamic pressure which appears in the Riemann problem in the hydro solver (so there are identically zero pressure forces) between any gas cells which reside between $10 < r < 100\,$kpc from the galaxy center. 	{\em Bottom:} Same, for {\bf m12i}, which in default runs transitions to smooth SF, restarted just before the transition would occur naturally ({\em left}) or well after ({\em right}). In both cases these extreme temperature/pressure modifications ensure it is impossible for the gas in the halo to be in virial/hydrostatic equilibrium. In all cases the SFR remains smooth where it would be smooth otherwise.
	\label{fig:smooth.bursty.tests.nopressure}}
\end{figure*}

In the main text, we survey a number of parameters and galaxy properties which do not appear to correlate well with the formation of disks nor with the transition from bursty to smooth SF. In particular, in \S~\ref{sec:alt.scaling.failures} we discuss a variety of models for ``smooth SF'' one might imagine based on more complicated variations of some of the arguments in the text or other extreme limits. In Fig.~\ref{fig:bursty.criteria.gallery} we show the radial profiles of a variety of these, to demonstrate explicitly (as we noted in the main text) that none of these quantities appears to provide a better separation between smooth and bursty SF, compared to the models we discuss more explicitly in the text. We also show one additional test in Fig.~\ref{fig:orr.fgas.criterion} for the disky versus non-disky distinction, which also does not provide a better explanation compared to the models in the main text. 

Fig.~\ref{fig:smooth.bursty.tests.nopressure} considers even more extreme variations of the sort of tests in \S~\ref{sec:bursty} showing that even radical modifications to the thermodynamics of the gas -- e.g.\ removing all thermal gas pressure from the CGM or not allowing any gas to reach the ``hot phase'' and virialize in the halos simulated here -- qualitatively changes the fact that sufficiently-deep potential wells maintain gas in the ``smooth'' SFR regime.

\end{appendix}

\end{document}